\documentclass[1p,final]{elsarticle}
\usepackage{amsmath}
\usepackage{amssymb}
\usepackage{amsfonts}
\usepackage{dsfont}
\usepackage{MnSymbol}
\usepackage{lineno}  
\usepackage[english]{babel}
\usepackage{slashed}
\usepackage{bm,braket}
\usepackage{verbatim}
\usepackage{booktabs}
\usepackage{mathtools}
\usepackage{graphicx}
\usepackage[pdftex,dvipsnames,table]{xcolor}
\usepackage{graphbox}
\usepackage{bigstrut}
\usepackage[T1]{fontenc}
\usepackage{nicefrac}
\usepackage{float}
\usepackage{braket}
\usepackage[export]{adjustbox} %
\usepackage{makecell}
\usepackage[normalem]{ulem}
\usepackage[section]{placeins}%
\usepackage{tabularx}%
\usepackage{xargs}      %
\journal{Progress in Particle and Nuclear Physics}
\biboptions{sort&compress}
\usepackage{titlesec}
\usepackage{sectsty}
\titleformat{\section}{\normalfont\Large\bfseries}{\thesection}{1em}{}
\titleformat{\subsection}{\normalfont\large\bfseries}{\thesubsection}{1em}{}
\titleformat{\subsubsection}{\normalfont\normalsize\bfseries}{\thesubsubsection}{1em}{}
\usepackage{hyperref}
\hypersetup{,  
	colorlinks= true,
	urlcolor  = blue,
	linkcolor = blue,
    urlcolor  = blue,
    citecolor = blue,
    anchorcolor = blue,
	pdftitle  = {Review on dynamical coupled-channel approaches},
	pdfauthor = {M. Doering, J. Haidenbauer, M. Mai, T. Sato},
	pdfsubject= {Review on dynamical coupled-channel approaches}
}
\usepackage{cleveref}
\crefformat{figure}{Fig.~#2#1#3}
\crefformat{table}{Tab.~#2#1#3}
\crefformat{equation}{Eq.~(#2#1#3)}
\crefformat{section}{Sect.~#2#1#3}
\numberwithin{equation}{section}
\graphicspath{{plots/}} %

\newcommand\vertarrowbox[3][8ex]{%
  \begin{array}[t]{@{}c@{}} #2 \\
  \left\updownarrow\vcenter{\hrule height #1}\right.\kern-\nulldelimiterspace\\
  \makebox[0pt]{\scriptsize#3}
  \end{array}%
}

\newcommand{\GeV}{{\rm GeV}}
\newcommand{\MeV}{{\rm MeV}}
\newcommand{\La}{{\Lambda}}
\newcommand{\Si}{{\Sigma}}

\renewcommand{\Re}{{\rm Re}\,}
\renewcommand{\Im}{{\rm Im}\,}
\newcommand{\barr}[1]{\bar{\bar{#1}}}
\newcommand{\dv}[1]{\mathrm{d} #1}
\newcommand*\dif{\mathop{}\!\mathrm{d}}
\newcommand{\non}{\nonumber \\}
\newcommand{\po}{{\rm P}}
\newcommand{\npo}{{\rm NP}}
\newcommand{\be}{\begin{eqnarray}}
\newcommand{\ee}{\end{eqnarray}}
\newcommand\TT{\rule{0pt}{2.6ex}}       %
\newcommand\BB{\rule[-1.2ex]{0pt}{0pt}} %
\newcommand\BBB{\rule[-1.6ex]{0pt}{0pt}} %
\newcommand{\cg}[3]{(#1~#2|#3)}
\newcommand{\D}{\bar D}

\newlength{\feynwidthb} 
\makeatletter
\def\widebreve{\mathpalette\wide@breve}
\def\wide@breve#1#2{\sbox\z@{$#1#2$}%
     \mathop{\vbox{\m@th\ialign{##\crcr
\kern0.08em\brevefill#1{0.8\wd\z@}\crcr\noalign{\nointerlineskip}%
                    $\hss#1#2\hss$\crcr}}}\limits}
\def\brevefill#1#2{$\m@th\sbox\tw@{$#1($}%
  \hss\resizebox{#2}{\wd\tw@}{\rotatebox[origin=c]{90}{\upshape(}}\hss$}
\makeatletter

\begin{document}
\begin{frontmatter}
\title{Dynamical coupled-channel models for hadron dynamics}

\author[1,2]{Michael D\"oring\corref{cor1}}
\ead{doring@gwu.edu}

\author[3]{Johann Haidenbauer}
\ead{j.haidenbauer@fz-juelich.de}

\author[4,1,5]{Maxim Mai}
\ead{maxim.mai@faculty.unibe.ch}

\author[6]{Toru Sato}
\ead{tsato@rcnp.osaka-u.ac.jp}
\affiliation[1]{
    organization={Department of Physics, The George Washington University},%
    addressline={725 21st St, NW}, 
    city={Washington},
    postcode={20052}, 
    state={District of Columbia},
    country={USA}}
\affiliation[2]{
Theory Center, Thomas Jefferson National Accelerator Facility, Newport News, VA 23606, USA
}
\affiliation[3]{
    organization={Institute for Advanced Simulation (IAS-4),  Forschungszentrum Jülich GmbH},%
    city={Jülich},
    postcode={52428}, 
    state={NRW},
    country={Germany}}
\affiliation[4]{
    organization={Albert Einstein Center for Fundamental Physics, Institute for Theoretical Physics, University of Bern},
    addressline={Sidlerstrasse 5}, 
    city={Bern},
    postcode={3012},
    country={Switzerland}}
\affiliation[5]{
    organization={Helmholtz-Institut f\"ur Strahlen- und Kernphysik (Theorie) and Bethe Center for Theoretical Physics},
    addressline={Universit\"at Bonn}, 
    city={Bonn},
    postcode={53115},
    country={Germany}}
\affiliation[6]{
    organization={Research Center for Nuclear Physics (RCNP), Osaka University},%
    addressline={10-1 Mihogaoka}, 
    city={Ibaraki},
    postcode={567-0047}, 
    state={Osaka},
    country={Japan}}

\cortext[cor1]{Corresponding author}

\begin{abstract}
Dynamical coupled-channel (DCC) approaches parametrize the interactions and dynamics of two and more hadrons and their response to different electroweak probes. The inclusion of  unitarity, three-body channels, and other properties from scattering theory allows for a reliable extraction of resonance spectra and their properties from data. We review the formalism and application of the ANL-Osaka, the Juelich-Bonn-Washington, and other DCC approaches in the context of light baryon resonances from meson, (virtual) photon, and neutrino-induced reactions, as well as production reactions, strange baryons, light mesons,  heavy meson systems, exotics, and baryon-baryon interactions. Finally, we also provide a connection of the formalism to study finite-volume spectra obtained in Lattice QCD, and review applications involving modern statistical and machine learning tools.\\~\\
Preprint number: JLAB-THY-25-4298
\end{abstract}

\begin{keyword}
Dynamical coupled-channel approaches \sep 
Baryon spectroscopy \sep 
Meson spectroscopy \sep 
Amplitude analysis \sep
S-matrix theory \sep
Photoproduction reactions\sep
Baryon-baryon interactions

\end{keyword}
\end{frontmatter}
\tableofcontents
\section{Introduction}
\label{sec:intro}
\subsection{What are Dynamical Coupled-Channel approaches?}
Scattering processes, including production reactions in which final states re-scatter, represent one of the few avenues to experimentally access the properties of the sub-atomic world. Together with the study of decay processes they allow to study strong interactions, or the response of strongly interacting systems to electroweak probes. Often, the desired output extracted from scattering consists in partial-wave projected amplitudes and the hadronic resonance spectrum contained therein, that allow for comparison with theory and models. This includes quark models, unitarized chiral approaches, Dyson-Schwinger, and other functional methods. Some theory approaches, including modern Lattice QCD-based hadron spectroscopy efforts (for reviews see \cite{Mai:2022eur, Aoki:2016frl, Prelovsek:2013cta, Mohler:2012nh, Lin:2011ti, Briceno:2017max}), allow in some cases to directly compare to amplitudes, while other models might only predict the spectrum of hadronic resonances, but not their widths or branching ratios.

Amplitude analyses aim to reconstruct scattering and production amplitudes, resonances, and their properties from experimental data to allow for such comparisons. S-matrix principles,  unitarity and analyticity, help constrain this endeavor, which often requires the parametrization of the amplitude in kinematic variables like energy, sub-energies, and scattering angles. Dynamical coupled-channel (DCC) approaches are a class of amplitude parametrizations formulated in terms of mesons and baryons as degrees of freedom that are stable under strong interactions (see \cref{sec:DCC-formalism} for the formalism). The ``coupled-channel'' aspect refers to the fact that strongly interacting systems can transform back and forth to different states as allowed by conserved quantum numbers. Conversely, the coupled-channel framework enables one to synthesize independent sources of information of a strongly interacting system manifested in its different decay channels and excitation mechanisms.

The ``dynamical'' part of DCC approaches is less well defined. 
Historically it is attributed to the explicit meson and baryon exchanges in the included interactions in all three Mandelstam variables $s$, $t$, and $u$, or other kinematic variables for reactions involving more than two-body systems. This is to be seen in contrast to  modern approaches guided by the idea of scale separation \cite{Weinberg:1990rz} in which, at a given order, some particle exchange processes could be explicitly included while others are simply absorbed into contact terms. Although the explicit modeling of strong interactions through exchange processes undoubtedly forms an aspect of DCC approaches, the use of contact terms is common practice, as well. Explicit particle exchange is also a necessity in terms of unitarity once three-body channels are considered as discussed in \cref{sec:unitarity}. In that sense, the dynamical exchange of mesons and baryons is less a model ingredient and more a requirement of S-matrix properties.

A technical aspect of ``dynamical'' concerns the relativistic scattering equation of DCC approaches that always involves at least one explicit momentum integration, rendering it a genuine integral equation. For two-body systems one can restrict the driving interaction to on-shell kinematics, simplifying the integral equation to an algebraic ``K-matrix''(-like) expression. After all, only on-shell quantities are accessible experimentally and off-shell extensions are ambiguous, as discussed in \cref{sec:compare}. 
However, for three-body channels the integration over a ``spectator'' momentum is always required even if the two-body sub-channels can still be expressed in terms of the on-shell amplitudes, see Sects.~\ref{sec:DCC-formalism} and \ref{sec:unitarity} for a discussion. In that case one is stuck with an integral equation. 

Indeed, many DCC approaches were formulated for systems including  three-body states because the DCC parametrization allows to relatively easily incorporate coupled-channel two- and three-body unitarity, at least in principle. The prime example is the pion-nucleon 
($\pi N$) sector with the ubiquitous $\pi\pi N$ channel into which all excited baryon resonances except for the $\Delta(1232)$  decay\footnote{The PDG notation includes the spin-parity assignment $J^P$ in resonance names that we often omit for better readability.}. The three-body aspect of DCC approaches is discussed in depth in Sects.~\ref{sec:unitarity} and \ref{sec:threemesons}.

Sometimes, the only aspect kept from DCC approaches for three-body physics is the unitarity structure and the pertinent particle exchange processes, while the two-body sub-amplitudes, referred to as ``isobars'', can be provided in any parametrization respecting two-body unitarity. In such cases, no attempts are made to understand interactions in terms of microscopic exchange processes (except for processes required by unitarity). Instead, DCC approaches serve for a relatively clean amplitude parametrization to analyze large data sets for the purpose of determining the spectrum of excited states and their properties. 

Indeed, since DCC approaches have a substantial theoretical underpinning, they are often used to analyze meson-baryon scattering (\cref{sec:mesonproduction}), real-photon induced reactions (\cref{sec:photoproduction}), virtual-photon induced reactions~(\cref{sec:electroproduction}), neutrino-induced reactions~(\cref{sec:neutrinoproduction}), multi-meson production reactions~(\cref{sec:threemesons}), and lattice QCD data~(\cref{sec:LQCD}). There are also baryon-baryon systems where coupled-channel aspects play an important role, for example for the $\Lambda N$-$\Sigma N$ interaction (\cref{sec:Baryon-Baryon}).
And coupled-channel dynamics is one of the starting points for interpreting
and explaining the various structures observed in the heavy sector,
i.e., in systems that involve charmed and bottom hadrons, the 
XYZ states and pentaquarks (\cref{sec:heavy-sector}).

While DCC approaches are extensively applied for data analysis and respect S-matrix principles, their origin lies in the field-theoretical description of particle interactions. In their original formulation, an effective Lagrangian using meson and baryon degrees of freedom is expanded to provide the vertices for exchange processes that are then re-summed. We start this discussion in \cref{subsec:ANL-Osaka} that is focused on the ANL-Osaka formulation and subsequently compare it to the Juelich model formulation in \cref{sec:JB}. These two approaches are highlighted throughout this review, but there are many more approaches pursued in the last decades as summarized in \cref{sec:historyandapproaches}.

Finally, we note that 
despite the advantages mentioned above DCC approaches require substantial formal and computational resources to unite aspects of field theory with a plethora of coupled channels and different excitations of the strongly interacting system by hadrons and/or electroweakly interacting particles. Such models are then used to analyze tens of thousands of data of different final states. Therefore, DCC approaches tend to be refined over many years meaning that there is also a historical aspect to this class of approaches discussed in the next subsection. Statistical aspects are discussed in \cref{subsec:statistics} and extensions to finite-volume physics relevant for lattice QCD in Sect.~\ref{sec:LQCD}.

As for notation, the total energy of the produced baryon-meson, meson-meson, or baryon-baryon systems is traditionally named differently. Throughout this review, we use
\begin{align}
    E=\sqrt{s}=W=z \ 
\label{eq:ESWZ}
\end{align}
except for Sect.~\ref{sec:analytic} where $z$ stands for an uniformization variable.
The notation $E$ refers to the energy in the context of quantum mechanics, $\sqrt{s}$ is the Mandelstam invariant, $W$ is traditionally used in baryon spectroscopy for the center-of-mass (c.m.) energy, and $z$ is sometimes used when the range of the energy is extended to complex values.

\subsection{Brief overview of DCC approaches for the light baryon sector}
\label{sec:historyandapproaches}
The concept of coupled channels has a rich history and broad scope of applications~\cite{Oller:2025leg, Badalian:1981xj}. While the term ``dynamical coupled channel approaches'' was coined in Refs.~\cite{Sato:1996gk, Matsuyama:2006rp}, earliest examples for DCC approaches in the literature are studies of the $\bar KN$ interaction~\cite{Alberg:1976xm, Henley:1980dp, Landau:1982mi, Siegel:1988rq, Mueller-Groeling:1990uxr}. In those works the coupling to the $\pi \Sigma$ and (in some cases) also to $\pi\Lambda$ channels were included. Note, however, that initially separable potentials were used. Only in later works \cite{Siegel:1988rq,Mueller-Groeling:1990uxr} the dynamics is based on the meson-exchange picture. In Ref.~\cite{Siegel:1988rq} separable potentials as well as meson exchange are considered. 

Regarding the $\pi N$ system an early example for a study in the spirit of the DCC approach is an off-shell isobar model from 1985 that includes the coupling to $\eta N$ and to effective two-body channels representing $\pi\pi N$ \cite{Bhalerao:1985cr}. See also the dynamical model including pion photoproduction of Ref.~\cite{Nozawa:1989pu}. The 
Dubna-Mainz-Taipei (DMT) model also builds on a long history~\cite{Lee:1991dd,Hung:1994hg,Hung:2001pz}. It was extended to include the $\eta N$ channel besides the $\pi N$ channel, plus imaginary parts for resonance propagators modeling decay into other channels~\cite{Chen:2007cy}. The analysis was extended up to $F$-waves and a center-of-mass (c.m.) energy of $\sqrt{s}\approx 2\,\GeV$. The DMT partial waves were analytically continued to complex scattering energies to extract resonance poles and residues~\cite{Tiator:2010rp}. In the same paper, this direct determination of resonance singularities was compared to approximate methods relying on information on the real-energy axis, like speed plots~\cite{Hohler:1992ru} and the regularization method~\cite{Ceci:2006zw}. 

While the relativistic $\pi N$ model of Gross and Surya is restricted to the elastic $\pi N$ channel~\cite{Gross:1992tj}, its treatment of unitarized dynamics retaining the Dirac structure of the amplitude allows for the gauge invariant coupling of the photon~\cite{Surya:1995ur}. One of the surprising findings of Ref.~\cite{Gross:1992tj} is that setting the pion on its mass shell and iterating the nucleon exchange once produces a good approximation of the box and crossed box Feynman diagrams. This is taken as motivation for the three-dimensional reduction of the four-dimensional Bethe-Salpeter equation
(BSE) using a prescription of Pearce and Jennings~\cite{Pearce:1990uj}. In 
the latter work, different reduction schemes are compared. See also Ref.~\cite{Hung:2001pz} for a systematic investigation of different reduction schemes. Similarly, the covariant $\pi N$ model of Pascalutsa and Tjon~\cite{Pascalutsa:2000bs} is based on the solution of a dimensionally reduced (quasipotential) Bethe-Salpeter equation, later expanded to electroproduction reactions~\cite{Caia:2005hz}. 
A more thorough discussion on the different schemes for the three-dimensional 
reduction of the BSE, leading to scattering equations like the ones by
Blankenbecler-Sugar~\cite{Blankenbecler:1965gx}, 
Kadyshevsky~\cite{Kadyshevsky:1967rs}, or Thomson~\cite{Thompson:1970wt} 
that are commonly used in studies of meson-baryon and/or baryon-baryon interactions, can be found in 
Refs.~\cite{Woloshyn:1973mce,Erkelenz:1974uj,Wu:2012md}. 

Solving the full, four-dimensional BSE is necessary to avoid dependence on reduction schemes or when including the $\pi N$ amplitude in a bigger setting of, e.g., photoproduction amplitude while maintaining unitarity and gauge invariance at the same time~\cite{vanAntwerpen:1994vh,Borasoy:2007ku,Bruns:2010sv,Bruns:2020lyb,Bruns:2022sio,Mai:2012wy,Ruic:2011wf} although gauge invariance can be reconstructed~\cite{Huang:2011as}. So far, such four-dimensional BSE have been solved analytically when using only contact terms~\cite{Borasoy:2007ku,Bruns:2010sv} or with non-local terms using Wick rotation~\cite{Lahiff:1999ur}. In both approaches, the complexity of overlapping singularities due to $u$-channel baryon exchange has been circumvented for the price of giving up exact three-body dynamics. In \cref{sec:unitarity} we discuss one reduction scheme of the BSE in more detail for illustration.

Starting point of the Juelich-Bonn coupled-channel meson-baryon interaction is the simple $\pi N$ potential by Sch\"utz et al.~\cite{Schutz:1994ue,Schutz:1994wp}
where, as the main new aspect, the 
phenomenological $\sigma$- and $\rho$-exchanges were
eliminated and the corresponding contributions in 
the scalar-isoscalar ($0^+$) and vector-isovector ($1^-$) channels were 
fixed from a microscopic evaluation of correlated $2\pi$-exchange. 
It was derived within relativistic time-ordered perturbation
theory (TOPT), in close analogy to the famous Bonn potential for 
nucleon-nucleon (NN) scattering~\cite{Machleidt:1987hj}.
The range of validity was up to
$p_{\text{lab}}\approx 500\,\MeV/c$ ($\sqrt{s}\approx 1.35\,\GeV$). 
In the next step \cite{Schutz:1998jx} the $\eta N$
channel and effective $\pi\pi N$ channels (in the 
form of $\sigma N$ and $\pi \Delta$) were added and the dynamical origin of the $N(1535)$ and $N(1440)$ resonances were investigated. Specifically, it was argued that there are strong hints that the $N^*(1535)$ is a genuine resonance in the sense of the quark model while the $N(1440)$ is very likely dynamically generated. The energy range was still limited, reaching up to $\sqrt{s}\approx 1.4\,\GeV$ in general and up to $\sqrt{s}\approx 1.6\,\GeV$ for specific partial waves like $S_{11}$ and
the $P_{11}$, in the common notation $L_{2I2J}$ with orbital angular momentum $L$ ($S$, $P$, $D$, $\dots$), isospin $I$ and total angular momentum $J$.

In the work of Krehl et al.~\cite{Krehl:1999ak,Krehl:1999km} some theoretical refinements were implemented and the dynamics of the
$N^*(1440)$ (Roper resonance) was examined thoroughly.
In that model, besides two physical channels ($\pi N$, $\eta N$), now three effective channels were included
($\sigma N$, $\rho N$, $\pi \Delta$) in order to represent the physics of the $\pi\pi N$ channels.
Finally, in Ref.~\cite{Gasparyan:2003fp} further
improvements in the treatment of the dynamics were
made. Specifically, a consistent description of the
$\pi N$ phase shift in the $S_{11}$ partial wave
and of the $\pi N \to \eta N$ transition cross section
near the threshold was achieved. 
Later milestones in the development of the Juelich-Bonn model included the extension to channels with strangeness~\cite{Ronchen:2012eg}, photoproduction~\cite{Ronchen:2014cna, 
Ronchen:2015vfa,
Ronchen:2018ury, Ronchen:2022hqk}, and inclusion of the $\omega N$ channel~\cite{Wang:2022osj}. The extension to electroproduction reactions~\cite{Mai:2021vsw,Mai:2021aui,Mai:2023cbp} is referred to as Juelich-Bonn-Washington model (JBW). In this review, we often refer to all different sub-approaches as ``JBW'' for simplicity. Related formalisms were used to explore the nature of states with hidden charm~\cite{Shen:2017ayv, Wang:2022oof, Shen:2024nck}.

The ANL-Osaka DCC model shares some but also differs in some construction principles from the JBW model. In the ANL-Osaka model, a Hermitian effective Hamiltonian that is independent of the scattering energy is constructed from a field theoretical model Lagrangian. Because of the energy-independent Hamiltonian,
unitarity of the scattering amplitude is almost trivially satisfied within the Fock space in which one works.

The method to obtain the Hermitian Hamiltonian was first  developed for nuclear potentials by Fukuda et al.~\cite{fukuda1954construction}, Okubo~\cite{Okubo:1954zz} and Nishijima~\cite{nishijima1951adiabatic,nishijima1951adiabaticII}
(FST-Okubo-Nishijima). In this method, the Fock space is truncated to only nucleons by eliminating the meson degrees of freedom. This allows one not only to determine the Hamiltonian but also the electroweak currents of the system. A consistent application of the model for all generators and currents
produces approximate Lorentz invariance of the form factors and currents conservation~\cite{Sato:1990mj,Tamura:1991gj}.
This transformation method can be understood as a systematic way of a three-dimensional reduction of the Bethe-Salpeter amplitude~\cite{Nambu:1950dpa} similar to the Folded Diagram Method~\cite{Johnson:1976uq}. However, the straightforward application of the FST-Okubo-Nishijima method gives singular non-static two-boson exchange potential in momentum space. In the ANL-Osaka model, a modified FST-Okubo-Nishijima method, the ``unitary transformation method''~\cite{Kobayashi:1997fm,Julia-Diaz:2009nkc}, is utilized to construct a Hermitian Hamiltonian which describes the system below and above the meson production threshold.

The first application of the method was the analysis of single pion photoproduction and electroproduction around the $\Delta$ resonance region~\cite{Sato:1996gk,Sato:2000jf}. In addition to the magnetic transition form factor, electric and Coulomb quadrupole form factors of $N\Delta$ transition were predicted within the dynamical model. The cross sections were found to agree well with the experimental data. The accumulated precise $p(e,e'\pi)N$ data was then re-analyzed and $N\Delta$ transition form factors were revised~\cite{Julia-Diaz:2006ios}. The coupled channel dynamics was studied
extending the approach for the $KY \oplus \pi N$ DCC model of the $\gamma p \rightarrow KY$ reaction~\cite{Julia-Diaz:2006rvy}. 

To investigate the meson production reactions beyond the $\Delta(1232)$ energy region, the effective Hamiltonian and scattering equations for the two-body states $\pi N, \eta N, \pi\Delta, \sigma N, \rho N$  and the three-body state $\pi\pi N$ 
were formulated in Ref.~\cite{Matsuyama:2006rp}. The DCC model for $\pi N$ scattering up to $\sqrt{s} < 2\,\GeV$ including
$S-$, $P-$, $D-$ and $F$-waves was first studied in Ref.~\cite{Julia-Diaz:2007qtz} (JLMS model). Using the meson-induced reaction model, the helicity amplitudes of $\gamma N \rightarrow N^*$ were extracted from the analysis of the  photoproduction $(\gamma,\pi)$~\cite{Julia-Diaz:2007mae} data for $\sqrt{s}<1.65$~GeV. Then, the $Q^2$ dependence of helicity amplitudes for $S_{11}$, $P_{11}$ and $D_{13}$ were
obtained by analyzing electroproduction $p(e,e'\pi)N$~\cite{JuliaDiaz:2009ww} data. The validity of the model for the inelastic processes was tested by the analyses of  $\pi N \rightarrow \eta N$~\cite{Durand:2008es},
$\pi N \rightarrow \pi\pi N$~\cite{Kamano:2008gr} and 
$\gamma N \rightarrow \pi\pi N$~\cite{Kamano:2009im}. The results indicted a reasonable description of those processes, but suggested a need for the improvement in the higher energy region. Meanwhile, the analytic structure of the scattering amplitude, the way to access the unphysical Riemann sheets, and the characterization of resonances through their pole positions and residues within the DCC model were studied in Refs.~\cite{Suzuki:2008rp, Suzuki:2010yn}, while poles of the $P_{11}$ resonances were examined in Ref.~\cite{Suzuki:2009nj}.

Later, the JLMS model was extended to (1) include $K\Lambda$ and $K\Sigma$ channels and to (2) perform a combined analysis of  pion and photon-induced reactions  on the proton~\cite{Kamano:2013iva} for $\sqrt{s} < 2\,\GeV$, which is the basis of the current ANL-Osaka model. The pole positions, residues of the scattering amplitudes, and electromagnetic helicity amplitudes of $\gamma p \rightarrow N^*$ were extracted in Ref.~\cite{Kamano:2013iva}.
The helicity amplitude of $\gamma n \rightarrow N^*$ are provided in Ref.~\cite{Kamano:2016bgm, Nakamura:2018fad} allowing one to separate the iso-scalar and iso-vector helicity amplitudes. 
The elementary photoproduction reactions and meson-baryon scattering amplitudes also served as input for the calculation of $\eta$ photoproduction on the deuteron~\cite{Nakamura:2017sls}.
The DCC model was extended for the weak vector and axial vector currents
for $Q^2 < 3~(\GeV/c)^2$~\cite{Kamano:2012id, Nakamura:2015rta}, which provides cross sections for neutrino physics. The $Q^2$ dependencies of the helicity amplitudes in the ANL-Osaka model were reported in Ref.~\cite{Kamano:2018sfb}. The model was also extended to $\bar{K}N$ reactions for studying
hyperon resonances~\cite{Kamano:2014zba, Kamano:2015hxa}.

\section{Formalism of dynamical coupled-channel approaches}
\label{sec:DCC-formalism}

\subsection{Scattering equation}
Many approaches have been developed to extract the spectrum and properties of nucleon resonances from  pion, photon and electron-induced meson production reactions.  Dynamical coupled channel approaches provide a particularly suitable reaction framework to analyze the pertinent data. The scattering amplitudes of the meson production reactions are obtained  by solving a three-dimensional relativistic Lippmann-Schwinger equation of two- and three-particle states with the model interactions. The particles and their interactions are usually introduced via a model Lagrangian in relativistic quantum field theory. The form of the Lagrangian is constrained by the symmetries, such as Lorentz invariance, flavor and chiral symmetry, and their breaking where applicable.  The two approaches discussed in the following reflect two different approaches of how to cast this framework in practical applications. In the implementations some differences exist between them.

\bigskip

\begin{description}
  
\item[Approach I: The Argonne-Osaka/ANL-Osaka]
assumes that in the 'medium' energy region only 'few-body' degrees of freedom are active and that they must be treated explicitly instead of solving the
field theory which is an intrinsic many body problem. The effects due to  the
many-body states are included as an effective interaction calculated
perturbatively in coupling constants. The Fock-space of the model consists of ``bare'' resonances, two-body meson-baryon, and $\pi\pi N$ three-body states. The interactions among the particles are obtained by applying a unitary transformation perturbatively on the field theoretical Hamiltonian. A feature of the approach is that the resulting interactions do not depend on the scattering energy $\sqrt{s}$ and the single nucleon state decouples from meson-baryon scattering states. The scattering equation of the meson-baryon systems are obtained by applying the projection operator technique.
  
\item[ Approach II: Juelich-Bonn-Washington/JBW approach]
  is based on relativistic time-ordered perturbation theory (TOPT), with 
  the effective interaction derived from the Wess-Zumino Lagrangian and from explicit Lagrangians for excited baryons states. In this approach, retardation effects from the exchange diagrams are retained and as a consequence the interaction depends explicitly on the energy.
  The resulting pseudo-potential is iterated in a Lippman-Schwinger type scattering equation, ensuring unitarity. There is also a renormalization program for the nucleon bound state. The approach shares many properties with the ANL-Osaka approach as a comparison of the scattering equations below shows [\cref{eq:mext} vs. \cref{scattering}].

\end{description}

\subsection{ANL-Osaka model}
\label{subsec:ANL-Osaka}

\subsubsection{Effective Hamiltonian and the method of unitary transformation}

To illustrate the  unitary transformation method, 
we consider the simplest phenomenological Hamiltonian of $N,N^*$ and $\pi$,
\begin{eqnarray}
    H=H_0 + H_I\ . \label{eq:htot}
\end{eqnarray}
Here, $H_0$ is free energy of particles.
The interaction $H_I$ describes
pion production or absorption $B \leftrightarrow B' + \pi$,
where $(B,B')$ are $(N,N),(N, N^*)$ or $(N^*,N)$. In this section, 
both the excited baryons of isospin $I=1/2$($N^*$) and $I=3/2$($\Delta$) are
represented by $N^*$ while $\Delta$ represents the $\Delta(1232)3/2^+$ resonance.

The essence of the unitary transformation method is to extract 
an effective Hamiltonian in a "few-body" space defined by a
unitary operator $U$, such that the resulting scattering equations can be
solved in practice,
\begin{eqnarray}
    H^\prime &=& UHU^\dagger \ .
\end{eqnarray}
In the  approach of Ref.~\cite{Kobayashi:1997fm}, 
the first step is to decompose the
interaction Hamiltonian $H_I$ into two parts,
\begin{eqnarray}
    H_I &= &H_I^P + H_I^Q \ ,
    \label{eq:hpq}
\end{eqnarray}
where $H_I^P$ defines the process  which can take place in free space.
In this model, $H_I^P$ describes the process $N^* \rightarrow N + \pi$.
$H_I^Q$ defines the virtual process, which cannot occur in free space
because of energy-momentum conservation. $H_I^Q$ includes
$N \rightarrow N + \pi$ and  $N \rightarrow N^* + \pi$.
The virtual processes are eliminated by choosing an appropriate unitary transformation $U$.
This can be done systematically by using a perturbative expansion of $U$
in powers of coupling constants. As a result the effects
of ``virtual processes'' appear as effective interactions
in the transformed Hamiltonian. Defining the first transformation $U = \exp(-iS^{(1)})$ by a Hermitian operator $S^{(1)}$ and expanding $U = 1-iS^{(1)} + ...$\,, the transformed Hamiltonian can be written as
\begin{eqnarray}
    H' &=& UHU^\dagger 
    = H_0 + H^{P}_I + H^{Q}_I + [H_0 , iS^{(1)}\,] 
    + [H_I^{P} , iS^{(1)}\,] + \frac{1}{2!} \, \Big[ [H_0 , iS^{(1)}\,] , iS^{(1)}\,\Big] + \cdots \ .
    \label{eq:hpq1}
\end{eqnarray}
The virtual processes $H^Q_I$ which are of first-order in 
the coupling constant are eliminated from Eq.~(\ref{eq:hpq1})
by imposing 
\begin{equation}
    H^{Q}_{I} + [H_0 , iS^{(1)}\,] = 0 \ .
    \label{eq:hpq-s}
\end{equation}
Since $H_0$ is a diagonal operator in the Fock space, Eq.~(\ref{eq:hpq-s}) clearly implies that $iS^{(1)}$ must have the same
operator structure as $H^{Q}_{I}$ and is first order in the coupling constant. By using Eq.~(\ref{eq:hpq-s}), Eq.~(\ref{eq:hpq1}) can be written as
\begin{eqnarray}
    H^\prime = H_0 + H_I^P + [H_I^P,iS^{(1)}\,] + \frac{1}{2} \,[H_I^Q,iS^{(1)}\,] 
    + \mbox{higher order terms}\,.
    \label{eq:hpq2}
\end{eqnarray}
In this example, the effective Hamiltonian includes
decay and production of $N^*$ (the second term $H_I^P$)
and  s- and u-channel baryon-exchange, non-resonant $\pi B-\pi B'$ potentials
(the third and the fourth terms). 
The backward propagation of hadrons, for example, pair production, is also included in
$H_I^Q$ and therefore taken into account as part of the effective potential.
It is informative to quote the matrix element of the $\pi N$ potential with s-channel nucleon exchange in this approach,
\begin{eqnarray}
    \braket{\pi(\bm{p}')N(-\bm{p}')|V |\pi(\bm{p})N(-\bm{p})}  = 
   \Gamma(p') \frac{1}{2}\left(\frac{1}{E_N(\bm{p}')+\omega_\pi(\bm{p}') - M_N} + 
   \frac{1}{E_N(\bm{p})+\omega_\pi(\bm{p}) - M_N}\right)\Gamma(p),
   \label{eq:pin-s1}
\end{eqnarray}
where $\Gamma(p)$ is the $\pi NN$ vertex, $E_N(\bm{p})=\sqrt{\bm{p}^2+M_N^2}$
and $\omega_\pi(\bm{p})=\sqrt{\bm{p}^2 + m_\pi^2}$.
In TOPT, the corresponding matrix element of the driving term is given as
\begin{eqnarray}
    \braket{\pi(\bm{p}')N(-\bm{p}')|V |\pi(\bm{p})N(-\bm{p})}  = 
   \Gamma(p') \frac{1}{E - M_N}  \Gamma(p),
   \label{eq:pin-s2}
\end{eqnarray}
where $E$ is scattering energy. Eq.~\eqref{eq:pin-s1} and Eq.~\eqref{eq:pin-s2}
coincide at the on-energy shell matrix element. Note that up to the same order Eq.~(\ref{eq:hpq1}) should have
an additional term which is the one-pion-loop contribution to
the single nucleon state.
Such a mass renormalization term is dropped in practice,
since it is part of the physical nucleon mass in the resulting effective 
Hamiltonian. The interaction $MB \rightarrow \pi\pi N$ is obtained by
applying the next order unitary transformation $U'=\exp(i S^{(2)})$.

The effective electromagnetic current or weak current
which should be used together with the Hamiltonian $H'$  is
obtained by applying the same 
transformation  $U$ as
\begin{eqnarray}
    J^\mu_{\rm eff} = U J^\mu U^\dagger.
    \label{eq:unit-ext}
\end{eqnarray}
Here $J^\mu$ represents  electromagnetic current  $J^\mu_{EM}$
or weak charged/neutral current $J^\mu_{CC}$/$J^\mu_{NC}$.
The effective current gives, for example, $J^\mu_{em} + N \rightarrow \pi N$ or $J^\mu_{em} + N \rightarrow N^*$ processes.
They are used to describe meson photo-, electroproduction and neutrino reactions.

In summary, a feature of the method is that the effective Hamiltonian is hermitian and independent on the scattering energy $\sqrt{s}$.
As a result, the unitarity relation of the scattering amplitude is
almost trivially satisfied. Another feature which is convenient for the phenomenological description is that the effective Hamiltonian is parametrized in terms of the 'physical' nucleon. The single nucleon state  is diagonalized in the effective Hamiltonian and has no coupling with the $\pi B$ Fock space. Analytic properties of the reaction amplitudes based on the ANL-Osaka model and the relation to the three-dimensional reduction of the BSE and TOPT are discussed in Refs.~\cite{Julia-Diaz:2009nkc,Sato:2009de}.

\subsubsection{Model Hamiltonian}

Towards a realistic model, the starting point is a set of
Lagrangians describing the interactions between mesons 
($M =\pi, \eta ,\rho, \omega, \sigma, K$) and 
baryons ($B = N, \Lambda, \Sigma, \Delta, N^*$). 
We start from the Hamiltonian density ${\cal H}$ of the meson-baryon system,
which is derived by the standard quantization procedure from the
model Lagrangian density ${\cal L}$,
\begin{eqnarray}
{\cal H} = \sum_j  \dot{\Phi}_j\Pi_j - {\cal L} \ , 
\label{Legendre}
\end{eqnarray}
where $\Pi_j$ is the conjugate momentum of $\Phi_j$ given by 
$\displaystyle \Pi_j  =  \frac{\delta {\cal L}}{\delta \dot{\Phi}_j}$,
and we set equal time commutation relation between the field and its conjugate
momentum.

\begin{figure}[tb]
    \centering
    \includegraphics[width=0.7\linewidth, clip]{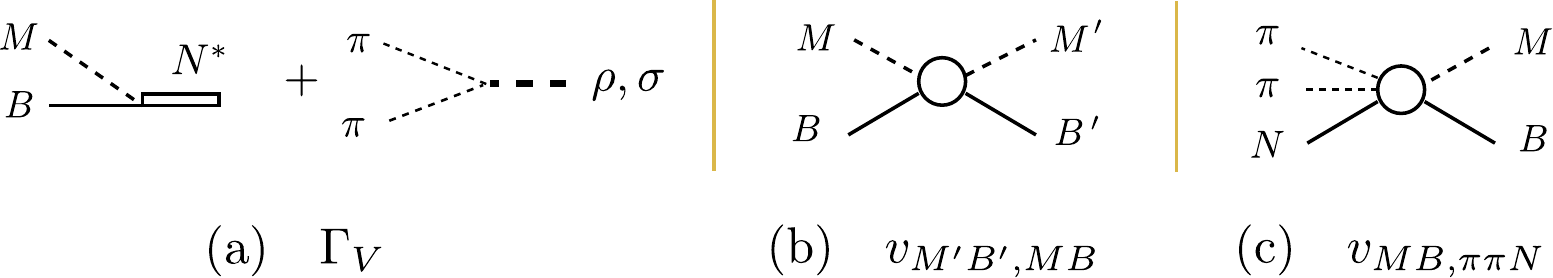}
    \caption{Basic mechanisms of the model Hamiltonian defined in Eq. (\ref{eq:H-1}).
    }
    \label{fig:h}
\end{figure}

By applying the unitary transformation 
method briefly explained in the previous section, we obtain the
effective Hamiltonian for the 
'bare' excited baryon states $P_1 = N^*$, 
the meson-baryon states
$P_2 = \pi N, \eta N, \pi\Delta, \rho N, \sigma N, K\Lambda, K\Sigma$ and
the $Q = \pi\pi N$ state. The Fock space of the model Hamiltonian
consists of $P + Q$ with $P= P_1 + P_2$.

The effective Hamiltonian of the model derived from the starting Lagrangian is given as~\cite{Kamano:2013iva,Matsuyama:2006rp}
\begin{eqnarray}
    H_\text{eff}= H_0 + V \,.
    \label{eq:H-1}
\end{eqnarray}
$H_0$ is the free energy operator. For example, $\bra{\alpha(\bm{p})}H_0\ket{\alpha(\bm{p})} =\sqrt{m_\alpha^2+\bm{p}^2} = E_\alpha(\bm{p})$ for particle $\alpha$ with  mass $m_\alpha$. The interaction Hamiltonian is 
\begin{eqnarray}
    V =\Gamma_V + \sum_{M'B',MB} v_{M'B',MB}   + \left(\sum_{MB} v_{\pi\pi N,MB} + (h.c.)\right) \,.
    \label{eq:H-2}
\end{eqnarray}
Here $h.c.$ denotes the hermitian conjugate. The interaction $\Gamma_V$ connects Fock-spaces $P_1 \leftrightarrow P_2$
and $P_2 \leftrightarrow Q$ given as
\begin{eqnarray}
    \Gamma_V&=& 
    \left(\sum_{N^*,MB} \Gamma_{N^* \rightarrow MB} +
     \sum_{M^*=\rho,\sigma} h_{M^*\rightarrow \pi\pi} \right) + (h.c.) \ . 
    \label{eq:gammav} 
\end{eqnarray}
The resonances associated with the $bare$ excited baryon states $N^*$ and meson states $M^* =\rho, \sigma$ are induced by the interactions $\Gamma_{N^* \rightarrow MB }$ and $\Gamma_{M^*\rightarrow \pi\pi}$. These interactions  are illustrated in Fig.~\ref{fig:h} (a). The masses $M^0_{N^*}$ and $m^0_{M^*}$ of those bare  states $N^*$ and $M^*$ are the parameters of the model which will be determined by fitting the
$\pi N$ and $\pi\pi$ scattering data.

\begin{figure}[tb]
    \centering
    \includegraphics[width=0.7\textwidth,angle=-0]{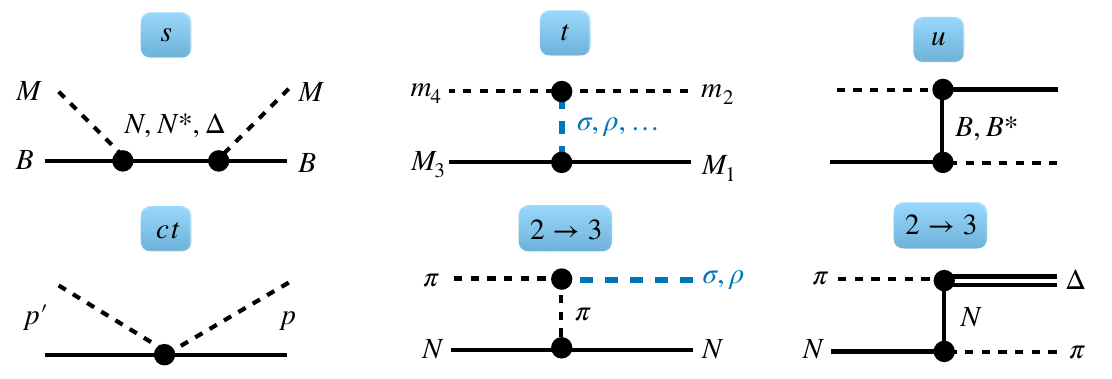}
    \caption{
    Meson ($M$) baryon ($B$) transitions through $s$-, $t$-, $u$-channel, and contact $(ct)$ processes. Also, the $2\to 3$ processes to populate the effective  $\sigma N$, $\rho N$, and $\pi\Delta$ channels are shown. The figure also labels incoming (outgoing ) c.m. momenta $p$ ($p'$) and baryon (meson) masses $M_1,\,M_3$ ($m_2,\,m_4$).}
    \label{fig:alldiagrams}
\end{figure}

The non-resonant meson-baryon interactions $v_{M'B',MB}$ and $MB \rightarrow \pi\pi N$ interaction $v_{\pi\pi N,MB}$
are illustrated in Fig.~\ref{fig:h}(b) and
Fig.~\ref{fig:h}(c), respectively. All of these interactions are defined by the tree-diagrams generated from the considered Lagrangians. 
The mechanism of the meson-baryon interaction 
consist of the $u$- and $s$- channel baryon exchange, $t$-channel meson exchange 
and the contact interaction shown in Fig.~\ref{fig:alldiagrams}.
Here, as an example, the $\pi N$ potential $v_{\pi N,\pi N}$ with s-channel nucleon exchange is given:
\begin{align}
    \bra{
    \pi(\bm{k}',i), N(\bm{p}',s')}
    v_{\pi N,\pi N}
    \ket{
    \pi(\bm{k},j), N(\bm{p},s)}
    =&
    \frac{1}{(2\pi)^3}\sqrt{\frac{M_N}{E_N(p')}}\sqrt{\frac{M_N}{E_N(p)}}\frac{1}{\sqrt{2\omega_\pi(k') 2\omega_\pi(k)}}
    \nonumber \\
    &\times 
     \left(\frac{f_{\pi NN}}{m_\pi}\right)^2\bar{u}_{s'}(p')\slashed{k}'\gamma_5 \tau_i\frac{1}{2}
    \left[
    \frac{1}{\slashed{k}' + \slashed{p}'+ M_N} + \frac{1}{\slashed{k} + \slashed{p}+ M_N}
    \right]
    \slashed{k}\gamma_5 \tau_j u(p)_s 
\label{ANL:potential-1}
\end{align}
where $k^\mu = (\omega_\pi(k),\bm{k}), p^\mu = (E_N(p),\bm{p})$.
In Eq.~\eqref{ANL:potential-1}, the nucleon propagator is similar to the covariant propagator except the time component of the momentum. The u-channel nucleon exchange potential is expressed in a similar form.
The explicit expressions of all interactions in Eq.~(\ref{eq:H-2}) can be found in Refs.~\cite{Kamano:2013iva,Matsuyama:2006rp}.

\begin{figure}[thb]
    \begin{tabularx}{\linewidth}{rX}
         $\pi N\to\pi \pi N$~~~~~~~~
         & \raisebox{-.5\height}{\includegraphics[width=0.85\linewidth]{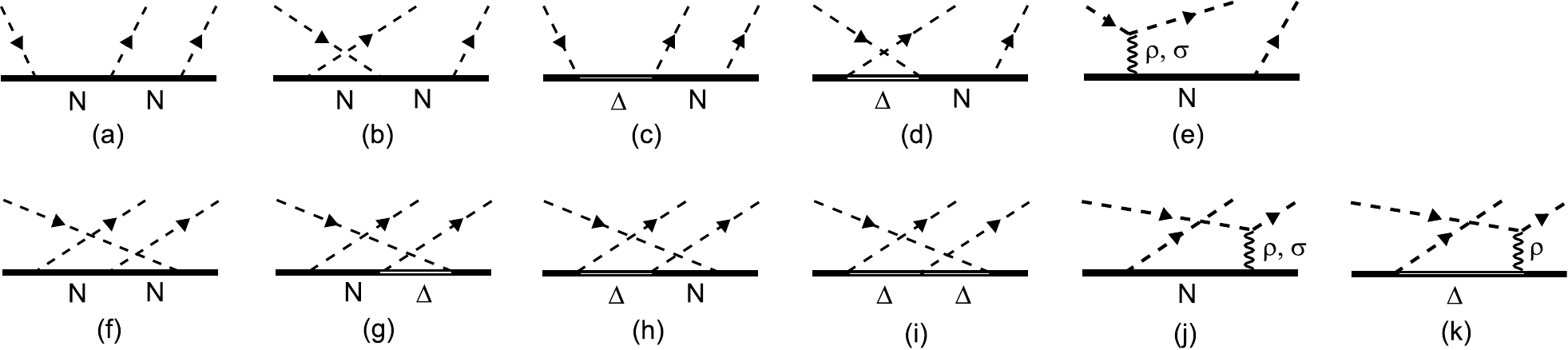}}\\
         ~&~\\
         $\gamma N\to \pi\pi N$~~~~~~~~
         & \raisebox{-.5\height}{\includegraphics[width=0.85\linewidth]{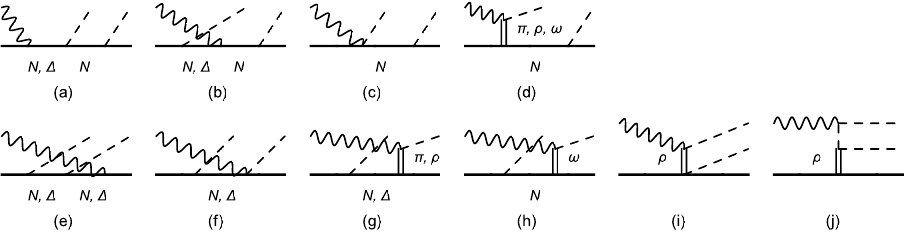}}
    \end{tabularx}
    \caption{The mechanisms for $v_{\pi\pi N,\pi N}$ and $v_{\pi\pi N,\gamma N}$ }
    \label{fig:v23pi}
\end{figure}

The non-resonant interaction $v_{\pi\pi N,MB}$ is calculated from Eq.~\eqref{eq:hpq2}.
The mechanisms for $v_{\pi\pi N,\pi N}$ are shown in Fig.~\ref{fig:v23pi} together with the mechanisms of $v_{\pi\pi N,\gamma N}$.

\subsubsection{Scattering equation}
\label{sec:ANLfull}

Our next task is to derive a set of dynamical coupled-channel equations for describing $\pi N, J^\mu N \rightarrow MB, \pi\pi N$ reactions within the model space $N^*\oplus MB \oplus \pi\pi N$.

The starting point is the Lippmann-Schwinger equation for the
T-matrix,
\begin{align}
    T(E)= V  + V\frac{1}{E-H_0 + i\epsilon} T(E)  \ ,
    \label{eq:loweq}
\end{align}
where $E=\sqrt{s}$ is the total energy. The interaction $V$ is defined from the effective Hamiltonian in Eq.~(\ref{eq:H-2}). The unitarity condition is given as
\begin{align}
    T(E) - T^\dagger(E) = - 2\pi i T^\dagger(E)\delta(E - H_0) T(E)\ .
\end{align}
This relation is satisfied since our model interaction does not depend on the scattering energy $E$. $V$ does not have discontinuities.

\begin{figure}[t]
\centering
    \includegraphics[width=10cm,angle=-0]{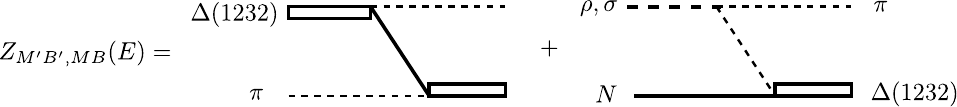}
    \caption{One-particle-exchange interactions $Z_{\pi\Delta,\pi\Delta}(E)$, $Z_{\pi\Delta,\rho N}(E)$ and $Z_{\pi\Delta,\sigma N}(E)$.}
\label{fig:z}
\end{figure}

We cast Eq.~(\ref{eq:loweq}) into a more convenient form for practical calculations. Applying the standard projection operator
techniques~\cite{fesbach1992theoretical}, the
coupled equations of P space ($N^*$ and MB) and Q space ($\pi\pi N$)
 are given as
\begin{align}
    T_{PP}(E) & =  \bar{V}_{PP}(E) + \bar{V}_{PP}(E) \frac{P}{E - H_0 + i\epsilon} T_{PP}(E)\ ,
    \\
    T_{QP}(E) & =  V_{QP} [ 1 + \frac{P}{E - H_0 + i\epsilon} T_{PP}(E)] \ , 
    \\
    \bar{V}_{PP}(E) & =  V_{PP} + V_{PQ} \frac{Q}{E - H_0 + i\epsilon}V_{QP}\ .
\end{align}
Here, for example, $T_{QP} = Q T P$, and $V_{QP} = Q V P$ using projection
operators $P$ and $Q$ of the Fock-spaces P and Q. The effective
interactions for $N^*$ and MB $\bar{V}_{PP}$ become energy dependent
due to the channel coupling.
The eliminated $\pi\pi N$ intermediate states generate the energy dependent particle exchange
interaction $Z_{\beta,\alpha}(E)$ shown in Fig.~\ref{fig:z}
and the self-energies of unstable particles for the $\pi\Delta, \rho N$ and $\sigma N$ Green's functions. The P space in the above equation contains $N^*$ and MB.
The equation for the two-body MB T-matrix is obtained by eliminating $N^*$ components 
or, equivalently, by the two-potential formula of Eq.~\eqref{deco1} for the JBW approach.
The  full description of the coupled-channels equation is given in Appendix B of Ref.~\cite{Matsuyama:2006rp}. 

The partial wave T-matrix  of
$M(\bm{p}) + B(-\bm{p}) \rightarrow M'(\bm{p}') + B'(-\bm{p}')$ is given as 
\begin{align}
    T_{\beta,\alpha}(p',p;E)
    =  t_{\beta,\alpha}(p',p;E)
    + t^R_{\beta,\alpha}(p',p;E) \ ,
    \label{eq:fullt}
\end{align}
where $\alpha, \beta$ represent the MB
states $\pi N,\eta N,\rho N, \sigma N, 
\pi\Delta$, in addition, external current $JN$ for $\alpha$ and angular momentum and isospin. 
The first term in Eq.~(\ref{eq:fullt}) is defined by the following coupled-channels equation
of non-resonant interaction (non-pole term),
\begin{align}
    t_{\beta,\alpha}(p',p;E)
    & =  \bar{V}_{\beta,\alpha}(p',p;E)+     \int_C dq q^2 \sum_{\gamma}    \bar{V}_{\beta,\gamma}(p',q;E)G_\gamma(q;E)
    t_{\gamma,\alpha}(q,p;E)\ ,
    \label{eq:mext}
\end{align}
where $\bar{V}_{\beta,\alpha}(p',p;E)$ is defined as
\begin{align}
    \bar{V}_{\beta,\alpha}(p',p;E) & =  v_{\beta,\alpha}(p',p) + Z_{\beta,\alpha}(p',p;E)\ .
    \label{eq:vvz}
\end{align}
Here, the energy dependent particle exchange interaction $Z_{\beta,\alpha}(p',p;E)$ is due to the coupling of MB states with the $\pi\pi N$ states. The integration path $C$, in principle ranging from $q=0$ to $\infty$, will be discussed later.

The meson-baryon Green's function for the stable particles ($\pi N,\eta N, K\Lambda, K\Sigma$) and unstable particles ($\pi \Delta, \rho N, \sigma N$) is given by
\begin{align}
    G_{MB}(q;E) & =  \frac{1}{E - \omega_M(q) - E_B(q) + i\epsilon}\ ,\\
    G_{MB}(q;E) & =  \frac{1}{E - \omega_M(q) - E_B(q) - \Sigma(q;E)}\ ,
    \label{unstableprop}
\end{align}
respectively. For example, the self-energy for $MB=\pi\Delta$ is given as
\begin{align}
    \Sigma_\Delta(p;E)= \frac{m_\Delta}{E_{\Delta}(k)}\int_{C'}  dq q^2 \frac{M_{\pi N}(q)}{E_{\pi N}(q,p)}\frac{\Gamma^2_{\Delta,\pi N}(q) }
    {E-\omega_\pi(p)- E_{\pi N}(q,p)}\ , 
    \label{eq-sigma-pid}
\end{align}
where $M_{\pi N}(q) = \omega_\pi(q) + E_N(q)$ and $E_{\pi N}(q,p)=\sqrt{M_{\pi N}^2(q) + p^2}$.
In Eq. (\ref{eq-sigma-pid}), the loop momentum is rewritten by the relative momentum of the $\pi N$ center of mass system based on the well-known instant form of relativistic quantum mechanics.  A detailed derivation of the formula is given in the appendix of Ref.~\cite{Kamano:2008gr}.

The resonant $T$-matrix of the second term on the r.h.s of Eq. (\ref{eq:fullt}) is given as
\begin{align}
    t^R_{\beta,\alpha}(p',p;E) = 
    \sum_{i,j}\bar{\Gamma}_{\beta,i}(p';E)
    [G_{N^*}(E)]_{i,j}
    \bar{\Gamma}_{\alpha,j}(p;E) ,
    \label{eq:tr}
\end{align}
with
\begin{align}
    [G_{N^*}^{-1}]_{i,j}(E)=
    (E - m_{N^*_i})\delta_{i,j} - \Sigma_{i,j} (E)\ .
    \label{eq:tr-g}
\end{align}
Here $i,j$ denote the bare $N^*$ states defined in the Hamiltonian.
$ m_{N^*_i}$ are their masses.
In general, the bare states mix with each other
through the off-diagonal matrix elements of the self-energies.
The dressed vertices and the energy shifts of the second term in 
Eqs.~(\ref{eq:tr})-(\ref{eq:tr-g}) are defined by
\begin{align}
    \bar{\Gamma}_{\alpha,j}(p;E) & = 
    \Gamma_{\alpha,j}(p) + \int_C dq q^2 \sum_\gamma
    t_{\alpha,\gamma}(p,q;E)G_\gamma(q;E)\Gamma_{\gamma,j}(q) \ ,
    \label{eq:dressf} 
    \\
    \Sigma(E)_{i,j} & =  \int_C dq q^2 \sum_\gamma
    \Gamma_{\gamma,i}(q)G_\gamma(q;E)\bar{\Gamma}_{\gamma,j}(q)\ ,
    \label{eq:selfe}
\end{align}
where $\Gamma_{\alpha,i}(p)$ defines the coupling of the $i$-th bare $N^*$ state to channel $\alpha$.

\begin{figure}[t]
    \centering
    \includegraphics[width=\linewidth]{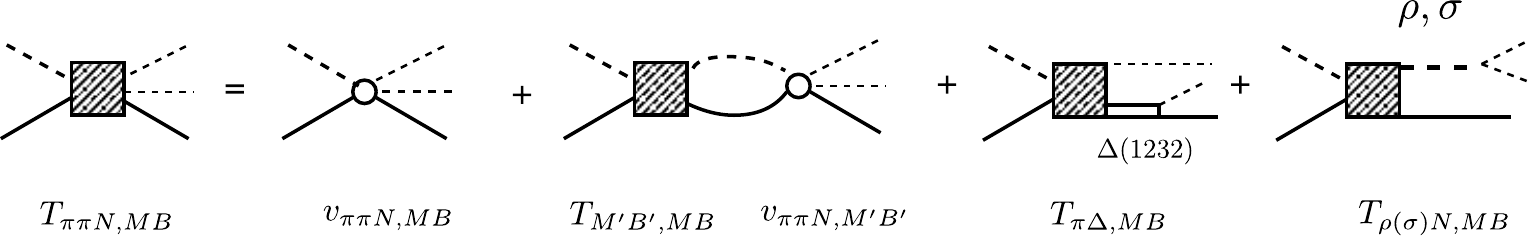}
    \caption{Graphical representations of  $T_{\pi\pi N,MB}$
    defined by Eq.~(\ref{eq:tpipin-1}).
    }
    \label{fig:mb-pipin-1}
\end{figure}

The amplitudes $T_{M'B',MB}=t_{M'B',MB}+t^R_{M'B',MB}$ defined by Eq.~(\ref{eq:fullt}) can be used directly to calculate the cross sections of the $\pi N \rightarrow MB$ and the $J N \rightarrow MB$ reactions. They are also the input to the calculations of the two-pion production amplitudes. The two-pion production amplitudes  are illustrated in Fig.~\ref{fig:mb-pipin-1}. They can be cast into the following form
\begin{align}
    T_{\pi\pi N,MB}(E) &= T^\text{dir}_{\pi\pi N,MB}(E)+ T^{\pi\Delta}_{\pi\pi N,MB}(E)+T^{\rho N}_{\pi\pi N,MB}(E) +T^{\sigma N}_{\pi\pi N,MB}(E)\ ,
    \label{eq:tpipin-1}
\end{align}
with
\begin{align}
T^\text{dir}_{\pi\pi N,MB}(E)
    & =  \sum_{M'B'}v_{\pi\pi N,M'B'}[\delta_{M'B',MB}   +G_{M'B'}(E) T_{M'B',MB}(E)] 
    \label{eq:tpipin-dir} \ , \\
    T^{\pi\Delta}_{\pi\pi N,MB}(E) & =  \Gamma^\dagger_{\Delta\rightarrow \pi N} G_{\pi\Delta}(E) T_{\pi\Delta, MB}(E)\ ,
    \nonumber  \\
    T^{\rho N}_{\pi\pi N,MB}(E) & =  h^\dagger_{\rho\rightarrow \pi\pi} G_{\rho N}(E) T_{\rho N, MB}(E)\ ,
    \nonumber \\
    T^{\sigma N}_{\pi\pi N,MB}(E) & =  h^\dagger_{\sigma \rightarrow \pi\pi} G_{\sigma N}(E) T_{\sigma N, MB}(E) \ .
    \label{eq:tpipin-sigman-ext}
\end{align}

The $\pi\pi N$ state is mainly produced through the unstable $\pi\Delta$, $\rho N$, and $\sigma N$ states illustrated in Fig.~\ref{fig:mb-pipin-1}. Each term has contributions from the non-resonant amplitude $t_{M'B',MB}(E)$ and the resonant term $t^R_{M'B',MB}(E)$. 
The mechanism of non-resonant interactions $v_{\pi\pi N, \pi N}$ and $v_{\pi\pi N,\gamma N}$ are shown in Fig.~\ref{fig:v23pi}.

\subsection{Meson-baryon interactions in the JBW approach and its SU(3) structure}
\label{sec:JB}

The interaction of mesons and baryons in the Juelich-Bonn (JB) approach is parametrized through the Wess-Zumino effective Lagrangian~\cite{Wess:1967jq}. Chiral symmetry of QCD is one of the guiding principles for its construction. The interaction also contains vector mesons. While there are general guiding principles to include them~\cite{Meissner:1987ge, Borasoy:1995ds}, the JB approach adopts the version of Ref.~\cite{Wess:1967jq}. The starting point is the nucleon and the pion field in a nonlinear $\sigma$ model. Vector mesons are then included by gauging the chirally invariant Lagrangian, i.e., by requiring the Lagrangian to be invariant under local transformations, leading to vector and axial gauge fields. The $\rho\pi\pi$ coupling strength is related to $f_\pi$ via the KSFR relation~\cite{Kawarabayashi:1966kd}. The Wess-Zumino Lagrangian provides $\pi NN$, $\rho NN$, $\rho\pi\pi$, and $a_1\rho\pi$ interactions; there are are also contact terms that contribute. See Appendix B of Ref.~\cite{Ronchen:2012eg} for a list.

The SU(3) extension of the approach is achieved by relating these interactions to other hadrons via flavor symmetry~\cite{Ronchen:2012eg}, based on Ref.~\cite{deSwart:1963pdg}. The coupling  of a given meson nonet with the octet baryon current $8_B\otimes 8_B$  yields two products, namely $8_B\otimes
8_B\otimes 1_M$ and $8_B\otimes 8_B\otimes 8_M$. The coupling to the meson octet $8_M$ depends on two parameters, $g_1, \,g_2$, corresponding to
the symmetric and antisymmetric representations of $8_B\otimes 8_B$. They can be re-written in terms of a coupling constant $g$ and a mixing
parameter $\alpha$ in the notation of Ref.~\cite{deSwart:1963pdg} that is used in the JBW as well, 	
\begin{eqnarray}
 g=\frac{\sqrt{30}}{40}\,g_{1}+\frac{\sqrt{6}}{24}\,g_{2},\quad \alpha = 
\frac{\sqrt{6}}{24}\,\frac{g_{2}}{g}\ . 
\end{eqnarray}
Here $g_{1}$ and $g_{2}$ can be also expressed in terms of the standard $D$ and $F$ axial couplings \cite{ParticleDataGroup:2024cfk}:
\begin{eqnarray}
 D=\frac{\sqrt{30}}{40}g_{1}\,  , \;\;\; F= \frac{\sqrt{6}}{24}\,g_{2}\ ,
\end{eqnarray}
leading to $\alpha=F/(D+F)$.  The coupling of the baryon octets to the meson singlet $1_M$ involves  a further independent coupling constant.  For the transitions $MB\to MB$, with the $J^P=0^-$ octet mesons $M$ and the $J^P=1/2^+$ octet baryons $B$, the JB model includes the full set of diagrams required by SU(3) symmetry. This applies to the classes of diagrams of the exchange of $J^P=1^-$ vector mesons, $J^P=1/2^+$ baryons, and $J^P=3/2^+$
baryons. In particular, non-diagonal transitions such as $\eta N\to KY$ ($Y=\La,\,\Si$) are considered.  See Appendix B2 of Ref.~\cite{Ronchen:2012eg} for further details. 

The same reference also provides information on how several processes not covered by SU(3) symmetry are modeled. For example, the coupling of scalar mesons to nucleons, $\sigma NN$, and among themselves, $\sigma\sigma\sigma$, is described by Lagrangians with couplings strengths fit to data. Similarly, $\pi\Delta\Delta$ and $\rho\Delta\Delta$ interaction are allowed for, with interaction strength fixed from fits to data or external information. The physical $\omega$ and $\phi$ states are defined by ideal mixing of $\omega_1$ and $\omega_8$ states.

Effective Lagrangians for higher spin nucleon and $\Delta$ excited states with $J\leq 3/2$ are listed in Appendix B of Ref.~\cite{Doring:2010ap}.
Resonance interactions with total spin $J \geq 5/2$ are not described by corresponding Lagrangians but their decay vertices are  related to those of lower spin with additional powers of momentum ensuring the correct centrifugal barriers. In general, $s$-channel resonances are allowed to couple to the MB (pseudo-scalar octet, baryon octet) channels $\pi N$, $\eta N$, $K\Lambda$ and $K\Sigma$ with independent coupling strength not related through SU(3) symmetry. Additionally, resonances can couple to $\pi\Delta$ and $\rho N$ states, but not to $\sigma N$. The argument is that, as long as the JBW model does not analyze  $\pi\pi N$ data, there is no need to populate each possible $\pi\pi N$ three-particle states. 
There is usually more than one independent way to combine spin and angular momentum of $\rho,\, N$ or $\pi,\,\Delta$ to a given $J^P$ (see Table~\ref{tab:couplscheme}), but the JB model uses only one resonance Lagrangian suitably projected into different $\pi\Delta$ and $\rho N$ channels, with only one coupling constant, for simplicity.

\subsection{From interactions to scattering in the JBW approach}
\label{sec:Juba}
For the construction of interactions from Lagrangians, the JB group uses 
TOPT~\cite{Schweber2005-hi}. The resulting pseudo-potentials are combined into
$s$-, $t$-, and $u$-channel diagrams which are summed up by solving 
a three-dimensional integral equation of Lippmann-Schwinger type with 
relativistic kinematics. Such a three-dimensional scattering equation
leads to a simplified singularity structure, as compared to that of a 
four-dimensional treatment in the context of the Bethe-Salpeter equation
(BSE) discussed in \cref{sec:unitarity}, although even in a three-dimensional formalism using TOPT one has ``moving`` singularities in exchange processes that depend on the incoming and outgoing momenta and the energy. More details are provided in \cref{sec:realmom}.

Through the Legendre transformation~\eqref{Legendre} the Wess-Zumino Lagrange density~\cite{Wess:1967jq} is converted to a Hamilton density
  ${\cal H}$ that  
is decomposed as
\begin{align}
{\cal H}={\cal H}^0-{\cal L}_\text{int}+{\cal H}_\text{ct}
\end{align}
with a non-interacting part ${\cal H}^0$, an interacting part, and contact interactions ${\cal H}_\text{ct}$. The latter originate if the interaction part contains explicit time derivatives of fields, or non-dynamical zero-components of vector fields~\cite{Weinberg_1995}.
Then, the contact terms restore the on-shell equivalence between the time-ordered and the covariant formalism. In practical terms, it is straightforward to construct the contact terms at the diagrammatic level as the difference of the covariant Feynman expression and both time orderings (TO)~\cite{Schweber2005-hi}, 
\begin{align}
    V_\text{ct}=V_\text{Feyn}-V_\text{TOPT}^\text{1st TO}-
    V_\text{TOPT}^\text{2nd TO} \ .
\end{align}
The Hamiltonian is obtained by integrating ${\cal H}$ over the spatial coordinates and can, again, be written in terms of a free and an interacting part, $H=H_0+H_I$. In terms of the Hamiltonian, the scattering operator reads 
\begin{align}
    S=\lim_{\substack{t\to\infty\\ t_0\to-\infty}} \hat U(t,t_0) \ ,
\end{align}
where $\hat U(t,t_0)=e^{iH_0 t}e^{-i H(t-t_0)}e^{-iH_0t_0}$ is the evolution operator in the interaction picture. The (onshell) T-matrix has the quantum mechanical normalization
\begin{align}
    S_{fi}=:\delta_{fi}-i(2\pi)\delta(E_f-E_i){\cal T}_{fi} \ ,
    \label{eq:qmT}
\end{align}
where $i (f)$ denote initial (final state) and $E_i, \,E_f$ are the respective energies. 
The on-energy shell T-matrix operator can be written as  
\begin{align}
{\cal T}=H_I+H_I\frac{1}{E-H+i\epsilon}\,H_I=H_I+H_I\frac{1}{E-H_0-H_I+i\epsilon}\,H_I \ .
\end{align}
One can expand ${\cal T}$ in a power series in $H_I$,
\begin{align}
    {\cal T}=H_I\frac{1}{1-{\cal G}H_I}=H_I+H_I{\cal G}H_I+\dots
    \label{tjb}
\end{align}
with the Green's function of the free particle states, 
\begin{align}
    {\cal G}=\frac{1}{E-H_0+i\epsilon}=\sumint_n\frac{\ket{n}\bra{n}}{E-E_n+i\epsilon} \ .
\end{align}
The sum/integral over the right-hand side formally extends over all multi-particle states allowed, and $n$ is a composite index for continuous and discrete variables. The next step towards writing a scattering equation consists in representing the perturbation series of Eq.~\eqref{tjb} through matrix elements obtained from diagrams derived from $H_I$ through TOPT rules that are similar to standard Feynman rules. The main differences consists in the occurrence of non-covariant TOPT contact terms that arise in addition to contact terms from the vector mesons from the Wess-Zumino Lagrangian itself as mentioned before. 
In the Green's function, one can explicitly separate 2-particle intermediate states off the rest, ${\cal G}=\tilde {\cal G}+{\cal G}^{(2)}$, arriving at 
\begin{align}
{\cal T}={\cal V}+{\cal V}{\cal G}^{(2)}{\cal T}
\end{align}
with an effective potential 
\begin{align}
    {\cal V}=H_I\frac{1}{1-\tilde {\cal G}H_I} \ ,
\end{align}
that contains the intermediate states with more than two particles. Furthermore, both the potential ${\cal V}$ and the ${\cal T}$-matrix can be generalized to the off-shell case $E\neq E_i\neq E_f$,
\begin{align}
  {\cal T}_{fi}={\cal V}_{fi}+\sumint_j\frac{{\cal V}_{fj}{\cal T}_{ji}}{E-E_j-\omega_j+i\epsilon}   \ ,
  \label{tfi}
\end{align}
for initial, final, and intermediate two-body states. The index $j$ indicates the sum over discrete quantum numbers like particle spins, two-body channels, and continuous three-momenta ${\bm q}_1$ and ${\bm q}_2$ with states normalized as $\braket{{\bm q}|{\bm q}'}=\delta({\bm q}-{\bm q}')$ and baryon and meson energies
\begin{align}
    E_j=\sqrt{M_B^2+{\bm q}_1^2}\ ,\quad \omega_j=\sqrt{m_M^2+{\bm q}_2^2} \ ,
    \label{eq:energies}
\end{align}
where $M_B$ and $m_M$ are the baryon and meson masses, respectively. In the center-of mass frame where the momenta of initial and final particles fulfill ${\bm p_1}+{\bm p_2}={\bm p_3}+{\bm p_4}$, the matrix elements of ${\cal A}\in\{{\cal T},{\cal V}\}$ can be written in terms of $A\in\{T,V\}$ as
\begin{align}
      \braket{{\bm p_3},{\bm p_4},\lambda_3,\lambda_4,\mu,I,I_3|{\cal A}(E)|{\bm p_1},{\bm p_2},\lambda_1,\lambda_2,\nu,I,I_3}
    =\delta({\bm p_3}+{\bm p_4}-{\bm p_1}-{\bm p_2})\,
    A_{\mu\nu}^I(E,\bm{p}',\lambda',\bm{p},\lambda) \ ,  
\end{align}
where ${\bm p}={\bm p}_1$, ${\bm p}'={\bm p}_3$, $\mu$ and $\nu$ are channel indices $(\pi N, \,\eta N,\,K\Lambda, \,K\Sigma,\,\omega N,\dots)$, and $I$ is the total isospin. Furthermore, $\lambda=\lambda_1-\lambda_2$ and $\lambda'=\lambda_3-\lambda_4$ are helicity indices for the helicities of incoming and outgoing states~\cite{Jacob:1959at}. With these simplifications, Eq.~\eqref{tfi} reads
\begin{align}
    T^I_{\mu\nu}(\bm{p}',\lambda',{\bm p},\lambda;E)=V^I_{\mu\nu}(\bm{p}',\lambda',{\bm p},\lambda;E)     +\sum_{\kappa,\lambda''} \int\mathrm{d}^3q\,V^I_{\mu\kappa}(\bm{p}',\lambda',{\bm q},\lambda'';E)\,\frac{1}{E-E_\kappa(q)-\omega_\kappa(q)+i\epsilon}\,T^I_{\kappa\nu}(\bm{q},\lambda'',{\bm p},\lambda;E) \ ,
    \label{eq:lse}
\end{align}
which formally resembles a Lippmann-Schwinger equation with relativistic energies. Note that this is similar to the Thompson equation~\cite{Thompson:1970wt, Erkelenz:1974uj,Wu:2012md}.
Three-body states can then be re-inserted as individual channels instead of effective potentials, ordered by quantum numbers. This is discussed in \cref{sec:tauiso}. Note that the above equation is identical to the one of the ANL-Osaka approach in Eq.~\eqref{ANL:potential-1}. 

If one aims at obtaining the interactions $V$ from Lagrangians leading to a vertex $\bar V$ by using TOPT rules, one 
starts out with the standard on-shell invariant scattering amplitude $\bar T$ defined through
~\cite{Schutz:1994ue}
\begin{align}
        S_{fi}=\delta_{fi}-i(2\pi)^{-2}\delta^4(P_f-P_i)\sqrt{\frac{M_3}{E_3}}\sqrt{\frac{M_1}{E_1}}\frac{1}{\sqrt{2\omega_4}}\frac{1}{\sqrt{2\omega_2}}\bar T_{fi} \ ,
        \label{eq:invaT}
\end{align}
which implies for the interactions that 
\begin{align}
    \bar V(E,\bm{p}',\lambda',{\bm p},\lambda)=\bar u(\bm{p}',\lambda')\Gamma u(\bm{p},\lambda)
\end{align}
with Dirac spinors normalized to $\bar uu=1$ and $\Gamma$ containing the entire spin and isospin structure of the transition.
The comparison of  Eqs.~\eqref{eq:invaT} and 
\eqref{eq:qmT}
reveals a conversion factor 
\begin{align}
A(\bm{p}',\bm{p})=\kappa'\,\bar A \ ,\quad 
\kappa'=
\frac{1}{(2\pi)^3}\sqrt{\frac{M_3}{E_3(\bm{p}')}}\sqrt{\frac{M_1}{E_1(\bm{p})}}\frac{1}{\sqrt{2\omega_4(\bm{p}')}}\frac{1}{\sqrt{2\omega_2(\bm{p})}}\,\ ,
\end{align}
where, again, $A$ stands for $V$ or $T$ and $\omega_2(\omega_4)$ are the incoming (outgoing) meson energies and analogously for baryon energies $E_1\,(E_3)$. In more recent versions of the JBW model, i.e., 
Ref.~\cite{Gasparyan:2003fp} and later, one has
\begin{align}
\kappa'\to\kappa=
\frac{1}{(2\pi)^3}\frac{1}{\sqrt{2\omega_4(\bm{p}')}}\frac{1}{\sqrt{2\omega_2(\bm{p})}}\,\ ,
\end{align}
due to a change of the spinor definition~\cite{Machleidt:1987hj}
resulting in $\bar uu=M/E$.

Note that separation of the identity and interacting part (the reaction amplitude $T$) from the S-matrix differ vastly in the literature. For example, in many meson-baryon studies (see, e.g.,  Ref.~\cite{Mai:2009ce,Bruns:2010sv,Mai:2012wy,Ruic:2011wf,Becher:1999he,Becher:2001hv}) canonical H\"ohler nomenclature~\cite{Hohler:1974ht} defines $\hat S=\mathds{1}+i\hat T$ on the level of operators and then utilizes spinors normalized as $\bar u_f(p)u_i(p)=2M\delta_{fi}$
while defining the general Lorentz invariant and parity conserving T-matrix element for a meson-baryon transition $M(q)B(p)\to M(q')B(p')$ as 
\begin{align}
\label{eq:Hohler-general-structure}
    M_{fi}=
    \frac{1}{8\pi\sqrt{s}}
    \bar u_f(p')
    \left(
        A(s,t)+\frac{\slashed{q}'+\slashed{q} }{2}B(s,t)
    \right)    
    u_i(p)\ .
\end{align}
Ultimately, the scalar quantities $A, B$ are used to define partial-wave amplitudes, relating also to observable quantities like phase-shifts or cross sections.

\subsection{Partial-wave projection}
\label{sec:pwa}

Dynamical coupled-channel models deal with $s$-channels containing spin-1/2 nucleons but also  the spin-3/2 $\Delta$ and spin-1 vector mesons $(\rho,\omega)$. The partial-wave projection of $A\in\{V,T\}$ in the scattering equation~\eqref{eq:lse} is a two-step process for the discussed DCC frameworks: first,  the scattering amplitude is projected to partial waves in the helicity basis, and then these partial waves are transformed to the JLS basis. Partial waves in the JLS basis can also be calculated directly~\cite{Oller:2019opk, Oller2019-xp}. 

As helicity and angular momentum commute, one can expand the helicity eigenstates in terms of eigenstates of total angular momentum. In particular, with incoming (outgoing) helicities $\lambda_1, \, \lambda_2\,(\lambda_3,\,\lambda_4)$, the quantity $A$ as written in the states $\ket{\bm{p}',\lambda_3,\lambda_4}$ and $\ket{\bm{p},\lambda_1,\lambda_2}$ can be decomposed as~\cite{Jacob:1959at, Chung:1971ri}
\begin{align}
    \braket{\bm{p}',\lambda'|A|\bm{p},\lambda}
    =\frac{1}{4\pi}\sum_J(2J+1)D^{J*}_{\lambda\lambda'}(\Omega_{\bm{p}'\bm{p}},0)\braket{p',\lambda'|A^J|p,\lambda} \ ,
\end{align}
that depends only on the helicity differences $\lambda:=\lambda_1-\lambda_2$ and $\lambda':=\lambda_3-\lambda_4$. Here, $D$ are  the Wigner-D functions with $\Omega_{\bm{p}'\bm{p}}$ the solid angle between $\bm{p}$ (incoming nucleon) and $\,\bm{p}'$ (outgoing nucleon). We choose $\Omega_{\bm{p}'\bm{p}}=(\phi_0,\theta_0)$ in the specific geometry where particle 1 (nucleon) comes in along the $+z$-direction and particle 3 (nucleon) leaves at angles $\phi_0$ and $\theta_0$~\cite{Chung:1971ri}. Note that in some papers one can find a slightly different PWA expression, with the connection to the above one discussed in Sect.~IIE of Ref.~\cite{Feng:2024wyg}. The inverse is given by
\begin{align}
  \braket{p',\lambda'|A^J|p,\lambda}=2\pi\int\limits_{-1}^1\dv(\cos\theta) \, d^J_{\lambda\lambda'} (\theta)\braket{\bm{p}',\lambda'|A|\bm{p},\lambda} \ ,
\end{align}
for which $\bm{p}=(0,0,p)^\intercal$ and $\bm{p}'=(p'\sin\theta,0,p'\cos\theta)^\intercal$ is usually chosen for simplicity. 
The second step concerns the transformation from helicity to JLS basis. For total spin $J$ with third component $M$ one has
\begin{align}
    \ket{JM\lambda_1\lambda_2}=\sum_{LS}\braket{JMLS|JM\lambda_1 \lambda_2}\,\ket{JMLS} 
\end{align}
with orbital angular momentum $L$ and total spin $S$ that is the sum of the individual spins. For example, the $\rho N$ states can have $S=1/2$ or $S=3/2$. The transformation matrix is given by Clebsch-Gordan coefficients~\cite{Chung:1971ri}, 
\begin{align}
  \braket{J'M'LS|JM\lambda_1 \lambda_2}=\sqrt{\frac{2L+1}{2J+1}}
  \braket{L0\,S\lambda|J\lambda}\braket{S_1\lambda_1\,S_2-\lambda_2|S\lambda}\,\delta_{JJ'}\delta_{MM'} \ ,
\end{align}
where, again, $\lambda=\lambda_1-\lambda_2$. After these transformations, the partial-wave projected scattering equation in the c.m. frame, for given isospin $I$ and spin-parity $J^P$, reads \cite{Krehl:1999km, Gasparyan:2003fp, Doring:2009bi, Doring:2009yv, Doring:2010ap} 
\begin{align}
    T_{\mu\nu}(p',p;E)=V_{\mu\nu}(p',p;E)
    +\sum_{\kappa}\int\limits_0^\infty \dv{q}\,
     q^2\,V_{\mu\kappa}(p',q;E)\,G_\kappa(q;E)\,T_{\kappa\nu}(q,p;E) \ , 
    \label{scattering}
\end{align}
where $p'\equiv|\bm{p}\,'|$ ($p\equiv |\bm{p}\,|$) is the modulus of the outgoing (incoming)  three-momentum that may be on- or off-shell, $E=\sqrt{s}$ is the scattering energy, and $\mu,\,\nu,\,\kappa$ are channel indices. In
Eq.~(\ref{scattering}), the propagator for a stable meson and a stable baryon is given by 
\begin{align}
    G_\kappa=\frac{1}{E-E_\kappa(q)-\omega_\kappa(q)+i\epsilon}\ ,
    \label{gkappa}
\end{align}
with energies from Eq.~\eqref{eq:energies}. The propagators for unstable channels are more complicated as discussed in \cref{sec:tauiso}. Note that there are different possibilities for the $\pi\Delta$ and $\rho N$ channels to couple to a given $J^P$, that are counted as individual channels. The complete coupling scheme (applying to $s$-, $t$-, and $u$-channel exchanges) is shown in Table~\ref{tab:couplscheme}. 
There, the total spin $S=|\bm{ S}_N+\bm{ S}_\rho|$ is given by the sum of the $\rho$ spin and the nucleon spin, and $L$ is the orbital angular momentum. 

\begin{table}[b]
\caption{Angular momentum structure of the coupled channels in isospin $I=1/2$ up to $J=9/2$. The $I=3/2$ sector is similar up to obvious isospin selection rules.
}
\begin{center}
\begin{tabularx}{\linewidth}{lX|ll|ll|ll|ll|ll}
    \hline \hline
    $\mu$	&\multicolumn{1}{r}{$J^P=$}
    &$\frac{1}{2}^-$&$\frac{1}{2}^+$&$\frac{3}{2}^+$&$\frac{3}{2}^-$&$\frac{5}{2}^-$&$\frac{5}{2}^+$&$\frac{7}{2}^+$&$\frac{7}{2}^-$&$\frac{9}{2}^-$&$\frac{9}{2}^+$\BBB \TT
    \\
    \hline
    $1$	&$\pi N$ 			&$S_{11}$		&$P_{11}$		&$P_{13}$		&$D_{13}$		&$D_{15}$		&$F_{15}$		&$F_{17}$		&$G_{17}$		&$G_{19}$		&$H_{19}$		\bigstrut[t]\\
    $2$	&$\rho N(S=1/2)$		&$S_{11}$		&$P_{11}$		&$P_{13}$		&$D_{13}$		&$D_{15}$		&$F_{15}$		&$F_{17}$		&$G_{17}$		&$G_{19}$		&$H_{19}$		\\
    $3$	&$\rho N(S=3/2, |J-L|=1/2)$	&---		&$P_{11}$		&$P_{13}$		&$D_{13}$		&$D_{15}$		&$F_{15}$		&$F_{17}$
    &$G_{17}$		&$G_{19}$		&$H_{19}$		\\
    $4$	&$\rho N(S=3/2, |J-L|=3/2)$	&$D_{11}$		&---		&$F_{13}$		&$S_{13}$		&$G_{15}$		&$P_{15}$		&$H_{17}$
    &$D_{17}$		&$I_{19}$		&$F_{19}$		\\
    $5$	&$\eta N$ 			&$S_{11}$		&$P_{11}$		&$P_{13}$		&$D_{13}$		&$D_{15}$		&$F_{15}$		&$F_{17}$		&$G_{17}$		&$G_{19}$		&$H_{19}$		\\
    $6$	&$\pi\Delta (|J-L|=1/2)$	&---		&$P_{11}$		&$P_{13}$		&$D_{13}$		&$D_{15}$		&$F_{15}$		&$F_{17}$
    &$G_{17}$		&$G_{19}$		&$H_{19}$		\\
    $7$	&$\pi\Delta (|J-L|=3/2)$	&$D_{11}$		&---		&$F_{13}$		&$S_{13}$		&$G_{15}$		&$P_{15}$		&$H_{17}$
    &$D_{17}$		&$I_{19}$		&$F_{19}$		\\
    $8$	&$\sigma N$			&$P_{11}$		&$S_{11}$		&$D_{13}$		&$P_{13}$		&$F_{15}$		&$D_{15}$		&$G_{17}$		&$F_{17}$		&$H_{19}$		&$G_{19}$		\\
    $9$	&$K\Lambda$ 			&$S_{11}$		&$P_{11}$		&$P_{13}$		&$D_{13}$		&$D_{15}$		&$F_{15}$		&$F_{17}$		&$G_{17}$		&$G_{19}$		&$H_{19}$		\\
    $10$	&$K\Sigma$ 			&$S_{11}$		&$P_{11}$		&$P_{13}$		&$D_{13}$		&$D_{15}$		&$F_{15}$		&$F_{17}$		&$G_{17}$		&$G_{19}$		&$H_{19}$
    \BBB
    \\
    \hline \hline
\end{tabularx}
\end{center}
\label{tab:couplscheme}
\end{table}

\subsection{Channels and transitions}
\label{sec:chatra}
Both the ANL-Osaka and the JB models include two-body channels with stable hadrons, $\pi N$, $\eta N$, $K\Lambda$,  $K\Sigma$, and $\omega N$, as well as $\pi\Delta$, $\sigma N$, and $\rho N$. The unstable ``isobars'' in the latter three channels stand for two-body sub-channels with sub-energy-dependent amplitude rather than simple resonances, see \cref{sec:tauiso}. The extension of the original JBW model to include the $K\Lambda$ and $K\Sigma$ channels was carried out in Ref.~\cite{Doring:2010ap}, while the $\omega N$ channel was added only recently by Wang et al.~\cite{Wang:2022osj}. In the same work, to make the dynamics of the $K\Sigma$ system complete, the exchange of $a_0$ was added for the $K\Lambda\to K\Sigma$ and $K\Sigma\to K\Sigma$ transitions.
For a summary of exchange diagrams see Fig.~\ref{fig:alldiagrams} and Table~\ref{tab:trans}, Ref.~\cite{Ronchen:2012eg}, or the supplementary material of Ref.~\cite{Wang:2022osj} for the current status.

In the ANL-Osaka model~\cite{Julia-Diaz:2007qtz,Kamano:2013iva}, the meson-baryon interaction is constructed as described in \cref{sec:ANLfull}. The hadron exchange mechanisms of this model are also summarized Table~\ref{tab:trans}.
$K\Lambda$ and $K\Sigma$ channels are included in ~\cite{Kamano:2013iva}.
Here, the interaction in channels involving unstable particles $\pi\Delta$, $\sigma N$ and $\rho N$ includes $Z$-diagrams discussed in section~\ref{sec:ANLfull}. The $\omega N$ channel was added to the ANL-Osaka model in Ref.~\cite{Paris:2008ig}.  

DCC approaches provide channel transitions through $s$-, $t$-, and $u$-channel exchanges as well as contact terms (see Fig.~\ref{fig:alldiagrams}). The interaction $V$ in Eq.~(\ref{scattering}) can be separated into different pieces according to
\begin{align}
V_{\mu\nu}=V^\npo_{\mu\nu}+\underbrace{\sum_{i=1}^{n} \frac{\gamma^a_{\mu;i}\,\gamma^c_{\nu;i}}{E-M_i^b}}_{V_{\mu\nu}^\po}+\underbrace{\frac{\gamma_\mu^{\text{CT};a}\gamma_\nu^{\text{CT};c}}{M_N}}_{V_{\mu\nu}^\text{CT}} \ .
\label{blubb}
\end{align}
The first term, $V^\npo$, contains non-separable $t$- and $u$-channel exchanges (see Fig~\ref{fig:alldiagrams}). The second term, $V^\po$, contains separable $s$-channel resonance contributions,
with $n$ being the number of bare $s$-channel states in a given partial wave. To simplify the notation, the explicit dependence on momenta and
energy are omitted here and in the following. In Eq.~(\ref{blubb}), $\gamma^c$ ($\gamma^a$) are the bare resonance creation (annihilation)
vertices, indicated by the subscript $c$ ($a$), of the $s$-channel states of bare mass $M_i^b$. 

The third term of Eq.~\eqref{blubb}, $V^\text{CT}$, contains phenomenological contact terms that are added to the interaction in a given partial wave~\cite{Ronchen:2015vfa}. These interactions are separable both in channel and momentum space; the correct threshold behavior is built into the $\gamma^\text{CT}$ for each channel $\mu$. These contact terms absorb contributions beyond the ones explicitly included in the model.
As discussed in \cref{sec:Juba}, there are also contact terms arising at the Lagrangian level or from the TOPT formalism requiring on-shell equivalence with Feynman formalism. These terms are associated with $V^\npo$. 

\subsubsection{The \texorpdfstring{$u$}{u}-channel transitions}
\label{sec:uandpheno}
\begin{table}[tb]
\caption{ The $t$- and $u$-channel transitions in the JBW and ANL Osaka models. Both models also include $s$-channel poles and contact interactions that are not listed here.
Black entries show common transitions, blue (green) transitions indicate processes only contained in the JBW model but not the ANL-Osaka model (ANL-Osaka model but not JBW model). The particle type implies which interaction channel ($t$ or $u$) is meant. Abbreviations used: $\Delta$ for $\Delta(1232)3/2^+$, $\sigma$ for $f_0(500)$, $\kappa$ for $K_0^*(700)$, $\rho$ for $\rho(770)$, $a_0$ for $a_0(980)$, $f_0$ for $f_0(980)$, $\Sigma^*$ for $\Sigma(1385)3/2^+$, $\Xi^*$ for $\Xi(1530)3/2^+$, and $C$ for contact term. The $(\pi\pi)_\sigma$ and $(\pi\pi)_\rho$ entries correspond to correlated two-pion exchange that is treated differently in both approaches. 
}
\begin{center}
\begin{tabularx}{\linewidth}{l|XXXXXXXX} 
\hline \hline
$\mu$ & $\pi N$ & $\eta N$ & $K\Lambda$ & $K\Sigma$ & ${\color{blue}\omega N}$ & $\pi \Delta$ & $\sigma N$ & $\rho N$\TT\\
\hline
\rowcolor{lightgray}
    $\pi N$ & \makecell{$N$, $\Delta$,\\
    {\color{blue}$(\pi\pi)_\sigma$}, {\color{teal} $\sigma, f_0$}\\ 
    {\color{blue}$(\pi\pi)_\rho$}, {\color{teal} $\rho$}}
    & {\color{blue} $a_0$}, $N$                        %
    & \makecell{$K^*$, $\Sigma$,\\ $\Sigma^*$
    ,\\ {\color{teal} $\kappa$} }\ \ \ %
    & \makecell{$K^*$,  $\Lambda$,\\ $\Sigma$, $\Sigma^*$
    ,\\ {\color{teal} $\kappa$} } \  %
    & {\color{blue} $\rho$, $N$}
    & $\rho$, $N$, $\Delta$ %
    & $\pi$, $N$ 
    & \makecell{$\pi$, $\omega$,\\ {\color{blue}$a_1$}, $N$
    ,\\ {\color{blue} $\Delta$}, $C$}
\\
$\eta N$ &&
    $N$, {\color{blue}$f_0$}
    & \makecell{$K^*$, $\Lambda$,\\ {\color{teal} $\kappa$}} %
    & \makecell{$K^*$, $\Sigma$
    ,\\ {\color{blue}$\Sigma^*$}, {\color{teal}$\kappa$}} %
    & {\color{blue} $\omega$, $N$} %
    &   %
    & {\color{teal}$N$}  %
    & {\color{teal}$N$}  %
\\ \rowcolor{lightgray}
$K\Lambda$ &&&
     \makecell{$\omega$, {\color{blue}$f_0$}, $\phi$,\\ $\Xi$, {\color{blue}$\Xi^*$} } %
    & \makecell{$\rho$, {\color{blue}$a_0$},\\ $\Xi$, {\color{blue}$\Xi^*$}}\ \ \ \ %
    & \makecell{ {\color{blue}$K$, $K^*$},\\ {\color{blue}$\Lambda$} } %
&&& \\
$K\Sigma$ &&&&  
     \makecell{$\rho$, $\omega$, $\phi$,\\ {\color{blue} $f_0$, $a_0$},\\ $\Xi$, {\color{blue}$\Xi^*$}} 
     & \makecell{ {\color{blue}$K$, $K^*$},\\{\color{blue} $\Sigma$, $\Sigma^*$} } %
\\ \rowcolor{lightgray}
{\color{blue}$\omega N$} &&&&&
    {\color{blue}$\sigma$, $N$} &&&
\\
$\pi\Delta$ &&&&&&
     $\rho$, {\color{blue} $N$}, $\Delta$ %
     & {\color{blue}$\pi$} %
     & $\pi$, $N$
\\ \rowcolor{lightgray}
$\sigma N$ &&&&&&&
    {\color{blue}$\sigma$}, $N$ &{\color{teal} $N$} %
\\
$\rho N$ &&&&&&&& \makecell{{\color{blue}$\rho$}, $N$,\\ $\Delta$, $C$}
\BBB
\\
\hline \hline
\end{tabularx}
\end{center}
\label{tab:trans}
\end{table}
 The  $t$- and $u$-channel exchanges included by the ANL-Osaka and JBW approaches are shown and compared in Table~\ref{tab:trans}.

The $u$-channel exchanges (or Z-diagrams) involve the exchange of a baryon.  For exchanged octet states, the JBW model employs TOPT nucleon propagators. For spin-$3/2$ propagators in $u$-channel transitions the choice 
\begin{align}
    P^{\mu\nu}(k)=\frac{\slashed{k}+M_\Delta}{k^2-M_\Delta^2+i\epsilon}\left[-g^{\mu\nu}+\frac{\gamma^\mu\gamma^\nu}{3}+\frac{2k^\mu k^\nu}{3M_\Delta^2}-\frac{k^\mu\gamma^\nu-k^\nu\gamma^\mu}{3M_\Delta}\right]
\end{align}
is made that suppresses off-mass-shell terms~\cite{Pearce:1990uj}.
The $u$-channel exchange
plays a special role in the context of three-body channels that are included both in the ANL-Osaka and JBW models. 
As discussed in \cref{sec:tauiso} three-body unitarity in the JBW approach requires $u$-channel exchange contributions in the potential $V$ leading to moving singularities that depend both on incoming and outgoing momenta and energy. Correspondingly, the solution of the integral equation~\eqref{eq:lse} requires a tailored treatment discussed in \cref{sec:analytic}. Figure~\ref{fig:alldiagrams} illustrates the nucleon $u$-channel exchange in the $\pi N\to \pi\Delta$ transition as one of several processes required by unitarity.
In the ANL-Osaka model, the $u$-channel exchange processes which develop $\pi\pi N$ cut are only the $Z$ diagrams shown in Fig.~\ref{fig:z}. The other $u$-channel exchange mechanisms are included in the potential $v_{M'B',MB}$ of Eq.~\eqref{eq:H-2} with no $\pi\pi N$ cut. The pion production processes associated with the $\pi NN$ vertex are included in $v_{\pi\pi N, MB}$ of Eq.~\eqref{eq:H-2}.

\subsubsection{The \texorpdfstring{$t$}{t}-channel transitions and correlated two-pion exchange}
In the JBW model, the $t$-channel exchanges are, on one hand, given by the exchange of scalar and vector mesons. As explained in the supplementary material of Ref.~\cite{Wang:2022osj} the exchange of the vector meson and scalar meson nonet except for the $\kappa$-meson is included. They provide realistic meson-baryon interactions that contribute to all partial waves simultaneously. 

Second, there are $t$-channel pion exchanges for non-diagonal transitions such as $\pi N\to\rho N$ or $\pi N\to \sigma N$. Unlike other $t$-channel exchanges, these nondiagonal transitions go on-shell in the inelastic physical region $\sqrt{s}>2m_\pi+M_N$ as required by unitarity, similar to the above-discussed $u$-channel exchanges (see \cref{sec:unitarity}).

\begin{figure}[t]
    \begin{center}
    \includegraphics[width=0.38\textwidth]{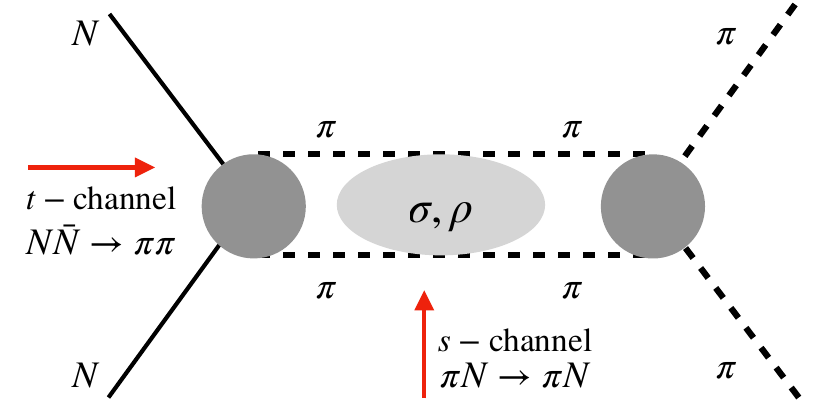}
    \end{center}
    \caption{Scalar-isoscalar and vector-isovector exchanges in the $t$-channel, determined by using dispersion techniques, crossing
    symmetry, and a fit to quasi-empirical data of the $N\bar N\to\pi\pi$ reaction. Figure adapted from Ref.~\cite{Schutz:1994ue}.
    }
    \label{fig:dispersion}
\end{figure} 
Third, in the JBW model the most prominent $t$-channel exchanges in $\pi N$ scattering, namely the scalar-isoscalar and the vector-isovector channels, are not mediated via explicit meson fields but by correlated $\pi\pi$ exchange as introduced in Ref.~\cite{Schutz:1994ue}.
The correlated two-pion exchange is related to the $N\bar N\to\pi\pi$ amplitude as schematically indicated in Fig.~\ref{fig:dispersion}.
The idea~\cite{Schutz:1994ue} is to relate a model that describes ``quasi-empirical'' data of the $N\bar N\to \pi\pi$ and $\pi\pi$ amplitudes to the $\pi N\to\pi N$ reaction through crossing symmetry. 
The $N\bar N\to \pi\pi$ were obtained by analytic continuation of $\pi N$ and elastic $\pi\pi$ data into the unphysical region below the $N\bar N$ threshold~\cite{Nielsen:1972ndh, Hohler:1974ht, Hohler:1984ux}. The amplitudes $f_+^0(t)$ and $f_\pm^1(t)$ describing them are related to the partial-wave projected invariant $A^{(+)}_\sigma$, $B^{(+)}_\sigma$ and $A^{(-)}_\rho$, $B^{(-)}_\rho$ amplitudes, respectively, where the upper index labels isoscalar $(+)$ and isovector $(-)$ amplitudes. For the Dirac structures associated with $A$ and $B$, see Eq.~\eqref{eq:Hohler-general-structure}.  The invariant amplitudes can be related to the $T$ matrix that can then be continued in different kinematic variables to the crossed reaction. See Ref.~\cite{Schutz:1994ue} for more details of the calculation, including a discussion of how to avoid double-counting of processes generated through iteration of the Lippmann-Schwinger equation. 

 It should be emphasized that the correlated two-pion exchange is implemented at the level of the interaction $V$ and not the $s$-channel unitarized amplitude $T$. More sophisticated dispersive treatments consider crossing symmetry at the amplitude level (Roy Steiner equations)~\cite{Hoferichter:2015hva, Hoferichter:2015tha, RuizdeElvira:2017stg} and have been recently continued to complex energies to, e.g., extract the pole of the Roper resonance~\cite{Hoferichter:2023mgy}.

\subsubsection{The \texorpdfstring{$s$}{s}-channel transitions, two-potential formalism, and renormalization}
\label{sec:schannel}

The $s$-channel processes play a special role as they are separable and allow for the separation into a ``pole part'' and a ``non-pole'' part through the two-potential formalism. This is illustrated for the ANL-Osaka model in \cref{sec:ANLfull}. For the JBW model, a very similar separation is discussed in Ref.~\cite{Ronchen:2012eg}. In DCC approaches, the backwards-going parts of $s$-channel processes and channel propagation are usually neglected, but their role has been  quantified for a mesonic systems some time ago~\cite{Zhang:2021hcl}.
In the ANL-Osaka model, the pair production processes are included in the potential as discussed on Eq.~\eqref{ANL:potential-1}, while the backwards-going propagation of the bare excited baryons are not included.

On the one hand, resonance poles in the T-matrix can arise from solving the nonlinear scattering equation with only $t$- and $u$-channel exchanges. Such states are usually referred to as ``dynamically generated''. In the JBW model, this occurs, e.g., for the Roper resonance~\cite{Krehl:1999km}. On the other hand, and 
in most of the cases, resonances are accounted for by explicit $s$-channel exchanges. These are simple singularities of $V$ in energy $E=\sqrt{s}$ that move into the complex energy plane once they are unitarized in the Lippmann-Schwinger equation~(\ref{eq:lse}). In the JBW model, $s$-channel baryonic states are included via Lagrangians up to spin $J=3/2$, see, e.g., Table B.8 of Ref.~\cite{Doring:2010ap}.  For higher spin, resonances are related to their lower-spin pendants by adding suitable barrier factors~\cite{Doring:2010ap}.  With this input, the partial-wave projected resonance creation and annihilation vertices $\gamma_\mu^a$ and $\gamma_\nu^c$ of Eq.~\eqref{blubb} are calculated containing coupling constants to be fit to data.

In the following we discuss the ``two-potential formalism'' in the JB model that allows one to separate off $s$-channel exchanges. See, e.g., Ref.~\cite{Oller:2025leg}, Sec. 6.1 for a recent derivation of the general formalism. In the ANL-Osaka model a similar decomposition is performed  as given in Eqs.~(\ref{eq:fullt}), (\ref{eq:mext}) and (\ref{eq:tr}).
For development and early application  of the two-potential formalism see Refs.~\cite{Ray:1980ck, Nakano:1982bc}. The two-potential formalism has been recently used in the context of connecting resonance lineshapes to resonance pole parameters and residues~\cite{Heuser:2024biq}.

The full scattering T-matrix of Eq.~(\ref{scattering}) can be written as the sum of ``pole'' and ``non-pole'' part~\cite{Ronchen:2012eg},
\begin{equation}
    T_{\mu\nu}=T_{\mu\nu}^\po+T_{\mu\nu}^\npo \, ,
    \label{deco1}
\end{equation} 
which is the same decomposition as the one into the ``resonant'' and ''non-resonant'' part in the ANL-Osaka formulation of Eq.~\eqref{eq:fullt}.
First, all $t$- and $u$-channel contributions, $V^\npo$ from Eq.~\eqref{blubb}, are unitarized in the non-pole part, 
\begin{align}
    T_{\mu\nu}^\npo= V_{\mu\nu}^{\npo}+ \sum_\kappa   V_{\mu\kappa}^{\npo}G_\kappa T_{\kappa\nu}^\npo \ ,
    \label{ttnpo}  
\end{align}
in operator notation implying integration over the second term.
This part can actually develop poles that are dynamically generated through the non-linearity in $V^\npo$. An example is given by the Roper resonance discussed in Sec.~\ref{sec:P11}.
Contributions from the $s$-channel exchanges, the pole part $T^\po$, can be evaluated from the non-pole part $T^\npo$ given in
Eq.~(\ref{ttnpo}).  For this, we define the following quantities using the bare vertices of Eq.~\eqref{blubb}:
\begin{eqnarray}
    \Gamma_{\mu;i}^c	=\gamma^c_{\mu;i}+\sum_\nu  \gamma^c_{\nu;i}\,G_\nu\,T_{\nu\mu}^\npo \ , \quad
    \Gamma_{\mu;i}^a	=\gamma^a_{\mu;i}+\sum_\nu  T_{\mu\nu}^\npo\,G_\nu\,\gamma^a_{\nu;i} \ , \quad
    \Sigma_{ij}		&=&\sum_\mu  \gamma^c_{\mu;i}\,G_\mu\,\Gamma^a_{j;\mu} \;,
    \label{dressed}
\end{eqnarray}
where $\Gamma^c$ ($\Gamma^a$) are the ``dressed resonance creation (annihilation) vertices'' and $\Sigma$ is the ``self-energy''. The indices $i,j$ label the $s$-channel state in the case of multiple resonances. For the two-resonance case with bare masses $m_1$ and $m_2$, the pole part reads explicitly~\cite{Doring:2009uc}
\begin{align}
    T^{\po}_{\mu\nu}&=\Gamma^a_{\mu}\, D^{-1} \, \Gamma^c_\nu \ ,    
\end{align}
where
\begin{align}
    \Gamma^a_\mu&=(\Gamma^a_{\mu;1},\Gamma^a_{\mu;2})\ , 
    \quad
    \Gamma^c_\mu=
    \begin{pmatrix}
    \Gamma^c_{\mu;1}\\
    \Gamma^c_{\mu;2}
    \end{pmatrix}
    \ , \quad
      D=\begin{pmatrix}
    E-m^b_1-\Sigma_{11}&&-\Sigma_{12}\\
    -\Sigma_{21}     &&E-m^b_2-\Sigma_{22}
    \end{pmatrix}
     \ ,
    \label{2res}
\end{align}
with inverse propagator $D$, from which the one-resonance case follows immediately. This decomposition is widely used in the literature,  see, e.g., Refs.~\cite{Matsuyama:2006rp,Afnan:1980hp, Lahiff:1999ur, Mikhasenko:2019vhk}. The non-pole part $T^\npo$ is sometimes referred to as {\it background}, although the unitarization of Eq.~(\ref{ttnpo}) may lead to dynamically generated poles in $T^\npo$. In turn, $T^\po$ contains constant and other terms beyond the pole contribution of the Laurent expansion~\cite{Doring:2009bi}. This demonstrates that a decomposition into resonant and non-resonant contribution according to the two-potential formalism is not identical to terms of the Laurent expansion, showing that this separation is not unique, and that it is difficult to attribute physical properties to them in a  model-independent way. 

However, there is a numerical advantage to this separation: All $s$-, $t$-, and $u$- channel exchanges have free fit parameters to be matched to data. Some parameters come through the form factors in $t$-, $u$- channel exchanges while others, the pole parameters, come through resonances coupling to the photon and hadronic channels. The two-potential formalism allows for hierarchical fit structures: One starts by evaluating  $T^\npo$ once, with a given set of $t$-, and $u$-channel ``slow'' parameters. Evaluating $V^\npo$ is the numerical bottleneck because it requires the partial-wave projection of all matrix elements in momentum and channel space. Once these transitions are evaluated, one optimizes all pole parameters plus the separable contact terms discussed in \cref{sec:uandpheno}. Indeed, due to their separable form, the latter can be organized in a structure very similar to the two-potential formalism for resonances~\cite{Ronchen:2015vfa} discussed before. Once these fast-evaluating parts of the amplitude are optimized, the next optimization step of slow parameters takes place, involving another evaluation of $V^\npo$.

There is one special $s$-channel exchange that accounts for the nucleon itself. Its spin-parity of $J^P=1/2^+$ corresponds to the $P_{11}$ partial wave that also hosts the Roper resonance $N(1440)$ and higher-lying states. Parameterizing the nucleon pole in terms of an $s$-channel singularity, $V\sim (f_N^b)^2/(E-M_N^b)$ makes it apparent that the bare $\pi NN$ coupling $f_N^b$ and bare mass $M_N^b$ undergo a renormalization such that the pole of the fully dressed T-matrix is at the physical nucleon mass, $M_N$, and its residue equals the residue induced by the physical $\pi NN$ coupling, $f_{\pi NN}$. The renormalization depends on the included number of channels, the non-pole (i.e., $t$-, $u$-and contact) interactions, and the regularization. In principle, one could numerically ensure the physical nucleon mass and coupling by fitting $M_N^b$ and $f_N^b$ to $M_N$ and $f_{\pi NN}$. 
 
It is also possible (and realized in the JBW model~\cite{Ronchen:2012eg}) to perform the renormalization analytically by separating the nucleon $s$-channel contributions from $t$- and $u$-channel contributions through the discussed two-potential formalism~\cite{Koch:1985bp,Lahiff:1999ur}.
 Following Eq.~(\ref{2res}), the pole part $T^\po$ in the one-resonance, one-channel ($\pi N$) case, with an $s$-channel nucleon exchange of bare mass $M_N^b$, can be
written as 
\begin{eqnarray}
     T^\po=\frac{\Gamma^a\,\Gamma^c}{E-M_N^b-\Sigma(E)} \ .
    \label{TP1chan}
\end{eqnarray}
We require that $T^\po$ has a pole at the physical nucleon mass $M_N$ or, equivalently, that 
\begin{eqnarray}
    M_N= M^{b}_N+\Sigma(M_N) \ .
    \label{renorma1}
\end{eqnarray}
To derive an expression for $M^b_N$ that respects the renormalization condition~(\ref{renorma1}) we expand the nucleon self-energy about
$E=M_N$,
\begin{eqnarray}
     \Sigma(E)=\Sigma(M_N)+(E-M_N)\frac{\partial \Sigma(E)}{\partial E}\bigg|_{E=M_N}+{\cal O}(E-M_N)^2 \ ,
    \end{eqnarray}
and define the reduced quantities $\tilde{\Sigma}$ and $\tilde{\Gamma}^{a,c}$ using that the bare coupling $f_{N}^b=x\,f_{\pi NN}$ can be factorized off $\Sigma$ and $\Gamma$,
\begin{eqnarray}
     \Sigma&=& (f_N^b)^{2}\Sigma^{\text{red}}=x^{2}f_{\pi NN}^{2}\,\Sigma^{\text{red}}=x^2\,\tilde{\Sigma} \ , \non
    \Gamma^{a,c}&=&f_{N}^b \Gamma^{\text{red}; a,c}=x\,f_{\pi NN}\, \Gamma^{\text{red}; a,c}=x\,\tilde{\Gamma}^{a,c}\, .
    \label{redquan1}
\end{eqnarray}
In Eq.~(\ref{redquan1}), $\tilde{\Sigma}$ is the nucleon self-energy calculated with the physical nucleon coupling $f_{\pi NN}$, instead of the bare coupling. The same applies to $\tilde{\Gamma}^{a,c}$, and $T^\po$ from Eq.~(\ref{TP1chan}) reads now
\begin{eqnarray}
     T^\po=\frac{1}{E-M_N}
     \left(\frac{x^2\tilde{\Gamma}^a\tilde{\Gamma}^c}{1-x^{2}\partial_{E}\tilde{\Sigma}}\right)_{E=M_N}+{\cal O}(E-M_N)^0\, .
    \label{tpopo}
\end{eqnarray}
To determine the bare coupling $f_N^b$, $x$ is calculated: the physical residue of the
nucleon pole in energy $E$ is given by
\begin{eqnarray}
    (a_{-1})_{\pi N\to \pi N}=\tilde\gamma^a\,\tilde\gamma^c\ ,
    \label{resin}
\end{eqnarray}
where $\tilde\gamma^{a,c}$ are the bare nucleon vertices calculated at $E=M_N$ with the physical nucleon coupling $f_{\pi NN}$ instead of the bare coupling (cf. Appendix B.1. of Ref.~\cite{Doring:2010ap}). The residue in Eq.~(\ref{resin}) has to agree with the residue of $T^\po$ from Eq.~(\ref{tpopo}) at the pole position:
\begin{eqnarray}
    \left(\frac{x^2\tilde{\Gamma}^a\,\tilde{\Gamma}^c}{1-x^{2}\partial_{E}\tilde{\Sigma}}\right)_{E=M_N}= 
    (a_{-1})_{\pi N\to \pi N}\, ,
\end{eqnarray}
leading to
\begin{eqnarray}
    \frac{1}{ x^{2}}=
    \left(\partial_{E}\tilde{\Sigma}+\frac{\tilde{\Gamma}^a\,\tilde{\Gamma}^c}{\tilde\gamma^a\,\tilde\gamma^c}
    \right)_{E=M_N} \ .
    \label{x1}
\end{eqnarray}
The bare mass $M_N^b$ is then
\begin{eqnarray}
    M^b_{N}=M_N-x^2\tilde{\Sigma}(M_N) \ . 
    \label{mb1}
\end{eqnarray}
This matching procedure is valid for the nucleon in case it couples only to the $\pi N$ channel. In Ref.~\cite{Krehl:1999km}, the
renormalization for one bare $s$-channel state coupling to more than one channel is developed. In Ref.~\cite{Ronchen:2012eg} the renormalization procedure in presence of an additional $s$-channel state, that may couple to all channels, is derived. This version is implemented in all versions of the JBW model thereafter. While the nucleon pole gets renormalized no attempt is made for a similar, self-consistent treatment of the nucleon appearing in exchange potentials $V_{\mu\nu}$ or the propagation functions $G_\kappa$ of Eq.~(\ref{scattering}). Instead, the physical nucleon mass is used in these instances.

 In the ANL-Osaka approach, the single nucleon state does not couple with MB states by construction and the nucleon state only appears in the $u$- and $s$-channel potential and Z-diagrams, where the \emph{physical} nucleon mass is used. The introduction of a bare nucleon state within the model is examined in Ref.~\cite{Kamano:2010ud} for the analysis of the $P_{11}$ partial-wave amplitude.

\subsection{Three-body unitarity and its implementation for the \texorpdfstring{$\sigma N,\,\rho N$}{sigmaN, rhoN} and \texorpdfstring{$\pi\Delta$}{piDelta} channels}
\label{sec:Three-body unitarity}

\subsubsection{The S-matrix and its unitarity}
\label{sec:unitarity}
We illustrate how unitarity relates imaginary parts of the isobar-spectator propagation to imaginary parts of their interaction via $u$-channel processes. For the ANL-Osaka and the JBW models, the isobars refer to the resonant $\pi N$ channel $P_{33}$ for the $\Delta$ and to the $\sigma$ and $\rho$ quantum numbers for the two-pion amplitudes as discussed in Sect.~\ref{sec:tauiso}. For three-meson systems (Sect.~\ref{sec:threemesons}), there are resonant channels but also non-resonant isobars~\cite{Feng:2024wyg}, for example the repulsive $3\pi^+$ system that does not host any resonances~\cite{Mai:2018djl}.

For simplicity, the argument in this subsection is outlined in terms of spinless particles and isobars following Ref.~\cite{Mai:2017vot}. 
A relativistic, infinite-volume three-body scattering amplitude  was formulated in Ref.~\cite{Mai:2017vot} following much earlier work by Aaron, Amado and Young~\cite{Aaron:1968aoz}. It can be expressed in terms of on-shell, two-body unitary $2\to 2$ amplitudes (the isobars), complex exchange processes, and genuine three-body interactions containing the microscopic dynamics, which are forced to be real-valued by three-body unitarity. 
In the following we collect only the main results of the derivation, relevant for this review, and refer the reader for details to the original work~\cite{Mai:2017vot}.

The central quantity of most theoretical approaches to the properties of excited hadrons is the S-matrix, introduced in the first half of $20^{\rm th}$ century, see, e.g., Refs.~\cite{Wheeler:1937zz,Heisenberg:1943zz}. It relates asymptotically free states in the far past to those in the distant future. Consider $n$ incoming and $m$ outgoing spinless states with three-momenta $\{\bm{p}\}=\{\bm{p}_1,...,\bm{p}_n\}$ and $\{\bm{p}'\}=\{\bm{p}'_1,...,\bm{p}'_m\}$, respectively. The operators for the S- and T-matrix are 
\begin{align}
    \langle \{\bm{p}'\}|\hat S|\{\bm{p}\}\rangle
    =
    \langle\{\bm{p}'\}|(\mathds{1}+\mathrm{i}\hat T)|\{\bm{p}\}\rangle\,.
\end{align}
The crucial point is that this allows one to formalize otherwise abstract principles of scattering theory in mathematical properties of the S-matrix. In particular, the S-matrix obeys \emph{crossing symmetry}, \emph{analyticity} and \emph{unitarity}, see for more details Ref.~\cite{Mai:2022eur} and references therein. In view of DCC approaches for hadron spectroscopy, analyticity and unitarity are of paramount importance. To be more specific, the S-matrix elements ($\in \mathds{C}$) are analytic (holomorphic) functions of $(3m+3n-10)$ kinematic variables ($\in \mathds{C}$). At the boundary of physical kinematics ($\in \mathds{R}$), the S-matrix elements are, indeed, related to the physical/observable quantities, such as, e.g., phase shifts $\delta_\ell=\arg S_l/(2i)$ for the $n=m=2$ case, projected to the $\ell$-th partial wave. The unitarity of the S-matrix,
\begin{equation}
    \hat S\hat S^\dagger=\mathds{1}
    ~\Leftrightarrow~
    \hat T-\hat T^\dagger=\mathrm{i}\hat T^\dagger \hat T \ ,
\end{equation}
puts another constraint on the imaginary parts of matrix elements and invariant matrix elements $(\mathcal{M})$
\begin{equation}
    {\mathcal{M}_{\{p\}\to \{p'\}}
    =
    (2\pi)^4\delta^{(4)}\left(\sum_{i=1}^n p_i-\sum_{i=1}^m p_i'\right)\langle \{\bm{p}'\}|\hat T|\{\bm{p}\}\rangle}\,.
\end{equation}
In the case of one particle type (scalar with mass $m$) this unitarity constraint reads 
\begin{align}
\label{T-matrix Unitarity}
    \langle \{\bm{p}'\}|(\hat T-\hat T^\dagger)| \{\bm{p}\}\rangle&=
    i\int\prod_{i=1}^N\frac{\mathrm{d}^4k_i}{(2\pi)^{4}}\,(2\pi)\delta^+(k_i^2-m^2)
    \langle \{\bm{p}'\}| \hat T^\dagger|\{\bm{k}\}\rangle
    \langle\{\bm{k}\}|\hat T| \{\bm{p}\}\rangle \,,\\
    \mathcal{M}_{\{p\}\to\{p'\}}-\mathcal{M}^*_{\{p'\}\to\{p\}}&=
    i\int\prod_{i=1}^N\frac{\mathrm{d}^4k_i}{(2\pi)^{4}}\,(2\pi)\delta^+(k_i^2-m^2)
(2\pi)^4\delta^{(4)}\left(P-\sum_{i=1}^N\,k_i\right)\mathcal{M^*}_{\{p'\}\to\{k\}}\mathcal{M}_{\{p\}\to\{k\}}\,.\nonumber
\end{align}
Here, $\delta^+(k_i^2-m^2)$ selects the positive energy solution only and $P$ denotes the total four momentum of the system. The first observation is that putting $N$ intermediate particles on-shell requires the system to have at least $E\ge N\cdot m$ total energy. The second observation is that when some set of intermediate particles can go on-shell, a discontinuity of the T-matrix elements is manifested along the real energy-axis -- a branch cut -- often referred to as the unitarity cut or right-hand cut.

In the following the $3\to 3$ case of the above unitarity condition is considered for identical, spinless particles of mass $m$ and out- and in-going four-momenta $p'_1,p'_2,p'_3$ and $p_1,p_2,p_3$, respectively.
The T-matrix consists of a fully connected ($\hat T_c$) and a once-disconnected piece ($\hat T_d$), related to the isobar-spectator scattering amplitude $T$ and isobar-propagator $\tau$ as
\begin{align}
    \label{eq:t33full}
    \langle p'_1,p'_2,p'_3|&\hat T| p_1,p_2,p_3\rangle
    =
    \langle p'_1,p'_2,p'_3|\hat T_c |p_1,p_2,p_3\rangle 
    +\langle p'_1,p'_2,p'_3|\hat T_d| p_1,p_2,p_3\rangle\\
    &=
    \frac{1}{3!}\sum_{n=1}^3\sum_{m=1}^3\,
    v(p'_{\bar{n}},p'_{\barr{n}})
    \Bigg(
    \tau(\sigma(p'_n))\,
    T(p'_n,p_m;s)\,
    \tau(\sigma(p_m))\, 
    -2E_{p_n'}\tau(\sigma(p'_n))(2\pi)^3\delta^3(\boldsymbol{p'}_n-\boldsymbol{p'}_m)
    \Bigg)
    v(p_{\bar{m}},p_{\barr{m}}) \,,\nonumber
\end{align}
where $P$ is the total four-momentum of the system, $s=E^2=P^2$. All four-momenta $p_1,p'_1,...$ are on-mass-shell, and the square of the invariant mass of the isobar reads 
\begin{align}
   \sigma(p'):=(P-p')^2=s+m_S^2-2\sqrt{s}E_{p'} 
   \label{eq:sigma}
\end{align}
for  spectator of mass $m_S=m$ and energy $E_{p'}$. We work in the total center-of-mass frame where ${\bm P}=\bm{0}$. The dissociation vertex $v(p,p')$ of the isobar decaying in asymptotically stable particles, e.g., $\rho\,(p+p')\to \pi(p)\pi(p')$, is chosen to be cut-free in the relevant energy region, which is always possible. The notation is such that, e.g., for a spectator momentum $p_n$, the isobar decays into two particles with momenta $p_{\bar{n}}$ and $p_{\barr{n}}$. 

\begin{figure}[tb]
\centering
\includegraphics[width=0.65\textwidth, keepaspectratio]{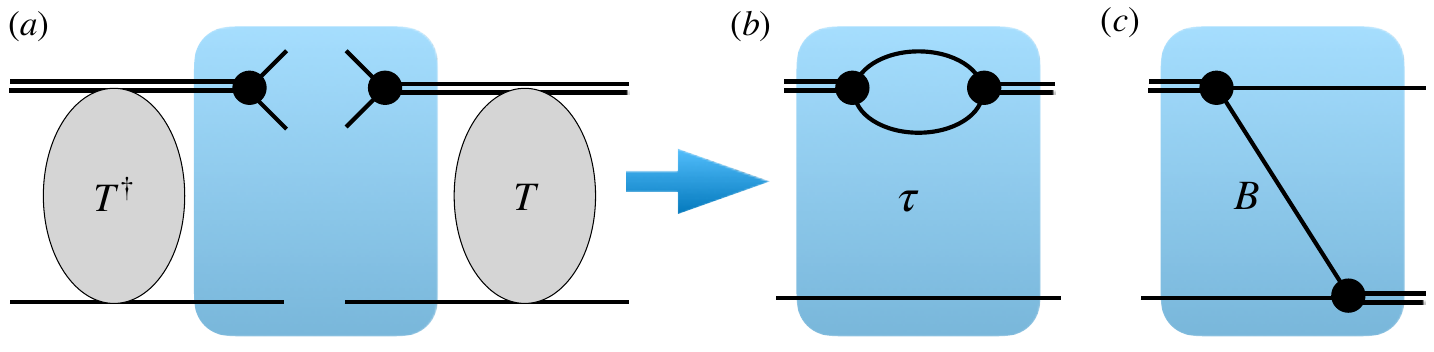}
\caption{(a) Unitarity relation for stable particles (single lines) with intermediate isobars (double lines). The on-shell, three-body, intermediate state is populated by decaying isobars (shaded highlights) in two distinct ways: (b) by combining the isobar decay products to form a self energy resulting in the propagator $\tau$, or (c) by combining a spectator and an isobar decay product to form a particle exchange $B$.
}
\label{fig:threee}
\end{figure}

Consider again the connected isobar-spectator amplitude $T$ from Eq.~\eqref{eq:t33full}. It appears twice on the right-hand side of the unitarity relation~\eqref{T-matrix Unitarity} as schematically depicted in Fig.~\ref{fig:threee} (a). The blue highlighted area shows the three stable intermediate particles that combine according to the integrations and $\delta$-functions in Eq.~\eqref{T-matrix Unitarity}. As inserts (b) and (c) show, there are two distinct ways for this to occur: (b) One can combine the isobar decay products to form a ``self energy'' or one can (c) combine a spectator with one of the isobar decay products. Next, one formulates a BSE ansatz for $T$ and realizes combinations (b) and (c), leading to 8 terms. In \cref{subsec:bseUCHPT} we quote the BSE in a different context.
Equating these terms to 8 diagrammatically equivalent terms on the left-hand side of \cref{T-matrix Unitarity} translates into 3 conditions: for the imaginary parts of $\tau^{-1}$, imaginary parts of $B$, while also putting the intermediate spectator on the energy-shell, reducing the BSE to a three-dimensional scattering equation that is manifestly unitary. The second condition leads through the use of an un-subtracted dispersion relation to 
\begin{align}
    \label{eq:B}
    B(p',p;s)
    =
    &\frac{-\lambda^2}
    {2E_{p'+p}\left(E-E_{p'}-E_p-E_{p'+p}+i\epsilon\right)}
    \,,
\end{align}
where $E=\sqrt{s}$, $E_{p'+p}^2=m^2+(\bm{p}+\bm{p}')^2$ etc., and $p$ and $p'$ denote the on-shell four-momenta of the in- and outgoing spectator, respectively. In addition, the dissociation vertex is chosen to be of simple scalar form, $v(p,p'):=\lambda$. 

The first condition, on Im~$\tau^{-1}$, leads to the following form of the isobar propagator~\cite{Mai:2017vot} using a twice subtracted dispersion relation, 
\begin{align}
    \label{eq:tau}
    \frac{1}{\tau(\sigma(q))}=
    \sigma(q)-M_0^2-\int \frac{\dv{^3k}}{(2\pi)^3} 
    \frac{\lambda^2}{2E_k(\sigma(q)-4E_k^2+i\epsilon)} \,,
\end{align}
where $M_0$ is a free parameter that can be used to fit (together with $\lambda$ and cutoff) the two-body amplitude corresponding to the considered isobar, $T_{22}:=v\tau v$. We refer to the exchange  process in Eq.~\eqref{eq:B} as $B$-term in the following, and to the integral term in Eq.~(\ref{eq:tau}) as self-energy. This formulation of two-body scattering for the isobars is used in both the ANL-Osaka and JBW approaches, see Eq.~\eqref{unstableprop} and the next subsection. It should be emphasized that there is no need for $\tau$ to be of this specific form. For $S$-wave isobars, an amplitude $\tau'$ with Im~$(\tau')^{-1}=\text{Im }v^2/\tau$ can also be used, removing $v$ factors from the $B$-term suitably. For $p$-wave isobars, a slightly different matching has to be performed because there is a genuine angular dependence in the decay vertices $v$~\cite{Sadasivan:2020syi,Feng:2024wyg,Yan:2024gwp}.

Finally, the last condition which puts the spectator on its energy-shell reduces the four-dimensional BSE to a three-dimensional scattering equation,
\begin{align}
    \label{eq:T}
    T(p',p;s)= 
    B(p',p;s)+C(p',p;s)-
    \int
    \frac{\dv{^3 q}}{(2\pi)^3}
    \left(B(p',q;s)+C(p',q;s)\right)\,
    \frac{\tau(\sigma(q))}{2E_q}\,
    T(q,p;s) \ .
\end{align}
Here we have added an additional real-valued term $C$ which does not impact the three-body unitarity constraints, parameterizing the so-called three-body force. Recently, the determination of this quantity through use of  lattice QCD and Effective Field Theory became quite an active field of research~\cite{Dawid:2025zxc, Noh:2024nvp,Draper:2023boj, Baeza-Ballesteros:2023ljl,Garofalo:2022pux,Brett:2021wyd,Sadasivan:2021emk,Alexandru:2020xqf,Capel:2020obz,Kamada:2019irm,Mai:2018djl}. 
Note however that $C$ is not an observable quantity, meaning that it will always depend on the details of the integral equation and its regularization.
Note also that
the distribution of signs in Eqs.~\eqref{eq:B} and \eqref{eq:T} is owed to the convention $\hat S=\mathds{1}+i\hat T$ in Ref.~\cite{Mai:2017vot}. In the JBW model for baryons, the ``Infinite-volume unitary'' (IVU) model for mesons discussed in \cref{sec:threemesons}), as well as the ANL-Osaka model, the convention $\hat S=\mathds{1}-i\hat T$ is adopted.

In the ANL-Osaka approach, expressions of the full amplitudes satisfying unitarity are given in ~\cite{Matsuyama:2006rp}. In the actual numerical analysis, the scattering amplitudes  given in section~\ref{sec:ANLfull} is used with the perturbative approximation of $v_{\pi\pi N,MB}$, i.e., the direct two-pion production mechanism. Apart from this, to satisfy  three-body unitarity, the interaction $Z$ in Eq.~\eqref{eq:vvz} illustrated in (c) of Fig.~\ref{fig:threee} and the self-energy $\Sigma(q,E)$ in Eq.~\eqref{unstableprop} illustrated in (b) of Fig.~\ref{fig:threee} play a crucial role. The unitarity of the model is tested numerically.

Finally, we also note that even if explicit isobar (or dimer) fields are used to parametrize the $2$-body subsystem, the isobar formulation is not an approximation but a re-parametrization of the full two-body amplitude as shown in Ref.~\cite{Bedaque:1999vb} and also discussed in Ref.~\cite{Hammer:2017kms}. 
There are some mandatory requirements on the two-body amplitude, though: it should be formulated in terms of the Lorentz scalar $\sigma$ because its center of mass is not at rest in the three-body system; another requirement is obviously two-body unitarity; in addition, absence of poles or cuts on the first Riemann sheet as required by causality~\cite{Gribov2012-yx}, and, in close connection to this, a regular behavior in the sub-threshold region. The latter two requirements are also connected to the contour deformation method for the SMC that extends to these regions as will be discussed in \cref{sec:cont3B} in relation to Fig.~\ref{fig:SMCSEC}.

\subsubsection{Three-body channels in the ANL-Osaka and JBW models}
\label{sec:tauiso}
\begin{figure}[tb]
    \centering
\includegraphics[width=1\textwidth,angle=-0]{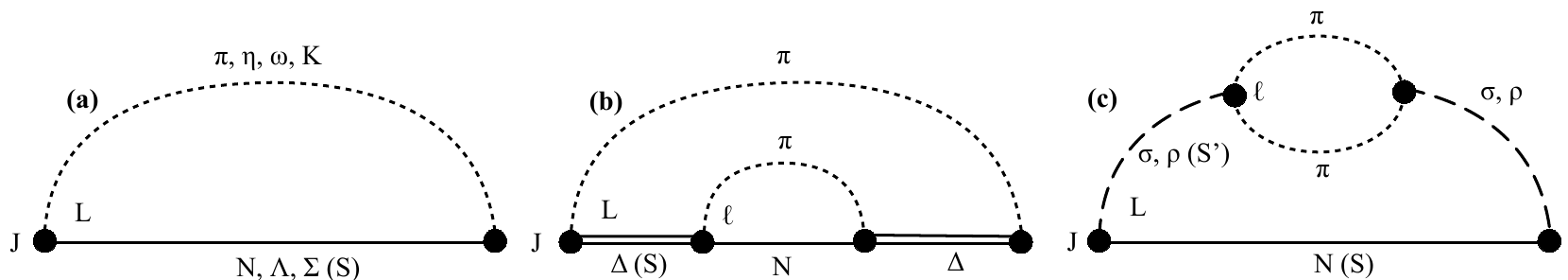}
    \caption{(a) Two-body channels, (b)$\pi\Delta$ channel, (c) $\pi\sigma$, $\pi\rho$ channels included in both the  ANL-Osaka and JBW models. See text for further details.}
    \label{fig:Sigma}
\end{figure}

Both the ANL-Osaka~\cite{Kamano:2013iva,Kamano:2019gtm} and JBW approaches~\cite{Ronchen:2012eg} incorporate three-body channels with the effective quantum numbers of $f_0(500)N$, $\rho(770) N$, and $\pi\Delta(1232)3/2^+$, in short, called $\sigma N$, $\rho N$, and $\pi\Delta$. As the $\rho$, $\Delta$, and nucleon have non-zero spin, some of these channels can couple to a given, overall spin-parity $J^P$ in more than one way as shown in Table~\ref{tab:couplscheme}. The isobars are shown in Fig.~\ref{fig:Sigma}.
The total angular momentum is indicated with $J$, the orbital angular momentum between isobar and spectator with $L$, the baryon spin with $S$, and the meson isobar spin (for the $\rho$ only) with $S'$. The orbital angular momentum between the isobar decay products, forming the self-energy $\Sigma$, is indicated with $\ell$.

On the one hand, unitarity requires the inclusion of the complex isobars shown in Fig.~\ref{fig:Sigma}. These structures simply express the physical process of realizing  three particles $\pi\pi N$. The three-body Green's function $G_\kappa$ for $\kappa\in\{\sigma N, \,\rho N,\,\pi\Delta\}$ is obtained by first fitting the isobar dissociation process $v$ to the respective two-body phase shifts. For the $\sigma$ and $\Delta$ quantum numbers in the JBW approach this was performed in Ref.~\cite{Schutz:1998jx}. The $\rho$ isobar was fit to $\pi\pi$ phase shifts in Ref.~\cite{Krehl:1999km}. The Green's functions $G_\kappa$ of the JBW are very similar to the ones of the ANL-Osaka approach. See Eq.~\eqref{unstableprop} and Refs.~\cite{Schutz:1998jx, Krehl:1999km} for explicit expressions, i.e., they are parametrized with an isobar propagator and a re-summed self energy as in Eq.~\eqref{eq:tau}.

On the other hand, as discussed in \cref{sec:unitarity}, the on-shell condition for $\pi\pi N$ must also be realized through complex-valued $3\to 3$ exchanges of isobar decay products. 
This reflects the fact that all three particles can go on-shell simultaneously in the exchange process, if the energy is large enough.
These transitions are shown in Fig.~\ref{fig:z}. In the context of the ANL-Osaka model, they are referred to as $Z$-diagrams; in the JBW model, these transitions are calculated using time-ordered perturbation theory, and in both approaches they provide the necessary imaginary parts. In both models, $s$-channel and  $t$-channel exchanges can be added to the amplitude to model the microscopic interactions as described in \cref{sec:chatra}.
There are also $2\to 3$ transitions required by unitarity in the JBW approach, as shown in the last two diagrams of Fig.~\ref{fig:alldiagrams}. In particular, there is an important pion exchange populating the $\sigma$ and $\rho$ isobars, and a nucleon exchange to populate the $\pi\Delta$ state. In \cref{sec:cont3B}  we discuss how the 3-body singularities from exchange processes require special care when continuing the amplitude to complex energies to search for resonance poles.

For the meson-exchange models discussed in Sect.~\ref{sec:threemesons}, three-body unitarity is guaranteed. Coupled-channel two-body unitarity is strictly fulfilled in the JBW model for light baryons, but only  approximately  for the three-body channels. On the one hand, the included nucleon $u$-channel diagram develops an imaginary part above $E=2m_\pi+M_N$, but the corresponding isobar-spectator channel (with quantum numbers $\pi P_{11}$) is neglected in the inelastic region. On the other hand, form factors are used that at, large momenta, deviate from unity to guarantee the convergence of the scattering equation. For unitarity, these form factors should appear consistently in the corresponding isobar self energies. Indeed, covariant form factors can be constructed for that purpose~\cite{Sadasivan:2020syi}.

\subsection{Solving the scattering equation}
\label{sec:realmom}
The integral equation~\eqref{scattering} can be solved along a complex spectator momentum contour (SMC) to avoid three-body singularities from particle exchange processes, as well as two-body singularities. This is a common solution strategy for the ANL-Osaka, JBW, and also the IVU approaches for excited baryons and mesons, respectively.  
For generic particles of mass $m_a$ and $m_b$,
the two-body singularities are given as the solution of $\sqrt{s}=E_a(p)+E_b(p)$,
\begin{align}
    p_\text{cm}=\frac{1}{2E}\sqrt{(E^2-(m_a-m_b)^2)(E^2-(m_a+m_b)^2)} \ .
    \label{pcm}
\end{align}
To solve the integral equation, the complex contour is discretized into a suitable set of mesh points $\{p_i\}$ and integration weights, $dp\to w_i$. With $p,\,p'$, and $q$ sampled at the same momenta $\{p_i\}$, the integral equation becomes a set of linear equations. The solution is then known along a set of complex momenta for fixed $\sqrt{s}$. This is sufficient if one is only interested in searching for resonance poles. For phase shifts and observables, however, one needs to know the solution for real on-shell momenta $p=p_\text{cm}(\sqrt{s})$ in case of stable two-body channels $\kappa$. For this, the set of mesh points is enlarged, $\{p_i\}\to\{p_i,\,p_\text{cm}\}$ which provides on-shell$\leftrightarrow$off-shell and the physical on-shell$\to$on-shell transitions~\cite{Hetherington:1965zza}. For $n_\kappa$ channels of two stable particles and $n_\text{SMC}$ mesh points, the dimension of the numerical T-matrix is then $n_\kappa\cdot (n_\text{SMC}+1)$.
In reduction to one channel, it is easy to see that with matrix elements $T^I_{lk}=T(p_l,p_k;E)$ with momenta from $\{p_i,\,p_\text{cm}\}$ and, analogously, $V_{lk}$, together with the discretized propagator matrix containing integration weights $w_i$,
\begin{align}
    G^I=\text{diag}\left(\frac{w_1\,p_1^2}{E-E_\kappa(p_1)-\omega_\kappa(p_1)},\dots,\frac{w_{n_\text{SMC}}\,p_{n_\text{SMC}}^2}{E-E_\kappa(p_{n_\text{SMC}})-\omega_\kappa(p_{n_\text{SMC}})},0\right) \ ,
    \label{discreteG}
\end{align}
 the T-matrix is obtained by matrix inversion of $T^I=V+VG^IT^I$, with the physical on-shell to on-shell scattering given by its $(n_\text{SMC}+1),(n_\text{SMC}+1)$ matrix element. The generalization to $n_\kappa$ coupled channels is obvious. The superscript $I$ serves to indicate the solution on the first Riemann sheet, in contrast to the solution on other sheets discussed in \cref{sec:contstable}. See also Ref.~\cite{Eichmann:2019dts} for the role of contour deformations of Lorentz-invariant scattering equations.

\begin{figure}[tb]
\centering
\includegraphics[align=c,width=0.32\textwidth, keepaspectratio]{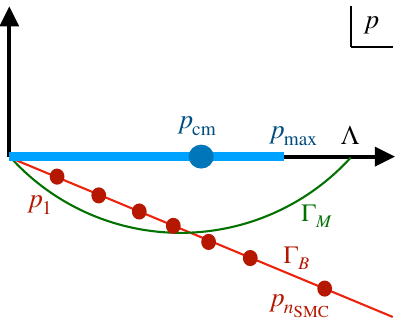}
\hspace*{0.4cm}
\includegraphics[align=c,width=0.45\textwidth, keepaspectratio]{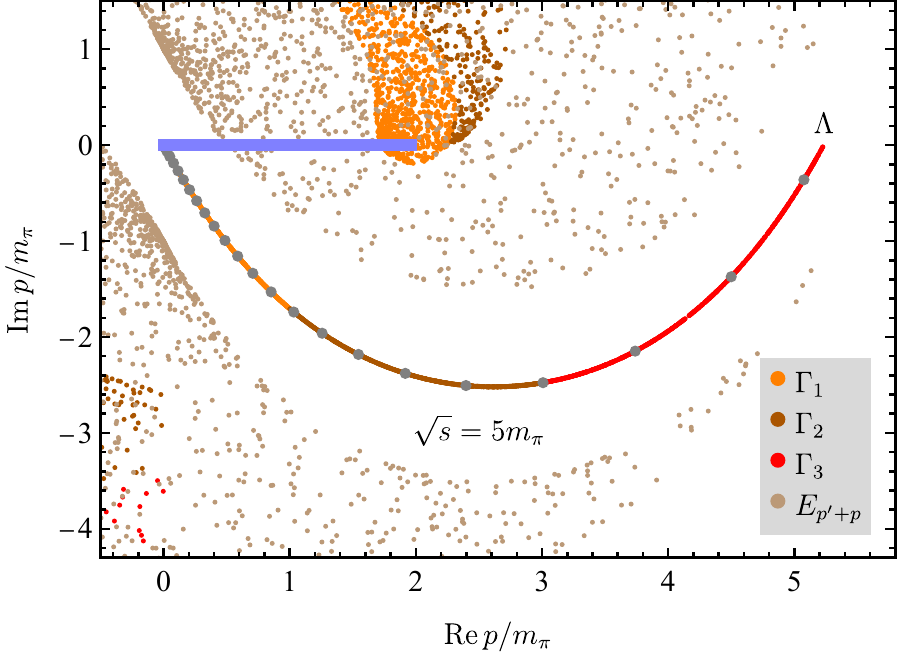}
\caption{Left: Complex spectator momentum plane showing an integration contour (SMC) $\Gamma_M$ for  regularization by cutoff $\Lambda$,  and another SMC $\Gamma_B$ with form factor regularization implying $\Lambda\to\infty$. Also shown are the integration mesh points $\{p_1,\dots,p_{n_\text{SMC}}\}$, the on-shell point for two-body channels of stable particles, $p_\text{cm}$, and the on-shell region for three-body channels, $p\in [0,p_\text{max}]$.
Right: Situation of three-body singularities (dots) color-coded for the corresponding parts of a typical SMC $\Gamma=\Gamma_1\cup\Gamma_2\cup\Gamma_3$ that appear in the same color. See text for further explanations.
}
\label{fig:contours}
\end{figure}

In Fig.~\ref{fig:contours} to the left, some commonly used contours are shown. The JBW model uses regularization through form factors, allowing for a straight integration path $\Gamma_B$ into the lower $p$ plane as the integral back to the real axis at infinity vanishes (red line). 
An alternative to render the integrand convergent (allowing for an integration path to infinity)
is by over-subtracting the dispersive propagation amplitude $\tau$ making it automatically convergent, see Eq.~(14) in Ref.~\cite{Mai:2017vot}. Similarly, unitarity only determines the imaginary part of the particle exchange $B$ from Eq.~\eqref{eq:B}, and over-subtracting it preserves unitarity while simultaneously guaranteeing the solution of the scattering equation~\cite{Mai:2017vot}.

One can also use a hard cutoff as shown with the green line $\Gamma_M$ and realized in Refs.~\cite{Sadasivan:2021emk, Feng:2024wyg} for excited meson systems (IVU model). The cutoff has to be large enough to not cut into the physically allowed state space, for all two- and three-body channels~\cite{Sadasivan:2021emk}.
In any case, the cutoff $\Lambda$ has to be chosen beyond all physically realizable momenta of the two-body and three-body channels~\cite{Sadasivan:2021emk, Feng:2024wyg}, symbolized with the blue dot and blue line in Fig.~\ref{fig:contours} to the left. The former stands for the two-body singularities at $p_\text{cm}$, while the latter indicates the range $[0,p_\text{max}]$ of physically allowed spectator momenta in three-body channels. 

The right-hand side of Fig.~\ref{fig:contours} shows the solution $p(p',\cos\theta)$ of $B^{-1}=0$ with the exchange term $B$ from Eq.~\eqref{eq:B} in the notation of momenta
\begin{align}
\sqrt{s}-E_{p'}-E_p-E_{p'+p}=0    \ .
\label{eq:ffunc}
\end{align}
 It is given by~\cite{Feng:2024wyg}
\begin{align}
p_{\pm}(m_1,m_3,m_u,x) 
=\frac{p'x(q'^2-\alpha^2+m_u^2-m_1^2)\pm\alpha\sqrt{\left(\beta-m_1\right)^2-m_u^2+p^{\prime 2}\left(x^2-1\right)}\sqrt{\left(\beta+m_1\right)^2-m_u^2+p^{\prime 2}\left(x^2-1\right)}}{2\beta^2}\ ,
\label{eq:singularity3B}
\end{align}
where
\begin{align}
\alpha=\sqrt{s}-\sqrt{m_3^2+p^{\prime 2}}\ ,\quad\beta^2=\alpha^2-p^{\prime 2}x^2 \
\end{align}
for incoming (outgoing) spectator of mass $m_1$ ($m_3$) and exchanged mass $m_u$. The color coding to the right in Fig.~\ref{fig:contours} indicates from which $p'$ of the SMC $\Gamma=\Gamma_1\cup\Gamma_2\cup\Gamma_3$ the solutions $p_\pm(p',\cos\theta)$ originate, while the brown points arise analogously from the pre-factor $E_{p+p'}=0$ in the $B$-term, see Eq.~\eqref{eq:B}. The chosen kinematics corresponds to pion spectators in incoming and outgoing states together with pion exchange. The figure demonstrates that the integration along $\Gamma_M$ is numerically stable because none of the solutions $p_{\pm}$ intersects with $\Gamma_M$, which would produce a pole in $B$. However, the figure also shows that one cannot simply add an on-shell point like in the previously discussed case of two stable propagating particles: Indeed, the region of real, on-shell spectator momenta, shown  with the blue horizontal line, intersects with the $p_\pm$. This is not a coincidence and cannot be circumvented with a simple modification of $\Gamma_M$. Instead, it represents a genuine problem requiring nontrivial solutions discussed in the following.

\subsubsection{Continuation to real momenta and triangle singularities}
\label{subsec:Cahill-Triangle}
To discuss the problem of continuation of the three-body amplitude to real spectator momenta, consider the case of a real-valued contour $\Gamma=\{p|p\in[0,\Lambda]\}$.
\begin{figure}[tb]
    \begin{center}
    \includegraphics[width=0.49\textwidth]{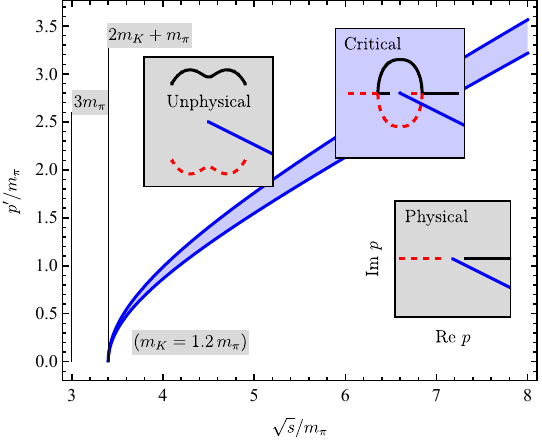}
    \includegraphics[width=0.49\textwidth]{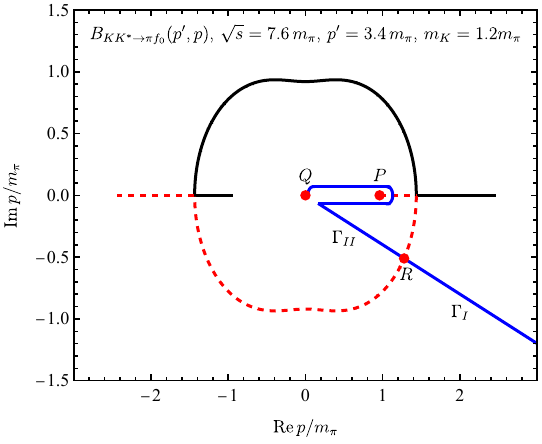}
    \end{center}
    \caption{Left: Singularity structures in the $s, \,p,\,p'$ variables. The insets show the position of three-body cuts in $p(p',x)$ for  three different $(\sqrt{s},p')$ regions (physical, critical, and unphysical), for $x\in [-1,1]$ for solutions $p_+$ (black lines) and $p_-$ (red dashed lines) according to Eq.~\eqref{eq:singularity3B}, for the kaon exchange in the triangle-singularity reaction $KK^*\to \pi f_0$. The insets also show a typical complex integration path with blue solid lines. The position of the insets corresponds to the regions in the $(\sqrt{s},p')$ plane. Right: Illustration of the integral path in the "critical region" for fixed outgoing $p'$, shown in the complex plane of incoming momentum $p$. The integration starts at the origin on the first Riemann sheet. Then, it goes around the logarithmic singularity at $P$, passing through the cut down onto the second Riemann sheet, and back to  $Q$, that indicates the origin on the second Riemann sheet. From there, the integration follows the original SMC $\Gamma$ with a smooth analytic transition of $B$ at $R$ from second sheet, $B^{II}$, back to the first sheet, $B^I$.}
    \label{fig:illustration}
\end{figure}
The solutions $p_\pm$ for this case are shown to the left in Fig.~\ref{fig:illustration}, for the unequal-mass case. For small values of $p'$ (inset ``Physical''), the singularities lie on the real axis, but a complex integration contour starting at $p=0$ (blue solid lines in insets) can easily avoid them. For intermediate values of $p'$ (inset ``Critical'' of Fig.~\ref{fig:illustration}), called ``critical region'' in Ref.~\cite{Feng:2024wyg}, there is no way for an integration contour to ``escape'' the closed cut. For $p'$ outside of the physical region, $p'>p_\text{max}$, the cut opens again and the integration contour can avoid the three-body singularities (inset ``Unphysical''). 

The problem of closed cut structure in the critical region can be solved by analytically continuing the $B$-term to the second sheet and an additional path deformation as shown in Fig.~\ref{fig:illustration} to the right, for the process $KK^*\to \pi f_0$ as indicated. This method proposed by Cahill and Sloan~\cite{Cahill:1971ddy, schmid1974quantum} was also applied recently in Ref.~\cite{Pang:2023jri}, see also Ref.~\cite{Zhang:2024fxy}. 
The caption of Fig.~\ref{fig:illustration} describes the path on first ($\Gamma_I$) and second ($\Gamma_{II}$) Riemann sheets with smooth transitions between them at $P$ and $R$. In particular, the logarithmic singularity at $P$ is exactly canceled in the chosen path which renders this method stable. Note that to perform this contour deformation, one first needs to know the solution on the interval $0<p<P$, which is usually possible because it corresponds to values of $p'$ below the critical region. One can then simply use a complex path as shown in the lower right inset in Fig.~\ref{fig:illustration} to the left. See Ref.~\cite{Feng:2024wyg} for more detailed explanations.

However, for certain mass combinations, the lower end of the critical region, $p'_{\rm min}$, is to the left of $P$, located at 
\begin{align}
    p_P=p_-(m_1,m_3,m_u,-1)
    \label{eq:trianglepoint}
\end{align}
 with $p_-$ from Eq.~\eqref{eq:singularity3B}, i.e., $p'_{\rm min}<p_P$. In Ref.~\cite{Zhang:2024fxy} and in Ref.~\cite{Fix:2019txp}, a path for this case is discussed. It passes right through the point where both solutions $p_\pm$ touch each other as shown to the right of point $P$ in Fig.~\ref{fig:illustration}. 

Note also the special kinematics when $P$ coincides with the quasi-on-shell momentum $p_\text{cm}$ from Eq.~\eqref{pcm} of an adjacent channel (e.g., $KK^*(892)$) that contains a narrow resonance. At the corresponding $\sqrt{s}$, the exchanged particle, the spectator and the very narrow resonance are all on-shell, leading to a ``triangle singularity''. Unitarized triangle singularities have been investigated in Ref.~\cite{Sakthivasan:2024uwd}, see also references therein for further information on this kinematic phenomenon. Sakthivasan et al.~\cite{Sakthivasan:2024uwd} also compare the Cahill and Sloan method with a direct extrapolation of the amplitude at complex momenta to the real ones through continued fractions. We summarize the findings in \cref{sec:triangles}.

\subsubsection{Production reactions and ``direct inversion''}
\label{sec:direct}
\begin{figure}[tb]
\begin{center}
\includegraphics[width=0.8\textwidth]{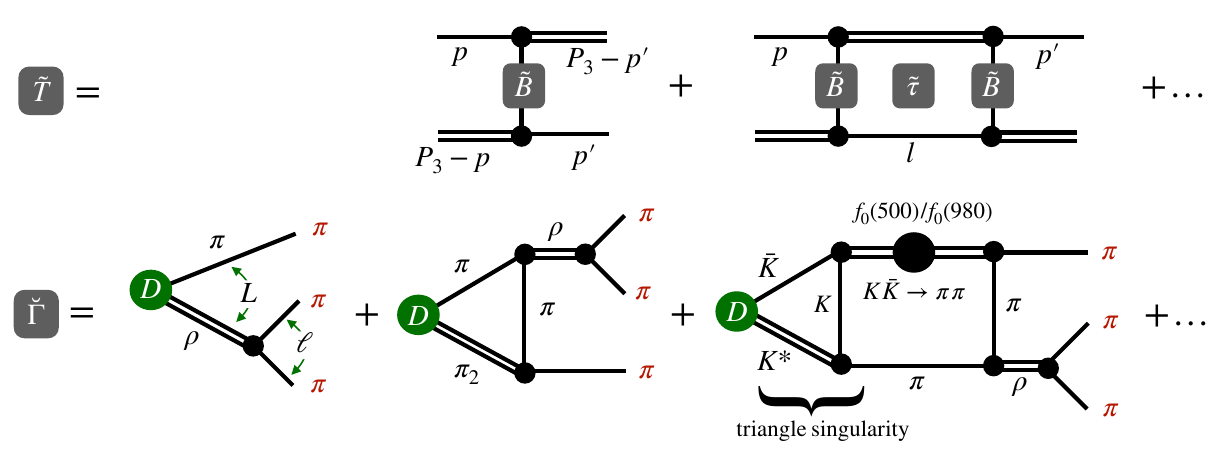}
\end{center}
\caption{Upper row: The rescattering series $\tilde T$ of isobars (double lines) and spectators (single lines) interacting through particle exchanges $\tilde B$, followed by propagation $\tilde \tau$. The amplitude $\tilde T(s,p',p)$ depends on the total Mandelstam $s=P_3^2$ and the incoming (outgoing) spectator momentum $p$ ($p'$), while the isobar propagation depends on $\sigma=(P_3-p)^2$, i.e., $\tilde\tau(\sigma(l))$ for the internal propagation shown.
This series forms the final-state interaction in the production reaction $\breve \Gamma(s,p')$ in the lower row. For $\breve \Gamma(s,p')$,  some coupled channels are indicated that contribute to a three-pion final state, as well as the orbital angular momentum $L$ and isobar spin $\ell$. Triangle singularities are naturally included in this rescattering series as indicated in an example. 
}
\label{fig:ccillu}
\end{figure}

The method by Cahill and Sloan discussed in the previous section is stable and computationally inexpensive. However, in the form presented here it only allows for continuation to real momenta of one of the two momenta of $T$ for $3\to 3$ scattering (either incoming or outgoing). It can still be used in the phenomenologically  relevant $2\to 3$ kinematics of the reaction $\pi N$ to $\pi\pi N$, though. It can also be used in production reactions in which $T$ provides the final-state interaction. The production reaction is referred to as $\breve \Gamma$ as shown in Fig.~\ref{fig:ccillu}.
It is obtained by multiplying the ``amputated'' production reaction $\tilde \Gamma^T$ with the propagator of the final isobar $\tilde \tau$ and its decay $\breve v_{j}$~\cite{Feng:2024wyg}, with sums over identical channel indices implied,
\begin{align}
    \breve\Gamma_{j}(s,p')=\breve v_{j}(\sigma(p'))\tilde\tau_{jk}(\sigma(p'))\left[\tilde\Gamma_{k}(s,p')+D_{k}(s,p')\right]=: \breve v_{j}(\sigma(p'))\tilde\tau_{jk}(\sigma(p'))\,\tilde \Gamma^T_k(s,p') 
    \label{eq:gammabrev}
\end{align}
with a new quantity $\tilde\Gamma^T$ consisting of a ``disconnected'' part $\breve v\tilde \tau D$ and the rescattering part $\tilde \Gamma$. The latter fulfills
\begin{equation}
\tilde\Gamma^T_j(s,p')=D_j(s,p')+\int_0^\Lambda \frac{\dv{q}\,q^2}{(2\pi)^3 2E_{q}}\,
\tilde B_{ji}(s,p',q)\,
\tilde\tau_{ik}(\sigma(q))\,
\tilde\Gamma^T_k(s,q) 
\label{dirprod}
\end{equation}
for channels $i,\,j,\,k$, outgoing spectator momenta $p'$, and exchange process $\tilde B$ similar to the expression given in Eq.~\eqref{eq:B}. For simplicity, contact terms are chosen $C=0$ in Ref.~\cite{Feng:2024wyg}. For the ``tilde'' notation and further details see \cref{sec:nine}. Note also the appearance of a second index in $\tau$ owing to the possibility of sub-channel transitions in the isobar itself, e.g., $\pi\pi\leftrightarrow K\bar K$ as illustrated in Fig.~\ref{fig:ccillu} to the lower right. 

One disadvantage of the contour deformation method by Cahill and Sloan is the dependence of the contour on the spectator masses (e.g., the position of point $P$) which can become complicated in the multichannel case with unequal masses (although this problem was solved in Ref.~\cite{Sakthivasan:2024uwd}). 

An alternative method is discussed in Ref.~\cite{schmid1974quantum}, based on the original reference~\cite{Sohre:1971hy}. We refer to it as direct inversion (DI) as compared to the contour deformation method (CD) discussed before. Its advantage is the independence from channel masses. The main idea is that the solution $\tilde \Gamma^T$ is suitably discretized in $q$ by a finite set of real $q_i\in[0,\,\Lambda]$, which can be well approximated by an interpolating polynomial connecting the discrete $\tilde \Gamma^T(q_i)$ (channel index and index $s$ omitted),
\begin{align}
  \tilde \Gamma^T(q)\approx
  \sum_{i=1}^N \tilde \Gamma^T(q_i)\,H_i(q) \ ,
  \label{ansatzDI}
\end{align}
using a set of interpolating polynomials $H_i$, e.g., Lagrange polynomials. Inserting this Ansatz into Eq.~\eqref{dirprod} immediately yields that the integral equation decouples into an algebraic equation, with the complicated three-body cut structure moved to integrals. The latter can be evaluated with standard numerical methods that are robust to the presence of logarithmic singularities. 

Alternatively to the use of Lagrange polynomials, one can approximate $\tilde\Gamma^T$ by splines of a given order, which requires a certain care in choosing appropriate grid points~\cite{Matsuyama:2006rp}.

\begin{figure}[tb]
    \begin{center}
    \includegraphics[width=6cm]{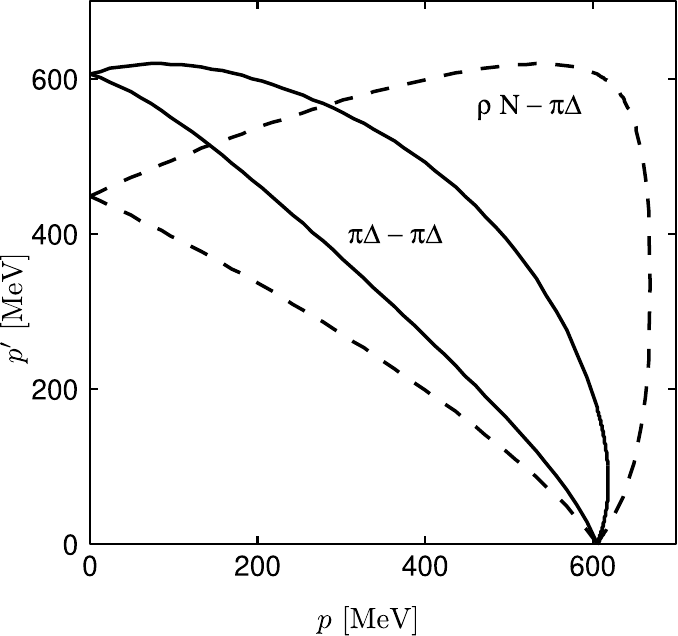}
    \caption{Logarithmically divergent regions of the matrix elements of $Z(p',p;E)$ for the $\pi\Delta-\pi\Delta$ and $\rho N -\pi\Delta$ transitions shown in Fig.~\ref{fig:z}.}
    \label{fig:moonpid}
    \end{center}
\end{figure}

In the ANL-Osaka approach, a crucial point of numerical calculation is how to choose the integration path $C, C'$ and grid points of momentum in Eq.~\eqref{eq:mext} and Eq.~\eqref{eq-sigma-pid}. 
The coupled integral equation is solved by using the matrix inversion method in momentum space. 
In order to calculate  observables, the integration paths $C$ and $C'$ are chosen from zero to infinity on the real axis without any contour deformation.   When only $v_{\beta,\alpha}$ in Eq. \eqref{eq:vvz} contributes,  the choice of the grid points of momentum is well known. Including the on-shell momentum of the two-body  Green's function, the integrand of principal value integral  is replaced by adding the subtraction term to obtain smooth momentum dependence and stabilize the numerical integration~\cite{Haftel:1970zz}.
The particle exchange interaction $Z(p',p;E)$ has logarithmic singularity above the $\pi\pi N$ three-body breakup threshold.  The logarithmically divergent moon-shape region is shown in Fig. \ref{fig:moonpid}. The imaginary part is non-zero inside the moon-shape region, which corresponds to the on-shell momentum for the two-body Green's function. The singularity depends on both $p'$ and $p$, which makes it difficult to solve the integral equation.
To overcome the difficulty, one writes the unknown half-off shell t-matrix in terms of an interpolation function $S_i(p)$ and
chooses appropriate grid points $\{ p_i \}$, 
\begin{eqnarray}
    t_{\beta,\alpha}(p,p_0;E) = \sum_i S_i(p)t_{\beta,\alpha}(p_i,p_0;E) \ .
\end{eqnarray}
Then, the required integration can be carried out as precisely as necessary
using the known function $S_i(p)$. Splines are chosen as interpolation functions, which depend locally on the grid points and which are  known to be less oscillating compared to the polynomial interpolation. Hermitian splines are used that only have the first derivative  continuous. Both the first and the second derivatives are continuous for natural splines. Furthermore, the appropriate choice of  the grid points is explained in Refs.~\cite{Matsuyama:1984qv,Matsuyama:2006rp} using the Amado model~\cite{Amado:1963twy}, as an example.  As a fast numerical calculation,  one can use a contour $C$ in the complex momentum plane as will be discussed in \cref{sec:technical} to calculate observables for the two-body final states.

\subsection{Analytic continuation to complex energies}
\label{sec:technical}
Resonance poles are situated at complex  energies $E=\sqrt{s}$ and, in sub-channels, at complex sub-energies or invariant masses $\sqrt{\sigma}$. Phenomenologically relevant poles are the ones close to the physical axis, situated on the analytically continued amplitude in the lower $\sqrt{s}$ half-plane. Technically, the analytic continuation to complex $\sqrt{s}$ with negative imaginary part can be achieved, for example, by a complex contour deformation~\cite{Doring:2009yv, Suzuki:2009nj, Sadasivan:2021emk, Dawid:2023jrj,Dawid:2023kxu} of the integration momentum running inside the loop integrals. Mesonic resonance poles in three-body amplitudes were only extracted recently~\cite{Sadasivan:2021emk, Mai:2021nul,Nakamura:2023obk,Yan:2024gwp}, while the extraction of exited baryon resonance poles, including three-body channels, has been carried out by various groups since a long time~\cite{Doring:2009yv, Suzuki:2009nj}. The first extraction of a mesonic resonance pole, the $a_1(1260)$, from data, with a three-body unitary amplitude, was achieved in Ref.~\cite{Sadasivan:2021emk}, including a quantification of the noticeable effect from the inclusion of particle exchange as required by unitarity. 
First Lattice-QCD-based determinations of pole positions of three-body resonances have also been pioneered recently, for the $a_1(1260)$~\cite{Mai:2021nul} and the $\omega(782)$~\cite{Yan:2024gwp}. For the mesonic amplitude discussed in Secs.~\ref{sec:a1} and \ref{sec:nine}, one can directly follow Ref.~\cite{Sadasivan:2021emk}. We summarize the main ideas of that study in the following. 

\subsubsection{Analytic structure of the multi-channel scattering amplitude}
\label{sec:analytic}
\begin{figure}[t]
    \centering
    \includegraphics[width=\linewidth]{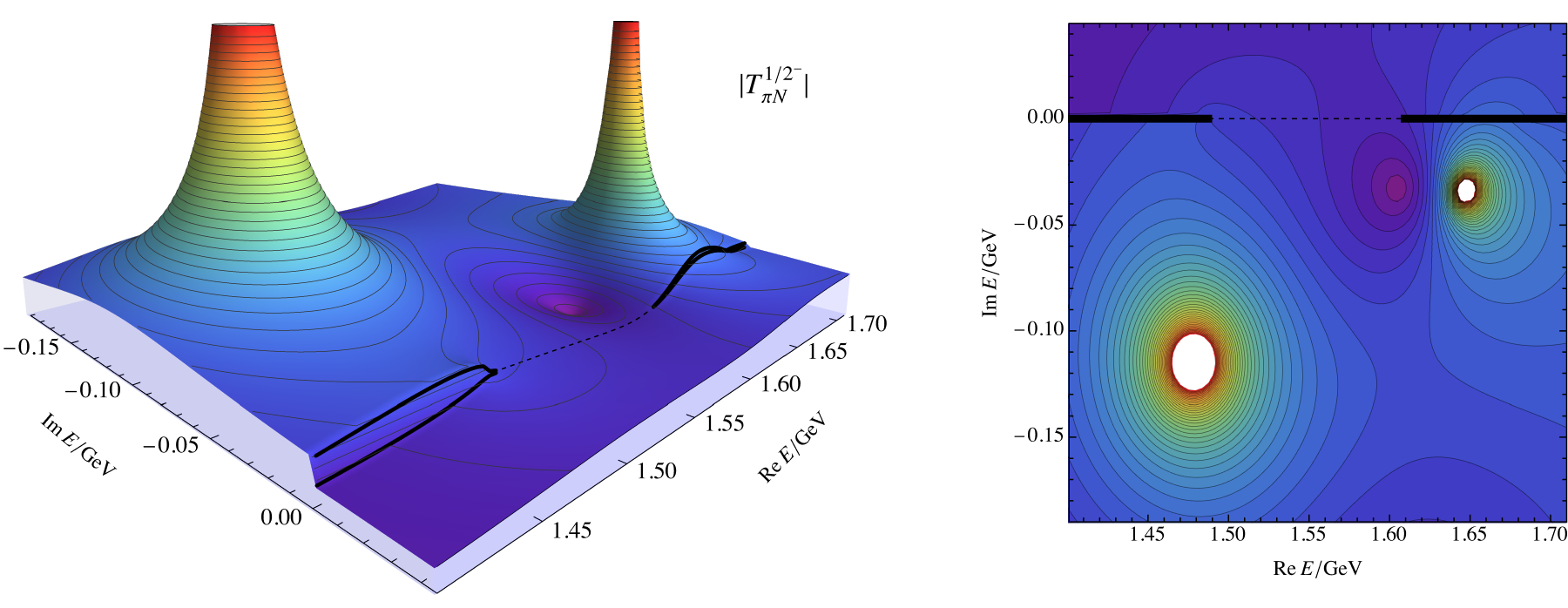}
    \caption{Example of Riemann sheets $[++++++]$ and $[---+++]$ with respect to $\{p\pi^0,n\pi^+,p\eta,\Lambda K^+,\Sigma^0K^+,\Sigma^-K^0\}$ meson-baryon two-particle channels. Both Riemann sheets are connected along the real energy axis between $p\eta$ and $\Lambda K^+$ thresholds. The amplitude is taken from Ref.~\cite{Bruns:2010sv} corresponding to $\pi N$ scattering for $I=1/2$, $J^P=1/2^-$. The poles correspond to the $N(1535)$ and a $N(1650)$ resonances.}
    \label{fig:RiemannSheets}
\end{figure}

The analytic structure of the scattering amplitude is characterized by resonance/bound state poles and branch points. Zeros of the amplitude are related to poles~\cite{Oller:1998zr}; their interplay in the meson-baryon sector has been illustrated in Ref.~\cite{Doring:2009uc}. In addition, there can be triangle singularities~\cite{Gribov2012-yx, Guo:2019twa}. In general, each branch point is associated with one or more cuts. Some branch points and cuts exist in the plane-wave amplitude (right and left-hand cut), but some cuts arise only after partial-wave projection, like the circular cut and the short nucleon cut~\cite{Hohler:1984ux}. Resonance poles and branching ratios have been related in Ref.~\cite{Heuser:2024biq}.

The multi-valuedness of the T-matrix, discussed in the previous section, leads to the concept of single-valued Riemann sheets~\cite{Hohler:1984ux} which, together, form the multi-valued Riemann surface. Any measured (experimentally or as a result of a numerical lattice calculation) values are located on the real energy-axis of the so-called physical sheet,  denoted most commonly by $[+\ldots+]$ referring to ${\rm Sgn}(\Im(q_i(s)))$ in each two-body channel where $q_i$ indicates the c.m. momentum in channel $i$ according to Eq.~\eqref{pcm}. The physical axis in the Mandelstam variable $s$ lies at $s+i\epsilon$ to make it clear that it is situated above the ``right-hand cut''.  Generalizing this notation, we denote all sheets by the same type of sequence, see, e.g., Ref.~\cite{Mai:2022eur}. Specifically, the unphysical sheet connected to the physical one along the right-hand cut, between the first and second threshold, is denoted as $[-+\ldots+]$, the one connected to the physical sheet between the second and third threshold by $[--+\ldots+]$ etc.. An example of Riemann sheets for a meson-baryon system is depicted in Fig.~\ref{fig:RiemannSheets}.

There is a broad $N(1535)$ pole and a narrower $N'(1650)$ pole. Note that the latter is ``hidden'' behind the $K\Lambda$ threshold (hence the prime notation). It is a replica of the actual $N(1650)$ resonance pole on the $[----++]$ sheet (not shown). Indeed, poles often repeat on different Riemann sheets. In particular, if the pole residue to a given channel is small, the pole repeats at almost the same complex energy $E$ on the two sheets corresponding to that channel.

The example of the $S_{11}$ partial wave is also illuminating because the other resonance in that partial wave, $N(1535)$, is located close to the $\eta N$ channel opening at  $E\approx 1486\,\MeV$, and it couples \emph{strongly} to that channel. This leads usually to an enhancement of threshold cusps~\cite{Doring:2009uc} as further discussed in Sec.~\ref{subsec:S11}. Another example is the large cusp in the $\Lambda N$ cross section at the opening of the $\Sigma N$ channel which is due to the appearance of an (unstable) $\Si N$ bound state close to the $\Si N$ threshold 
\cite{Haidenbauer:2021smk}.

\begin{figure}[tb]
\begin{center}
 \includegraphics[width=0.5\textwidth]{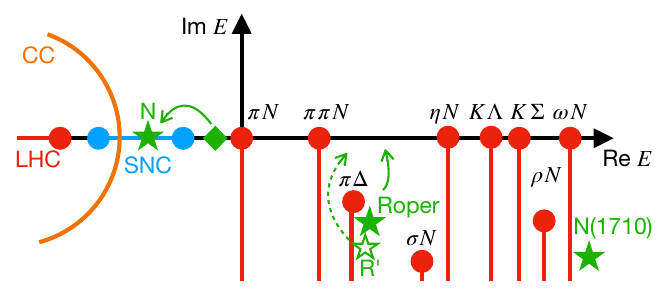}
\end{center}
\caption{Analytic structure of the $P_{11}$ partial wave, for complex scattering energies $E=\sqrt{s}$. The circular
cut is labeled ``CC'', the left-hand cut ``LHC''. The nucleon is labeled as $N$.  See text for further explanations. }
\label{fig:anap11}     
\end{figure}

The analytic structure of the $P_{11}$ partial wave is discussed in the following to illustrate three-body aspects. More phenomenological aspects of that partial wave are discussed in Sect.~\ref{sec:P11}. The $P_{11}$ wave ($J^P=1/2^+$) is particularly interesting because it hosts the nucleon as a bound state. Its analytic structure is displayed in Fig.~\ref{fig:anap11}.
Channels of two stable particles ($\pi N$, $\eta N,\dots$) have branch points (thresholds) on the real-energy axis. Each of these two-body channels is associated with one right-hand cut. The figure also shows the $\pi\pi N$ branch points on the real axis and in the complex plane from the $\pi\Delta$, $\sigma N$, and $\rho N$ channels as discussed in \cref{sec:cont3B}.  To search for poles on the most relevant Riemann sheets, the cuts are chosen to be oriented from the threshold openings into the negative imaginary $E$-direction as indicated in the figure. In addition, the figure shows the sub-threshold short nucleon cut (SNC), circular cut (CC) and left-hand cut (LHC). The structure of some of these cuts and branch points is discussed in Ref.~\cite{Hohler:1984ux}, see also~\cite{Doring:2009yv}. 

The (green) stars indicate the positions of the physical poles. The nucleon appears as a bound state pole on the first Riemann sheet in the amplitude, right on top of the SNC. The green diamond indicates the bare nucleon pole that moves to the physical position after renormalization (see \cref{sec:schannel}). 
There are two Roper poles on different $\pi\Delta$ sheets, labeled as $N(1440)$ and $N^\prime(1440)$ (not to be confused with the two-pole structure of the $\Lambda(1405)$ for which both poles are on the \emph{same} Riemann sheet). This two-pole structure is found in several analyses, such as GWU/SAID~\cite{Arndt:2006bf}, EBAC~\cite{Suzuki:2009nj}, and JB~\cite{Ronchen:2012eg}. It is due to the fact that poles can repeat on different sheets as discussed previously for the case of the $N(1535)$ and $N(1650)$. In addition, the proximity of the complex $\pi\Delta$ branch point to the resonance poles leads to a non-standard shape of the Roper resonance that is not of the Breit-Wigner type. Another reason for the unusual shape is the strong influence of the $\sigma N$ channel. All three particles $\pi$, $\pi$, $N$ are in relative $S$-waves for $P_{11}$. The absence of any centrifugal barriers then, indeed, leads to a large inelasticity at low energies, strongly distorting the Roper resonance shape. In addition, other resonances ($N(1710)1/2^+$ and $N(1750)1/2^+$) have been reported in some analyses as the figure indicates. The role of the $N(1710)1/2$ is further discussed in \cref{sec:P11}.

\begin{figure}[tb]
    \begin{center}
    \raisebox{-.5\height}{\includegraphics[width=6cm]{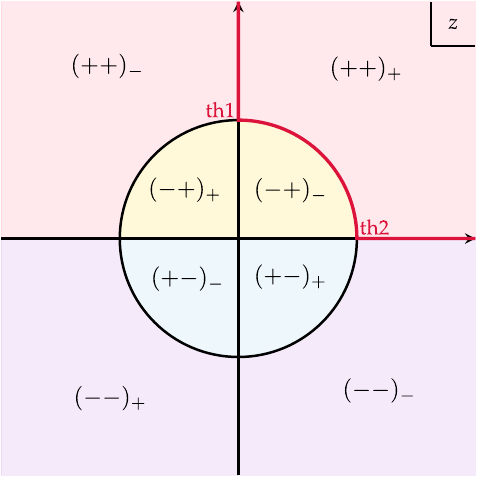}}
    ~~~~~
    \raisebox{-.5\height}{\includegraphics[width=8cm]{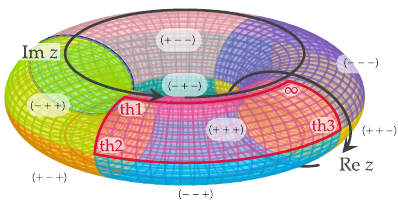}}
    \caption{Left: Riemann sheet of a two-channel system in terms of the uniformization variable $z$. 
    The subscript of the Riemann sheets denotes either the positive or negative complex half plane of the latter.
    Right: representation of a 3-channel system on a torus. The corresponding two-body thresholds are denoted by "th".}
    \label{fig:unif2}
    \end{center}
\end{figure}

Recently, an approach to extract resonance information from observables that incorporates the resonant and threshold
behavior was proposed by Morimatsu and Yamada~\cite{Yamada:2020rpd}.
The approach was applied to analyze  unstable states~\cite{Yamada:2021cjo,Yamada:2021azg, Yamada:2022xam,Yamada:2023wqp,Yamada:2025lkc}.
This approach is a tool for a data driven methodology extracting the resonance information without referring to a specific model.
The starting point is that one can unfold the Riemann surface choosing the appropriate
variable called uniformization variable $z$ in place of energy $\sqrt{s}$ (in all other parts of this review, except for this subsection, $z$ stands for $\sqrt{s}$ according to Eq.~\eqref{eq:ESWZ}).
For the single channel case, $z = p = \sqrt{s - \epsilon^2}/2$ with threshold energy 
$\epsilon$. For the two-channel case, the uniformization variable $z$ is given as~\cite{Kato:1965iee,Newton:1982qc},
\begin{align}
  z  =  \frac{p_1 + p_2}{\Delta}\ ,
\end{align}
where $p_i = \sqrt{s - \epsilon_i^2}/2$, $\epsilon_i = (m_i + m_i')$ and $\Delta = \sqrt{\epsilon_2^2 - \epsilon_1^2}/2$. Fig.~\ref{fig:unif2} shows the Riemann surface of the two-channel case with respect to  $z$ which represents four sheets $[++],[-+],[+-]$ and $[--]$. The red line corresponds to the real scattering energy $\sqrt{s}$ and the subscript $(\pm)$ indicates the sign of the imaginary part of $\sqrt{s}$. The lower(higher) threshold is at $z=i$($z=1$). 

The scattering amplitude ${\cal A}$ and the S-matrix become a single-valued function of uniformization variable $z$. The meromorphic function can be expressed by the sum of pole terms and an entire function, which is the Mittag-Leffler(ML) expansion~\cite{humblet1961theory,Nussenzveig:1972tcd,RamirezJimenez:2018dge}.
In the uniformized ML expansion approach,  the scattering amplitude is expressed as
\begin{align}
  {\cal A}(z) = \sum_{n=1} \left( \frac{r_n}{z - z_n} - \frac{r_n^*}{z + z_n^*} \right),
  \label{eq:ml-amp}
\end{align}
where $z_n$ and $r_n$ are the position and residue of a pole in the variable $z$.
The amplitude is expressed as a sum of the pair of the poles at $z_n$ and $-z_n^*$.
The symmetry imposes the series to obey the proper threshold behaviors.
The pole positions and residues are determined by fitting Im~${\cal A}$ to the experimental data  by a few pairs of the pole terms.
The method is applied to study the pole properties of the
$\Lambda (1405)$~\cite{Yamada:2021cjo} from the analysis of the
invariant mass distribution of  $\pi \Sigma$ in the
$\gamma p \rightarrow K^+ \pi \Sigma$ and $K^- p$ cross sections.
The data are well fitted as seen in Ref.~\cite{Yamada:2021cjo} with three pair of poles.
The pole values are $z_1= 0.524 \pm 0.006 + i(\,0.316 \pm0.006)\, (\sqrt{s_1}= (1.420 \pm 0.001 - i(\,0.048 \pm 0.002) \text{ GeV})$,
$z_2 = 1.64 \pm 0.07 - i\,(1.04 \pm 0.09)\,(\sqrt{s_2}= (1.43 \pm 0.01 -i\,(0.074 \pm 0.004)\text{ GeV})$ and $z_3 = 2.323 \pm 0.003 - i\,(0.069 \pm 0.003) (\sqrt{s_3}=1.5138 \pm 0.0003  -i\, (0.0068 \pm 0.0003) \text{ GeV})$. The conjugate poles are at $-z_1^*,-z_2^*$ and $-z_3^*$.
Pole 1 is on the $[-+]$ sheet just below the $\bar{K}N$ threshold and pole 2 and 3 are on the $[--]$ sheet. The broad peak between $\pi \Sigma$ and $\bar{K}N$ is sufficiently explained by the pair of poles $1 + 1^*$. 

\begin{figure}[t]
\begin{center}
    \raisebox{-.5\height}{\includegraphics[height=6cm]{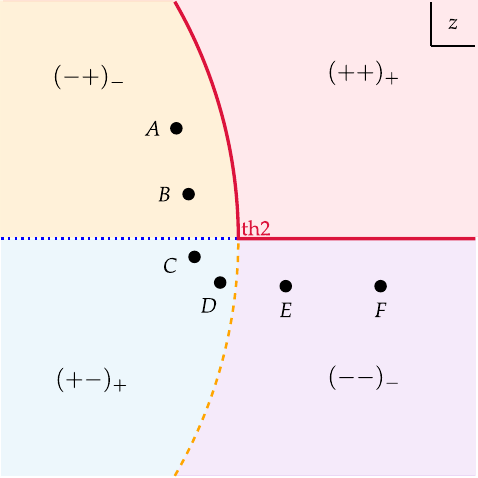}}
    ~~
    \raisebox{-.5\height}{\includegraphics[height=6cm]{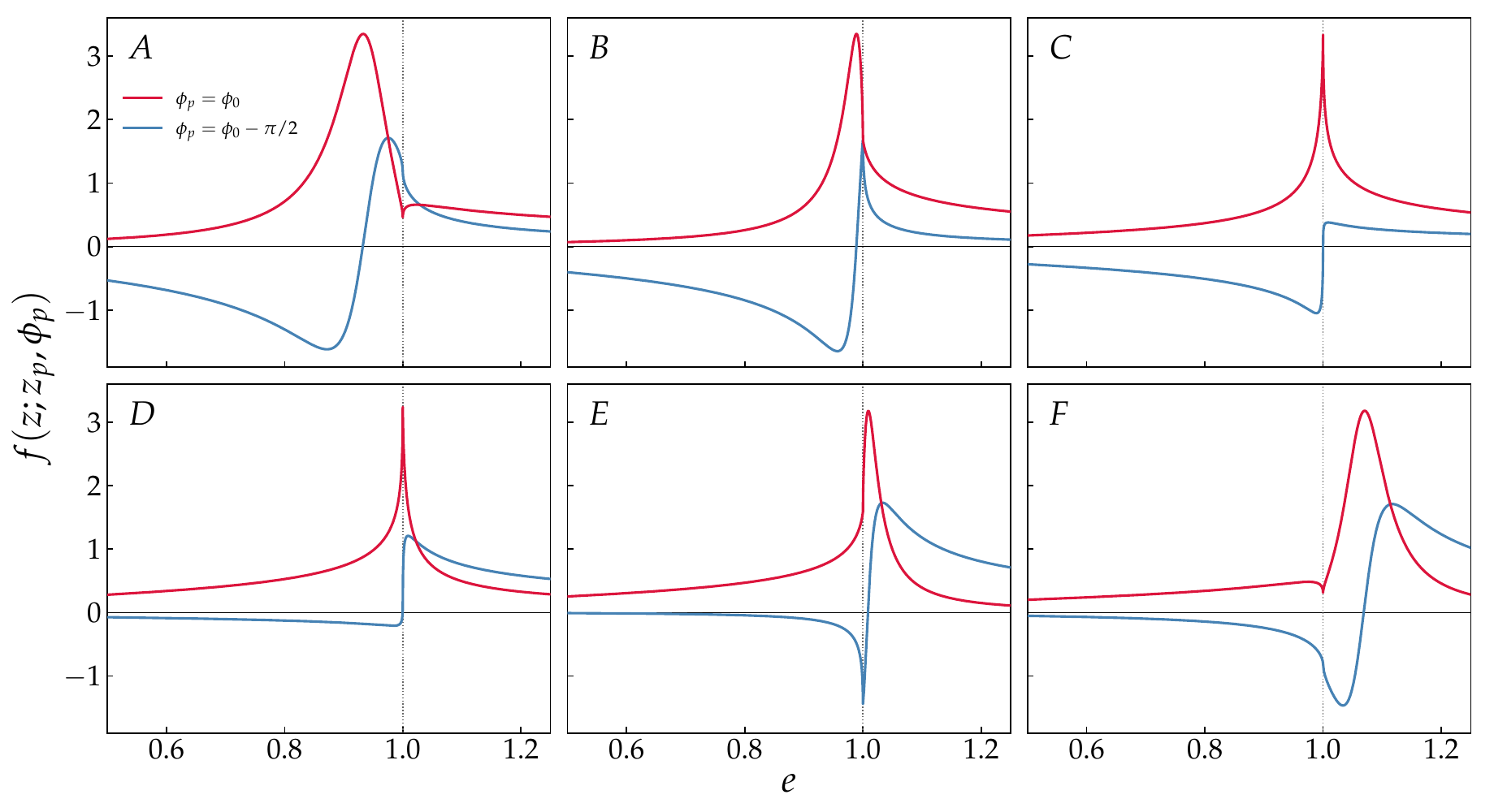}}
    \caption{\label{fig:unif2-amp}
    S-matrix pole close to the upper threshold(th2) in $z$ plane and the shape of the spectrum. On the left, the pole positions $A\to\ldots\to F$ are  shown, which also includes poles ($C,D$) being located on the more distant $[+-]_+$ 
    Riemann sheet. The effect of the pole on the imaginary part of the scattering amplitude for real energies \eqref{eq:variable-e} is depicted in the right panel. 
    }
    \end{center}
\end{figure}

The method is also applied to demonstrate the manifestation of S-matrix poles in the physical amplitude near an inelastic threshold~\cite{Yamada:2021azg}.  The left panel of \cref{fig:unif2-amp} shows six positions, A-F, of a pole near an inelastic two-body threshold. The right panel  shows imaginary parts of the pole-pair contribution of the normalized amplitude given as
\begin{align}
 f(z;z_p,\phi_p) = - \frac{1}{\pi}\text{Im} \left[ 
 \frac{\exp(i\phi_p)}{ z - z_p} - \frac{\exp(-i\phi_p)}{z + z_p^*}
 \right].
\end{align}
The two cases of  the phase of the residue, $\phi_p=\phi_0$ and $\phi_p=\phi_0 - \pi/2$ are shown.
 $\phi_0$ is chosen as $\phi_0=$Arg$(z_p)-\pi/2$, Arg$(z_p-1)-\pi/2$ and
 $0$ for $[-+]$, $[+-]$ and $[--]$ sheet respectively. See Ref.~\cite{Yamada:2021azg} for details.
The quantity $e$ is a normalized energy variable given as
\begin{align}
    e=\frac{s-\epsilon_1^2}{\epsilon_2^2-\epsilon_1^2}\, .
    \label{eq:variable-e}
\end{align}
The transition of the spectrum is continuous as the pole moves from A to F
crossing the boundaries of Riemann sheets when the pole position is sufficiently close to the threshold in the $z$ plane. The pole near the inelastic threshold exists irrespective of which Riemann sheet it is on, and the amplitude shows a peak shape.  Since there is no essential difference on the shape of the spectrum, for example 
B$[-+]_-$, C$[+-]_+$, D$[+-]_+$ and E$[--]_-$,  high precision data will be needed to determine to which Riemann sheet the pole belongs.  It is noticed, poles C, D with positive imaginary part of the energy are close to the upper threshold, while the conjugate poles are far from the upper threshold in  the $z$ plane.  The argument suggests that the peak  near the $\bar{D}D^*$ threshold found in the analysis of the HAL QCD~\cite{Ikeda:2017mee,HALQCD:2016ofq} is
most likely due to the existence of the pole near the $\bar{D}D^*$ threshold.

The uniformization variable $z$ itself has been used to analyze the
relation between the poles and the shape of the $\Lambda N$ cross
section~\cite{Miyagawa:1999zz} and the movement of poles in a two-channel
Breit-Wigner formula~\cite{Suzuki:2008rp}. An interesting observation of Ref.~\cite{Kato:1965iee}
is that, for the two-channel case, the minimum number of the set of poles is two at $z_1$ and $z_2$ and the positions of the two poles satisfy $|z_1 z_2|=1$.
Therefore, one of the poles is inside the unit circle of $z$, i.e., the $[+-]$ or $[-+]$ sheet and the other pole is outside the circle on the $[--]$ sheet.
One can also rewrite the Flatt\'e formula~\cite{Flatte:1976xu} in the form of Eq.~\eqref{eq:ml-amp} with two pairs of poles, whose explicit form is given in ~\cite{Yamada:2021azg}.
In general, the use of the uniformization variable can help understand the pole movement with respect to some internal parameters of the theory such as, e.g., interaction strength or quark mass in Lattice QCD calculations, see Refs.~\cite{Cieply:2016jby, Guo:2023wes, Doring:2016bdr}.

Clearly, one might be intrigued to extend this approach to the three-channel case. For that case, there are eight Riemann sheets. Note that the topological structure of the Riemann surface for two channels is equivalent to that of a sphere, whereas for three-channels it is equivalent to a torus~\cite{Weidenmuller:1964fbf}, see also a corresponding discussion in Ref.~\cite{Mai:2022eur}. The explicit construction of the uniformization variable
and the description of the Riemann surface is for the first time provided in Ref.~\cite{Yamada:2022xam}. It reads
\begin{align}
     z = \frac{1}{4 K(1/\gamma^2)} sn^{-1} (\gamma,z_{12},1/\gamma^2)\ .
\end{align}
Here, $z_{12}$ is the uniformization variable of channels 1 and 2, and $sn$ is the Jacobian elliptic function. The constants $\gamma$ and $K$ are given as
\begin{align}
    \gamma = \frac{\Delta_{13}+\Delta_{23}}{\Delta_{12}},
    \qquad
     K(k) = \int_0^1 \frac{ds}{\sqrt{(1-s^2)(1-k^2s^2)}}\ .
\end{align}
With this uniformization variable, the ML expansion of the amplitude is given as
\begin{eqnarray}
  {\cal A}(z) = \sum_i[ r_i (\zeta(z - z_i) + \zeta(z_i)) - r_i^* (\zeta(z + z_i^*) + \zeta(-z_i^*))]\ ,
\end{eqnarray}
where $\zeta(z)$ is Weierstra{\ss}'s zeta function. The use of the ML expansion is demonstrated with a simple example of nonrelativistic effective field theory with contact interactions
for the $0^+$, $\Lambda\Lambda-N\Xi-\Sigma\Sigma$ coupled-channel problem.

\subsubsection{Continuation for two-body channels}
\label{sec:contstable}
As discussed, resonance poles are situated on the unphysical sheet in the lower $\sqrt{s}\in\mathds{C}$ half plane that is analytically connected to the physical axis~\cite{Gribov2012-yx}.
To find resonance poles in $\sqrt{s}$ one cannot simply dial in complex energies because the resulting amplitude on the physical sheet has a right-hand unitarity cut along the real $\sqrt{s}$ axis that separates the lower $\sqrt{s}$ half plane from the physical axis at $\sqrt{s}+i\epsilon$. To illustrate this consider the integration path of Eq.~\eqref{scattering} or Eq.~\eqref{dirprod}. 

\begin{figure}[t]
    \centering
    \includegraphics[width=0.5\linewidth]{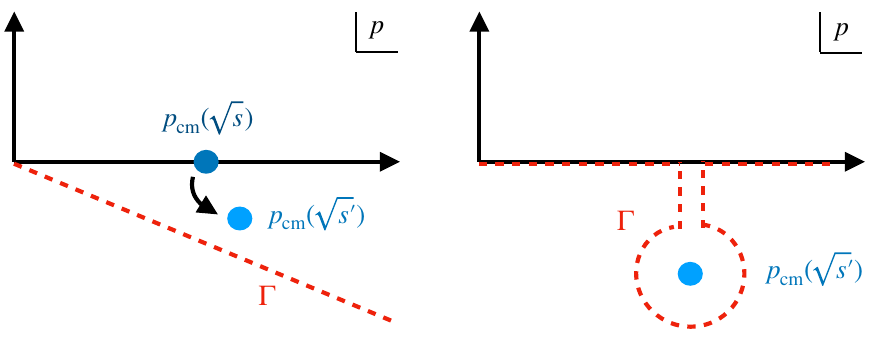}
    \caption{Left: Analytic continuation in $\sqrt{s}$ by deforming the integration contour $\Gamma$ (adapted from Ref.~\cite{Sadasivan:2021emk}). The figure indicates the position of the two-body singularity at $p_\text{cm}$ as the energy develops a negative imaginary part, Im~$\sqrt{s'}<0$. Right: Alternatively, one can integrate along the real axis and add the residue to the integral as contributions along the vertical parts of $\Gamma$ cancel (adapted from Ref.~\cite{Doring:2009yv}).
 }
 \label{fig:contours2}
\end{figure}

For one channel of two stable particles with masses $m_a$ and $m_b$, there is a two-body singularity given by Eq.~\eqref{pcm} at $p=p_\text{cm}$
that has to be avoided in the integration. Complex paths in the lower momentum plane as shown in Fig.~\ref{fig:contours} achieve this as $\sqrt{s}$ develops a negative imaginary part, $\sqrt{s}\to\sqrt{s'}$. Such a deformation allows to smoothly (``analytically'') continue into the lower $\sqrt{s'}$ plane onto the unphysical sheet, as long as $\Im \sqrt{s}$ does not become too negative making $p_\text{cm}(\sqrt{s})$  cross the integration contour. This situation is shown in Fig.~\ref{fig:contours2} to the left. The right-hand side of the figure illustrates an alternative: instead of deforming the path for analytic continuation, one can add the residue (given by the closed circle) to the integral of Eqs.~\eqref{scattering}  and proceed with a path along real momenta because the integrations along the vertical direction cancel~\cite{Doring:2009yv}:
\begin{align}
T_{\mu\nu}(E,p',p)&=V_{\mu\nu}(E,p',p)
+\sum_{\kappa}\int\limits_0^\infty \dv{q}\,
 q^2\,V_{\mu\kappa}(E,p',q)\,G_\kappa(E,q)\,T_{\kappa\nu}(E,q,p) +\sum_\kappa \delta G_\kappa\nonumber \ , \\ 
 \delta G_\kappa&=-\theta(E-m_\kappa-M_\kappa)V_{\mu\kappa}(E,p',q_\text{cm})\,\frac{2\pi i\,q_\text{cm}E_\kappa(q_\text{cm})\omega_\kappa(q_\text{cm})}{E}\,T_{\kappa\nu}(E,q_\text{cm},p) \ ,
\label{addingres}
\end{align}
with Im~$E<0$, $q_\text{cm}$ from Eq.~\eqref{pcm}, on-shell baryon and meson energies $E_\kappa(q_\text{cm}),\,\omega_\kappa(q_\text{cm})$, and the $\theta$ function ensuring that the continuation to the second sheet occurs only above the respective thresholds. Equivalently, one can say that this corresponds to rotating the right-hand cuts into the negative imaginary $E$-direction, see Fig.~\ref{fig:anap11}.
Adding the residue is, of course, equivalent to adding the discontinuity along the real axis. In a discretized matrix scheme for the integral equation, the residue term can simply be added to  $G^I$ of Eq.~\eqref{discreteG} leading to a new second-sheet propagator,
\begin{align}
  G^{II}_{lk}=\begin{cases}
      -\frac{2\pi i\,q_\text{cm}E(q_\text{cm})\,\omega(q_\text{cm})\,\theta(E-m_\kappa-M_\kappa)}{E}&\text{for }l=k=n_\text{SMC}+1\\
      G^I_{lk}&\text{otherwise.}
  \end{cases} 
\end{align}
The prescription of Eq.~\eqref{addingres} (adapted to possible differences in the normalization of the T-matrix) is often used in chiral unitary approaches~\cite{Inoue:2001ip, Doring:2009uc, Bruns:2010sv, Cieply:2016jby}, while contour deformation has the advantage that it can simultaneously work for two- and three-body channels if the path is chosen accordingly. The disadvantage of contour deformation is that the solution is only known along complex momenta on $\Gamma$; this is less relevant when searching for resonance, but one can still add the (now complex) ``on-shell'' point $p_\text{cm}$ of Eq.~\eqref{pcm} to a discretized integration scheme as discussed at the beginning of \cref{sec:realmom}. This is required when calculating pole residues related to branching ratios~\cite{Doring:2010ap}.

Also, for various channels with different masses, the contour deformation into the lower momentum plane allows to analytically continue the amplitude for all channels at once. In addition, it automatically ensures the continuation to the sheet closest to the physical axis in a given segment of $\sqrt{s}$ between two thresholds (see next section).

\subsubsection{Continuation for three-body channels}
\label{sec:cont3B}
For three-body channels, the contour deformation method can be generalized by taking into account the additional cuts, poles, and branch points from the three-body dynamics~\cite{Doring:2009yv, Suzuki:2009nj, Sadasivan:2021emk, Dawid:2023jrj,Dawid:2023kxu}. 
Instead of two stable particles, there is now a stable ``spectator'' and a two-body sub-amplitude called ``isobar'' or "dimer". That isobar may contain a resonance -- in the JBW and ANL-Osaka models, that is the case for the $\pi\Delta(1232)$, $f_0(500)N$ (usually called ``$\sigma N''$), and $\rho(770) N$ channels as discussed in \cref{sec:tauiso}. In case of mesonic amplitudes, the IVU model~\cite{Mai:2017vot, Mai:2018djl, Mai:2019fba, Alexandru:2020xqf, Feng:2024wyg} contains the $f_0(500)$, $f_0(980)$, $\rho(770)$, $K^*_0(700)$ (``$\kappa$''), and $K^*(892)$ resonances (apart from non-resonant, repulsive isobars) as discussed in \cref{sec:threemesons}.
\begin{figure}[t]
    \centering
    \includegraphics[width=0.5\linewidth]{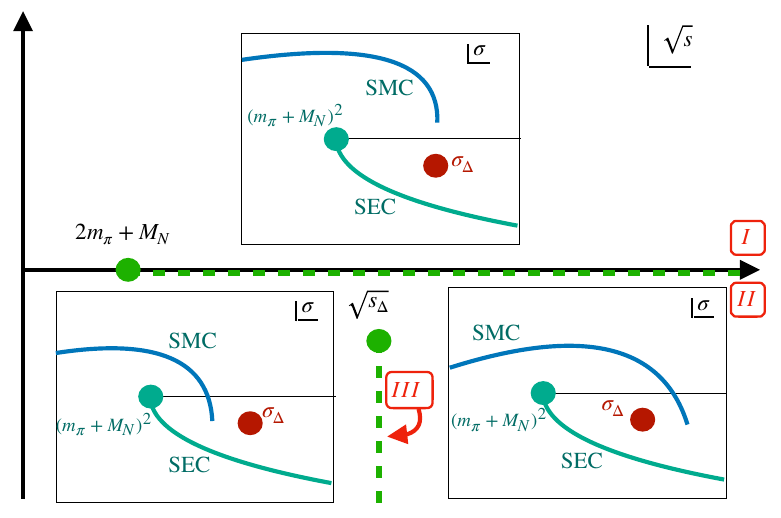}
    \caption{The inelastic cut connecting sheets I and II, and the complex branch point at $\sqrt{s_\Delta}$ inducing sheet III, for the example of the $\pi\Delta$ channel (figure adapted from Ref.~\cite{Sadasivan:2021emk}). Corresponding to their location in the $\sqrt{s}$ plane, the insets show the squared sub-energy plane $\sigma$ for different values of $\sqrt{s}$. They contain the $\Delta(1232)$ pole at $\sigma_\Delta$, the rotated two-body cut (SEC), and the spectator momentum contour (SMC) that avoids the SEC. The branch point at $\sqrt{s_\Delta}$ occurs when the SMC, mapped to the $\sigma$-plane, starts at $\sigma_\Delta$.}
    \label{fig:SMCSEC}
\end{figure}

In the following we first discuss the continuation in the JBW approach and then in the ANL-Osaka approach, pointing out the similarities even if the variables differ.
For complex energies $\sqrt{s}$, the SMC in the JBW approach still has to avoid the three-body cuts in the exchange diagrams as shown to the right in Fig.~\ref{fig:contours} which is one constraint on its form.
A new constraint arises by inspecting the sub-energy plane $\sigma$ given by \cref{eq:sigma}.
The insets in Fig.~\ref{fig:SMCSEC} show the SMC in the $\sigma$ plane for different $\sqrt{s}$ according to the approximate location of the insets. The additional constraint for the SMC arises from the threshold at $m_\pi+M_N$ and the associated right-hand cut that has to be avoided. Sometimes, for example in the JBW model, resonant two-body sub-amplitudes are parametrized with $s$-channel propagators and self-energies~\cite{Feng:2024wyg}, see \cref{sec:tauiso}. The latter can be evaluated with a selfenergy contour (SEC). In other instances, the isobar amplitude is expressed by the Inverse Amplitude Method~\cite{Truong:1988zp, GomezNicola:2001as, Nebreda:2010wv} as in Refs.~\cite{Alexandru:2020xqf, Mai:2018djl}, or by chiral unitary amplitudes~\cite{Oller:1997ti, Oller:1998hw} or simple contact terms as in Ref.~\cite{Feng:2024wyg}, that all can generate two-body resonances through the two-body unitarization. In all cases, one chooses the right-hand cut to be rotated into the lower $\sigma$ half plane as shown because the SMC contour starts at a $\sigma$ with negative imaginary part if $\sqrt{s}$ has a negative imaginary part.
Once both of the discussed constraints are fulfilled for the SMC, the amplitude is automatically continued from sheet I to sheet II as indicated in the figure, just as in case of two stable particles discussed before.

There are two occasions in which the SMC cannot avoid non-analyticities in the $\sigma$ plane: If $\sqrt{s}=\sqrt{s_\Delta}=m_\pi+\sqrt{\sigma_\Delta}$ the SMC, mapped to the $\sigma$ variable, starts at the resonance pole at $\sigma_\Delta$. This induces a branch point at $\sqrt{s_\Delta}$ and an associated cut connecting the amplitude to a new Riemann sheet $III$ as indicated in the figure. As one aims at searching for poles on the sheets most closely connected to the physical axis, one intends to rotate that cut into the negative imaginary direction as shown. 

The second case occurs when the SMC starts at $\sqrt{\sigma}=m_\pi+M_N$. This induces the branch point at $\sqrt{s}=2m_\pi+M_N$ as expected. Depending on the orbital angular momentum between isobar and spectator, $L$, and the decay angular momentum of the isobar, $\ell$, the barrier factors at the branch points can be calculated~\cite{Ceci:2011ae}. 

For a more technical discussion of analytic continuation see Ref.~\cite{Feng:2024wyg}. For the JBW model an approximation to the discussed contour deformation method is taken as discussed in Ref.~\cite{Doring:2009yv}. In short, the amplitude on the unphysical sheet $II$ is evaluated by adding the discontinuity along the righthand cut to sheet I (with the naming as shown in Fig.~\ref{fig:SMCSEC}), just as it can be done for the case of two stable particles discussed in \cref{sec:contstable}. The approximation consists in replacing the momentum-dependent nature of this process (for three-body channels) by choosing approximate on-shell momenta assuming the $\sigma$, $\rho$ and $\Delta$ resonances are stable. The resulting amplitude on sheet II still exhibits the full analytic structure with complex and real branch points discussed before. Ref.~\cite{Doring:2009yv} contains also suitable prescriptions for integration paths to continue to sheet III. In practical terms the poles are searched in the upper $\sqrt{s}$ planes on the corresponding sheet which is equivalent due to Schwartz' reflection principle~\cite{Doring:2009yv}.

\begin{figure}[tb]
\begin{center}
    \includegraphics[width=6cm]{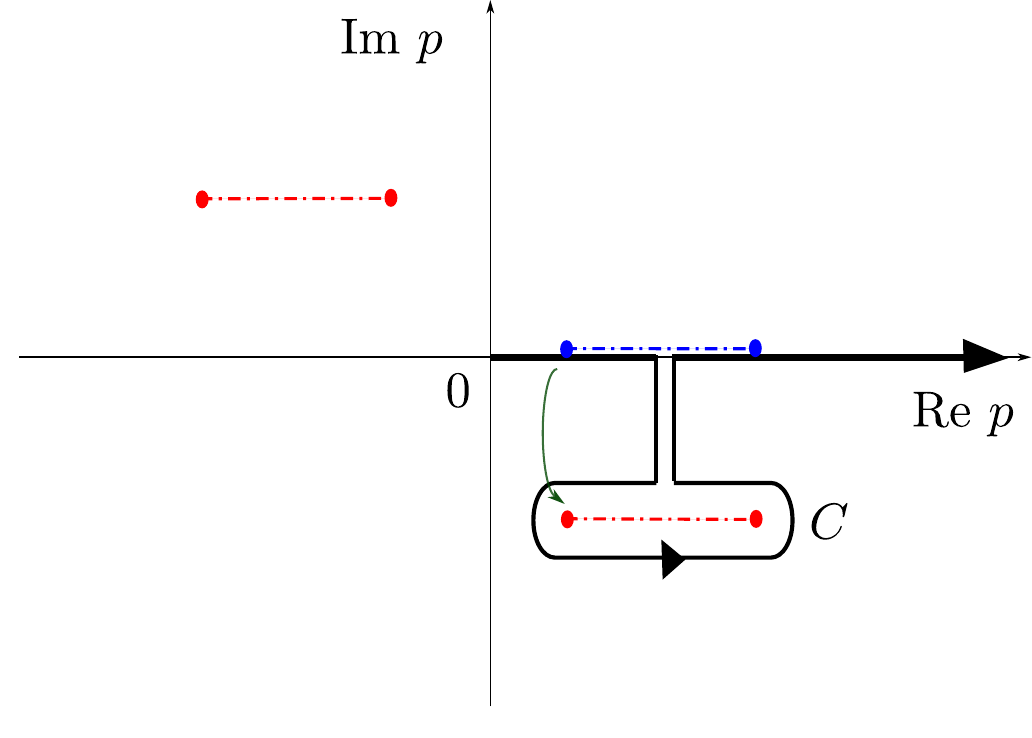}\hspace*{1cm}
    \includegraphics[width=6cm]{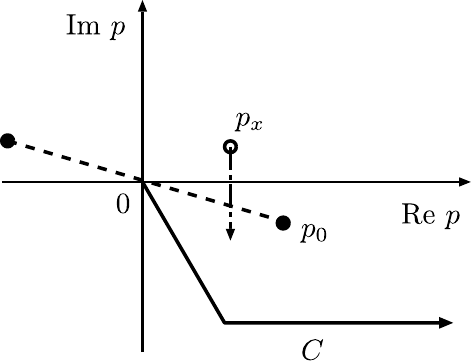} 
  \end{center}
  \caption{Integration contour $C$. Left: logarithmic singularity of $Z(p',p;E)$ for fixed $p'$
  from ~\cite{Ikeda:2007nz}. Right: Singularity of $G_{\pi\Delta}(p;E)$
  from ~\cite{Suzuki:2010yn}. 
  }
    \label{fig:path-ANLOSK}
\end{figure}

In the ANL-Osaka model, all the right hand cuts in Fig.~\ref{fig:anap11} are parallel to the real axis. Analytic continuation of the T-matrix to the unphysical sheet is performed by deforming the integral contour. For stable meson-baryon (MB) channels such as $\pi N, \eta N, KY$, only the potential $v_{\beta,\alpha}(p',p)$ contributes. For the T-matrix on unphysical sheet, the integration contour $C$ in Eq.~\eqref{eq:mext} is deformed so that the on-shell momentum does not cross $C$ in a similar way to the left panel of Fig.~\ref{fig:contours2}.
For unstable MB such as $\pi\Delta, \rho N,\sigma N$, the contour $C$ is chosen by taking into account the $\pi\pi N$ discontinuity of $Z(p',p;E)$ \cite{Ikeda:2007nz,Glockle:1978zz} and the singularity of  the MB Green's function~\eqref{unstableprop}~\cite{Suzuki:2010yn}.
For real energy $E$, $Z(p',p;E)$ develops a logarithmic singularity along the real $p$ axis for fixed $p'$ as illustrated with the blue dotted line in the left panel of Fig.~\ref{fig:path-ANLOSK}. For $E$ with negative imaginary part, the singularity
moves and becomes the red dotted line, and one can access the T-matrix of the nearest Riemann sheet 
across the $\pi\pi N$ cut by deforming the contour as $C$, similar to the stable MB case. Since $p'$ is also varying on $C$, the allowed path $C$ is similar to the one in right panel of Fig.~\ref{fig:contours}. One has to also take into account the singularity of the Green's function. As an example, $G_{\pi \Delta}(p;E)$ has a pole corresponding ''on-shell'' momentum $p_x$ (right panel of Fig.~\ref{fig:path-ANLOSK}) given as
\begin{eqnarray}
 E - \omega_\pi(p_x) - E_\Delta(p_x) - \Sigma_\Delta(p_x,E) = 0 \ .
 \end{eqnarray}
 In the JBW approach, $p_x$ corresponds to the $\Delta$-pole in the $\sigma$ variable at $\sigma_\Delta$, discussed above.
The Green's function also has a discontinuity associated with  $\pi\pi N$ through the self-energy indicated with the dashed line in the right panel of Fig.~\ref{fig:path-ANLOSK}, that has to be avoided by the contour.  Note that the integration contour $C'$ of Eq.~\eqref{eq-sigma-pid} has to be deformed~\cite{Suzuki:2010yn}.
In the JBW approach, the self-energy contour (SEC) is similarly deformed as indicated in Fig.~\ref{fig:SMCSEC}.
For unstable MB, the contour $C$ is determined by the singularity of $Z(p',p;E)$
and $G_{MB}(p;E)$ simultaneously.

\subsection{Related approaches and aspects}
\label{sec:compare}
The challenge to construct coupled-channel scattering amplitudes from general properties of the scattering S-matrix has received much interest in recent years as reviewed, e.g., by Oller~\cite{Oller:2019opk}, JPAC~\cite{JPAC:2021rxu} and, and Mai et al.~\cite{Mai:2022eur}.
In the following we discuss a particular approaches related to the previously introduced DCC models and point out similarities and differences.

\subsubsection{The Bethe-Salpeter equation in coupled channels}
\label{subsec:bseUCHPT}
We start the discussion with the general Bethe-Salpeter equation (BSE) from which most DCC approaches arise through different schemes for three-dimensional reductions as discussed in Sec.~\ref{sec:unitarity}. 

In the following we sketch a coupled-channel, chiral unitary formalism (UCHPT) which is related to DCC approaches but relies on chiral Lagrangians and is \emph{not} reduced to three dimensions. Various versions of UCHPT formalisms exist which differ in the form and iteration prescription of the chiral potential, for a review see~\cite{Mai:2020ltx}. Here we demonstrate the methodology~\cite{Borasoy:2007ku,Bruns:2010sv,Mai:2012dt} which is most closely related to an  expansion in Feynman diagrams allowing, e.g., for extensions of the approach to  manifestly gauge-invariant photoproduction amplitudes~\cite{Borasoy:2007ku,Mai:2012wy,Mai:2013cka}.

The formalism relies on the fully covariant BSE~\cite{Salpeter:1951sz}. We denote the in- and outgoing meson momenta by $q_1$ and $q_2$, and the overall four-momentum by $p=q_1+p_1=q_2+p_2$, where $p_1$ and $p_2$ are the momenta of in- and out-going baryon, respectively. For the meson-baryon scattering amplitude $T(\slashed{q}_2, \slashed{q}_1; p)$ and chiral kernel $V(\slashed{q}_2, \slashed{q}_1; p)$ the integral equation to solve reads 
\begin{align}
\label{eqn:BSE}
	T(\slashed{q}_2, \slashed{q}_1; p)= V(\slashed{q}_2, \slashed{q}_1; p) 
	+
	i\int\frac{d^d l}{(2\pi)^d}V(\slashed{q}_2, \slashed{l}; p) 
	S(\slashed{p}-\slashed{l})\Delta(l)T(\slashed{l}, \slashed{q}_1; p) \ ,
\end{align}
where $S$ and $\Delta$ represent the baryon (of mass $m$) and the meson (of mass $M$) propagator respectively, that are given by $iS(\slashed{p}) ={i}/({\slashed{p}-m+i\epsilon})$ and $i\Delta(k) ={i}/({k^2-M^2+i\epsilon})$. 
Note that, in contrast to the rest of this review and more in line with baryon CHPT, baryon (mesons) masses are indicated with lowercase $m$ (capitalized $M$) in this subsection.
The suppressed channel indices in the above formulas include all relevant (meson-baryon) coupled channels, such that $T$, $V$, $S$ and $\Delta$ are matrices in channel space. For strangeness $S=0$ and electric charge $Q=+1\,$ one is left with the following channels: $\{p \pi^0,~n \pi^{+},~p\eta,~\Lambda K^+,~\Sigma^0 K^+,~\Sigma^+ K^0\}$.

The interaction kernel to be iterated in \cref{eqn:BSE} includes only contact-term contributions from leading and next-to-leading order (NLO) chiral Lagrangians. Baryon exchange diagrams (Born terms) are omitted in this approach. Note that  one-baryon exchange diagrams are often included in potential formalisms, see, e.g., Refs.~\cite{Oller:2000fj,Borasoy:2005ie,Khemchandani:2018amu, Guo:2023wes} leading to potentially unphysical singularities related to meson-meson-baryon intermediate on-shell states. 
For a related discussion for the negative $S=-1$ meson-baryon scattering see Refs.~\cite{Borasoy:2005fq,Oller:2006ss}. For the explicit form and position of singularities see Ref.~\cite{Pittler:2025upn}.
See also Refs.~\cite{Entem:2025siq} and Ref.~\cite{Oller:2018zts} discussing approaches in coupled channels that would allow to include $u$-channel processes while strictly respecting the cut structure associated with the right- and crossed-channels.
Above procedure provides the only way to treat chiral NLO corrections in the covariant BSE, without making use of the on-shell approximation or $S$-wave projection of the chiral potential, so that also a $P$-wave can be iterated. Separating the momentum space from channel space structures the chiral potential considered here takes the form:
\begin{align}
\label{eqn:coupling}
	V(\slashed{q}_2, \slashed{q}_1; p)=
    \underbrace{
        A_{WT}(\slashed{q_1}+\slashed{q_2})}_{\text{leading order CHPT}}
	+\underbrace{
        A_{14}(q_1\cdot q_2)+A_{57}[\slashed{q_1},\slashed{q_2}]+A_{M}(q_1\cdot q_2)+A_{811}\big(\slashed{q_2}(q_1\cdot p)+\slashed{q_1}(q_2\cdot p)\big)}_{\text{next-to-leading order CHPT (local terms)}}
    \,,
\end{align}
where the first matrix ($A_{WT}$) only depends on the meson decay constants $F_\pi ,\,F_K,\,F_{\eta}$, whereas  $A_{14}$, $A_{57}$, $A_{811}$ and $A_{M}$ also depend on the NLO low-energy constants as specified in the appendix of Refs.~\cite{Mai:2012wy,Mai:2013cka}. The loop diagrams appearing in the BSE Eq.~(\ref{eqn:BSE}) are in general divergent and require renormalization. In case of a strict chiral perturbation expansion, the terms can be renormalized  in a quite straightforward way, order by order, including, at a given order, all the counterterms absorbing the loop divergencies. On the other hand, the treatment of the divergencies of the BSE is known to be a complicated issue, see, e.g., Refs.~\cite{Nieves:1999bx, Borasoy:2007ku}, which is why here the loops are regularized by treating the regularization scale as a fitting parameter.

Finally, to solve the BSE one embraces the fact that the invariant Dirac-structures of \cref{eqn:coupling} induce a closed set (under iterations) of Dirac-structures. Thus, the iteration through the BSE can be solved exactly embedding the T-matrix in the latter space as 
\begin{align}
	T(\slashed{q}_2, \slashed{q}_1; p)=\sum_{i=1}^{20} \aleph_i \cdot \text{T}_i \ , \quad
	V(\slashed{q}_2, \slashed{q}_1; p)=\sum_{i=1}^{20} \aleph_i \cdot \text{V}_i\,,
\end{align}
where the coefficients ${\rm T}_i$ and $V_i$ are $6\times 6$ matrices in channel space, which only depend on the c.m. energy $\sqrt{s}$ after fixing the low-energy constants, and $\aleph:=\Big(\slashed{q_1}${}, $\slashed{p}\slashed{q_1}${}, 
$\slashed{q_2}\slashed{p}\slashed{q_1}${}, $\slashed{q_2}\slashed{q_1}${}, 
$\slashed{p}\slashed{q_1}(q_2\cdot p)${}, $\slashed{q_1}(q_2\cdot{}p)${}, 
$\slashed{q_2}(q_1\cdot p)${}, $\slashed{q_2}\slashed{q_1}${}, 
$(q_1\cdot p)(q_2\cdot p)${}, $\slashed{p}(q_1\cdot~p)(q_2\cdot~p)$, 
$(q_1\cdot p)${}, $\slashed{p}(q_1\cdot p)${}, $(q_2\cdot q_1)${}, 
$\slashed{p}(q_2\cdot q_1)${}, $\slashed{q_2}\slashed{p}${},
$\slashed{q_2}${}, $\slashed{p}(q_2\cdot p)${}, $(q_2\cdot p)${}, $\mathds{1}${}, 
$\slashed{p}\Big)$ is a vector in the $20$--dimensional space of invariant structures. Note that the scalar products are listed here as independent structures because we include the full off-shell dependence of the chiral potential in the BSE, which prevents us from writing them as simple functions of the Mandelstam variables $s$ and $t$. Reinserting now the latter decomposition into the BSE~\eqref{eqn:BSE} and collecting all ${\rm T}_i(s)$ on the r.h.s. one obtains the following expression
\begin{align}
\label{eqn:solBSe}
     \sum_{i=1}^{20} {\rm ~V}_i~\aleph_i(q_2, q_1; p)=
     \sum_{i=1}^{20} \Big(\aleph_i(q_2, q_1; p) - \sum_{j=1}^{20} {\rm ~V}_j \Big( i\int\frac{d^d l}{(2\pi)^d} \frac{\aleph_j({q}_2, {l}; p)
     (\slashed{p}-\slashed{l}+m) \aleph_i({l}, {q}_1; p)}{({l^2-M^2+i\epsilon})({(p-l)^2-m^2+i\epsilon})}\Big) \Big) {\rm T}_i(s)\,,
\end{align}
where $V$ is expressed as coefficients of the vector elements of  $\aleph$. The term in the inner parentheses has the crucial property that, due to Lorentz invariance, it is also an element of the Dirac-momentum subspace spanned by the elements of the vector $\aleph$,
\begin{align}
\label{eqn:property}
    \underset{a\in\aleph(q_2, l; p), b\in\aleph(l, q_1; p)}{\forall} : \underset{C\in\mathds{C}^{20}}{\exists} \qquad \int\frac{d^d l}{(2\pi)^d}  \frac{a(\slashed p-\slashed l+m)b}{\big((p-l)^2-m^{2}\big)\big(l^2-M^2\big)} = \sum_{k=1}^{20}C_k~\aleph_k(q_2, q_1; p)~.
\end{align}
The coefficients $C_i$ are here complex-valued functions of scalar quantities only, i.e., $s, M, m$ and scalar loop integrals determined utilizing Passarino-Veltman reduction. Consequently, the r.h.s. of Eq.~\eqref{eqn:solBSe} can be rewritten as 
\begin{align*}
    \sum_{i=1}^{20} {\rm ~V}_i~\aleph_i(q_2, q_1; p)=\sum_{i=1}^{20}\sum_{j=1}^{20} \Big(\delta_{ij} -  \tilde{C}_{ij} \Big) {\rm ~T}_j(s)\aleph_i(q_2, q_1; p)~,
\end{align*}
where the coefficients $\tilde{C}_{ij}$ are matrices in channel space, obtained from Eq.~\eqref{eqn:property} replacing $a=V$ and $b=\aleph_i$. In fact, the latter expression is a system of $20$ coupled linear equations which we can rewrite in an even more elegant way as
\begin{align}
\label{eqn:MATRsol}
    \mathbf{X}^{~j}_i{\rm ~T}_j(s) = V_i~,
\end{align}
where we have utilized the Einstein summation convention for $i,j\in[1,20]$ and $\mathbf{X }\in Mat^{20\times20}$, whereas each element is a complex valued channel space matrix defined by $\mathbf{X}_{ij}=(\delta_{ij} -\tilde{C}_{ij})$. 

If one is interested in scattering quantities, one can solve  equation~\eqref{eqn:MATRsol} for all 20 structures reducing them in the final step to two on-shell structures $\{\slashed{p},\mathds{1}\}$, c.f., Eq.~\eqref{eq:Hohler-general-structure}, ultimately reducing to partial-wave amplitudes etc.~\cite{Mai:2012wy}. In \cref{subsec:S11} we discuss such an application. 

\subsubsection{Statistical aspects and machine learning tools}
\label{subsec:statistics}
The interpretation of near-threshold enhancement phenomena is one of the central issues of resonance searches. Common procedures, sometimes referred to as top-to-bottom approaches, start from a model such as UCHPT, or DCC which include symmetry assumptions. Then, comparing the model to the data or fitting free parameters and performing analytic continuation to complex energies, resonance parameters are extracted, see \cref{sec:analytic} or Ref.~\cite{Mai:2025wjb} for an introductory and historical review. Alternatively, one can ask  if observed line shapes actually encode some information about the nature of the enhancement inducing resonances as, e.g., virtual, resonance, or bound states. Recently, there has been  notable progress in this direction, sometimes referred to as bottom-to-top approaches, often relying on the development of machine learning tools. The general workflow is that minimal assumptions are made for the form of the amplitude producing a large set of synthetic data,  being variations of line shapes with known resonance pole class, training a neural network. A trained network can recognize pattern in the line shapes associated with the pole category. It is then applied to the real experimental data providing probabilities for the presence of a resonance, bound states, etc..

\begin{figure}[tb]
    \centering
    \raisebox{-.5\height}{\includegraphics[width=0.58\linewidth]{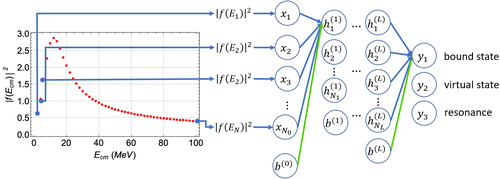}}
    \raisebox{-.5\height}{\includegraphics[width=0.38\linewidth]{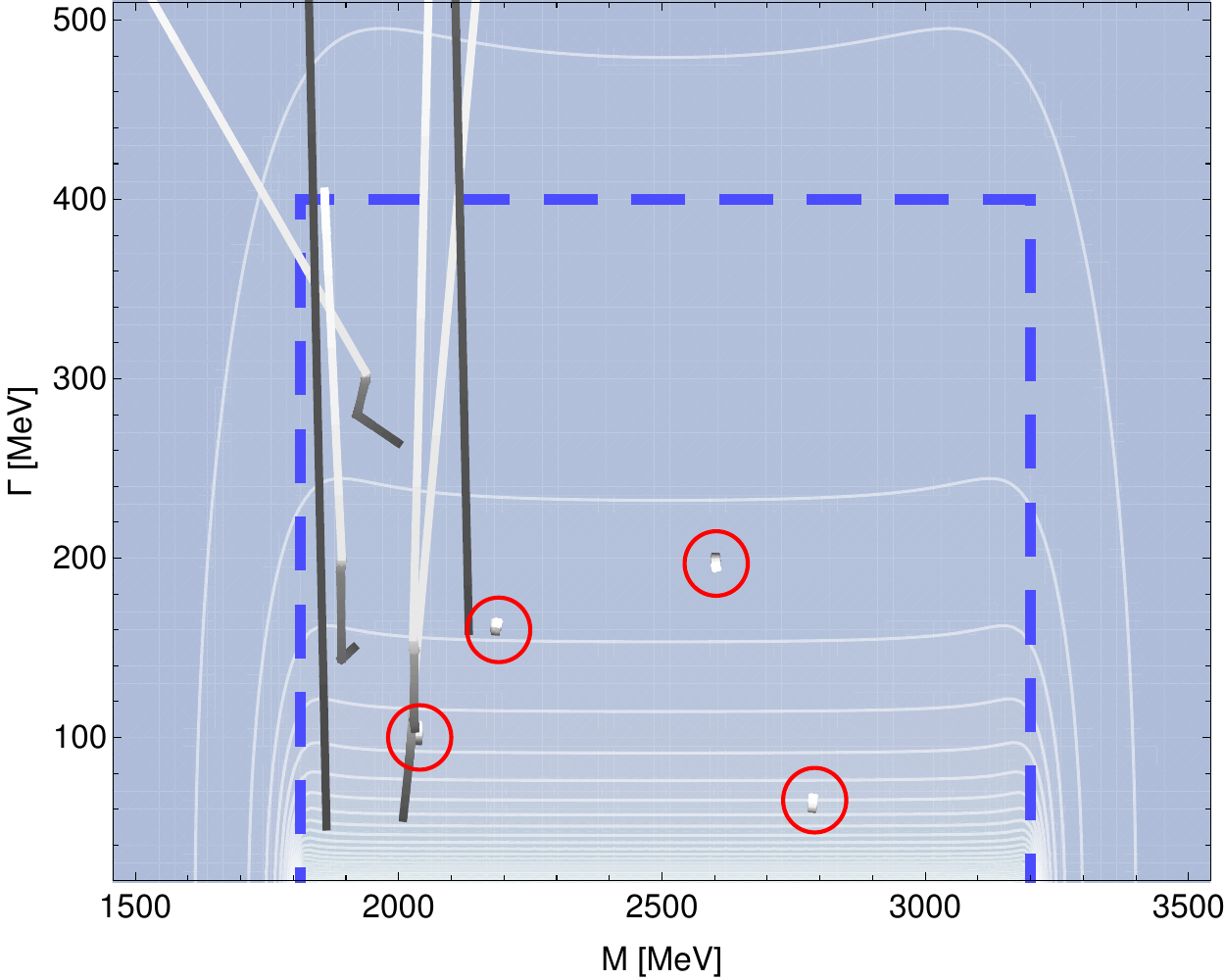}}
    \caption{{\bf Left}: Schematic picture of a deep neural network for S-matrix pole classification. Figure adapted from Ref.~\cite{Sombillo:2020ccg}. {\bf Right}: Resonance trajectories (thick solid lines) in the $M/\Gamma$-plane as function of the penalty parameter $\lambda$ penalizing an over-structured amplitude. The very short trajectories of significant resonances are highlighted by red circles. Some extra resonances are initially present in the chosen mass/width window, describing insignificant fluctuations leading to an improved $\chi^2$. Once the penalty is increased, the spurious resonances are pushed far into the complex plane, as the long trajectories show, while the data description does not suffer significantly. Figure adapted from Ref.~\cite{Landay:2018wgf}.}
    \label{fig:depsom}
\end{figure}

A program utilizing deep neural networks (DNN) was carried out in Ref.~\cite{Sombillo:2020ccg}. There, a deep feed-forward neural network (for a schematic representation of the latter see \cref{fig:depsom}) in classifying objects was utilized and shown to have high level of accuracy ($>90\%$) which however can drop with sizable background effects. 
DNN were also explored by JPAC~\cite{Ng:2021ibr}. In this proof of concept study, a neural network was trained on a synthetic data set containing $10^5$ line shapes produced with a generic form of an amplitude for  $J/\psi-p$ scattering. Each of these line shapes was labeled by the characteristics of the pole being a virtual or bound state, and the corresponding Riemann sheets. Through this (including the necessary validation procedures), the DNN learned how the nature of the state is related to the details of the pertinent line shape. Applying this so-called DNN classifier to the experimental data one can, indeed, obtain a probability for the nature of the state as favored by the data. This was done for the LHCb data for the $P_c(4312)$ signal, leading to a virtual state interpretation favored also by the previous traditional amplitude study~\cite{Fernandez-Ramirez:2019koa}. Studies of coupled-channel effects within  DNN-type  approaches can be found in Refs.~\cite{Sombillo:2021rxv}, see also Refs.~\cite{Santos:2024bqr,Ng:2024xye,Darulis:2022brn} for further applications and Ref.~\cite{Alghamdi:2023emm} for an example of how detector effects can be removed from the identification of multi-particle final states. Related approaches~\cite{Malekhosseini:2024eot,Co:2024bfl} aim at distinguishing  kinematic triangle singularities (see 
\cref{subsec:Cahill-Triangle}) from genuine resonance effects. For related reviews see  Refs.~\cite{JPAC:2021rxu,Aarts:2025gyp}.

In the context of spectroscopy, there are various ``blindfolded'' ways to test new resonances, without the need of manual intervention. These methods are robust in the sense that they allow for a relative model comparison even if the overall $\chi^2$ is not excellent. 
Bayesian inference to determine the resonance spectrum was introduced into baryon spectroscopy by the Ghent group~\cite{DeCruz:2012bv,DeCruz:2011xi}. In a related context, the necessary precision of data to discriminate models was determined in Ref.~\cite{Nys:2016uel}. 
Another method for the partial-wave analysis of mesonic systems was presented in Ref.~\cite{Guegan:2015mea}, see also Ref.~\cite{Williams:2017gwf}. The Least Absolute Shrinkage and Selection Operator (LASSO) allows one to generate a series of models with varying partial-wave content, for an application at COMPASS, see Ref.~\cite{Ketzer:2019wmd}. 
Ideally, LASSO tests all combinations of partial waves for all numbers of partial waves. Realistically, this is not achieved as fits are usually nonlinear, have different local minima etc.
The non-convexity of the problem tied to this has been recently addressed in Ref.~\cite{Dong:2023yad} in connection with ways to make pole searches in partial waves more efficient.
The best model among all the ones generated by LASSO can then be selected in an additional step through cross validation or various criteria from information theory (BIC, AIC,\dots)~\cite{10.1111/j.1467-9868.2011.00771.x, Hastie2009}.

 Specifically, in Ref.~\cite{Landay:2018wgf} the procedure was applied to the  $\bar K N\to K\Xi$ world data base, including polarized and unpolarized differential cross sections, to determine the minimally needed set of resonances to describe the data. For photoproduction data in Ref.~\cite{Landay:2016cjw}, minimal models compatible with data were automatically identified with the same technique, minimizing the number of significant partial waves and the number of parameters within the nonzero waves. The basic idea is to provide a very flexible model based on general properties of scattering theory allowing for a very large number of resonances, but penalizing overfitting data fluctuations through unphysical resonances. Several scenarios are also tested on synthetic data providing more certainty in these novel techniques. As an example, in one of the approaches~\cite{Landay:2018wgf} the penalty procedure smoothly ``pushed'' the resonances not necessarily demanded by the data, out of the relevant mass/width window as depicted in the right panel of~\cref{fig:depsom}. 
Related forms of penalty functions are utilized in Refs.~\cite{Bydzovsky:2021rog, Petrellis:2022eqw} in combination with LASSO, Ridge regression and other methods for kaon photoproduction reactions. In Ref.~\cite{Fernandez-Ramirez:2008ixe} Genetic Algorithms (evolution inspired update framework) are applied within a realistic Lagrangian model of the pion photoproduction reaction. This may, indeed, have an advantage over traditional $\chi^2$ minimization methodology when multiple local minima are present.

On a more general footing, machine learning tools aim at replacing microscopic models by data driven methods extracting some fundamental characteristics of underlying dynamics. Clearly, this procedure is very much sensitive to the quality and quantity of data, and it suffers  under possible systematic uncertainties in the data. An example of possible pitfalls is given in Ref.~\cite{Bydzovsky:2006wy} using kaon photoproduction data with a $\Lambda$-baryon in the final state. In this regard, the famous d'Agostini bias is worth mentioning, which systematically favors smaller model values when a normalization error in the experimental data is implemented in the naive way~\cite{DAgostini:1993arp}.
See also Ref.~\cite{Ball:2009qv} in which a way is discussed to avoid bias when fitting multiple normalization constants from different experiments.

As a forward-looking perspective for necessary development we emphasize that besides controlled systematics --such as choice of architecture-- a major challenge is the question of how to include symmetries in machine learning approaches. The use of observed symmetries is, indeed, what helped  particle physics to achieve a high level of predictability in the first place. In DCC models this is addressed through implementation of phenomenological Lagrangians. In some  idealized
physics problems neural networks can indeed learn symmetries by themselves~\cite{Liu:2021azq,Iten:2020ohp}. Still the question is how to implement this in a more data-sparse, realistic scenario for the example of meson-photoproduction off the nucleon discussed in this review. Perhaps one practical approach is to ``bake-in'' the symmetries directly into the architecture~\cite{Muller:2022grp,Mattheakis:2019tyi} possibly by utilizing existing techniques of DCC models.

\subsection{Finite-volume extensions}
\label{sec:LQCD}
\subsubsection{Two-body quantization condition and L\"uscher formula}

Lattice QCD provides information about the spectrum, structure, and interactions of hadrons as they emerge from quark-gluon dynamics. This numerical approach relies on Monte-Carlo sampling of correlation functions on Euclidean space-time evaluated in a finite (space-time) volume. Due to this, real-time observables such as scattering amplitudes cannot be accessed directly. Instead, as pioneered by L\"uscher for two-body systems~\cite{Luscher:1985dn, Luscher:1986pf, Luscher:1990ux}, the dependence of energy eigenvalues on the volume size can be used to single out dynamics of the system for not too small volumes. In this section, we briefly motivate the L\"uscher method suitable for two-body systems and discuss how it is related to DCC approaches. Then we show the connection between DCC models and the Finite-Volume Unitarity (FVU) approach for three-body systems. For specialized reviews on the topic see Refs.~\cite{Briceno:2017max,Hansen:2019nir,Mai:2021lwb,Mai:2022eur}.

Lattice QCD calculations are performed in a finite volume because of the finiteness of computational resources. This necessitates the need for boundary conditions, for simplicity most frequently periodic boundary conditions are chosen equating wave functions $\psi(x)=\psi(x+L)$ in each space dimension. The system, therefore, exhibits a discrete set of real energy eigenvalues. An equation which maps the latter to complex-valued infinite-volume scattering amplitudes is referred to as \emph{quantization condition}. 

To illustrate this in an example of two-to-two scattering, we follow the simplified derivation of Ref.~\cite{Doring:2011vk}. Consider two stable scalar particles interacting through some potential $V$ projected to the $S$-wave.  Suppressing for the following discussion the momentum dependence of the potential, $V=V(E^2)$, the scattering amplitude can be obtained  
as a solution of the scattering equation,
\begin{align}
\label{eq:fv-scattering}
    T&=V+VGT
     =\frac{1}{V^{-1}-G}
    \quad
    \text{for}
    \quad
    G=\int\limits^{|\bm{q}|<q_{\rm max}}
    \frac{\dv{^}3\bm{q}}{(2\pi)^3}\frac{1}{2\omega_1(\bm{q})\,\omega_2(\bm{q})}
    \frac{\omega_1(\bm{q})+\omega_2(\bm{q})}
    {E^2-(\omega_1(\bm{q})+\omega_2(\bm{q}))^2+i\epsilon}\,,
\end{align}
for total energy $E$ and particle masses $m_1,\,m_2$ with corresponding energies $\omega_{1,2}(\bm{q})=\sqrt{m_{1,2}^2+\bm{q}^2}$. Here the two-particle loop integral is regularized with some sharp cutoff $q_{\rm max}$. We note that the loop function $G$ is a relativistic form of the integral over the non-relativistic Green's function $1/(E-H_0)$ and thus $T\sim 1/(E-H)$. The imaginary part of the loop function is given by $\Im G=-q_{\rm cm}/(8\pi\sqrt{s})$ with $q_{\rm cm}$ from Eq.~\eqref{pcm}. This amplitude fulfills two-body unitarity and can be written in terms of the phase-shift as 
\begin{align}
    T=\frac{-8\pi E}{q_{\rm cm}\cot\delta-iq_{\rm cm}}
    \quad
    \text{for}
    \quad
    q_{\rm cm}\cot\delta=-8\pi E(V^{-1}-\Re G) \ .
\end{align}

Consider now the same system in a finite volume or, more specifically, a 3-dimensional cube with side length $L$ with periodic boundary conditions. In that system momenta become discretized, i.e., $\bm{q}\in\mathds{R}^3\to \bm{q}\in\aleph_L:=\{(2\pi/L)\bm{n}|\bm{n}\in\mathds{Z}^3\}$ such that the loop function in \cref{eq:fv-scattering} becomes
\begin{align}
    G\to \tilde G=\frac{1}{L^3}\sum_{\bm{q}\in\aleph_L}^{|\bm{q}|<q_{\rm max}}
    \frac{1}{2\omega_1(\bm{q})\,\omega_2(\bm{q})}\,\,
    \frac{\omega_1(\bm{q})+\omega_2(\bm{q})}
    {E^2-(\omega_1(\bm{q})+\omega_2(\bm{q}))^2}\,.
    \label{eq:fv-scattering-3}
\end{align}
Note that $\tilde G$ is a real but singular function, with poles emerging whenever the on-shell condition $E=\omega_1(\bm{q})+\omega_2(\bm{q})$ is fulfilled for any $\bm{q}\in\aleph_L$.
Furthermore, the above equation explicitly provides the volume dependence of $\tilde G$.
In contrast to the Green's function $G$, the interaction $V$ does not contain any configurations in which the two-particle system is on-shell, i.e., $V\in \mathds{R}$ in the physical region. In fact, this is a prerequisite for two-body unitarity. Therefore, the infinite-volume interaction equals the finite-volume interaction, $V\approx\tilde V$ up to small contributions. One can think of $V$ containing, e.g., tadpole loops for which the integration is replaced by a summation in finite volume, or $t$-channel exchange of off-shell particles. In contrast to $G$, these off-shell terms do not exhibit poles in the physical region; the sum that replaces the integral, therefore, approximates the integral well. The differences are \emph{exponentially suppressed} according to the regular summation theorem, $\sim e^{-ML}$ where $M$ is the characteristic energy scale of the problem, e.g., the pion mass.

With this information we can use the finite-volume loop function $\tilde G$ to define a finite-volume quantity $\tilde T$
\begin{align}
    T\to \tilde T=\frac{-8\pi E}{q_{\rm cm}\cot\delta-(-8\pi E)G_L}
    \quad
    \text{with}
    \quad
    G_L=\tilde G-\Re G\, ,
    \label{eq:fv-scattering-4}
\end{align}
with a real-valued, volume-dependent function $G_L$. Just like in the infinite volume, the discrete energy eigenstates $\{E_L\}$ of the Hamiltonian correspond to poles of $\tilde T\sim 1/(E-H)$ except for the modified boundary conditions. Consequently, for an energy eigenvalue $E^*$ the phase shift at this energy is obtained as
\begin{align}
    q_{\rm cm}(E^*)\cot\vertarrowbox{\delta}{
            \fbox{\begin{tabular}{c}
                Experiment\\Effective Field Theories\\\ldots
            \end{tabular}
        }}(E^*)=(-8\pi E^*)G_L(E^*)
        ~~\text{for}~~
        E^*\in\vertarrowbox{\{E_L\}}{\fbox{lattice QCD}}
\label{eq:fv-scattering-5}
\end{align}
and $\hat q=q_{\rm cm}(E^*)L/(2\pi)$. It is notable that the right-hand side is proportional to the Lüscher zeta-function ${\cal Z}_{00}$ up to regular and, thus, exponentially suppressed terms $e^{-ML}$,
\begin{align}
    (-8\pi E^*)G_L(E^*)
    =
    \frac{2}{\sqrt{\pi}L}\,{\cal Z}_{00}(1,\hat q^2)+\mathcal{O}(e^{-ML})\,,
\label{eq:fv-scattering-6}
\end{align}
see Refs.~\cite{Beane:2003yx,Doring:2011vk} for an explicit derivation. The quantization condition~\eqref{eq:fv-scattering-5} allows one to either predict the finite volume spectrum provided knowledge of the phase shift (e.g., from experiment or an effective theory), or, given a set of eigenenergies $\{E_L\}$ determined by lattice QCD, to calculate a set of phase shifts at these energies as schematically indicated by the arrows in \cref{eq:fv-scattering-5}. Note that there is no ultraviolet divergence for $G_L$ as $q_\text{max}\to \infty$. If one operates with an explicit cutoff regularization the integrand/summand is usually smoothened to make the cutoff effect exponentially suppressed with the volume~\cite{Doring:2011vk}.

The present discussion skips the special role of the angular momenta leading in the finite volume to the decomposition into irreducible representations of the cubic group, and we also skip extensions to multichannel systems. For more details on this we refer the reader to Refs.~\cite{He:2005ey,Gockeler:2012yj,Morningstar:2017spu,Doring:2011vk,Doring:2012eu,Lee:2020fbo,Pelissier:2011ib,Culver:2019qtx,Lee:2017igf}. The key takeaway point of the above discussion is, however, that unitarity identifies all on-shell configurations of two particles which then allows one to single out volume independent terms connecting quantities accessible on the lattice and the infinite volume/experiment. This methodology is also at the core of the FVU approach in deriving the three-body quantization condition proposed in Ref.~\cite{Mai:2017bge} and described in \cref{3BQC}.

\subsubsection{DCC approaches for two-body systems in finite volume}
Before turning to three-body systems we summarize the findings of Ref.~\cite{Doring:2011ip} on DCC approaches in the finite volume. This is closely related to the previous introductory discussion in that it also relies on explicit two-body unitarity. In addition, we discuss different boundary conditions and higher angular momentum potentials.
Starting with the partial-wave projected DCC equation in the meson-baryon normalization~\eqref{scattering} and replacing the integral with a sum as in Eq.~\eqref{eq:fv-scattering-3}, one obtains the scattering equation in discretized momentum space,
\begin{align}
    T^{\rm (BC)}(E,p',p)=V(E,p',p)+\frac{2\pi^2}{L^3}
    \sum_{i=0}^\infty\,\vartheta^{\rm (BC)}(i)
    \frac{V(E,p',q_i)\, T^{\rm (BC)}(E,q_i,p)}{\sqrt{s}-E(q_i)-\omega(q_i)},\quad 
    q_i=f^{\rm BC}(i) \ ,
    \label{scattdisc}
\end{align}
where $q_i$ is
the distance from the origin  to the $i$-th neighbors, called ``shell'', depending on the imposed boundary conditions (BC). 
All quantities in this equation are real and the potential $V$ automatically contains the correct threshold behavior due to its partial-wave projection. The  $\vartheta^{\rm (BC)}$ are the BC-dependent multiplicities counting the number of lattice points on shell $i+1$. In comparison to Eq.~\eqref{scattering}, the integration measure is replaced as $\dv{q}q^2\to 2\pi^2\,\vartheta^{\rm (P)}(i)/L^3$ in the finite volume. For periodic (P) boundary conditions, the $\vartheta^{\rm (P)}$ series (1,6,12,\dots) is given by the coefficients of the Taylor expansion around $x=0$ of the function
$g^{\rm (P)}(x)=[\vartheta_3(0,x)]^3$, 
\begin{align}
   g^{\rm (P)}(x)=\sum_{i=0}^\infty \vartheta^{\rm (P)}(i)\, x^i \ ,\quad \text{where}\quad
   \vartheta_3(0,x)=\sum_{k=-\infty}^\infty\,x^{k^2} \,,
\label{taylor}
\end{align}
denoting the elliptic $\vartheta$ function~\cite{OeisNumbers}. The distance to shell $i$ is simply $f^{\rm P}(i)=2\pi \sqrt{i}/L$. Another choice is given by antiperiodic (A) boundary conditions~\cite{Doring:2011vk}. In that case, the multiplicities, $\vartheta^{(\rm A)}= 8,24,24,\dots$ are given by the coefficients of the Taylor expansion [cf. Eq.~(\ref{taylor})] around $x=0$ of the function $g^{\rm (A)}(x)=[\vartheta_2(0,\sqrt x)]^3\,x^{-3/8}$, where $\vartheta_2$ is the elliptic $\vartheta$ function~\cite{OeisNumbers}
\begin{align}
    \vartheta_2(0,\sqrt x)=2\,x^{1/8}\sum_{k=0}^\infty\,x^{k(k+1)/2} \ ,  
\end{align}
and the distance from origin to shell $i$ is now given as $f^{\rm A}(i)=\frac{2\pi}{L}\,\sqrt{\frac{8i+3}{4}}$ with $i=0,1,\dots$, provided that the origin is now in the center of a momentum cell and no longer at the edge as for periodic BC. 

Comparing Eq.~\eqref{scattdisc} and Eq.~\eqref{eq:fv-scattering-4} we note that the former allows for explicit momentum dependence in $V$ while the latter is on-shell by construction rendering the scattering equation algebraic. In fact, if $V$ is derived from a Langrangian, its momentum is replaced by the on-shell momentum, $p\to p_\text{cm}$ for this purpose. We conclude that this replacement of putting momenta on-shell changes both the amplitude in the infinite volume and the energy spectrum in the finite volume. However, the difference between both formalisms, when it comes to mapping between finite and infinite volume, is exponentially suppressed as the approaches coincide on-shell. As exponentially suppressed contributions are model-dependent, anyway, there is no advantage of one formalism over the other. An exception is the case of a close-by left-hand cut. Such structures can be built explicitly into DCC approaches with their full momentum dependence. See
Refs.~\cite{Du:2023hlu,Meng:2023bmz,Hansen:2024ffk} for related approaches. 

Technically, the solution of Eq.~\eqref{scattdisc} proceeds just like the solution of the infinite-volume scattering equation~\eqref{scattering}
by discretization as explained in \cref{sec:realmom}. In fact, the only difference lies in the distribution of sample points and integration weights, that are replaced by multiplicities.

Note also that the situation radically changes when considering three-body systems discussed in \cref{3BQC}. Then, the momentum dependence of $V$ is compulsory due to three-body unitarity, and there is a whole range of on-shell momenta. In that case, the scattering equation can never be recast as a geometric series and remains a true integral equation. As a last remark, the determination of the two-body finite-volume spectrum from a Hamiltonian discussed here resembles the effective Hamiltonian approach for the same purpose~\cite{Hall:2013qba, Hall:2014uca, Liu:2015ktc, Liu:2016wxq}.

In the following, two predictions from Ref.~\cite{Doring:2011ip} of the finite-volume spectrum from existing DCC approaches are presented, for the $\Lambda(1405)$ (baryonic) and for the  $f_0(500)$ (mesonic) quantum numbers and the respective irreps $G_{1u}(0)$ and $A_1^+$. For  the $\Lambda(1405)$ and the $f_0$ several lattice QCD collaborations have produced data and finite-volume analyses leading to the first determinations of the respective resonance poles from QCD, see Refs.~\cite{BaryonScatteringBaSc:2023ori,BaryonScatteringBaSc:2023zvt, Menadue:2011pd, Hall:2014uca, Molina:2015uqp, Guo:2023wes, Zhuang:2024udv, Bruns:2024yga} and Refs.~\cite{LatticeStrongDynamics:2023bqp,  Rodas:2023nec, Rodas:2023gma, Mai:2019pqr, Culver:2019qtx, Guo:2018zss,Fu:2017apw,Doring:2016bdr,Briceno:2016mjc}, respectively. 

\begin{figure}[tb]
    \centering
    \includegraphics[width=0.37\textwidth]{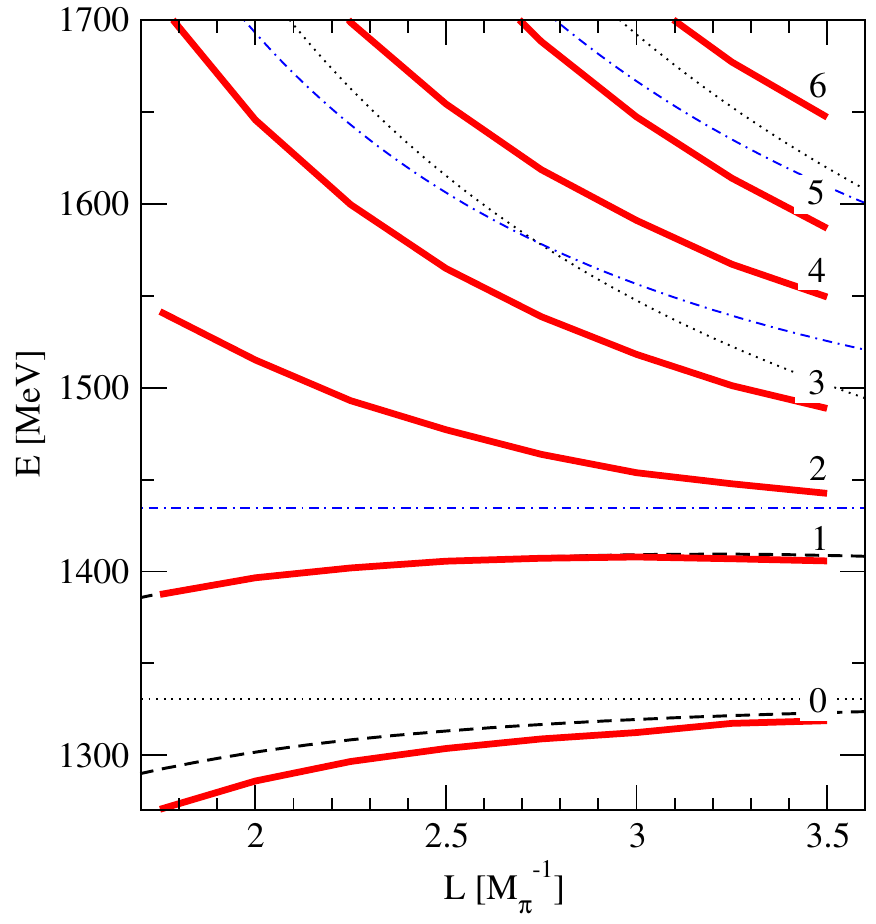}
    \hspace{0.03\textwidth}
    \includegraphics[width=0.43\textwidth]{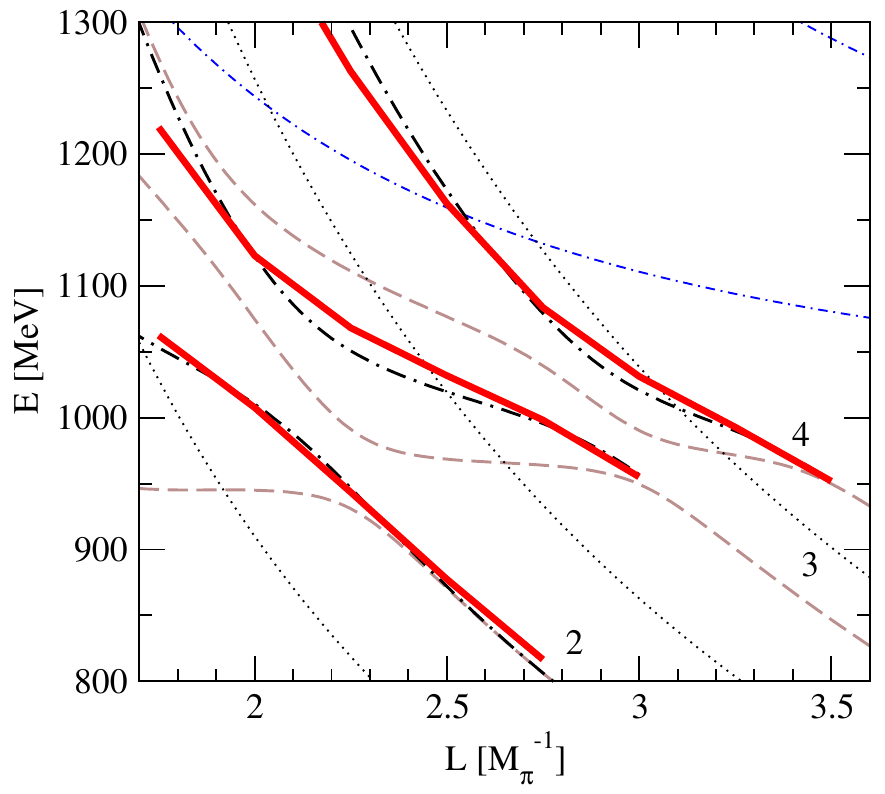}
    \caption{{\bf Left}: Spectrum $E(L)$ of the $I=0, \,S=-1$, $J^P={\frac{1}{2}}^-$ meson-baryon sector
    [$\Lambda(1405)$] for periodic boundary conditions in the $G_{1u}(0)$ irrep. Solid lines: results for the $\bar KN$ DCC Juelich model~\cite{Mueller-Groeling:1990uxr,Haidenbauer:2010ch}. 
    Dashed lines: For comparison, 
    the first two levels as obtained from the chiral unitary approach of Ref.~\cite{Oset:1997it}. 
    Dotted (dash-dotted) lines: free $\pi\Sigma\,(\bar KN)$ levels.
    {\bf Right}:  Finite-volume spectrum of the scalar-isoscalar sector with
    anti-periodic boundary conditions ($\theta=\pi$) in the $\bar KK$ channel, for the
    DCC model of Refs.~\cite{Janssen:1994wn,Krehl:1996rk} (solid 
    lines). For comparison, the result with anti-periodic boundary conditions for the chiral
    unitary approach of Ref.~\cite{Oller:1997ti} is shown with thick dash-dotted lines. Also, the result with periodic boundary conditions in the $A_1^+$ irrep is shown (dashed lines). Dotted (thin dash-dotted) lines: free $\pi\pi$\,(free anti-periodic $\bar KK)$ levels.
    }
    \label{fig:lamandf0}
\end{figure}

To produce the finite-volume spectra in the $I=0,\,S=-1$ meson-baryon sector shown to the left in Fig.~\ref{fig:lamandf0}, the dynamical coupled-channel model 
of Refs.~\cite{Mueller-Groeling:1990uxr,Haidenbauer:2010ch} is considered. It
includes the channels $\pi \Lambda$, $\pi \Sigma$, and $\bar K N$. The attraction in coupled channels is so strong that the $\Lambda(1405)$ appears dynamically generated.
In this approach, also the channels  $\bar K \Delta$, $\bar K^* N$, and $\bar K^*
\Delta$ are included.  However, these channels contribute only effectively to the direct $\bar K N$ interaction in form of box diagrams \cite{Mueller-Groeling:1990uxr}. Below the three-body thresholds, these channels are physically closed, the box diagrams are real, and one can refrain from  explicitly discretizing the integrals related to these box diagrams; the three-body system is more intricate, anyway, as outlined below. The level structure below the $\bar KN$ threshold is interesting because of a conjectured two-pole structure of the $\Lambda(1405)$ \cite{Oller:2000fj, Jido:2003cb}. Indeed, also the hadron-exchange approach of Ref.~\cite{Mueller-Groeling:1990uxr}, considered here, contains two poles below the  $\bar KN$ threshold as pointed out in Ref.~\cite{Haidenbauer:2010ch}. The two different approaches both predict an attractive $\pi\Sigma$ threshold level (level 0) and another level close to but below the $\bar K N$ threshold (level 1). The non-interacting threshold levels are given by the horizontal lines. Note that while both amplitudes contain two $\Lambda(1405)$ poles, the latter are not manifested in additional energy eigenvalues. 

For the scalar-isoscalar finite-volume spectrum shown to the right in Fig.~\ref{fig:lamandf0}, the 
interaction potential developed in Refs.~\cite{Janssen:1994wn,Krehl:1996rk} is used for which the coupling between the $\pi\pi$ (or $\pi\eta$) and $\bar KK$ channels is taken into account. The $\sigma\equiv f_0(500)$, $f_0(980)$,  and $a_0(980)$ resonances  all appear dynamically generated, i.e., without the need of corresponding $s$-channel exchanges. On the other hand, the model does include also a genuine resonance in form of an $s$-channel pole diagram, tentatively called $f_0(1400)$ in Ref.~\cite{Janssen:1994wn}, whose bare mass is at $1520\,\MeV$. The finite-volume spectrum exhibits a characteristic avoided level crossing at around  $980\,\MeV$ shown with the dashed lines. In general, avoided level crossing is not only a signal for a resonance but also for $S$-wave thresholds. Here the $f_0(980)$ coincides with the $K\bar K$ threshold, making it difficult to disentangle both effects. 
It is, therefore, useful to modify the BC to obtained more information on the system. Using the multiplicities $\vartheta^{\rm A}$ and shell radii $f^{\rm A}$ quoted above, the resulting energy eigenvalues are shown with the thick red solid lines. Indeed, the avoided level crossing disappears with anti-periodic BC~\cite{Bernard:2010fp, Doring:2011vk}. 

To conclude this part, it should be mentioned that chiral unitary approaches in which the full, four-dimensional BSE~\eqref{eqn:BSE} is solved, can also be matched to the finite volume, allowing for predictions or data analyses. This has been demonstrated in Ref.~\cite{Doring:2013glu}: For various coupled channels, extensions to twisted boundary conditions, and different hadron masses, the spectrum of the $N(1535)$ and $N(1650)$ was predicted.

\subsubsection{Connection of DCC approaches to three-body finite-volume systems}
\label{3BQC}

Since a few years, it has become possible to calculate three-body systems from LQCD due to two closely intertwined factors. First, increased computational capacities and algorithmic developments have made the implementation of multi-hadron operators possible, see  
Refs.~\cite{Lang:2016hnn,Kiratidis:2016hda,Liu:2016uzk,Horz:2019rrn,Culver:2019vvu,Fischer:2020jzp,Hansen:2020otl,Alexandru:2020xqf,Blanton:2021llb,NPLQCD:2020ozd,Buhlmann:2021nsb,Mai:2021nul,Garofalo:2022pux,Draper:2023boj, Yan:2024gwp, Dawid:2025doq,Dawid:2024dgy}
for recent studies including lattice calculations. Such programs provide a discrete finite-volume energy spectrum, which encodes the full QCD dynamics. Relating this spectrum to the infinite-volume (coupled-channel) amplitude is accomplished by the previously introduced Quantization Conditions (QCs). See 
Refs.~\cite{Hansen:2019nir,Mai:2021lwb} for dedicated reviews in the three-hadron sector and Ref.~\cite{Mai:2021nul} for a first application to a three-body resonance using two channels. 
As mentioned in \cref{sec:threemesons}, there are also recent extensions to unequal masses and coupled channels~\cite{Draper:2023boj,Draper:2024qeh,Hansen:2020otl} in the context of an alternative approach.

Here, we discuss the connection of three-body quantization conditions in form of the FVU approach to DCC approaches. The infinite-volume aspect of three-body unitarity (IVU)~\cite{Mai:2017vot} is reviewed in \cref{sec:unitarity}. The FVU approach was developed in Ref.~\cite{Mai:2017bge}. Following these works, the three-body scattering amplitude of identical spinless particles, $\Phi\Phi\Phi\to\Phi\Phi\Phi$, is given by Eq.~\eqref{eq:t33full} as the sum of a connected piece and a once-disconnected piece.
The connected piece contains the three-dimensional isobar-spectator re-scattering piece $T$ that obeys Eq.~\eqref{eq:T}, which resembles a Lippmann-Schwinger equation with relativistic kinematics. The re-arrangement interaction $B$ is complex-valued as demanded by unitarity and given in Eq.~\eqref{eq:B}. Together with a real-valued term $C$ they determine the dynamics of the three-body scattering. 

The two-body Green's function is given in Eq.~\eqref{eq:tau}. We rewrite it in a slightly more general form using a K-matrix-like $\tilde K$ as 
\begin{align}
    \tau(\sigma_p)=
    \frac{1}{\tilde K^{-1}(\sigma_p)-N_{2}\underbrace{\int\frac{dkk^2}{E_k}\frac{v(k,P-k)v^*(k,P-k)}{\sigma_p-4E_k^2}}_{=:\Sigma}}
    \quad\text{for}\quad
    N_2,N_3,v,C,\tilde K \in\mathds{R}\,.
    \label{eq:fv-3body-5}
\end{align}
The normalization constants $(N_2, N_3)$ address possible isospin or other combinatorial coefficients while $v$ carries angular dependence of the isobar decaying to the asymptotically stable final states. Neither of these have free parameters which can all be absorbed into the remaining unspecified real-valued terms $C, \tilde K$. Since $\tau$ is proportional to the two-body scattering amplitude $\Phi\Phi\to\Phi\Phi$ up to angular factors, one can, indeed, exactly relate $\tilde K\sim q_{cm}\cot\delta_{\phi\phi\to\phi\phi}$. Similarly, one could hope to identify also $C$ with some sort of three-body K-matrix, see, e.g., Refs.~\cite{Hansen:2014eka}. However, this is in general not possible since any three-body observables can only be obtained from the knowledge of the $C$-term via the integral equation~\eqref{eq:T} plugged into \cref{eq:t33full}. Thus, this relation will always depend on the details of the integration like cutoff-dependence, which, for example, leads to a cycling behavior of this three-body force becoming infinite at specific choices of the cutoff~\cite{Bedaque:1998kg, Hammer:2017kms, Doring:2018xxx}. Therefore, this term has more analogy with a coupling constant than a K-matrix.

In view of the relation between the finite-volume spectrum accessible in lattice QCD calculations and observables such as line shapes or Dalitz plots, the key point of identifying on-shell configurations is accomplished by invoking unitarity, leading to on-shell configurations not only in the isobar-spectator propagation but also in the particle exchange interaction. This enables one to follow the same steps as used in deriving the two-body quantization condition~\cref{eq:fv-scattering-5} and replace all (intermediate) momenta as $\bm{q}\in\mathds{R}\to \bm{q}\in\aleph_L:=\{(2\pi/L)\bm{n}|\bm{n}\in\mathds{Z}^3\}$. In the denominator of \cref{eq:fv-3body-5} this induces a sum over relative momenta in the two-body subsystem ($\Sigma\to\tilde\Sigma$) similarly to \cref{eq:fv-scattering-3}. This transforms the integral equation~\cref{eq:T} to a real-valued matrix equation with singularities located at the eigen-energies (set $\{E_3^L\}$) of the Hamiltonian in the finite volume borrowing the same logic as the one used to derive two-body quantization condition before. Overall, and noting that for any matrix the singularities are given by the zeros of the determinant, we obtain the FVU three-body quantization condition
\begin{align}
    &\operatorname{det}[(\vertarrowbox{\tilde K^{-1}}{
    \fbox{
        \begin{tabular}{c}
        2-body input\\ 
        Eq.~\eqref{eq:fv-3body-5}
        \end{tabular}
    }}(E^*)-\Sigma^{FV}(E^*))E_L-( B(E^*)+\vertarrowbox{C}{
        \vertarrowbox{
            \fbox{
                \begin{tabular}{c}
                    Scattering equation\\
                    $T_3=v(\tau T\tau+\tau)v$\\
                    $T=(B+C)+\int (B+C)\tau T$
                \end{tabular}}
            }{
            \fbox{
                \begin{tabular}{c}
                    Line shapes\\
                    Dalitz plots\\
                    ...
                \end{tabular}}
            }
        }(E^*))]=0
    \quad
    \text{for}
    \quad
    E^*
    \in \vertarrowbox{\{E_3^L\}.}{\fbox{lattice QCD}}
    \label{eq:fv-3body-6}
\end{align}
Here, the determinant is taken with respect to all possible momentum configurations. This is an infinite-dimensional space $\aleph_L$ since the spectator can take on any of these momenta. Similarly to  \cref{eq:fv-scattering-5}, we ignore here the angular momentum decomposition and the corresponding decomposition into the irreducible representations of the cubic group in the finite volume. In any case, the methodology of relating lattice determined energy eigenvalues to observable scattering quantities is here not one-to-one one as in \cref{eq:fv-scattering-5}. Instead, the energy eigenvalues determine the roots of the above equation, thus, constraining the volume-independent quantities $C$ and $\tilde K$. In fact, $C$ is fit to the three-body energies $E_3^L$, while input from the two-body lattice QCD system enters through $\tilde K$. It is only in the second step, by putting these quantities in the corresponding infinite-volume scattering equation, that the experimentally accessible quantities (line shapes, Dalitz plots, pole positions) can be retrieved as indicated in Eq.~\eqref{eq:fv-3body-6}. This is of course valid for all currently available and employed approaches to the three-body quantization condition~\cite{Hansen:2014eka,Hammer:2017kms,Mai:2017bge}.

The discussion above serves as an anchor point relating methods central to this review with  results of  modern lattice QCD approaches. Most technical details are left out for the sake of clarity. For more dedicated reviews on this topic we refer the reader to Refs.~\cite{Briceno:2017max,Hansen:2019nir,Mai:2021lwb,Mai:2022eur}. In summary, DCC approaches can be extended to tackle finite-volume multi-hadron systems.
The main connection, to, e.g., the FVU approach in the finite volume becomes, however, clear when considering three-body systems. Indeed, 
three-body unitarity serves as the prime tool to identify on-shell configurations in interaction and propagation, both guiding the construction of infinite-volume amplitudes but also finite-volume quantization conditions. First Lattice QCD based studies of resonant three-meson systems are available in Ref.~\cite{Mai:2021nul} for the $a_1(1260)$ and in Ref.~\cite{Yan:2024gwp} for the $\omega(782)$.
Extensions of the method to excited baryon systems are underway~\cite{Severt:2022jtg}.

\section{Light and strange baryons}
\label{sec:nonstrange}
\subsection{Background}
The determination of the light baryon spectrum, its properties, and its interpretation has received increased attention in the past two decades. Situated in the intermediate energy region between low and high-energy QCD,  light baryonic resonances represent the excited states of the proton and neutron, making them a bridge between matter as we know it, and the regime where QCD is understood perturbatively~\cite{Gross:2022hyw, Deur:2023dzc}. 

Theory predictions from quark models, dynamical quark and gluon 
approaches/functional methods, effective field theories, and lattice QCD were and are tested in large experimental campaigns at facilities like JLab, ELSA, MAMI, LEPS and other experiments around the world. So much experimental and theoretical work has been invested in the study of light baryons that we can only aim at providing a complete overview of reviews on the topic, with an incomplete list of papers tied to particular issues, in addition. 

The canonical literature for meson-baryon scattering and resonances contains the book by Bransden and Moorhouse~\cite{Bransden2015-vz}. The book by H\"ohler includes a technical discussion of (pion-nucleon) scattering theory~\cite{Hohler:1979yr} and is a widely used reference. Similarly, the book by Ericson and Weise~\cite{Ericson:1988gk} contains many aspects on pion-nucleon scattering, and the lecture series by Levi-Setti and Lasinski~\cite{alma991040892069706011} has many derivations of commonly used formulas in scattering with spin. An introductory work to K-matrix formalism was written by Chung et al.~\cite{Chung:1995dx}. A lecture series by Dalitz was published in Ref.~\cite{Dalitz_R_H_1962} discussing hadronic reactions and resonances, together with an introduction to dispersion theory. See also a monograph~\cite{Close:2007zzd} containing different contributions on ``Electromagnetic interactions and hadronic structure''.

First reviews on meson-baryon scattering often highlighted aspects of duality, see, e.g., Rosner~\cite{Rosner:1974sn}, Donnachie's review on partial-wave analysis~\cite{donnachie1973partial}, and a discussion by Dalitz~\cite{Dalitz:1970ga}, ``What is  resonance?'' A more recent review covering both experimental aspects but also models and theories for light  baryons was published by 
Klempt and Richard 15 years ago~\cite{Klempt:2009pi}. This review is a good starting point for the topic of light baryons;  it also contains a pedagogic introduction to quark models, see also Ref.~\cite{Richard:2012xw} by Richard.
An introduction to quark models was provided by Eichmann recently, as well, containing aspects of the flavor structure of baryons and functional methods, in addition~\cite{Eichmann:2022zxn} (see also an update in Ref.~\cite{Eichmann:2025wgs}). Mai recently provided a pedagogical introduction to resonances in Ref.~\cite{Mai:2025wjb} following a major review on the same topic as a key to understanding QCD~\cite{Mai:2022eur}. See also a discussion summary in Ref.~\cite{Lutz:2015ejy}.
Burkert, Eichmann, and Klempt most recently reviewed our current understanding of the nucleon excitation spectrum and the role of photo and electroproduction reactions~\cite{Burkert:2025coj} in that challenge.

Quark models for excited baryons and other hadrons~\cite{Isgur:1977ef, Isgur:1978wd, Isgur:1978wd, Capstick:1986ter, Ferraris:1995ui, Glozman:1996wq, Glozman:1997ag} were discussed in a dedicated review by Capstick 20 years ago~\cite{Capstick:2000qj}. 
Over the years, the approach has been further refined  in  relativistic quark models with
instanton-induced quark forces   ~\cite{Loring:2001kv, Loring:2001kx, Loring:2001ky, Ronniger:2012xp},  quark-diquark models~\cite{Santopinto:2004hw, DeSanctis:2014ria, Santopinto:2014opa, Barabanov:2020jvn}, hypercentral constituent quark models~\cite{Giannini:2001kb, Giannini:2015zia, Menapara:2020dhy, Menapara:2022ksj}, in covariant models~\cite{Gross:2006fg, Ramalho:2008ra, Ramalho:2009df, Ramalho:2010xj, Ramalho:2011ae, Marcucci:2015rca, Ramalho:2015qna} with predictions of electromagnetic properties including timelike form factors~\cite{Ramalho:2012ng}. Few-quark dynamics was even extended to five-body systems~\cite{Eichmann:2025gyz}. See also Refs.~\cite{ Vijande:2004he, Huang:2005gw, Garcilazo:2007eh, Bicudo:2016eeu, Zhong:2024mnt} for other quark models and related approaches for excited baryons.
A recent review~\cite{Ramalho:2023hqd} by Pe\~na and Ramalho is dedicated to form factors, their parametrization~\cite{Ramalho:2017muv, Ramalho:2016buz} and their connection to quark models. Hadronic reactions have been described in the framework of suitably extended constituent quark models, as well~\cite{Zhong:2011ht, He:2008uf, Golli:2011jk, An:2011sb, Golli:2009uk, Golli:2013uha, Golli:2016dlj, Xiao:2016dlf}, including even photoproduction reactions of baryon resonances~\cite{Bermuth:1988ms, Scherer:1987hr}.
The $1/N_c$ expansion of baryons and its role in spectroscopy was discussed in Refs.~\cite{Jenkins:1995td, Carlson:1998vx, 
Carlson:1998gw, 
Schat:2001xr, Goity:2002pu, Goity:2004ss, Garcia-Recio:2006uwg, Scoccola:2007sn, 
Cherman:2009fh,
Matagne:2016gdc, Matagne:2012tm} and reviewed in Ref.~\cite{Matagne:2014lla} by Matagne.

The relativistic light-front quark model~\cite{Aznauryan:2017nkz, Aznauryan:2012ec, Aznauryan:2012ba, CLAS:2009ces, Aznauryan:2009mx, Aznauryan:2004jd} and its application to electroproduction reactions was revied by Azauryan and Burkert~\cite{Aznauryan:2011qj} and by Aznauryan, Bashir, Braun, Brodsky, and Burkert in Ref.~\cite{Aznauryan:2012ba}. See also \cref{sec:electroproduction} that contains a more detailed introduction to electroproduction reactions with its own literature overview. We mention here only the short reviews by Carman and Mokeev~\cite{Carman:2023zke, Mokeev:2022xfo}, and also by Dai, Haidenbauer, and Mei{\ss}ner for timelike form factors~\cite{Dai:2024lau}.  New perspectives of exploring hadron structure through transition GPDs were recently summarized in a whitepaper~\cite{Diehl:2024bmd}.

Holographic QCD is an approach with extra spatial dimensions to model hadrons, as explained in an introductory text by Erlich~\cite{Erlich:2014yha}. A suggestive example of a particle propagating in a compact extra dimension is provided to illustrate that excited  hadrons can be interpreted as Kaluza-Klein modes. 
Holographic QCD for baryon resonances has been proposed in Refs.~\cite{Erlich:2005qh, deTeramond:2005su}, see also Refs.~\cite{Karch:2006pv, deTeramond:2008ht}. The
approach of Ref.~\cite{deTeramond:2005su} is highly successful in organizing the hadron
spectrum, although it underestimates the spin-orbit separations of the $L = 1$ orbital states. In Ref.~\cite{Forkel:2008un} a two-parameter mass formula from quark-diquark correlations emerging from AdS/QCD was derived which reproduces the baryon mass spectrum with surprising precision.
In Ref.~\cite{Brodsky:2014yha} this and related approaches were reviewed by Brodsky, de Teramond, Dosch, and Erlich.

Dyson-Schwinger approaches and functional methods have been developed to understand hadrons as reviewed by Roberts and Maris already 20 years ago~\cite{Maris:2003vk, Roberts:2007ji}. These methods were recently reviewed by Ding, Roberts, and Schmidt~\cite{Ding:2022ows}, following reviews by Roberts, Richards, Horn, and Chang~\cite{Roberts:2021nhw}, Burkert and Roberts~\cite{Burkert:2017djo}, and Bashir et al.~\cite{Bashir:2012fs} reflecting the progress in the baryon sector~\cite{Roberts:2011cf, Chen:2012qr, Chen:2019fzn} including the prediction of Transition Form Factors (TFFs)~\cite{Wilson:2011aa, Chen:2018nsg, Lu:2019bjs, Raya:2021pyr}, see \cref{sec:electroproduction} of this review. Furthermore, there is the review by Clo\"et and Roberts~\cite{Cloet:2013jya} with emphasis on the connection of the Schwinger-Dyson formalism to observables like form factors of pion and nucleon, and TFFs of excited baryons. 

There are reviews by Fischer and Eichmann~\cite{Fischer:2006ub,Eichmann:2016yit,Eichmann:2025wgs} for functional methods, and Refs.~\cite{Eichmann:2016hgl, Sanchis-Alepuz:2014sca, Sanchis-Alepuz:2014xua, Eichmann:2009qa, Chen:2023zhh, Yin:2023kom, Eichmann:2018adq, Segovia:2015hra} in the context of excited baryons. The role of diquarks was reviewed by Barabanov et al.~\cite{Barabanov:2020jvn}, see also Refs.~\cite{Francis:2022stl,Francis:2022fdj} for connections to Lattice QCD.

(Excited) Meson and (excited) baryon degrees of freedom form the basis for  Chiral Perturbation Theory (CHPT) approaches~\cite{Scherer:2009bt} to excited baryons~\cite{Pascalutsa:1998pw, Pascalutsa:1999zz, Fettes:2000xg, Fettes:2000bb, Fettes:2000gb, Becher:2001hv, Pascalutsa:2002pi, Fettes:2001cr, Borasoy:2006fk, Pascalutsa:2007yg, Baru:2007wf, Bernard:2009mw, Siemens:2014pma, Gegelia:2016xcw, Gegelia:2016pjm, Vonk:2022tho, Alharazin:2022wjj,Li:2025man}. See, for example, the reviews by Bernard, Kaiser, and Mei{\ss}ner \cite{Bernard:1995dp},  by Pascalutsa, Vanderhaeghen, and Yang~\cite{Pascalutsa:2006up} on the $\Delta(1232)$ resonance, and by Bernard ~\cite{Bernard:2007zu}. 

Through the unitarization of a chiral (or other) interactions to a given order~\cite{Guo:2017jvc}, light baryons~\cite{Feijoo:2024bvn, He:2017aps, He:2015yva, Roca:2015tea, Garzon:2014ida, Oset:2012ap, Garzon:2012np, Khemchandani:2011mf, Xie:2011uw, Bruns:2010sv, Sarkar:2010saz, Xie:2010ig, Mai:2009ce, Gonzalez:2008pv, Lutz:2005ip, Sarkar:2004jh, Kolomeitsev:2003kt, Garcia-Recio:2003ejq, Inoue:2001ip, Nieves:2001wt, Lutz:2001dr, Lutz:2001yb, Kaiser:1995cy, Siegel:1988rq} or light-strange baryons~\cite{Guo:2023wes, Sadasivan:2022srs, Sadasivan:2018jig, 
Feijoo:2018den, Cieply:2016jby, Khemchandani:2016ftn, Ramos:2016odk, Khemchandani:2013nma, 
Nakamura:2013boa,
Mai:2012dt, Jido:2003cb, Garcia-Recio:2002yxy, Oset:1997it, Oset:2001cn, Oller:2000fj, Kaiser:1995eg} can emerge (see remarkable progress for unifying these sectors in Ref.~\cite{Lu:2022hwm}). Their description as dynamically generated states can be tested by coupling the photon to the constituents~\cite{Borasoy:2007ku} and predicting electromagnetic properties~\cite{Bruns:2022sio, Kim:2021wov, Mai:2014xna, Sharma:2012jqz, Mai:2012wy, Ruic:2011wf, Gasparyan:2010xz, Sun:2010bk, Doring:2010fw, Doring:2010rd, Doring:2009qr, Doring:2009uc, Ajaka:2008zz, Jido:2007sm, Geng:2007hz, Doring:2007rz, Roca:2006pu, Doring:2006pt, Doring:2006ub, Doring:2005bx, Kaiser:1996js}.
Analogously, coupling vector mesons to constituents allows for predictions of the corresponding reaction cross sections~\cite{Khemchandani:2012ur, Geng:2008cv, Doring:2008sv}.
Rescattering through chiral unitary potentials can be extended to three particles~\cite{Malabarba:2021taj, Khemchandani:2020exc, MartinezTorres:2010zv, MartinezTorres:2009cw, MartinezTorres:2008kh, Torres:2008qkx, Khemchandani:2008rk, MartinezTorres:2007sr}. There are reviews on coupled-channel scattering by Hyodo~\cite{Hyodo:2011ur}, Mai~\cite{Mai:2020ltx}, Mei{\ss}ner~\cite{Meissner:2020khl}, Oller~\cite{Oller:2000ma}, and Oset~\cite{Oset:2016lyh} with emphasis on chiral unitary methods and, at least partially, covering light baryons. 
In this context, one has to mention the role of anomalous thresholds~\cite{Lutz:2018kaz} and their effect in unitarization. In certain kinematic regions they becomes so strong that even the dynamical generation of $P$-wave resonances becomes possible~\cite{Korpa:2022voo}.
Compositeness criteria to assess the nature of light baryons in terms of meson and baryon components were formulated and applied in Refs.~\cite{Wang:2025ecf,Wang:2023snv, Sekihara:2021eah, Molina:2015uqp}. Some baryonic resonances/structures in specific reactions have also been explained as enhancements due to triangle singularities~\cite{Wang:2016dtb, Samart:2017scf, Roca:2017bvy, CBELSATAPS:2021osa}. 

The Roy-Steiner set of equations were applied to the $\pi N$ and other sectors in Ref.~\cite{Ditsche:2012fv, Hoferichter:2015dsa, Hoferichter:2015tha, Hoferichter:2015hva} and also used to extract resonance pole parameters including the $\Delta(1232)$ and Roper resonances~\cite{Hoferichter:2023mgy}. Furthermore, the authors confirm a sub-threshold singularity on the unphysical sheet in $S_{11}$ close to the circular cut, see \cref{subsec:S11} for a more detailed discussion.  

As for lattice QCD approaches for baryon spectroscopy, see Refs.~\cite{Khan:2020ahz, Engel:2013ig, Dudek:2012ag, Edwards:2011jj, Bulava:2010yg, Melnitchouk:2002eg} and approaches in which phase shifts, scattering lengths and other properties are extracted~\cite{BaryonScatteringBaSc:2023ori, BaryonScatteringBaSc:2023zvt, Bulava:2022vpq, Silvi:2021uya, Pittler:2021bqw, Lang:2016hnn, Lang:2012db, Lin:2008qv}, as well as pioneering studies on higher excited baryons~\cite{Owa:2025mep, Zhuge:2024iuw, Liu:2023xvy, Hockley:2023yzn, Abell:2023nex, Wu:2017qve, Wu:2016ixr, Liu:2016wxq, Liu:2016uzk, Liu:2015ktc, Kiratidis:2015vpa}. We refer to dedicated reviews and whitepapers for more details~\cite{Bulava:2022ovd, Padmanath:2018zqw, Briceno:2017max}. For the extraction of Low Energy Constants from global fits to baryon lattice QCD data, see Refs.~\cite{Lutz:2024ubv, Hudspith:2024kzk} as well as the dedicated sections in the FLAG report~\cite{FlavourLatticeAveragingGroup:2019iem}.

The phenomenology and experimental aspects of baryon spectroscopy were reviewed by Crede and Roberts in 2013~\cite{Crede:2013kia}, including also aspects of heavy baryons, quark models, and a discussion of the impact of modern experiments on the baryon spectrum, especially from photoproduction reactions. 

Most recently, in 2022, Thiel, Afzal, and Wunderlich~\cite{Thiel:2022xtb} reviewed the light baryon spectrum including a consistent compendium of experimental results, discussion of complete experiments, different analysis frameworks (including DCC approaches), and impact on the baryon spectrum and its properties. Notably, this review provides also short introductions to theories and models for baryons like lattice QCD, quark models, Dyson-Schwinger methods etc. and represents a good starting point to obtain an overview of these methods.

There are also specialized reviews on the data aspects,  specific reactions,  and specific resonances in light baryons spectroscopy. This includes a 2019 overview by Ireland, Pasyuk, and Strakovsky~\cite{Ireland:2019uwn}, an older review by Schadmand and Krusche on meson photoproduction from 2003~\cite{Krusche:2003ik}, an $\eta$ physics handbook from 2002~\cite{Bijnens:2002zy}, and an updated review on $\eta$ and $\eta'$ production on nucleons and nuclei by Krusche from 2014~\cite{Krusche:2014ava}. One of the most prominent states of the baryon spectrum, the Roper resonance, $N(1440)1/2^+$, was reviewed by Burkert and Roberts in Ref.~\cite{Burkert:2017djo}, especially regarding the evidence of this resonance being the first radially excited state of the nucleon itself~\cite{Segovia:2015hra}. See also Ref.~\cite{Alvarez-Ruso:2010ayw} by \'Alvarez Ruso for a short review on the Roper resonance and Refs.~\cite{BESIII:2024vqu,Ping:2004wz, Alvarez-Ruso:1998puq} for examples of the Roper resonance appearing in other reactions than $\pi N$ and $\gamma N$-induced ones, including measurements and partial-wave analyses from BESIII. 

Regarding the role of amplitude analysis and high-energy parametrizations for light baryons, we mention work from JPAC and others~\cite{Stamen:2024gfz, JPAC:2018zjz, Mathieu:2018xyc, Mathieu:2018mjw, Nys:2018vck, JPAC:2016lnm, Mathieu:2015gxa, 
Lutz:2015lca,
Danilkin:2010xd} for aspects of S-matrix theory, their connection to Regge theory, finite-energy sum rules, and the formulation of new reaction amplitudes guided by these principles, as reviewed by JPAC in Ref.~\cite{JPAC:2021rxu}. In the context of amplitude analysis for resonance physics, the review by Yao, Dai, Zheng, and Zhou~\cite{Yao:2020bxx} discusses the role of chiral effective field theory and dispersion relations in the determination of partial waves and their resonance content, for a wide range of systems including $\pi N$.

Finally, we mention the role of meson-baryon dynamics in related contexts. Meson and photon-induced reaction are discussed in depth in the following because DCC approaches have been developed for their analysis. However, light excited baryons and, sometimes, their timelike electromagnetic form factors were also measured in $pp$-induced reactions at experiments like HADES~\cite{HADES:2017jkt, Agakishiev:2014wqa, HADES:2012aa} and COSY-TOF~\cite{COSY-TOF:2013uqx, TOF:2012tfr, TOF:2010ygk} and were analyzed by the 
Bonn-Gatchina (BnGa) group, Sibirtsev et al., and others~\cite{Ramalho:2016zgc, HADES:2015pgs, Ramalho:2012ng, Ermakov:2011at, Xie:2011me, Valdau:2010kw, Cao:2007md, Xie:2007vs, Anisovich:2007zz, Roca:2006pu, Sibirtsev:2007sk, Sibirtsev:2006uy, Sibirtsev:2005mv, Teis:1996kx, Gasparian:1999jj, Baru:2002rs}. 
See the review by Salabura~\cite{Salabura:2020tou}.

Laget reviewed exclusive meson photoproduction at high energies in the so-called partonic non-perturbative regime~\cite{Laget:2019tou} that is particularly relevant for current and future experiments at JLab and the EIC.
The in-medium properties of excited baryons were reviewed by Lenske~\cite{Lenske:2018bgr}, and, in close relation, by Metag~\cite{Metag:2017yuh}. 
Hadronic reactions close to thresholds have simple kinematics and  allow for tests of different models as reviewed by Moskal~\cite{Moskal:2002jm}, see also Ref.~\cite{Gonzalez:2008en}. Photoproduction reactions at low energies was reviewed in Ref.~\cite{Drechsel:1992pn}.
The field of light-strange baryons, including the $\Lambda(1405)$ is vast; the contribution from DCC approaches in discussed in \cref{sec:strangebaryons} and we mention here only the reviews by Hyodo~\cite{Hyodo:2011ur, Hyodo:2020czb}, Mai~\cite{Mai:2020ltx}, 
Mei\ss ner on two-pole structures~\cite{Meissner:2020khl}, and Tolos~\cite{Tolos:2020aln}.

\subsection{Meson-induced reactions}
\label{sec:mesonproduction}

Meson beams are versatile probes for hadron spectroscopy. Elastic pion-nucleon scattering was the first reaction providing access to the baryon spectrum in the classical partial-wave analyses by H\"ohler,  Koch, and Cutkoski leading to the establishment of many states by the 90's~\cite{Cutkosky:1979zv, Cutkosky:1979fy, Koch:1980ay, Hohler:1984ux,  Koch:1985bp}. See also the analysis by Manley et al. that includes elastic $\pi N$ scattering and the $\pi\pi N$ final state~\cite{Manley:1984jz, Manley:1992yb} and the Carnegie-Mellon-Berkeley (CMB) model by Vrana, Dytman, and Lee~\cite{Vrana:1999nt} and by the Zagreb group~\cite{Batinic:1995kr, Batinic:1997gk, Ceci:2006ra, Batinic:2010zz}. 

The analyses of $\pi N$ scattering including the reaction $\pi N\to\eta N$ and pion photoproduction on proton and neutron have been further refined to
this day by the GWU/SAID analysis group~\cite{Arndt:1985vj, Arndt:1995bj, Arndt:1998nm, Arndt:2006bf,
Dugger:2007bt,
Workman:2008iv, Arndt:2008zz, 
CLAS:2009tyz,
Workman:2011vb,
Workman:2012jf,
Workman:2012hx,
CLAS:2013pcs,
A2:2015mhs,
CLAS:2015ykk,
Anisovich:2016vzt,
A2:2017gwp,
CLAS:2017kua, 
A2:2019yud,
Briscoe:2020qat, Strakovsky:2022tvu,
Briscoe:2023gmb}. 
The SAID pion and photon-induced formalisms were unified into a Chew-Mandelstam parametrization in Ref.~\cite{Workman:2012jf}. 
The partial-wave dispersion relations by Frazer and Fulco~\cite{Frazer:1960zz} derived from the Mandelstam representation~\cite{Mandelstam:1958xc} provided useful constraints for many of the $\pi N$ analyses.  Notably, all other modern partial-wave analysis efforts rely on the SAID or other  analyses for elastic pion scattering, although the ANL-Osaka analysis~\cite{Julia-Diaz:2007qtz} follows a two-step procedure: (1) SAID input is used to locate the range of the parameters of the model; (2) The obtained parameters are  refined and confirmed by directly comparing with the original $\pi N$-scattering data.

In Ref.~\cite{Doring:2016snk} the covariance matrices between different partial waves of the SAID Single-Energy Solutions (SES) were provided, that allow, in principle, to reconstruct a meaningful $\chi^2$ representing the actual underlying data. However, this means one implicitly accepts the correction of systematic errors as applied by SAID. Systematic uncertainties and their correction are not small, at all, due to the problematic data situation of elastic $\pi N$ scattering. It is, therefore, better (albeit much more tedious) to fit the underlying $\pi N$ data base with a procedure to correct for various normalization factors~\cite{Ball:2009qv} as practiced in Ref.~\cite{Hoferichter:2015hva} in which the $\pi\pi \to \bar NN$ process is related to the $s$-channel $\pi N\to\pi N$ reaction. 

Meson-induced reactions are central in any modern analysis aiming at a complete data description, but the available data are problematic. In Ref.~\cite{Briscoe:2015qia} a case for new measurements using meson beams was made with minor updates regarding the planned Electron Ion Collider in Ref.~\cite{Briscoe:2021cay}. Historically, the argument for using photons in the discovery of excited baryons (instead of meson beams) was the sensitivity to states that couple only weakly to $\pi N$. A resonance that couples substantially to the photon but weakly to the $\pi N$ channel could be easier to detect in photoproduction reactions. Reactions like $\gamma N\to \eta N,\,K\Lambda,\,K\Sigma,\dots$ are ideal for this purpose. The downside of photoproduction is the larger number of amplitudes. In general, a resonance with given spin-parity $J^P$ couples to both an electric and a magnetic multipole, i.e., more degrees of freedom are present in photoproduction reactions than in pion-induced reactions. This substantially increases the number of necessary measurements to determine an amplitude. See next section for a discussion on ``complete experiments''. 

In addition, modern baryon spectroscopy analysis frameworks add the real or virtual photon to an existing hadronic rescattering part in one way or another~\cite{Surya:1995ur, Drechsel:1998hk, Anisovich:2004zz, Matsuyama:2006rp,   Drechsel:2007if, Julia-Diaz:2007mae, Durand:2008es, Anisovich:2011fc, Ronchen:2014cna, Ronchen:2015vfa, Ronchen:2018ury, Kamano:2019gtm, Mai:2021vsw, Mai:2021aui, Ronchen:2022hqk, Mai:2023cbp, Wang:2024byt} as discussed in \cref{sec:photoproduction}. The rescattering part itself is constrained from pion-induced reactions. In that, any data problems in pion-induced reactions will have consequences for the analysis of photon-induced reactions, especially when it comes to resonances that are tied to the strongly interacting rescattering part. This is particularly the case for pion-induced inelastic reactions such as $\pi N\to \eta N$, $K\Lambda$, $K\Sigma$, $\omega N$ or three-body channels. On one hand, data in these reactions are known to be problematic as discussed below. On the other hand, new resonances coupling weakly to $\pi N$ are still more visible in these reactions than in elastic $\pi N$ scattering. Summarizing, there are four reasons why re-measuring inelastic, pion-induced reactions can significantly contribute to an improved and more quantitative understanding of the excited baryon spectrum, and the discovery of new states: (a) the poor data situation, (b) reducing the bias for photoproduction analysis, (c) fewer 
amplitudes than in $\gamma^{(*)}$-induced reactions, and (d) sensitivity to channels other than $\pi N$ (for inelastic reactions).

Of the possible final states, the reactions $\pi N\to\eta N,\,K\Lambda$ and $\pi^+ p\to K^+\Sigma^+$ have to be highlighted because they isospin filter $I=1/2$ $N^*$'s~\cite{Ronchen:2012eg} and $\Delta$'s~\cite{Doring:2010ap}, respectively. The reactions $\pi N\to K\Lambda$ and $\pi N\to K\Sigma$ provide unique information in that the outgoing baryon is self-analyzing allowing for an easy access to spin observables. Take, for example, the reaction $\pi^+p\to K^+\Sigma^+$ measured in Ref.~\cite{Candlin:1982yv}. The measurements cover a substantial angular and energy range, and they include data on differential cross sections and recoil polarization $P$. Together with much less precise data on the spin-rotation parameter $\beta$ from Ref.~\cite{Candlin:1988pn} the set of observables is almost complete (see, e.g., the discussion in Ref.~\cite{Klempt:2009pi}). 

Regarding the overall data situation, an overview of inelastic pion-induced reactions is provided in Table~\ref{tab:datainelastic}.
The table is taken from Ref.~\cite{Ronchen:2022hqk} and amended by the available $\pi N\to\omega N$ data collected and analyzed in Ref.~\cite{Wang:2022osj}. The number of data points should be compared to the tens of thousands of data available in photoproduction reactions~\cite{Ronchen:2022hqk}, or the hundreds of thousands of data in electroproduction~\cite{Mai:2023cbp}. 

\begin{table}[b!]
    \caption{Inelastic pion-induced reaction data. The quantity $\beta$ is the spinrotation parameter. A full list of references to the different experimental publications can be found online~\cite{Juelichmodel:online}.
    }
    \label{tab:datainelastic}
\begin{center}
    \begin{tabularx}{\linewidth}{XXXXXX} 
    \hline \hline
    Reaction & \multicolumn{4}{c}{Observables ($\#$ data points)} & Total\BB\TT \\
    \hline
    $\pi^-p\to\eta n$ &&$d\sigma/d\Omega$ (676)&  $P$ (79)&  
    & 755 \TT\\
    $\pi^-p\to K^0 \Lambda$ 
    &&$d\sigma/d\Omega$ (814)& $P$ (472)& $\beta$ (72) 
    & 1358\\
    $\pi^-p\to K^0 \Sigma^0$ 
    &&$d\sigma/d\Omega$ (470)&  $P$ (120)&  
    & 590\\
    $\pi^-p\to K^+ \Sigma^-$ 
    &&$d\sigma/d\Omega$ (150)& & 
    & 150\\
    $\pi^+p\to K^+ \Sigma^+$ 
    &&$d\sigma/d\Omega$ (1124)&  $P$ (551) &  $\beta$ (7)
    & 1682\\ 
    $\pi N\to \omega N$ 
    & $\sigma(54)$&  $d\sigma/d\Omega$ (124)&&  
    & 178
    \BB\\
    \hline \hline
    \end{tabularx}
\end{center}
\end{table}

In addition, the pion-induced reaction data often carry large uncertainties and are, in many cases, contradictory due to underestimated systematic effects. This leads to  notoriously large $\chi^2$ values in global fits, making the application of traditional statistical criteria to, e.g.,  assess the significance of a signal for a new resonance, very difficult. The aim of a new pion-induced experiment would  not only be to provide more data, but, for the first time, provide {\it consistent} data. This is indeed possible: for example, the EPECUR experiment has drastically improved the data situation in elastic pion-nucleon scattering~\cite{EPECUR:2014wrs} (see, e.g., Fig.~4 in Ref.~\cite{Briscoe:2021cay}). See also Ref.~\cite{Shirotori:2012ka} for a demonstration how drastically the measurements of the $K^+\Sigma^+$ final state can be improved with modern experimental techniques. Planned reactions with meson beams at J-PARC are discussed in Ref.~\cite{Ohnishi:2019cif}; for the study of pion-induced reactions at HADES, see Ref.~\cite{Salabura:2015yot}.

\begin{figure}[tb]
    \centering
    \includegraphics[width=0.222\textwidth]{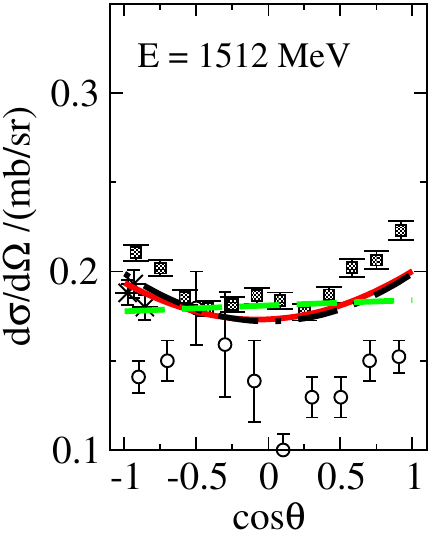}
    \hspace*{0.1cm}
    \includegraphics[width=0.167\textwidth]{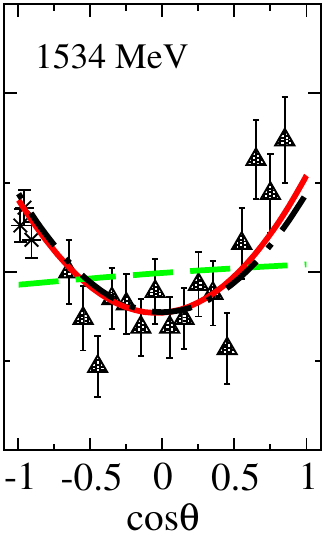}
    \hspace*{0.1cm}
    \includegraphics[width=0.34\textwidth]{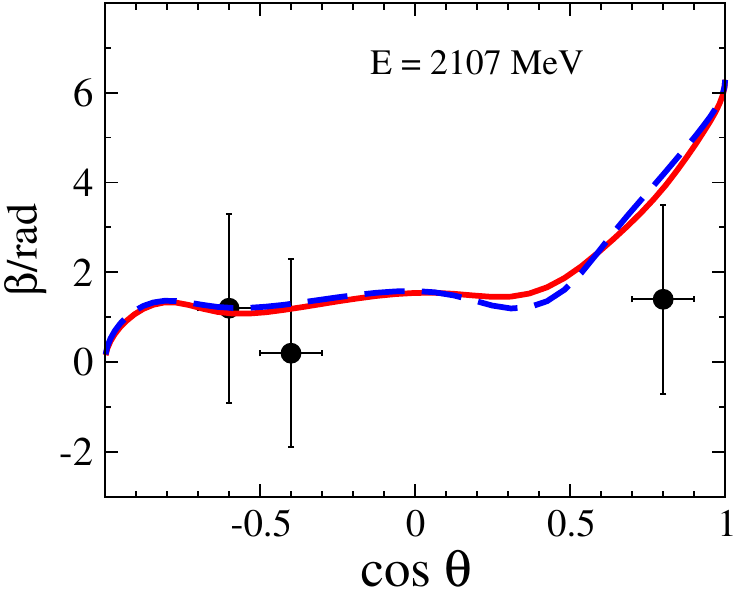}
    \hspace*{0.cm}
    \includegraphics[width=0.205\textwidth]{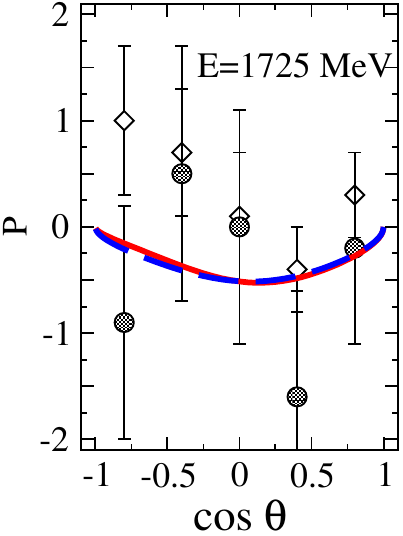}
    \caption{Left two figures: Differential cross section of the reaction $\pi^- p\to \eta n$ at different scattering energies $E$.
    The u-shape in fits and data reveals the $S$-$D$-wave interference of $N(1535)1/2^+$ and $N(1520)3/2^-$.
    Central figure: Spinrotation parameter $\beta$ for $\pi^+p\to K^+\Sigma^+$. Right figure: Polarization $P$ for the reaction $\pi^-p\to K^0\Sigma^0$. 
    Solid red (dashed blue) lines: JBW solutions A (B) from Ref.~\cite{Ronchen:2012eg}.
    For  data references see Ref.~\cite{Ronchen:2012eg}.}
    \label{fig:etaN_s+dwave}     
\end{figure}
The situation for the inelastic reactions is illustrated in Fig.~\ref{fig:etaN_s+dwave}.
 \begin{figure}[tb]
     \centering
     \includegraphics[width=0.48\linewidth]{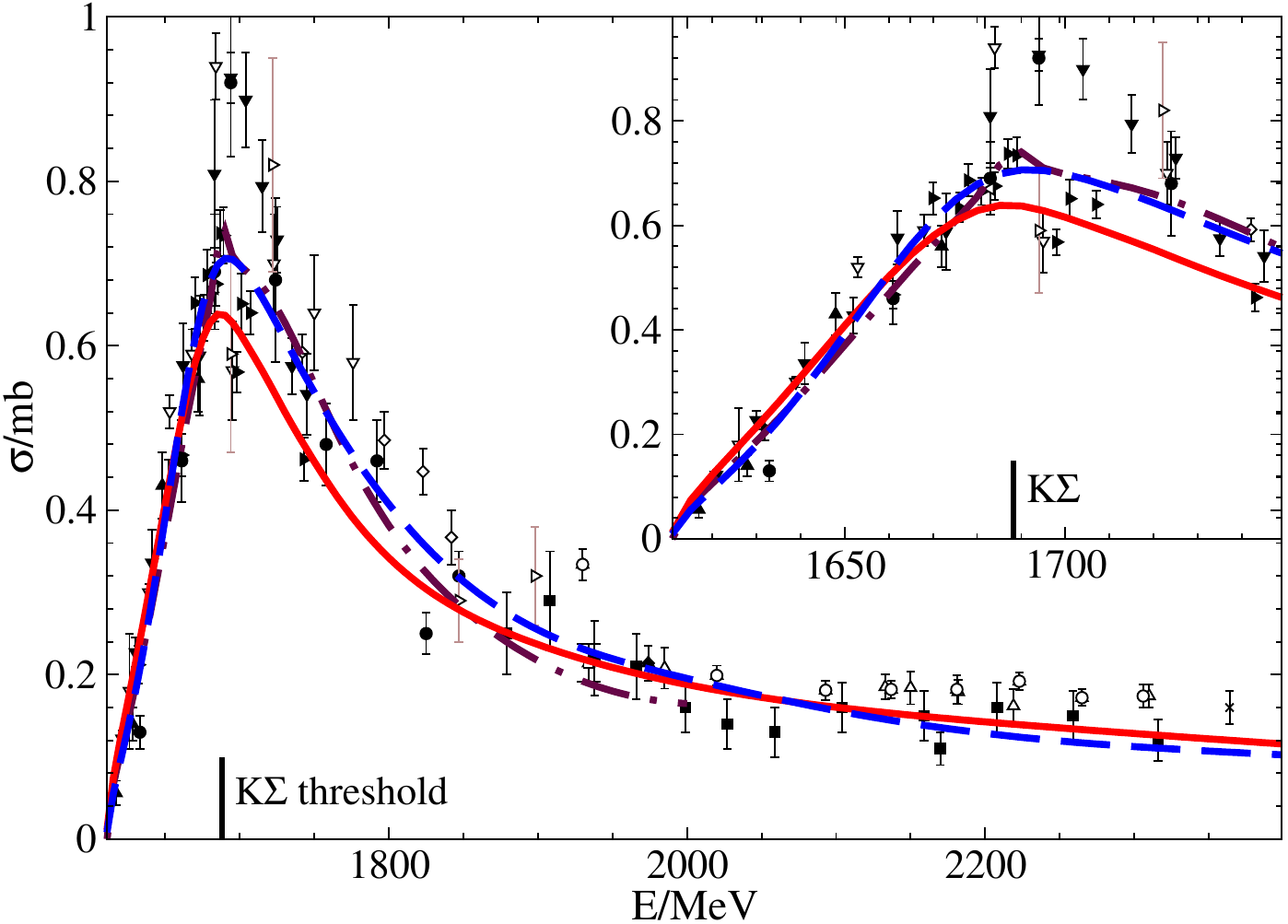}
     \includegraphics[width=0.48\linewidth]{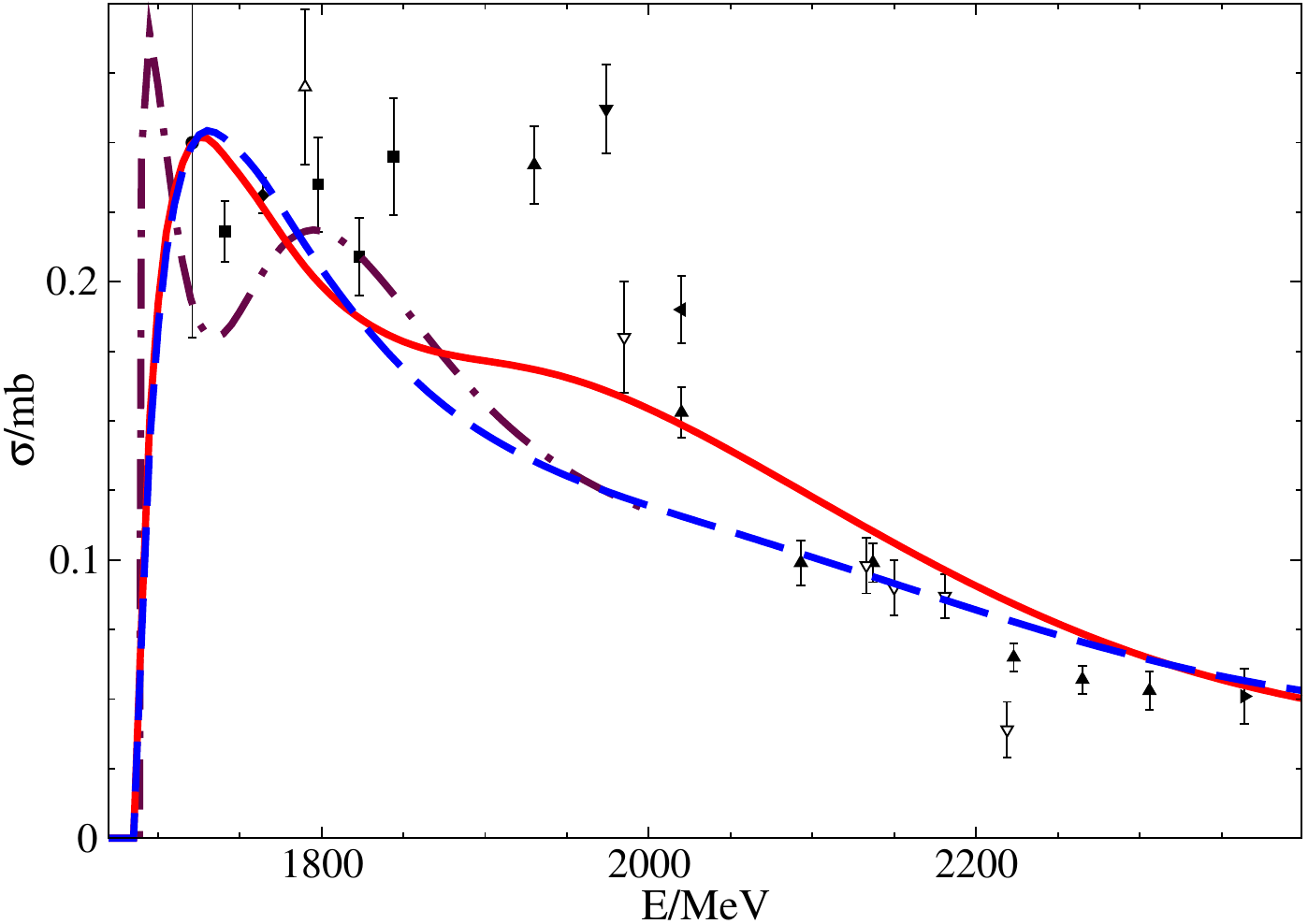}
     \caption{
     Total cross sections of the reactions $\pi^-p\to K^0\Lambda$ (left) and $\pi^-p\to K^+\Sigma^-$ (right). Red solid (blue dashed)  lines: JBW  solutions A (B) from Ref.~\cite{Ronchen:2012eg}. Dash-dotted dark brown lines: ANL-Osaka solution from Ref.~\cite{Kamano:2019gtm}. See Ref.~\cite{Ronchen:2012eg} for data references.}
     \label{CH4:figMM-3}
 \end{figure}
The differential cross section data demonstrates that the systematic problems are not just given by simple multiplicative factors that might be adjustable~\cite{Ball:2009qv} --the shapes of the cross sections do not match. The plot of the spinrotation parameter $\beta$ demonstrates how few data are available for this most valuable experimental quantity; and the right-hand plot shows the typical quality for polarization data  (that are supposed to be limited to $|P|\leq 1$).

The total cross sections shown in Fig.~\ref{CH4:figMM-3} exhibit similar problems, i.e., clearly contradictory data. For example, the total cross section of the reaction $\pi^-p\to K^+\Sigma^-$ is almost impossible to describe due to a supposed structure around a scattering energy of $E=z\approx 2\,\GeV$, which is, however, only due to one or two data points. Having more conclusive data would be particularly valuable because this reaction does not allow for $t$-channel meson exchange and is, thus, of special interest to study hadron dynamics. There is no polarization available for this reactions, at all. 

The $\pi N\to\pi\pi N$ reaction also provides valuable information on the baryon spectrum and its few-body decays. Such reactions have been considered in the ANL-Osaka DCC approach as described below. Although the $\pi N\to\pi\pi N$ data are few and similarly problematic as the data discussed before, new experimental efforts are being made at J-PARC in the E45 experiment to drastically increase the 
data base~\cite{Kamano:2013ona, Sako:2017xsw}.
Inelastic reactions like $\pi N\to \pi\eta N$ have also been used to gain information on the dynamical nature of resonances like $\Delta(1700)$~\cite{Doring:2006pt}, see also recent progress by JPAC/GlueX/COMPASS~\cite{Gleason:2022pry, JPAC:2018zyd, JPAC:2017dbi, COMPASS:2014vkj} in the context of searching for exotic mesons in this reaction.

The $\pi\pi N$ states are known to have strong couplings with most of the high-mass nucleon resonances and, thus, are relevant for spectroscopy. They are not directly included in the analysis of the nucleon resonances in the ANL-Osaka model. As a first step of a complete analysis to include two-body and three-body final states for the pion and photon induced reactions, the $\pi N \rightarrow \pi\pi N$~\cite{Kamano:2008gr, Kamano:2013ona} and $\gamma N \rightarrow \pi\pi N$~\cite{Kamano:2009im} reactions are studied within the  DCC model~\cite{Julia-Diaz:2007qtz,Kamano:2013iva}. The situation of the theoretical approaches and the experimental data on $\pi N \rightarrow \pi\pi N$ are briefly summarized in Ref.~\cite{Kamano:2008gr}.

\begin{figure}[thb]
    \begin{center}
    \includegraphics[width=16cm]{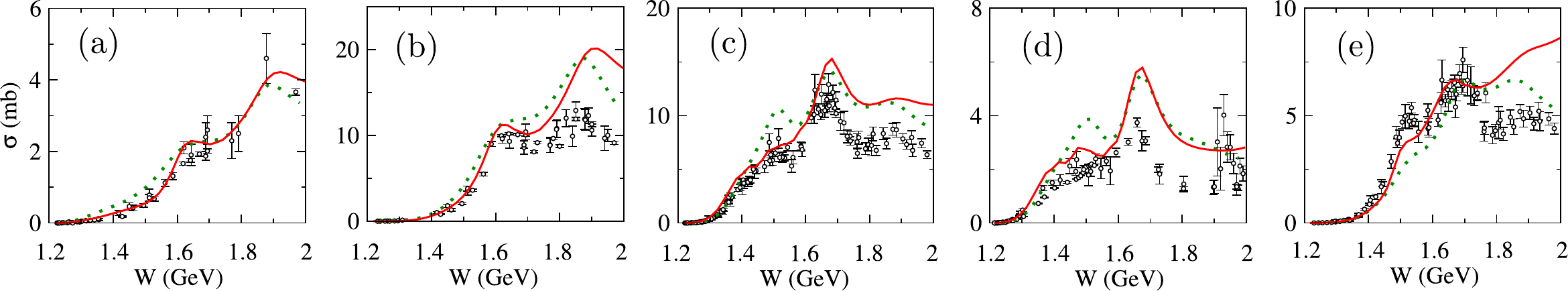}
    \end{center}
    \caption{The $\pi N \rightarrow \pi\pi N$ total cross section
    predicted by the ANL-Osaka models from Refs.~\cite{Kamano:2013iva} (red solid) and \cite{Julia-Diaz:2007qtz} (green dotted lines) as a function of energy $E=W$.
    The reaction channels are
    (a) $\pi^+ p \rightarrow \pi^+\pi^+ n$,
    (b) $\pi^+ p \rightarrow \pi^+\pi^0 p$,
    (c) $\pi^- p \rightarrow \pi^+\pi^- n$,
    (d) $\pi^- p \rightarrow \pi^0\pi^0 n$, and
    (e) $\pi^- p \rightarrow \pi^-\pi^0 p$.  
    }
    \label{fig:pi2pitcrs}
\end{figure}

With the approximation of $T_{\pi\pi N,\pi N}^{dir}$ (Eq.\eqref{eq:tpipin-dir}) neglecting the final state interaction on the mechanisms (f)-(k) of $v_{\pi\pi N,\pi N}$ in Fig.~\ref{fig:v23pi}~\cite{Kamano:2008gr},
the $\pi N \rightarrow \pi\pi N$ cross section is calculated using information of the ANL-Osaka model~\cite{Julia-Diaz:2007qtz,Kamano:2013iva} with no additional parameters. The total cross sections for five channels are compared with the data as shown in Fig.~\ref{fig:pi2pitcrs}. The red solid curves are from a recent version of the ANL-Osaka approach~\cite{Kamano:2013iva}
while the green dotted curves are from the previous JLMS model~\cite{Julia-Diaz:2007qtz}.  Both the magnitude and the energy dependence of the data for
five $2\pi$ production can be reproduced to a very large extent.
The direct mechanism $T_{\pi\pi N,\pi N}^{dir}$ plays an important role
especially for $\pi^+ p \rightarrow \pi^+\pi^+ n$ and
$\pi^- p \rightarrow \pi^- \pi^0 p$. In the low-energy region $W<1.4\,\GeV$,
the calculation is comparable to the calculation of chiral perturbation theory.
This suggests that the model of $v_{\pi\pi N,\pi N}$
is reasonable and the discrepancies with the data in the higher $W$ region
are more likely results from the uncertainties in the contribution
for $\pi \Delta, \rho N$ and $\sigma N$ transitions.

\begin{figure}[thb]
    \centering
    \includegraphics[width=0.75\linewidth]{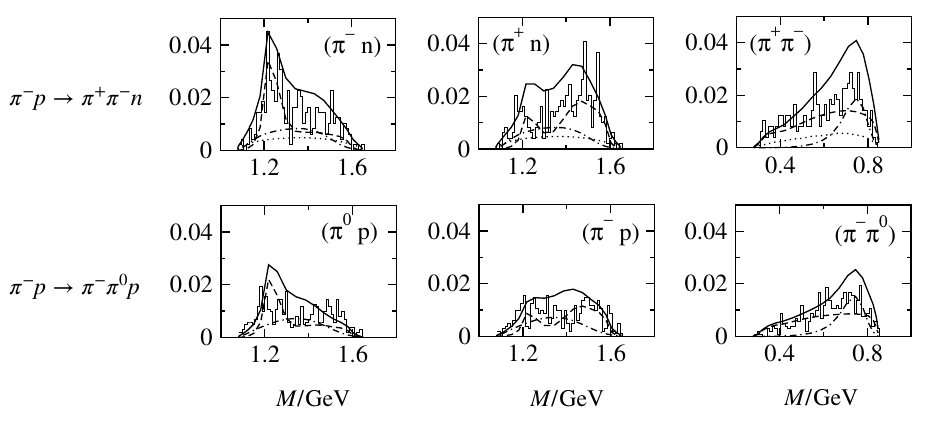}
    \caption{
    Predicted contributions from the $\pi \Delta$ (dashed), $\sigma N$ (dotted), and $\rho N$ (dot-dashed) channels to the invariant mass distributions  of $\pi^- p \rightarrow \pi\pi N$ at $W=1.79\,\GeV$. The data are from Arndt~\cite{Arndt-privatecomm}, with magnitudes determined by normalizing them to the total $\pi N \rightarrow \pi\pi N$ cross section.
    }
    \label{fig:pi2pim}
\end{figure}

The $\pi\pi$ and $\pi N$ invariant mass distributions at $W=1.79\,\GeV$ and the contributions of  each of the $\pi \Delta, \rho N, \sigma N$ channels are shown in Fig.~\ref{fig:pi2pim}. The DCC model is able to reproduce the main features of the data. The DCC model shows the strong interference among those channels for the invariant mass distributions. Future measurements of the $\pi N \rightarrow \pi\pi N$ reactions are important for the extraction of 
higher-mass nucleon resonances and for an improved description of the 
$N^* \rightarrow \pi\Delta, \rho N, \sigma N$ couplings.

In summary, inconsistencies dominate the data uncertainties for pion-induced reactions. This has severe consequences for baryon spectroscopy 
affecting the reliability of determining the spectrum, especially when it comes to hard statistical criteria. Indirectly, these problems also affect the analysis of photoproduction reactions, in which so much experimental and analysis effort has been invested in the past decades. Having a large set of consistent data on pion-induced reactions, especially inelastic ones, would allow one to take advantage of the inherent upsides of meson beams: Fewer amplitudes than in photoproduction (for single-meson final states), easier to achieve ``complete experiment'', larger cross sections and, to some extent, still having sensitivity to resonances that couple only weakly to $\pi N$.
The $\pi\pi N$ final states
were predicted by the ANL-Osaka approach demonstrating that DCC models can analyze these states. The impact of new data on different analyses, such as J-PARC E45~\cite{jparc:e45}, once available, could be quantified in common efforts across different analysis groups as practiced in Ref.~\cite{Anisovich:2016vzt} for photoproduction.

\subsection{Photoproduction reactions}
\label{sec:photoproduction}
\subsubsection{Overview of different analysis frameworks}

In the last three decades, a plethora of final states, differential cross sections, and spin observables were measured in photoproduction reactions at JLAB/CLAS~\cite{CLAS:2006pde, CLAS:2009rdi, CLAS:2015pjm, CLAS:2015ykk, CLAS:2017rxe, CLAS:2024bzi}, CB-ELSA/TAPS~\cite{CB-ELSA:2003rxy,  CBELSATAPS:2007oqn, CBELSA:2008epm, CBELSATAPS:2009ntt, Thiel:2012yj, CBELSATAPS:2014wvh, CBELSATAPS:2020cwk, CBELSATAPS:2021osa} (continued at the new BGOOD detector~\cite{ 
Rosanowski:2024rww, Figueiredo:2024zmh, Jude:2022atd, BGOOD:2021oxp, BGOOD:2021sog, Alef:2020yul, 
Jude:2020byj,
BGO-OD:2019utx}),
MAMI/A2/CrystalBall~\cite{A2:2017gwp, CrystalBallatMAMI:2013iig, A2:2014pie, CrystalBallatMAMI:2010slt, CrystalBallatMAMI:2009lze, GDH:2001zzk}, LEPS~\cite{LEPS:2017jqw, LEPS:2017pzl, LEPS:2009pib, LEPS:2005hji}, 
ELPH/FOREST~\cite{Ishikawa:2019rvz},
GRAAL~\cite{GRAAL:2008jrm,Ajaka:2008zz}, and other facilities. Sometimes observables are tied to specific choices of reference frames that are not unique across different experiments and analyses. The ``Rosetta Stone Paper'' provides an overview~\cite{Sandorfi:2011nv}. 
 
The data and their analyses provided unprecedented information on baryonic resonances, not only for their electromagnetic properties, but also for the spectrum itself. Indeed, many new states were added to the Particle Data Book~\cite{ParticleDataGroup:2024cfk} in the photoproduction era. States found before that were promoted in the PDG star rating while a few other ones descended or disappeared, see, e.g., Table 9 of Ref.~\cite{Crede:2013kia}. In the reviews~\cite{Klempt:2009pi, Crede:2013kia, Ireland:2019uwn, Thiel:2022xtb}, comprehensive compilations of data and their features can be found. We do not aim at replicating these works here and just pick out some interesting findings in the context of DCC approaches and general hadron dynamics, see Sects~\ref{subsec:S11} and \ref{sec:P11}. There are other topics beyond the scope of this review like the efforts to experimentally confirm the GDH sum rule~\cite{Strakovsky:2022tvu, GDH:2001zzk, GDH:2000tuw} or Compton scattering reviewed by Grie{\ss}hammer, McGovern, Phillips, and Feldman~\cite{Griesshammer:2012we}, including the role of the $\Delta(1232)$~\cite{McGovern:2012ew} in the low-energy chiral expansion.

There are also extended discussions on specific structures that have attracted much attention such as a peak in $\eta $ photoproduction on the neutron at around $W=1675\,\MeV$ that is absent at the proton target~\cite{CBELSA:2008epm, A2:2014pie}. This has been interpreted as an interference effect~\cite{Shklyar:2006xw}, a coupled-channel  effect~\cite{Doring:2009qr}, or a new resonance~\cite{A2:2016bij}, possibly a pentaquark~\cite{Diakonov:1997mm}. The need for a new narrow resonance was disputed in Ref.~\cite{Anisovich:2017xqg} by BnGa, and the Giessen group did not find new narrow states for photoproduction on the proton~\cite{Shklyar:2012js}.

The extraction of multipoles from photoproduction data is a challenge. There are continuous and discrete ambiguities~\cite{Svarc:2017yuo}. Due to the presence of electric and magnetic multipoles, the extraction requires in general more observables than in reactions induced by spinless mesons. 
Such questions tied to ``complete experiments'' have been discussed in Refs.~\cite{Barker:1975bp, Sandorfi:2010uv, Wunderlich:2013iga, Workman:2016irf, Wunderlich:2021xhp, Kroenert:2023ovd} by Wunderlich et al.. There are two variants of this issue: How many measurements are needed to determine the amplitude  at a given energy and scattering angle~\cite{Kroenert:2020ahf}, and, second, the ``truncated partial-wave'' complete experiment~\cite{Wunderlich:2013iga, Workman:2016irf, Tiator:2017cde} that is somewhat more relevant when it comes to baryon spectroscopy of resonances with given $J^P$. See Refs.~\cite{Svarc:2020cic, Svarc:2025zhb} for recent comparisons of these types of complete experiment and a discussion of the role of uncertainties.
In relation to this, one of the tasks in  spectroscopy is to determine which partial waves actually contribute in a given kinematic region. This question has  been addressed in Ref.~\cite{Wunderlich:2016imj} for the baryon sector and in Ref.~\cite{Landay:2016cjw} as a problem for which machine learning techniques can be used.

In practice, one needs assumptions on the amplitude to extract energy-dependent multipoles to reconstruct resonances and their electromagnetic properties. Assumptions on the smoothness of multipoles in energy are indeed mandatory as the set of observables is never complete in all kinematic regions. Starting from such energy-dependent solutions, single-energy solutions (SES) can be derived in which the solutions are allowed to vary in an energy bin describing only the data in that bin. One needs subsidiary conditions for this such as fixing multipole phases to avoid the SES to vary uncontrolled from bin to bin. This allows to scan for narrow structures and can also help detect systematic shortcomings of the global parametrization. See Refs.~\cite{Svarc:2024phb, Svarc:2021gcs, Anisovich:2017bsk, Svarc:2015usk, Svarc:2014sqa} for recent progress in this direction using the Laurent-Pietarinen expansion, as well as Refs.~\cite{Osmanovic:2021rck, Osmanovic:2019mux} in which $t$-channel analyticity and unitarity have been used to constrain the determination of single-energy solutions. Resonance pole parameters in the complex plane can be determined by fitting them with meromorphic functions~\cite{Svarc:2015usk}. The Mittag-Leffler expansion is used in ~\cite{Yamada:2020rpd,Morimatsu:2019wvk} by introducing  the uniformization variable discussed in section~\ref{sec:analytic}.

Many frameworks have been developed for the analysis of meson photoproduction reactions - sometimes, the efforts are bundled, e.g., to study the impact of new polarization data on the different analysis frameworks~\cite{Anisovich:2016vzt}. We discuss here only approaches that do \emph{not} fall into the DCC category. The ANL-Osaka and JBW approaches have been discussed in depth in this review, while other DCC approaches have been summarized in  \cref{sec:historyandapproaches}. Furthermore, we discuss here approaches that, in many cases, not only analyze photoproduction but also pion-induced reactions.

The SAID group including Arndt, Briscoe, Schmidt, Strakovsky,  Workman et al. developed one of the earlier approaches for the analysis of photoproduction on the proton~\cite{Briscoe:2023gmb, Workman:2011vb, CrystalBallatMAMI:2010slt, CLAS:2009tyz, Arndt:2002xv} but also neutron~\cite{Briscoe:2020qat, A2:2019yud, CLAS:2017kua} as reviewed in Ref.~\cite{Briscoe:2021siu}.  For the latter case, mostly referring to photoproduction on the deuteron, there are strong corrections (Fermi motion and final state interactions) that have to be taken into account~\cite{Tarasov:2011ec, Tarasov:2015sta}. As discussed in \cref{sec:mesonproduction}, the photoproduction and pion-induced reactions were formulated in a unified SAID framework in Ref.~\cite{Workman:2012jf}.

The MAID unitary isobar approach to meson photoproduction has also evolved over time. After initial studies using effective Lagrangian approaches~\cite{Knochlein:1995qz, Tiator:1994et} and fixed-t dispersion relations~\cite{Hanstein:1997tp}, Drechsel, Hanstein, Kamalov, Tiator et al. developed the unitary isobar MAID framework in Ref.~\cite{Drechsel:1998hk} for pion photoproduction. In another single-channel analysis, the framework was extended to $\eta$ photo and electroproduction~\cite{Chiang:2001as} with a reggeized version for the background presented in Ref.~\cite{Chiang:2002vq}. 
Kaon photoproduction was studied in KAON-MAID~\cite{Lee:1999kd, Bennhold:1999mt} by Bennhold, Haberzettl, Lee, Mart, and Wright, within a tree-level model that incorporates
hadronic form factors consistent with gauge invariance.
The MAID 2007 approach is a unified analysis of pion photo and electroproduction~\cite{Drechsel:2007if} using suitably unitarized Breit-Wigner forms for the resonances. The paper also contains an overview of the historic evolution of the MAID approach. See also \cref{sec:electroproduction} for a more specific literature review on electroproduction and the respective analysis frameworks.
An analysis of low-energy $\gamma^{(*)}$-induced data with chiral perturbation theory was performed under the chiral MAID label in Refs.~\cite{Hilt:2013uf, Hilt:2013fda}.
An update on $\eta$ and $\eta'$ photoproduction was published in Ref.~\cite{Tiator:2018heh}. Finally, Ref.~\cite{Tiator:2018pjq} summarizes the legacy of MAID and points towards future developments of the approach.

In the BnGa approach both pion and photon-induced reactions are analyzed simultaneously~\cite{Anisovich:2012ct}, like in the ANL-Osaka and JBW approaches. While electroproduction reactions are not yet considered, the range of analyzed photoproductions is larger than in the DCC approaches. 
Apart from the reactions analyzed in the mentioned DCC approaches, the BnGa
group analyzes also three-body final states of reactions like $\gamma p \to \pi^0\eta p$ or $\pi^0\pi^0 p$~\cite{CBELSATAPS:2015taz, CBELSATAPS:2014wvh, Oberle:2013kvb}. See also recent effort on two-pion photoproduction  by the JPAC group~\cite{JointPhysicsAnalysisCenter:2023gku, Mathieu:2019fts, JointPhysicsAnalysisCenter:2017del, CLAS:2008ycy} and Haberzettl~\cite{Haberzettl:2018hcl}, as well as tests of chiral unitary models in two-meson photoproduction~\cite{Doring:2010fw, Doring:2005bx} and other two-meson isobar models~\cite{Nacher:2001yr, Nacher:2000eq, Nacher:1998hh, GomezTejedor:1995pe}.

The BnGa amplitude was formulated in Ref.~\cite{Anisovich:2004zz} with main results published in Ref.~\cite{Anisovich:2011fc}, 
which have been updated based on new data and reactions over the years~\cite{Sarantsev:2025lik, Anisovich:2013jya, Thiel:2012yj}, such as 
$\omega$ photoproduction~\cite{Denisenko:2016ugz, CBELSATAPS:2015ftl}. 
Evidence for several new excited baryons and new decay channels of known baryons was found~\cite{Anisovich:2015gia, CBELSATAPS:2015taz, Burkert:2014wea, Anisovich:2013vpa, Burkert:2012ee, 
Anisovich:2010mks, Sarantsev:2005tg, Anisovich:2005tf} leading to  substantial updates of the Particle Data Book, for example for a $N(1900)3/2^+$~\cite{Nikonov:2007br, CBELSATAPS:2015tyg} and a $N(1895)1/2^-$~\cite{CBELSATAPS:2020cwk}. The case of the latter resonance is particularly intriguing due to the closeness of the $\eta' p$ threshold~\cite{CBELSATAPS:2020cwk}. In Ref.~\cite{Anisovich:2011ye} BnGa explored the region of high-lying states and found evidence for four new states by performing mass scans. In this technique, resonances are inserted into the amplitude as Breit-Wigner forms with their masses held fixed and other parameters allowed to re-adjust with the data. The $\chi^2$ is plotted as a function of the mass, preferably for different final states. Simultaneous minima of the $\chi^2$ for different final states are taken as a signal for a new resonance at that mass. The four new states were interpreted in light of the general mass ordering expected in quark models from the spin-flavor symmetry leading to the group SU(6) that is decomposed into 70-plets etc., which can then further be decomposed into different SU(3) groups with different spin multiplicities. An organization of states in spin doublets and quartets is plausible. If that is the case, ``the diquark hypothesis
- which freezes one pair of quarks into a quasi-stable $S$-wave flavor diquark - is ruled out as
explanation of the missing resonance problem''~\cite{Klempt:2012gu}. Dynamical diquarks behave differently~\cite{Barabanov:2020jvn, Chen:2019fzn}. 

The Giessen coupled-channel approach by Lenske, Mosel, Penner, Shklyar et al. employed a coupled-channel K-matrix formalism for the combined analysis of pion and photon-induced reactions~\cite{, Shklyar:2014kra, Cao:2013psa, Shklyar:2012js,  Shklyar:2006xw,
Shklyar:2005xg,
Shklyar:2004ba, Penner:2002md, Penner:2002ma} including vector meson and $\pi\pi N$ final states in form of the partial-wave results by Manley~\cite{Manley:1992yb}. 
The Kent state analyses by Manley et al. addressed a variety of final states including $\pi N$~\cite{Manley:1992yb, Shrestha:2012ep, Hunt:2018wqz, Hunt:2018mrt} but also inelastic, pion and photon-induced reactions~\cite{Shrestha:2012ep}, using a product S-matrix approach. Single-energy solutions~\cite{Shrestha:2012va} provide additional insight into structures not uncovered by global, energy-dependent multi-channel fits. 

Effective Lagrangian or Regge plus resonance approaches have also served as a tool to extract information on excited baryons from  photoproduction reactions by Kim, Mart, Petrellis, Skoupil,  Wang, Xie,  and many others~\cite{Wang:2025gfv, Wang:2025rvr, Kim:2024mqx, Wei:2024lne, Zhuge:2024iuw, 
Petrellis:2024ybj, Ben:2023uev, Petrellis:2022eqw, Wei:2022nqp, 
Strakovsky:2022jnx,
Clymton:2021wof,
Luthfiyah:2021yqe,
Wei:2019imo, 
Mart:2019fau,
Mart:2019mtq,
Bydzovsky:2019hgn,
Skoupil:2018vdh,
Clymton:2017nvp,
Wang:2017tpe,
Skoupil:2016ast,
Mart:2015jof, 
Wang:2015hfm,
Mart:2014eoa, Cao:2014mea, Kim:2014hha, He:2013ksa, Xie:2013mua, 
Mart:2013ida, 
Mart:2013fia,
Xie:2010yk,
Huang:2008nr,
Bydzovsky:2006wy,
Mart:2006dk,
Mart:1999ed, Mart:1995wu},
see also the work by the Ghent group~\cite{DeCruz:2012bv,DeCruz:2011xi} in combination with statistical tools as discussed in \cref{subsec:statistics}.
In addition, effective Lagrangian approaches have been used to analyze  related reactions simultaneously~\cite{Huang:2012xj} exploiting the universality of resonance couplings.
For such approaches the question of maintaining gauge invariance becomes relevant; Regge amplitudes with gauge invariance have been discussed by Haberzettl~\cite{Haberzettl:2024rai,Haberzettl:2021wcz} and JPAC~\cite{JointPhysicsAnalysisCenter:2024kck}; a related issue is the construction of gauge invariant form factors~\cite{Haberzettl:2019qpa, Haberzettl:1998aqi}. See also Refs.~\cite{Mathieu:2015eia, Mathieu:2017jjs, JPAC:2016lnm} for $\pi^0,\,\eta$, and $\eta'$ photoproduction at high energies using Regge theory and finite-energy sum rules, in connection with the 
first GlueX experiments~\cite{GlueX:2017zoo}.

\subsubsection{Photoproduction in the JBW approach}
\label{sec:photoJBW}
\begin{figure}[tb]
  \centering
\includegraphics[width=0.7\columnwidth,clip=]{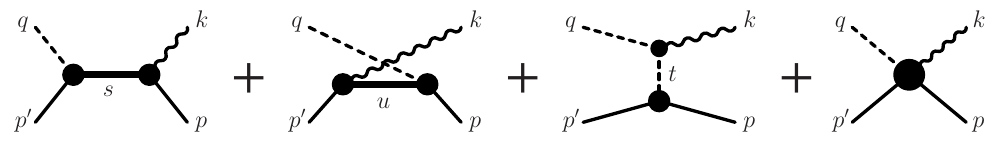}
  \caption{
  Generic structure of the pion photoproduction current $M^\mu$ for $\gamma N
  \to \pi N$. Time proceeds from right
  to left. Nucleons and pions are depicted by solid and dashed lines,
  respectively, and the photon is shown as a wavy line. The first three diagrams 
  depict the $s$-, $u$-, and $t$-channel pole diagrams $M^\mu_s$, $M^\mu_u$,
  and $M^\mu_t$, respectively. The
  last diagram shows the contact-type
  interaction current. Figure taken from Ref.~\cite{Huang:2011as}.}
  \label{fig:currents}
\end{figure}
The interaction of the photon with the hadronic interactions is constrained by gauge invariance in terms of the Ward-Takahashi identity~\cite{BOYER19671, Haberzettl:1997jg, Haberzettl:2018hcl, vanAntwerpen:1994vh}. Based on the formalism of Refs.~\cite{Haberzettl:2006bn, Haberzettl:2011zr}, single meson photoproduction was studied within the JBW approach by Huang et al. in Ref.~\cite{Huang:2011as}. The hadronic part of the JBW model was taken as final state interaction for a gauge invariant coupling scheme of the photon to $s,\,t,$ and $u$-channel processes, as well as a generalized, gauge-restoring contact term, to compute the pion photoproduction current $M^\mu$. This is indicated in Fig.~\ref{fig:currents}. The starting point is the topology of the $\pi NN$ vertex with the photon coupling to all legs. In addition, the photon can couple to the internal components of the complicated rescattering part which is parametrized with the contact term taking into account the internal structure. One obtains a tower of coupled non-linear Dyson-Schwinger-type equations that can only be solved iteratively. In practice, it is subject to a gauge-invariant truncation for the dressed nucleon current and the contact-type five-point current~\cite{Huang:2011as}. For this purpose, the non-covariant normalization of the scattering equation~\eqref{eq:lse} is transformed into the Kadyshevsky equation~\cite{Kadyshevsky:1967rs} which is a covariant three-dimensional reduction of the Bethe-Salpeter equation that preserves elastic unitarity. The photoproduction current is then constructed
within the same covariant, three-dimensional reduction.
With an efficient parametrization of the contact term in terms of a few parameters, neutral and charged pion photoproduction data were fitted up to $E=1.65\,\GeV$ because only resonances up to spin $J=3/2$ were included in Ref.~\cite{Huang:2011as}. Some coupling constants were taken from the literature, and only about 40 parameters (most of them resonance couplings to the photon) were varied to obtain an excellent description of differential cross sections and photon spin-asymmetry data on a proton as well as a neutron target~\cite{Huang:2011as}. This demonstrates that gauge invariance is not only a theoretical requirement but can be helpful in guiding the construction of efficiently parametrized amplitudes.

In subsequent analyses of photoproduction reactions, the JBW amplitude was technically simplified by switching to a partial-wave-wise parametrization. The photoproduction multipole amplitude in terms of a photoproduction kernel $V_{\mu\gamma}$ is given by~\cite{Ronchen:2014cna} 
\begin{align}
    M_{\mu\gamma}(q,E)=V_{\mu\gamma}(q,E) +\sum_{\kappa}\int\limits_0^\infty dp\,
     p^2\,T_{\mu\kappa}(q,p;E)G^{}_\kappa(p,E)V_{\kappa\gamma}(p,E)\ .
    \label{m2}
\end{align}
Here, the index $\gamma$ is used exclusively for the $\gamma N$ channel. 
Note that in the second term $V_{\kappa\gamma}$ produces a meson-baryon pair in channel $\kappa$ with
off-shell momentum $p$ that rescatters via the hadronic (partial-wave projected) half-offshell  T-matrix from Eq.~\eqref{scattering}, producing the final $\pi
N$ state (more generally, channel $\mu$) with momentum $q$. The meson-baryon propagator $G_\kappa$ is defined in Eq.~\eqref{gkappa} for two-body intermediate state; for three-body intermediate states see Sect.~\ref{sec:tauiso}.
The photoproduction kernel can be written as 
\begin{align}
    V_{\mu\gamma}(p,E)=\alpha^\npo_{\mu\gamma}(p,E)+\sum_{i} \frac{\gamma^a_{\mu;i}(p)\,\gamma^c_{\gamma;i}(E)}{E-M_i^b} \ .
    \label{vg}
\end{align}
Here, $\alpha^\npo_{\mu\gamma}$ represents the photon coupling to
$t$- and $u$-channel diagrams and to contact diagrams. These diagrams together form the non-pole part of the full photoproduction kernel as can bee seen from field-theoretical considerations~\cite{Haberzettl:1997jg}. Note that the form of the photoproduction kernel allows one to vary the strength of resonances and background independently without violating Watson's theorem~\cite{Razavi:2019vcr}. The summation in Eq.~(\ref{vg}) is 
over the resonances $i$ in a multipole, and the $\gamma^c_{\gamma;i}$ are the real-valued tree-level $\gamma NN^*_i$ and $\gamma N\Delta^*_i$ photon couplings. It is crucial that the resonance annihilation vertex $\gamma^a$ in Eq.~(\ref{vg}) is precisely the same as in the hadronic part of Eq.~(\ref{blubb}) so that the explicit singularity at $E=M_i^b$ cancels.

The two-potential formalism allows one to rewrite the photoproduction amplitude $M$ as
\begin{align}
    M_{\mu\gamma}&=\alpha^\npo_{\mu\gamma}+ \sum_{\kappa}\; T^\npo_{\mu\kappa}
     G^{}_\kappa\alpha^\npo_{\kappa\gamma}+\Gamma^a_{\mu;i}\,(D^{-1})^{}_{ij}\,\Gamma^c_{\gamma;j}\non
    \Gamma^c_{\gamma;j}&=\gamma^c_{\gamma;j}+ \sum_{\kappa}\;\Gamma^c_{\kappa;j} G^{}_\kappa\alpha^\npo_{\kappa\gamma}
    \label{twopotfinal}
\end{align}
with the dressed resonance propagator $D$ from Eq.~\eqref{2res} and with the dressed resonance-creation photon-vertex $\Gamma^{c}_{\gamma; j}$ which is a vector in resonance space, like the strong dressed vertex
$\Gamma^c_{\mu;i}$ in Eq.~(\ref{2res}). This standard result
has been derived, e.g., in Ref.~\cite{Doring:2009uc}. In the form of Eq.~(\ref{twopotfinal}) it becomes apparent
that in $M_{\mu\gamma}$ all singularities due to the bare resonance propagators of Eq.~(\ref{vg}) have canceled.

Alternatively, one can write the amplitude simply in terms of the full hadronic T-matrix as 
\begin{align}
    M_{\mu\gamma}= \sum_{\kappa} \left(1-VG\right)_{\mu\kappa}^{-1}\,
    V^{ }_{\kappa\gamma} \ .
    \label{final}
\end{align}
In principle, any of the forms (\ref{m2}), (\ref{twopotfinal}), or (\ref{final}) can be used in practical
calculations. 
In the form of Eq.~(\ref{final}), which resembles the one of Ref.~\cite{Hanhart:2012wi}, the similarity with the CM12
Chew-Mandelstam parameterization of the SAID group~\cite{Workman:2012jf} becomes apparent, in which the
hadronic kernel $\bar K_{\kappa\nu}$  of their hadronic T-matrix,
\begin{align}
    T_{\mu\nu}= \sum_{\kappa}(1-\bar K C)_{\mu\kappa}^{-1}\,\bar K^{}_{\kappa\nu}\ ,
\end{align}
is replaced by a photoproduction kernel, $\bar K_{\kappa\gamma}$,
\begin{align}
    M_{\mu\gamma}= \sum_{\kappa}(1-\bar K C)_{\mu\kappa}^{-1}\,\bar K^{}_{\kappa\gamma}\ .
    \label{saidamp}
\end{align}
Here, $C$ is the complex Chew-Mandelstam function that guarantees two-body unitarity and even the complex thresholds for effective three-body channels~\cite{Workman:2012jf}.
While Eq.~(\ref{saidamp}) is formally identical to Eq.~(\ref{final}), there is a practical difference:
Eq.~(\ref{final}) implies an integration over intermediate off-shell momenta, while the quantities $\bar K$
and $C$ in Eq.~(\ref{saidamp}) factorize. In both approaches the dispersive parts of the intermediate loops
$G$ and $C$ are maintained. In the JBW approach, the terms $\alpha_{\mu\gamma}^\npo$ and $\gamma^c_{\gamma;i}$ in Eq.~(\ref{vg}) are approximated by polynomials $P$,
\begin{align}
    \alpha^\npo_{\mu\gamma}(p,E)= \frac{ \tilde{\gamma}^a_{\mu}(p)}{\sqrt{M_N}} P^{\text{NP}}_\mu(E)\ ,  \quad
    \gamma^c_{\gamma;i}(E)= \sqrt{M^{}_N} P^{\text P}_i(E)\ ,
    \label{vg_poly}
\end{align} 
where $\tilde{\gamma}^a_{\mu}$ is a vertex function equal to $\gamma^a_{\mu; i}$ but stripped of any dependence on the
resonance number $i$. Equation~(\ref{vg_poly}) means that one has $n+m$ polynomials per multipole with $n$ resonances $i$ and $m$ hadronic channels $\mu$. With this
parameterization, non-analyticities from left-hand cuts, like the one from the pion-pole term, are approximated. As the distance to the physical region is quite large, such an approximation can be justified.
 In any case, it is clear that in this multipole-wise parametrization one needs substantially more fit parameters than in the plane-wave approach of Ref.~\cite{Huang:2011as} in which gauge invariance was used to guide the parametrization.

This review is focused on summarizing and explaining the formal aspect of DCC approaches. Therefore, we keep the discussion of applications of the JBW model to photoproduction very short. Pion photoproduction on the proton up to an energy of $E\approx 2.3\,\GeV$ was analyzed in Ref.~\cite{Ronchen:2014cna}, including also isospin breaking at low energies due to hadronic mass differences within the pion and nucleon multiplets. About 20,000 data for differential cross sections and all polarization observables available at that time were analyzed, with some exclusion of data in the extreme forward direction because the approach was truncated at a total angular momentum of $J=9/2$.
Photocouplings were extracted at the resonance poles and compared to values of the SAID and BnGa approaches. Notably, these helicity amplitudes are complex and not real as they would be for a Breit-Wigner formulation. In the pole definition, photocouplings are, by definition, independent of the final state because coupled-channel amplitude factorize at resonance poles. The photoproduction data of the $\eta N$, $K\Lambda$, and $K\Sigma$ final states were successively added in the respective analysis updates in Refs.~\cite{Ronchen:2015vfa}, \cite{Ronchen:2018ury}, and \cite{Ronchen:2022hqk}, with an accumulated data set of more than 70,000 points. All fit results to all data (pion- and photon-induced) are documented on a web site~\cite{Juelichmodel:online}. 
Results of electroproduction fits are collected on a different, interactive website~\cite{JBW-homepage}, see \cref{sec:electroproduction}. 
Many resonance pole positions, branching ratios, and electromagnetic couplings determined in the JBW approach are quoted in the PDG ``above the line''~\cite{ParticleDataGroup:2024cfk}.

\subsubsection{The weak \texorpdfstring{$\Lambda$}{Lambda} decay parameter \texorpdfstring{$\alpha_-$}{alpha-} from CLAS data}
\begin{figure}[tb]
    \centering
    \includegraphics[width=0.55\linewidth]{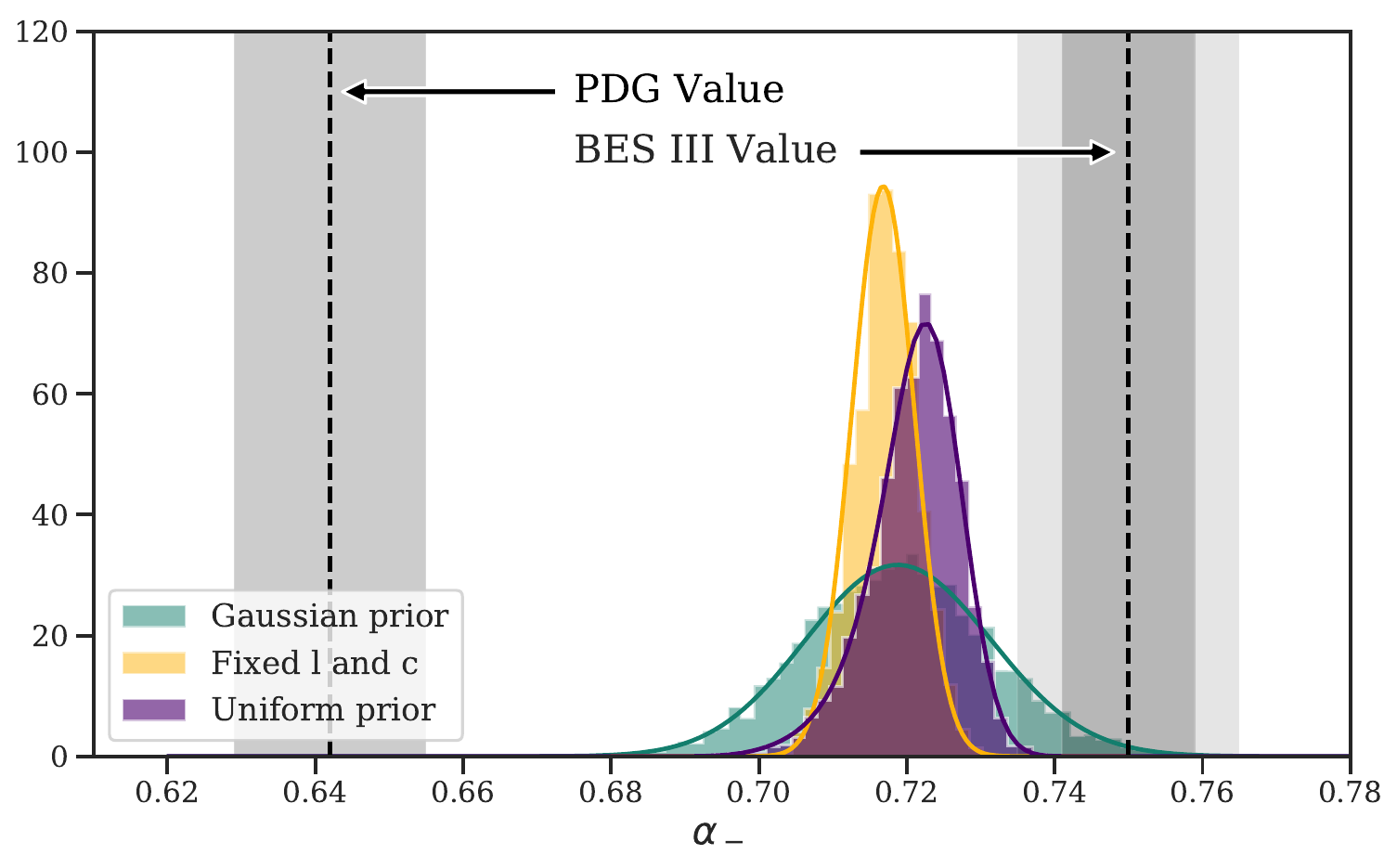}
    \caption{\label{fig:posteriors}Posterior densities for the weak $\Lambda$ decay parameter $\alpha_-$ from different priors for the beam polarization calibration constants $P_C^{\gamma}$ (scaling parameter $c$) and $P_L^{\gamma}$ (scaling parameter $l$), as indicated with the colored histograms/lines~\cite{Ireland:2019uja}. The gray vertical bands represent statistical uncertainty for the old PDG~\cite{ParticleDataGroup:2018ovx} and the new BESIII~\cite{BESIII:2018cnd} values. Figure taken from Ref.~\cite{Ireland:2019uja}.}
\end{figure}

For reactions with $\Lambda$ and $\Sigma$ final states, it should be noted that in experimental analyses the weak asymmetry parameter $\alpha_-$ scales the values of polarization observables. The value of this fundamental constant was questioned through new experimental evidence at BESIII~\cite{BESIII:2018cnd}. In Ref.~\cite{Ireland:2019uja}, Ireland and JBW independently determined $\alpha_-$ using photoproduction data in combination with Fierz identities, 
\begin{align}
    O^2_x + O^2_z + C^2_x + C^2_z + \Sigma^2 - T^2 + P^2 &= 1 
    \label{eq:Fierz1}
    \\
    \Sigma P - C_x O_z + C_z O_x - T &= 0 \ .
    \label{eq:Fierz2}
\end{align}
As all observables in Eqs.~(\ref{eq:Fierz1}) and (\ref{eq:Fierz2}) were measured by CLAS~\cite{CLAS:2009rdi, CLAS:2006pde, CLAS:2016wrl},  these Fierz identities can be used to estimate the calibration parameters $l$,  $c$, and $a$, where $l$ and $c$ are the relative systematic uncertainties of the linear and circular photon polarization, while $a$ varies the value of the weak decay parameter $\alpha_-$. It is known that  $O_x,\,O_z,\, T$ scale with $al$, $C_x$ and $C_z$ scale with $ac$, $\Sigma$ scales with $l$, and $P$ with $a$. If one has sufficiently many measurements, this set of scaling conditions together with the Fierz identities of Eqs.~(\ref{eq:Fierz1}, \ref{eq:Fierz2}) allow to extract the scaling factor $a$, and, hence, a re-determination of $\alpha_-$. Special care has to be paid to statistical aspects as this problem bears a few intricacies: the Fierz identities are to be fulfilled in a statistical sense; the estimation of normalization factors is notoriously prone to biases that have to be avoided; and the distribution of the systematic uncertainties for linear and circular polarization are unknown. Using different priors for these distributions, a new value of $\alpha_-=0.721(6)(5)$ was found which is close to, but not compatible with, the new BESIII value of $\alpha_-=0.750(9)(4)$. In contrast, the previous PDG value~\cite{ParticleDataGroup:2018ovx} of $\alpha_-=0.642(13)$ had been accepted and used for over 40 years, affecting all hyperon polarization data.
The results are summarized in Fig.~\ref{fig:posteriors}.
In Ref.~\cite{Ireland:2019uja}, the data affected by the change in $\alpha_-$ data were re-fitted with the JBW approach. The re-analysis shows that the lowest $\chi^2$ is indeed obtained with the $\alpha_-$ value obtained from CLAS data, pointing towards internal consistency of the new value, that is now quoted in the PDG ``above the line''~\cite{ParticleDataGroup:2024cfk}.

\subsubsection{Photoproduction in the ANL-Osaka approach}
\label{sec:ANLPhoto}
\begin{figure}
    \centering
    \includegraphics[width=\linewidth]{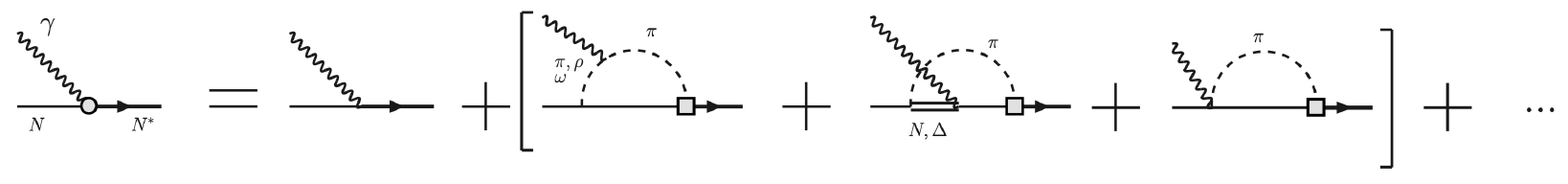}
    \caption{Graphical illustration of the contribution of the $\pi N$ intermediate state to the dressed $\gamma N \rightarrow N^*$~\cite{Julia-Diaz:2006ios} transition. The term in brackets illustrates the meson-cloud effect.}
    \label{fig:ANL-photoproduction}
\end{figure}

The partial wave T-matrix element for the meson photoproduction
$\gamma(\bm{p}) + N(-\bm{p}) \rightarrow M'(\bm{p}') + B'(-\bm{p}')$
  is given by Eq.~\eqref{eq:fullt} with
  $\alpha=\gamma N,J,\lambda_\gamma,\lambda_N$  
  and $\beta =M'B',LSJ$. In the following,
  we simply write $\alpha=\gamma N$
  and $\beta =M'B'$ to make clear the Fock-space component of the states. The photoproduction amplitude is given by a non-resonant and a resonant part,
\begin{align}
    T_{M'B',\gamma N}(p',p;E)
    =  t_{M'B',\gamma N}(p',p;E)
    + t^R_{M'B',\gamma N}(p',p;E).
    \label{eq:fullgam}
\end{align}
In the following, we assume only the first order term of the electromagnetic
coupling constant $e$  and the processes which need
higher order terms like the $\gamma-\gamma$ amplitude of
Compton scattering or $\gamma - Z$ amplitude of  parity violating processes are not considered here.
The non-resonant photoproduction T-matrix $t_{M'B',\gamma N}(p',p;E)$ is given
as follows by rewriting Eq.~\eqref{eq:mext},
\begin{align}
    t_{M'B',\gamma N}(p',p;E)
    & =  v_{M'B',\gamma N}(p',p)+     \int_C dq q^2 \sum_{M'',B''}
    t_{M'B',M''B''}(p',q;E)\ G_{M''B''}(q;E)\ 
    v_{M''B'',\gamma N}(q,p)\ ,
    \label{eq:mextgam}
\end{align}
which includes all order of the strong interaction within the model and the
first order electromagnetic interaction $v_{MB,\gamma N}$. The potential
$v_{MB,\gamma N}$ is constructed by the $s$-, $t$-, and $u$-channels hadron exchange processes
applying Eq.~\eqref{eq:unit-ext}.
The resonant T-matrix $t^R_{M'B',\gamma N}(p',p;E)$ is given by Eq.~\eqref{eq:tr}
with the dressed electromagnetic helicity amplitude $\bar{\Gamma}_{\gamma N,j}(p;E)$ of  $N_j^*$ as
\begin{align}
    t^R_{M'B',\gamma N}(p',p;E) = 
    \sum_{i,j}\bar{\Gamma}_{M'B',i}(p';E)
    \ [G_{N^*}(E)]_{i,j}\ 
    \bar{\Gamma}_{\gamma N,j}(p;E).
    \label{eq:tr-gam}
\end{align}
The dressed helicity amplitude $\bar{\Gamma}_{\gamma N,j}$ is given as
\begin{align}
    \bar{\Gamma}_{\gamma N,j}(p;E) & = 
    \Gamma_{\gamma N,j}(p) + \int_C dq q^2 \sum_{MB}
    \Gamma_{MB,j}(q) \ G_{MB}(q;E) \ t_{MB,\gamma N}(q,p;E) \ .
    \label{eq:dressf-gam} 
\end{align}
This formula corresponds to Eq.~\eqref{twopotfinal} of the JBW model.
Due to the non-resonant electromagnetic meson production process
$t_{MB,\gamma N}$, the bare helicity amplitude $\Gamma_{\gamma N,j}$
is dressed $\bar{\Gamma}_{\gamma N,j}$ as illustrated in
Fig.~\ref{fig:ANL-photoproduction}. The contribution of non-resonant processes is called the ``meson cloud effect''.

The pion photoproduction reactions are investigated up to the second resonance region $1.1 < W < 1.6  $ GeV in Ref.~\cite{Julia-Diaz:2007mae}.  
Starting from the strong interaction model JLMS~\cite{Julia-Diaz:2007qtz} of the pion induced reaction, the bare resonance helicity amplitudes are
the only free parameters adjusted to describe the data of pion photoproduction. Large effects of coupled channel in the second resonance region are found. The $Q^2$ dependence of the meson could effects are studied, which are important in the low-$Q^2$ region. 

The current ANL-Osaka DCC model~\cite{Kamano:2013iva}
is an extension of the JLMS model~\cite{Julia-Diaz:2007qtz}.
In the extension 
(1) $K\Lambda$ and $K\Sigma$ states are included
in addition to the  $\pi N, \eta N, \pi\pi N$ and
$\pi \Delta, \rho N, \sigma N$ states,
(2) a unified analysis of $\pi N$ and $\gamma N$ reactions is performed,
(3) the analysis is extended to include higher-energy $\sqrt{s} < 2.1$~GeV reactions.
The single energy solutions of the SAID partial wave amplitudes
are fitted up to $\sqrt{s} < 2.3$~GeV while $\pi N$ and $\gamma N$ reactions
are analyzed below $\sqrt{s} < 2.1$GeV. 
Lists of the fitted observables
and the number of data points are given in Ref.~\cite{Kamano:2013iva}.
The bare helicity amplitudes
$A_\lambda^{N^*}$ are parametrized  using  magnetic $M_{l\pm}^{N^*}$ and
electric $E_{l\pm}^{N^*}$ multipole amplitudes as 
\begin{eqnarray}
  A_{3/2}^{N^*} & = & \frac{\sqrt{l(l+2)}}{2}[- M_{l+}^{N^*} + E_{l+}^{N^*}],\ \ 
  A_{1/2}^{N^*}  = - \frac{1}{2}[l M_{l+}^{N^*} + (l+2) E_{l+}^{N^*}]
  \ \ \mbox{for} \ \ j=l+1/2 \\
  A_{3/2}^{N^*} & = & -\frac{\sqrt{(l-1)(l+1)}}{2}[ M_{l-}^{N^*} + E_{l-}^{N^*}],\ \ 
  A_{1/2}^{N^*}  =  \frac{1}{2}[(l+1) M_{l-}^{N^*} - (l -1) E_{l-}^{N^*}]
  \ \ \mbox{for} \ \ j=l-1/2.
  \label{eq:hel-bare}
\end{eqnarray}
In Ref.~\cite{Kamano:2013iva}, 
$M_{l\pm}^{N^*}$ and $E_{l\pm}^{N^*}$ are parametrized by the low-energy photon
momentum dependence of the magnetic and electric multipole amplitudes
and include a damping factor for the large momentum behavior. The 
bare helicity amplitudes were treated as constant in the previous analysis~\cite{Julia-Diaz:2007mae}. The results of the model for observables and the multipole amplitudes can be found on a web site~\cite{anl:online}.

\subsubsection{Summary of the light baryon spectrum}
\label{sec:specres}
\begin{figure}[t]
    \centering
    \includegraphics[width=0.41\linewidth]{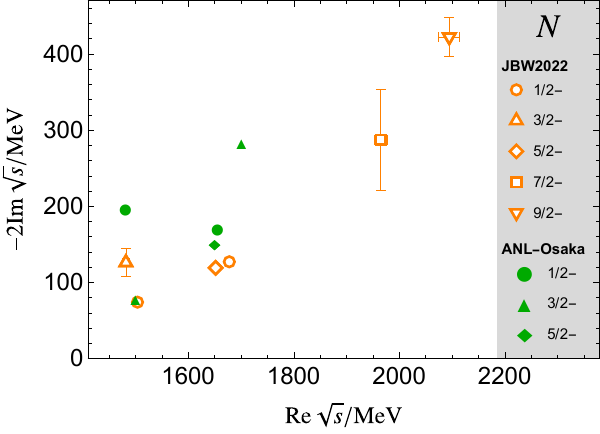}
    ~~~~
    \includegraphics[width=0.41\linewidth]{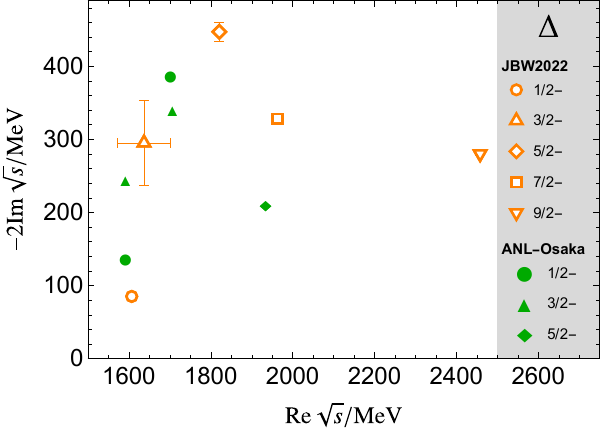}
    \\[-5mm]
    \includegraphics[width=0.41\linewidth]{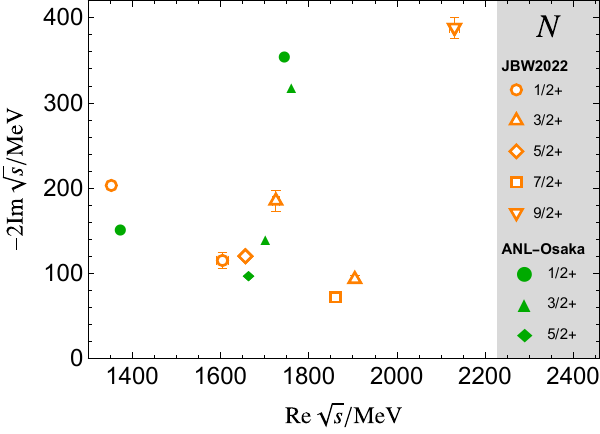}
    ~~~~
    \includegraphics[width=0.41\linewidth]{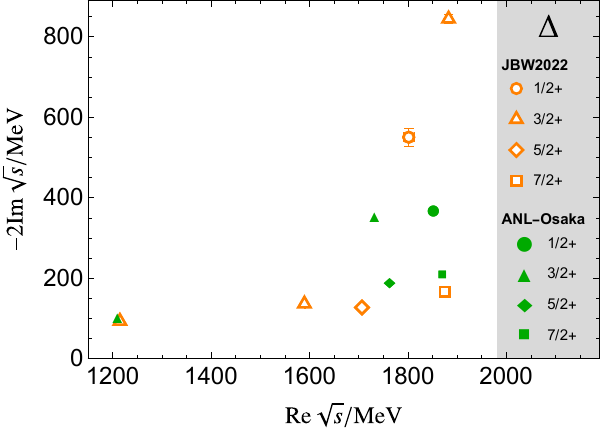}
    \\[0.3cm]
    \includegraphics[width=0.41\linewidth]{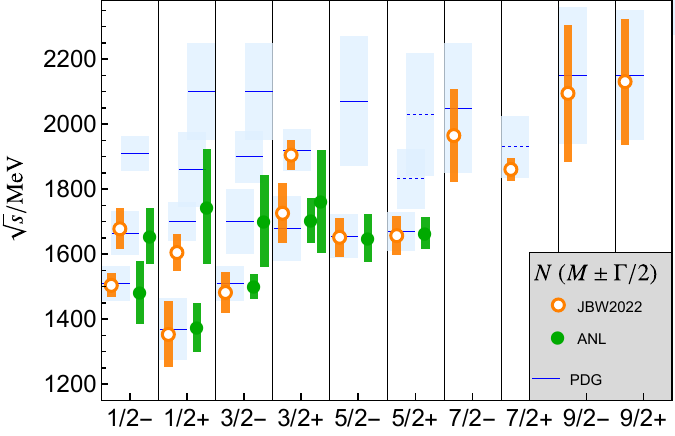}
    ~~~~
    \includegraphics[width=0.41\linewidth]{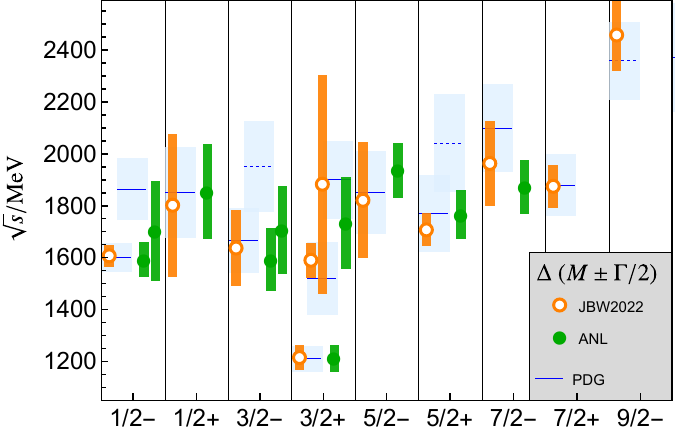}
    \caption{Summary of results for the baryon parameters obtained in the ANL-Osaka~\cite{Kamano:2013iva} and JBW~\cite{Ronchen:2022hqk} approaches. The error bars in the upper four plots show uncertainties, the lengths of the bars in the two lower plots correspond to the resonance widths. For these cases, PDG information~\cite{ParticleDataGroup:2024cfk} is displayed in blue. Note that for the ANL-Osaka model, only resonances up to $2\,\GeV$ in mass and $400\,\MeV$ in width are quoted.
    }
    \label{fig:BaryonsSummary}
\end{figure}

In Fig.~\ref{fig:BaryonsSummary} the baryon spectrum is shown as determined by the ANL-Osaka~\cite{Kamano:2013iva}
and the JBW~\cite{Ronchen:2022hqk} approaches. The upper set of pictures shows the pole positions with uncertainties ordered by isospin and parity. The two lower pictures display similar information in bar plots, but with PDG states in addition. Here, the heights of the bars correspond to the resonance widths.
The resonance spectrum in the JBW approach has changed over the years, with an example discussed in \cref{sec:P11}.

Overall, both approaches exhibit a similar spectrum although there are remarkable differences, especially for some of the widths. Other than $S_{11}$ and $P_{11}$, both approaches find two $N^*$ with $J^P=3/2^+$, but the heavier one differs in mass and width. For $J^P=1/2^-$ and $J^P=3/2^-$, the JBW approach has only one $\Delta$ resonance, while the ANL-Osaka has two. For the positive parity $J^P=3/2^+$, both approaches have, of course, the $\Delta(1232)$. The JBW approach has a $\Delta(1600)$ and a very wide, heavy $\Delta$ resonance. It is debatable how stable  such wide resonances are under re-parameterizations of the amplitude, which is why the ANL-Osaka approach only quotes states up to 400 MeV width. See also \cref{subsec:statistics} where stability tests for resonance signals are discussed. Another question is the bias induced by fitting to the SAID partial-wave analysis. Like in many other analyses, information from SAID on elastic $\pi N$ scattering is contained in both analyses, and many states of the SAID analysis are also quoted in the PDG. This might induce a confirmation bias. The solution lies in directly fitting the elastic $\pi N$ data. However, this is complicated because of a very inhomogeneous data set with systematic uncertainties that are often unknown. See \cref{sec:mesonproduction} for a discussion.

\subsubsection{Highlight: The \texorpdfstring{$S_{11}$}{S11} partial wave}
\label{subsec:S11}
The $J^P=1/2^-$ partial wave features the excited states $N(1535)$, $N(1650)$ and, possibly, an $N(1895)$. As Fig.~\ref{fig:BaryonsSummary} shows, the ANL-Osaka and JBW approaches find the first two resonances but not the third state. The latter state has been been found in the BnGa analysis that also contains the $\eta' N$ channel~\cite{CBELSATAPS:2020cwk} (it is also found in the Kent State analysis~\cite{Hunt:2018wqz}). Once the DCC approaches contain this channel, a need for that state may arise.

Theoretical interest in studying the first state, $N(1535)$, roots in the fact that it is heavier than the second state of the parity-partner channel $P_{11}$, the Roper resonance $N(1440)$, which challenges the expectation from  quark models
~\cite{Isgur:1977ef, Isgur:1978wd, Loring:2001kx}. Furthermore, inelastic meson-baryon thresholds ($\eta N$, $K\Lambda$, and $K\Sigma$) are located close to the first two $J^P=1/2^-$ states, such that coupled channel effects need to be taken into account. 
In Ref.~\cite{Liu:2005pm} it was argued that the $N(1535)$ couples even stronger to the $K\Lambda$ than to the $\eta N$ channel by examining BES data, see also Ref.~\cite{Sibirtsev:2006ia}. 

Within chiral unitary models, it was realized early~\cite{Kaiser:1995cy} that low-lying $S_{11}$ resonances might not be a three-quark (pre-existing) states but rather be generated by strong channel dynamics, with a substantial $K\Sigma - K\Lambda$ component in their wave functions.  This idea was extended in Ref.~\cite{Inoue:2001ip}, where within certain approximations the effects of 3-body $\pi\pi N$ channels were also included. In Ref.~\cite{Nieves:2001wt},  the $S_{11}$ phase shift was fitted from threshold to about $\sqrt{s} \simeq 2\,\GeV$ together with cross section data for $\pi^- p\to \eta n$ and $\pi^- p\to K^0 \Lambda$ in the respective threshold regions. This led to a satisfactory description of the $S_{11}$ phase. Two poles were found corresponding to the $S_{11}(1535)$ and the $S_{11}(1650)$ resonances together with a close-by unphysical  pole on the first Riemann sheet. More recently, it was pointed out in a  meson-exchange model that there is indeed  strong resonance interference between the two $S_{11}$ resonances, as each of these resonances provides an energy-dependent background in the region of the other~\cite{Doring:2009yv}.

In \cref{subsec:bseUCHPT} a scheme to solve the four-dimensional Bethe-Salpeter equation with chiral LO and NLO contact terms was discussed. The resulting coupled-channel, chiral unitary amplitude was derived in Ref.~\cite{Bruns:2010sv}, and the low-energy constants were fitted to the SAID partial waves providing a compelling picture of a dynamically generated $N(1535)$ as a pole on the second Riemann sheet shown in \cref{fig:RiemannSheets}. Additionally, a second pole is recorded outside of the fitted energy region. It is quite tempting to assume this to be a prediction of the $N(1650)$ state from chiral dynamics and coupled-channel unitarity. In Ref.~\cite{Garzon:2014ida} it was found that this resonance has a strong coupling to vector-meson baryon components.

\begin{figure}[t]
    \centering
    \includegraphics[width=0.5\linewidth]{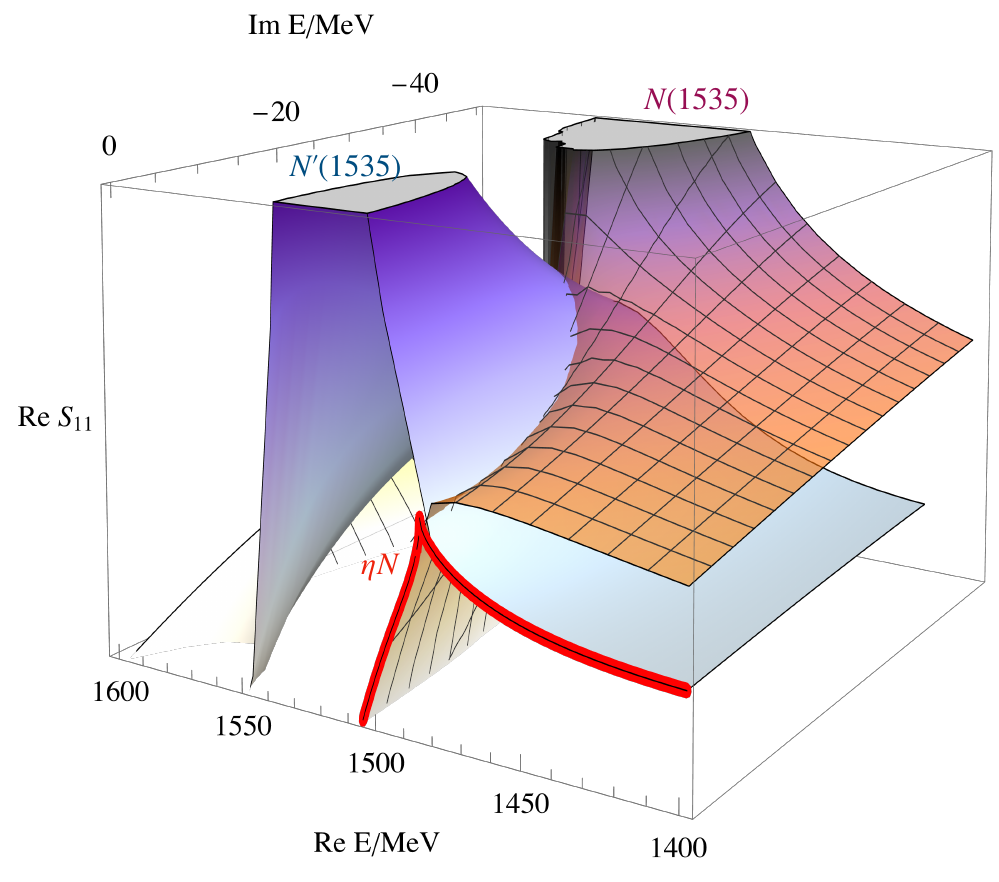}
    \includegraphics[width=0.42\linewidth]{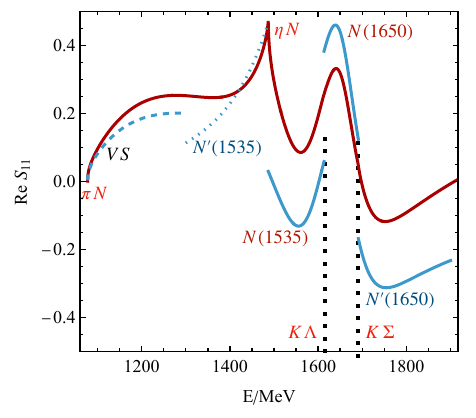}
    ~~~~~
    \caption{{\bf Left}: In orange, the $N(1535)$  on the sheet connected to the physical axis (red line) above the $\eta N$ cusp;   in blue, a hidden replica $N'(1535)$ on the sheet connected to the physical axis below the  $\eta N$ cusp. Together, they generate the particular cusp structure and unusual $N(1535)$ shape of the $S_{11}$ amplitude (red solid line). See also Fig.~\ref{fig:unif2-amp} for a discussion on cusp structures.
    {\bf Right:}
    The real part of the $S_{11}$ amplitude (dark red) with two-body thresholds and residue terms $a_{-1}/(E-E_0)$ where $a_{-1}$ is the residue of the pole at $E=E_0$ (blue dashed, dotted, and solid lines). The enhancement of Re~$S_{11}$ at the $\pi N$ threshold is due to a virtual state (VS) as the blue dashed line shows. See text for further explanations.
    Figures adapted from Ref.~\cite{Doring:2009uc}.}
    \label{fig:S11cusp}
\end{figure}

Coupled-channel dynamics is an integral aspect of the $N(1535)$, irrespective of its nature. This resonance is located very closely to the $\eta N$ threshold to which it also couples very strongly~\cite{ParticleDataGroup:2024cfk}. 
As resonance poles usually repeat on different Riemann sheets (see discussion in Sec.~\ref{sec:analytic}), one has here the situation of a regular $N(1535)$ and a hidden $N'(1535)$ on a different sheet that, together, generate the particular $\eta N$ cusp structure of the $S_{11}$ partial wave as illustrated in Fig.~\ref{fig:S11cusp}.

It should be noted that this is not what is usually referred two as ``two-pole nature'' of a resonance. In the limit of vanishing coupling of a resonance to an inelastic two-body channel, it (trivially) repeats on both sheets of that channel at the exact same position. If the channel coupling is stronger, as it is the case for the $N(1535)$ and the $\eta N$ channel, the poles repeat at substantially different positions. In contrast, the $\Lambda(1405)$, which is believed to have a true two-pole nature, has both its poles on the same Riemann sheet~\cite{Oller:2000fj, Jido:2003cb}.

To the right in Fig.~\ref{fig:S11cusp},
the real part of $S_{11}$ from the fit of Ref.~\cite{Doring:2009uc} is shown (dark red). In addition we show the ``pole approximations'' (PA) of the $N(1535)$ and the hidden $N'(1535)$ with the solid and dotted blue line as indicated in the figure. These curves are given by $a_{-1}/(E-E_0)$ where $a_{-1}$ is the residue of the pole at $E=E_0$. Consequently, the dotted curve only shows the ``shoulder'' of the hidden $N'(1535)$. Indeed, the shapes of these two PA's explain the cusp quantitatively. Still the PA's of the $N(1535)$ and $N(1650)$ show a large offset with the actual partial wave - this reflects the fact that each of these resonances provides a substantial background for the other one. 

In addition to these PA's, there is another PA from a virtual state (VS) at $E_0=(1031-i\,203)$~MeV shown with the dashed line, that explains the sharp rise of the $S_{11}$ partial wave at threshold. While that state was found in Ref.~\cite{Doring:2009uc} in 2009, no claims for its existence were made as it is located far from the physical axis and the amplitude parametrization of Ref.~\cite{Doring:2009uc} was rather simple. However, this state has recently received substantial interest. In Ref.~\cite{Wang:2017agd} (see also Ref.~\cite{Wang:2018nwi} and Ref.~\cite{Li:2025man} for a review), Wang et al. discovered it at $ E_0=(861 \pm 53)-i(130 \pm 75)\,\MeV$ using the general Peking University (PKU) representation~\cite{Xiao:2000kx, He:2002ut, Zheng:2003rw, Zhou:2006wm} in connection with relativistic baryon chiral perturbation theory to estimate the contributions from the left-hand cut. In Ref.~\cite{Cao:2022zhn} the sub-threshold singularity was found again at $E_0= (918 \pm 3)-i(163 \pm 9)\,\MeV$ in a Roy-Steiner analysis that was checked numerically with the previous framework of Hoferichter et al.~\cite{Hoferichter:2015hva}. 
Resonance poles in the latter analysis were searched in
Ref.~\cite{Hoferichter:2023mgy}. Apart from a determination of the $\Delta(1232)$ and Roper poles, the subthreshold pole at $E_0=(913.9\pm 1.6)-i(168.9\pm 3.1) \,\MeV$ was found again. There have also been dispersive studies of low-energy pion photoproduction~\cite{Cao:2021kvs, Ma:2020hpe} extracting the photocoupling of that state which was found to be substantial. The pole trajectory of the state with pion mass has been studied in a linear $\sigma$ model~\cite{
Li:2025fvg}. While there are now strong indications for this state, ``the physical interpretation of this singularity is far from obvious, as its position, far in the complex plane, casts doubts on its relevance for observables in the physical region''~\cite{Hoferichter:2023mgy}.

\subsubsection{Highlight: The \texorpdfstring{$P_{11}$}{P11} partial wave}
\label{sec:P11}
In the nucleon channel $J^P=1/2^+$, the properties, interactions, and nature of the $N(1440)$ or Roper resonance~\cite{Roper:1964zza, Arndt:1985vj} have been discussed extensively. The reversed mass pattern between $N(1535)$ and $N(1440)$ appeared as a puzzle in quark models~\cite{Isgur:1977ef, Isgur:1978wd, Loring:2001kx}. A proper description in the constituent quark model can be achieved by including Goldstone bosons as effective degrees of freedom~\cite{Glozman:1995fu}. 
Note that most recent functional methods do indeed reproduce the right ordering~\cite{Eichmann:2016hgl}. For further details see the review~\cite{Burkert:2017djo}.

Phenomenological analyses reveal an intricate analytic structure of the Roper~\cite{Arndt:2006bf, Doring:2009yv, Alvarez-Ruso:2010ayw} including strong coupling to the three-body channels. Indeed, the $P_{11}$ partial wave has an early and strong onset of inelasticity~\cite{Arndt:2006bf}. This is related to the fact that for the $J^P=1/2^+$ quantum numbers three particles $\pi,\,\pi,\, N$ can all be in a relative $S$-wave, i.e., there is no centrifugal barrier that could suppress the influence of three-body channels in the $\sigma N$ channel (and other effective two-body channels such as a $\pi N$ system with $J^P=1/2^-$ and a third pion). In addition, the Roper pole is close to the complex $\pi\Delta$ branch point as shown in Fig.~\ref{fig:anap11}. As poles repeat on different sheets, there is a complicated interaction of two poles and a complex branch point as first noted by the SAID group~\cite{Arndt:2006bf}. On the physical axis, this leads to the unusual, non-Breit-Wigner shape of the Roper. As discussed before, one would \emph{not} refer to this as a two-pole resonance like the $\Lambda(1405)$, 
since for the latter both poles lie on the same Riemann sheet.

As there is a $\sigma N$ channel in relative $S$-wave, where $\sigma$ stands for the $f_0(500)$, the ``threshold'' of that pseudo-two-body system lies, indeed, around the Roper $N(1440)$ mass. Dynamically generated states are often situated close to two-body $S$-wave thresholds, and, indeed, the Roper resonance has been found as dynamically generated from this channel in the JB approach without a $qqq$ core~\cite{Schutz:1998jx,Krehl:1999km}.
Mass trajectories in specific amplitude parametrization, that fit QCD lattice eigenvalues, seem to support the dynamical generation~\cite{Owa:2025mep}.
A substantial ingredient in this picture is also the renormalized nucleon pole that has a pole part reaching to energies above the $\pi N$ threshold (see \cref{sec:schannel} and Fig. 38 in Ref.~\cite{Ronchen:2012eg}). The Roper resonance was also dynamically generated in the multichannel model of Refs.~\cite{Golli:2017nid, Golli:2017bai} in which quark model predictions were used to fix coupling constants. 

\begin{figure}
    \begin{center}  \raisebox{-.5\height}{\includegraphics[width=0.4\textwidth]{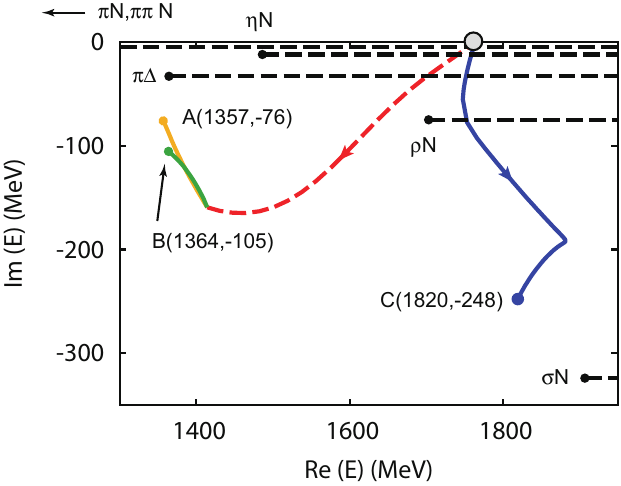}}~~~~~~
\begin{minipage}{0.5\textwidth}
  \hspace*{-0.1cm}
    \includegraphics[width=1\textwidth]{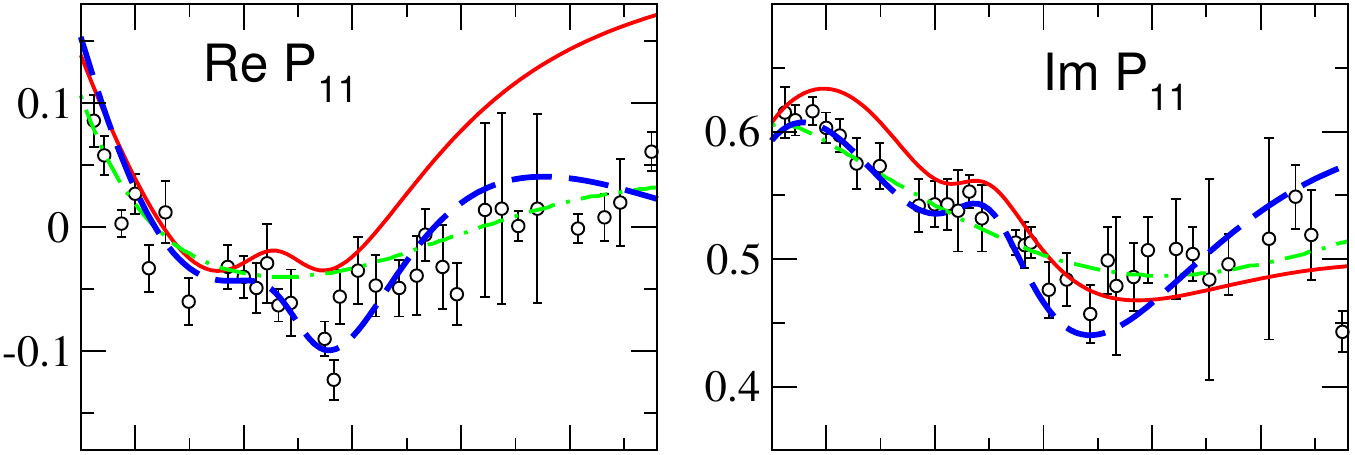}\\
    \vspace*{-0.5cm}\\
\includegraphics[width=1\textwidth]{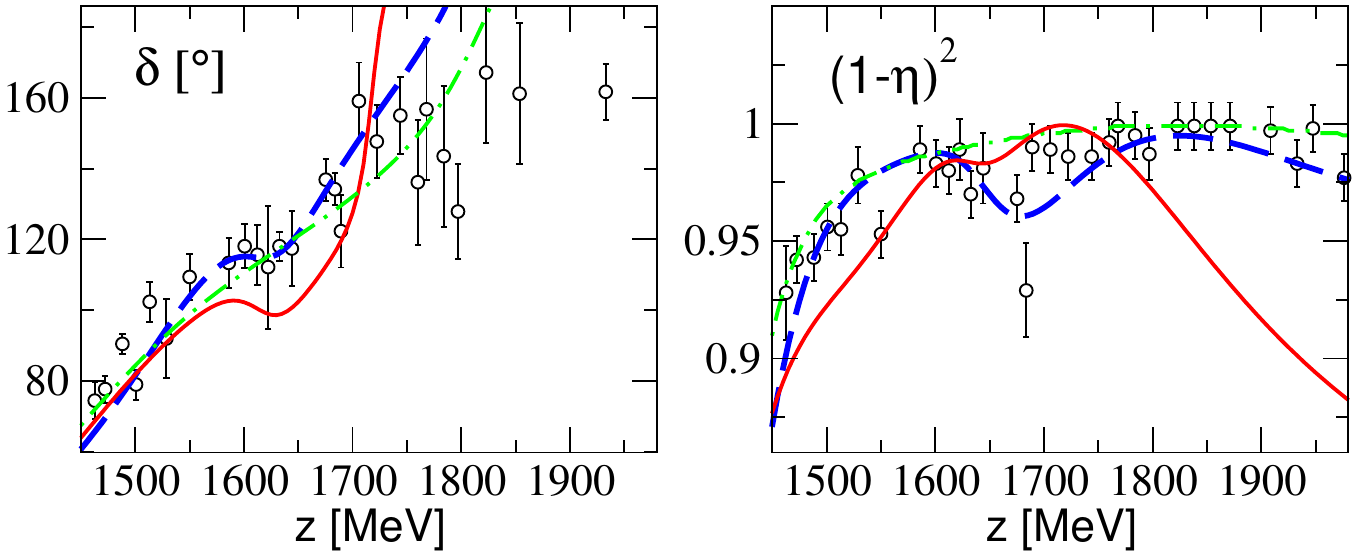}
\end{minipage}
    \end{center}
    \caption{{\bf Left}: The Roper poles A, B and the $N(1710)$ pole C in the ANL-Osaka parametrization of Ref.~\cite{Suzuki:2009nj}, on different Riemann sheets. The trajectories are obtained by slowly switching off the self energies of a genuine $s$- channel state showing that, in this parametrization, the resonances originate from the same bare state. Here, the cuts are run into the real-energy direction in contrast to Fig.~\ref{fig:anap11} where they are run into the negative Im~$E$ direction.
    {\bf Right}: Detail of the partial-wave amplitude (top) and phase shift/inelasticity (bottom) of the elastic $\pi N\to\pi N$ $P_{11}$ wave in the JB solution of 2012~\cite{Ronchen:2012eg}, as a function of energy $z=E$. Shown are fit A (solid red lines) and fit B (dashed blue lines) that represent fits to the energy-dependent GWU/SAID
    solution~\cite{Arndt:2006bf} (dash-dotted green lines). The single-energy solution~\cite{Arndt:2006bf} (data points) is {\it not} included in
    the fitting procedure but predicted.
    Figures taken from Refs.~\cite{Suzuki:2009nj, Ronchen:2012eg}.
    }
    \label{p11detail}  
\end{figure}

In the ANL-Osaka model, the Roper resonance resonance appears from dressing a bare state (``genuine resonance'') through coupled channels~\cite{Suzuki:2009nj}. In fact, even the next-higher state, the $N(1710)1/2^+$ originates from the same bare state as shown in Fig.~\ref{p11detail} to the left. The corresponding trajectories are obtained by slowly switching on the bare pole self energies in the different channels. 
While both pictures of the Roper are intriguing and allow for their own interpretation, they also demonstrate 
the dependence of the interpretations (of the same data) on the underlying parametrization.
In Fig. 43 of Ref.~\cite{Ronchen:2012eg} the dependence of the bare mass on the bare resonance couplings and included channels was studied. It was shown that one can obtain very similar descriptions of the  $\Delta(1620)$ amplitude in the $S_{31}$ wave for variations of the bare mass over more than a GeV (with corresponding variations of bare couplings). In contrast, as expected, resonance residues change very little. 

 The dependence of the extracted resonance parameters on the reaction model and the accuracy of the empirical partial wave amplitude are commonly
 investigated issues.
In the ANL-Osaka model, the analytic structure of the $P_{11}$ amplitude near the nucleon pole is rather different from the JBW model.  In Ref.~\cite{Kamano:2010ud}, a bare nucleon state is included  with mass and vertex renormalization. Then, the SAID amplitude is re-fitted and the CMB amplitude is examined for comparison.
The results indicate that the pole structure of the Roper resonance is rather insensitive to the analytic structure below the $\pi N$ threshold. In addition, the study showed the need for more accurate data in the $  \sqrt{s} > 1.6$~GeV region for precise extraction of poles.

Many attempts have also been made to tackle the Roper channel through Lattice QCD, but no definite conclusions have been drawn despite significant efforts~\cite{Alexandrou:2013fsu, Roberts:2013ipa, Alexandrou:2014mka, Liu:2014jua, Engel:2013ig, Edwards:2011jj, Mathur:2003zf}. More sophisticated lattice studies~\cite{Lang:2016hnn, Kiratidis:2016hda} followed later, consistently ruling out the exclusive $qqq$ interpretation of the Roper-resonance. Instead the sizable coupling to the five-quark operators (meson-baryon states) is noted in Ref.~\cite{Kiratidis:2016hda}. In a recent lattice calculation, a multi-two-body-channel model was set up to extend the energy range up to the next excited state beyond the Roper~\cite{Owa:2025mep}. Owing to different available pion masses, resonance pole trajectories were mapped out. 

Notably, there is a missing level~\cite{Lang:2016hnn,Padmanath:2017oya,Leskovec:2018lxb} when comparing to the expectation from elastic $\pi N$ scattering alone. There are also further attempts based on Hamiltonian effective field theory (HEFT)~\cite{Wu:2017qve} to interpret the measured finite-volume spectrum~\cite{Lang:2016hnn}. The clean solution to analyzing the finite-volume spectrum of the Roper resonance lies in correctly including the three-body dynamics in form of the  $\pi\pi N$ inelastic channels, see Ref.~\cite{Severt:2020jzc} for progress in this direction. For progress of finite-volume three-body physics in general, see \cref{3BQC} and reviews~\cite{Hansen:2019nir,Mai:2020ltx}. 

In Fig.~\ref{p11detail} to the right, the region around $\sqrt{s}=E\approx 1.7\,\GeV$ is displayed in more detail.
The figure shows, apart from solutions A and B from the 2012 JBW analysis~\cite{Ronchen:2012eg}, the energy-dependent $P_{11}$ partial wave from Ref.~\cite{Arndt:2006bf} that was fitted (green
dash-dotted lines). The SES (data) were \emph{not} fitted. Nonetheless,  fit B, and to some extent also A, do exhibit the structure of the SES from Ref.~\cite{Arndt:2006bf} as the $N(1710)$ moved to that position.
In the 2012 JB analysis, the inclusion of the $N(1710)$ became necessary for an entirely different reason, i.e., to improve the description of the $\pi^- p\to K^0\Lambda$ differential cross
section and polarization data. In Ref.~\cite{Ronchen:2012eg}, the unexpected repercussion for elastic $\pi N$ scattering was taken as strong 
evidence for a resonance at this energy, and as  a powerful demonstration how DCC approaches connect different reactions. 

The existence of the $N(1710)$ resonance has also been advocated for based on $\eta N$ data~\cite{Ceci:2006ra}. 
While the existence of this resonance is well established, nowadays the $P_{11}$ partial
wave around that energy has served as a testing ground to discuss the role of complex threshold openings from three-body channels in baryon spectroscopy~\cite{Ceci:2011ae}. The starting point of this exercise is the smooth, energy-dependent $P_{11}$ amplitude from SAID or an older version of the JBW amplitude that both do not contain an $N(1710)$ pole, but which both \emph{do} contain the complex threshold opening of the $\rho N$ channel (see Fig.~\ref{fig:anap11}) at around the same energy, $E=M_N+E_\rho$ where $E_\rho$ is the complex $\rho(770)$ pole position. This amplitude is then fit with an early version of the Carnegie-Mellon-Berkeley type model developed by the Zagreb group~\cite{Batinic:1995kr, Batinic:1997gk, Batinic:2010zz, Ceci:2006ra} which does not contain this branch point, but which does contain resonances to fit amplitude structures. Indeed, the fit demands a resonance at $E=(1698- i130)\,\MeV$ which simulates the branch point missing in that model. The exercise demonstrates the need to explicitly include known non-analyticities like complex threshold openings in amplitude parametrizations, to avoid false positives in baryon detection.

\subsection{Electroproduction reactions}\label{sec:electroproduction}
Electromagnetic probes of strongly interacting matter provide independent access to emergent phenomena of QCD such as resonances.  Photoproduction reactions have been used to determine the spectrum and properties of excited baryons~\cite{Ireland:2019uwn, Thiel:2022xtb} as analyzed by different groups~\cite{Shklyar:2005xg, Drechsel:2007if, Anisovich:2011fc, Workman:2012jf, Kamano:2013iva, Ronchen:2014cna, Hunt:2018mrt}. These analyses allow for a comparison to theory like lattice QCD~\cite{Burch:2006cc, Bulava:2010yg, Engel:2010my, Edwards:2011jj, Menadue:2011pd, Edwards:2012fx, Dudek:2012ag, Alexandrou:2013ata, Alexandrou:2015hxa, Stokes:2019zdd, RQCD:2022xux} or quark models~\cite{Ferraris:1995ui, Glozman:1995fu, Glozman:1997ag, Loring:2001kx, Giannini:2001kb, Santopinto:2004hw, Bijker:2009up, Ronniger:2011td, Klempt:2012fy, Liu:2022ndb}. See Ref.~\cite{Mai:2022eur} for a review. Notably, first calculations of meson-baryon scattering amplitudes in lattice QCD have appeared recently, some of them containing the $\Delta(1232)3/2^+$ resonance~\cite{Meissner:2010ij,Andersen:2017una, Silvi:2021uya, Pittler:2021bqw, Bulava:2022vpq}.
Complementary to photoproduction reactions, radiative decays of excited baryons, such as measured by CLAS~\cite{CLAS:2002zlc,CLAS:2006ogr,Nasseripour:2008aa,CLAS:2009sbn,CLAS:2005bgo,Carman:2012qj,CLAS:2014udv},  can reveal information about the nature of the $\Delta(1700)$ resonance~\cite{Doring:2007rz} and the $\Lambda(1405)$ and $\Lambda(1520)$ hyperons~\cite{Geng:2007hz,  Doring:2006ub}. 

In addition, the momentum transfer of the probe can be tuned once the photon is allowed to become virtual, testing strong interactions at different scales. 
Indeed, electroproduction reactions are a prime tool to study the structure of excited baryons~\cite{Carman:2020qmb, Mokeev:2022xfo, Ramalho:2023hqd, 
Proceedings:2020fyd,
Carman:2023zke}, but they also serve to discover new baryonic states~\cite{Mokeev:2020hhu} as discussed below. One cannot directly test the response of a resonance to a virtual photon, but  determine transition form factors (TFFs) in the electro-excitation of the resonance from the nucleon. One can also map out the transverse charge density by using electromagnetic form factors~\cite{Carlson:2007xd}. The $Q^2$-dependent multipoles can also be used to test calculations within chiral perturbation theory~\cite{Bernard:1992ms, Bernard:1992rf, Bernard:1993bq, Bernard:1994dt, Bernard:1996bi, Steininger:1996xw, Bernard:2000qz, Krebs:2004ir} and unitary extensions~\cite{Jido:2007sm, Doring:2010rd, Mai:2020ltx}, chiral resonance calculations~\cite{Gail:2005gz, Bauer:2014cqa}, quark models~\cite{Capstick:1992uc, Merten:2002nz, Gross:2006fg, Ramalho:2008dp, Ramalho:2011ae, Santopinto:2012nq, Golli:2013uha, Aznauryan:2017nkz, Obukhovsky:2019xrs, Ramalho:2020nwk} and other theory predictions~\cite{Williams:1992tp, Cohen:2002st, Jayalath:2011uc, Anikin:2015ita}. Notably, a chiral unitary framework for kaon electroproduction was developed in Ref.~\cite{Borasoy:2007ku} and extended later~\cite{Ruic:2011wf, Mai:2012wy}. Transition form factors also serve as point of comparison for dynamical quark calculations referred to as Dyson-Schwinger approaches~\cite{Cloet:2008re, Wilson:2011aa, Segovia:2014aza, Segovia:2015hra, Eichmann:2016hgl, Chen:2018nsg, Qin:2019hgk, Chen:2019fzn, Lu:2019bjs}. In this context, remarkable agreement of the lower-lying baryon spectrum with predictions has been achieved~\cite{Eichmann:2016hgl, Qin:2019hgk}, showing little evidence for a ``missing resonance'' problem at lower energies. See Refs.~\cite{Aznauryan:2011qj, Aznauryan:2012ba, Bashir:2012fs, Eichmann:2016yit, Eichmann:2022zxn} for reviews.
Methods to study the $Q^2$-dependence of resonance couplings in lattice QCD were proposed in Ref.~\cite{Agadjanov:2014kha}.
A pioneering lattice calculation was carried out recently in the meson sector~\cite{Radhakrishnan:2022ubg}. 

Transition form factors have been defined in different ways~\cite{Aznauryan:2011qj}, but the only reaction-independent definition is given in terms of $Q^2$-dependent couplings at the resonance pole, to be determined by an analytic continuation of electroproduction multipoles~\cite{Tiator:2016btt}. We review the steps taken in the JBW model towards determining TFFs at resonance poles.

The multipoles themselves are determined by analyzing the exclusive electroproduction of one or more mesons. The advantage of simultaneously analyzing different final states in a coupled-channel approach lies in the factoriziation of the amplitude at the pole, i.e., the fact that the resonance TFF is the same for any final state, by construction. 

Single-channel analyses of single-meson electroproduction data have a long history; one of the first approaches is MAID for pion photo- and electroproduction~\cite{Tiator:2003uu, Drechsel:2007if}, later complemented by a chiral-MAID approach at low energies~\cite{Hilt:2013fda}. There is also the etaMAID2001 analysis on eta electroproduction~\cite{Chiang:2001as}. See Ref.~\cite{Tiator:2011pw} for a review. The CLAS collaboration extracted helicity amplitudes for several resonances from their experiment using two different methods, namely a unitary isobar model and an amplitude constructed with fixed-t dispersion relations, developed by Aznauryan~\cite{Aznauryan:2014xea, CLAS:2009ces,CLAS:2008roe, Aznauryan:2002gd}. One of the findings was the unusual zero in the $A_{\nicefrac{1}{2}}$ Roper form factor.
See also the analysis of Refs.~\cite{Mokeev:2023zhq, HillerBlin:2022ltm, HillerBlin:2019jgp} on the extraction of TFFs and their role in different reactions. 
See also Refs.~\cite{Petrellis:2024ybj, Skoupil:2018vdh, Maxwell:2016hdx, Maxwell:2015psa, Maxwell:2012zz} for electroproduction analyses using tree-level effective Lagrangian frameworks.

Questions on efficient parametrizations of electroproduction amplitudes and transition form factors are discussed in Refs.~\cite{Ramalho:2016zzo, Ramalho:2017muv, Ramalho:2017xkr, Ramalho:2019ocp}.
The two-pion electroproduction reaction has also been measured at CLAS~\cite{CLAS:2018drk, CLAS:2018fon, CLAS:2017fja, CLAS:2008ihz, CLAS:2002xbv, Trivedi:2018rgo} and analyzed with the JM reaction model~\cite{CLAS:2012wxw, Mokeev:2015lda, Mokeev:2020hhu, Mokeev:2023zhq}, see also Ref.~\cite{Bernard:1994ds}.
Consistent results on $N^*$ electroexcitation amplitudes obtained from independent studies of two major exclusive meson electroproduction channels in the resonance region, with different non-resonant contributions $\pi N$ and $\pi^+\pi^-p$, support the reliable extraction of these quantities.

In Ref.~\cite{Mokeev:2020hhu} the existence of a second $N(1720)3/2^+$ is claimed which would be the first resonance exclusively discovered in electroproduction. Notably, much higher $Q^2$ values for resonance transition form factors become accessible in ongoing CLAS12 experiments~\cite{Carman:2020fsv, Aznauryan:2012ba}. Most relevant for the present analysis of $K\Lambda$ electroproduction is KAON-MAID~\cite{Bennhold:1999mt, Mart:2002gn}, an analysis using an effective Lagrangian approach~\cite{Maxwell:2012zz}, and the more recent analyses using a Regge-plus-resonance amplitude~\cite{Corthals:2007kc, DeCruz:2012bv}. See Ref.~\cite{Carman:2018fsn} for an overview of kaon electroproduction reactions and Refs.~\cite{Haberzettl:1998aqi, Mart:1999ed, Mart:2006dk, Mart:2011ez, JPAC:2016lnm, Blin:2021twt, Haberzettl:2021wcz} for related analyses and theoretical developments by JPAC and others.

The data situation in electroproduction reactions is more challenging than in photoproduction. On the one hand, this is due to the presence of another kinematic variable in addition to the energy $W$, namely the virtuality of the photon $Q^2=-q^2$, where $q$ is the transferred four-momentum of the photon. Even though the number of data points is larger in electro- than in photoproduction, the data are still sparser due to this additional variable. On the other hand, there are longitudinal multipoles to be determined from data, in addition to the electric and magnetic ones that parametrize the photoproduction amplitudes. While there are in principle more spin observables than in photoproduction~\cite{Knochlein:1995qz}, fewer have been measured than in photoproduction. 

However, the data situation is improving rapidly through new measurements with the CLAS12 detector~\cite{Accardi:2023chb, Carman:2020fsv, Aznauryan:2009da}. 
CLAS12 is the only available facility in the world capable of extending information on the evolution of the $N^*$ electroexcitation amplitudes towards $Q^2=10\,(\text{GeV}/c)^2$. The first measurements with the CLAS12 of inclusive $(e,e'X)$ cross sections~\cite{CLAS:2025zup} have revealed the presence of resonance structure in the second and third resonance regions, suggesting promising prospects for the extraction of $N^*$ electroexcitation amplitudes within the range of $Q^2\leq 10\,(\text{GeV}/c)^2$.  The development of  coupled channel approaches capable of extracting $N^*$ electroexcitation amplitudes within that range represents an important and very needed direction in theory support of these ongoing experimental efforts.

The related question of how many measurements are necessary to determine a truncated partial-wave expansion of the electroproduction amplitude is discussed in Refs.~\cite{Tiator:2017cde, Wunderlich:2021xhp}. 

\begin{figure}[tb!]
    \centering
    \includegraphics[width=0.41\linewidth]{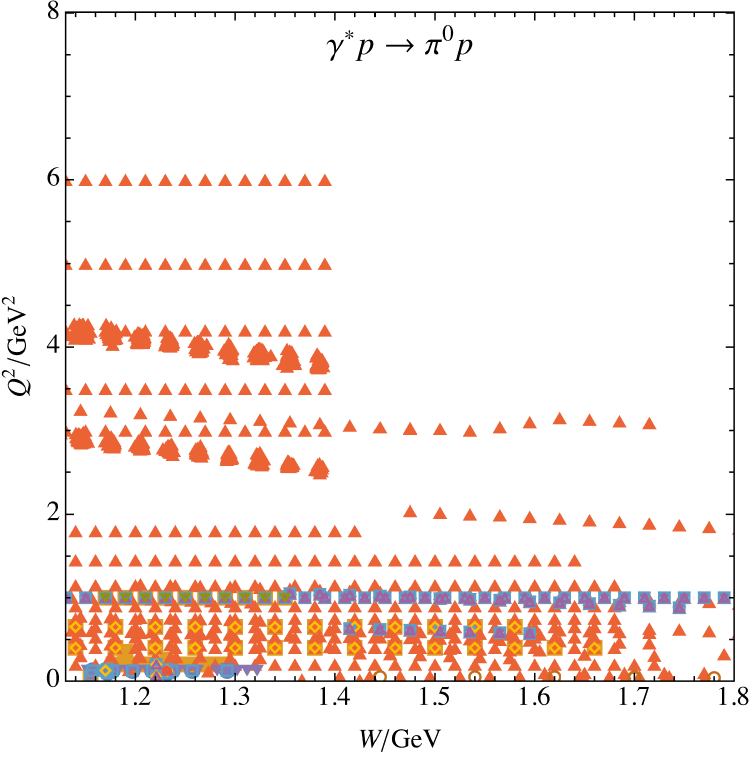}
    ~~~
    \includegraphics[width=0.41\linewidth]{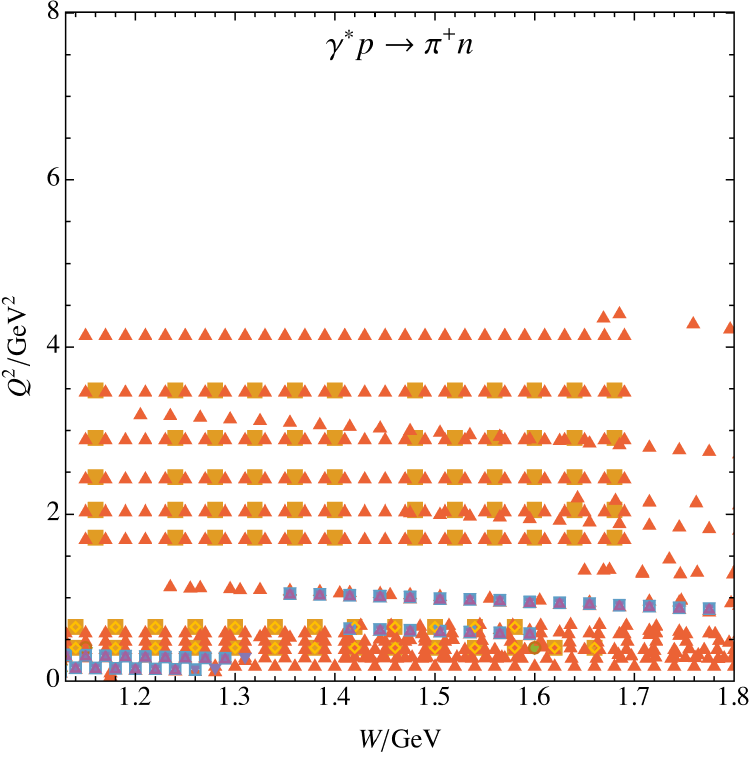}\\
    \includegraphics[width=0.41\linewidth]{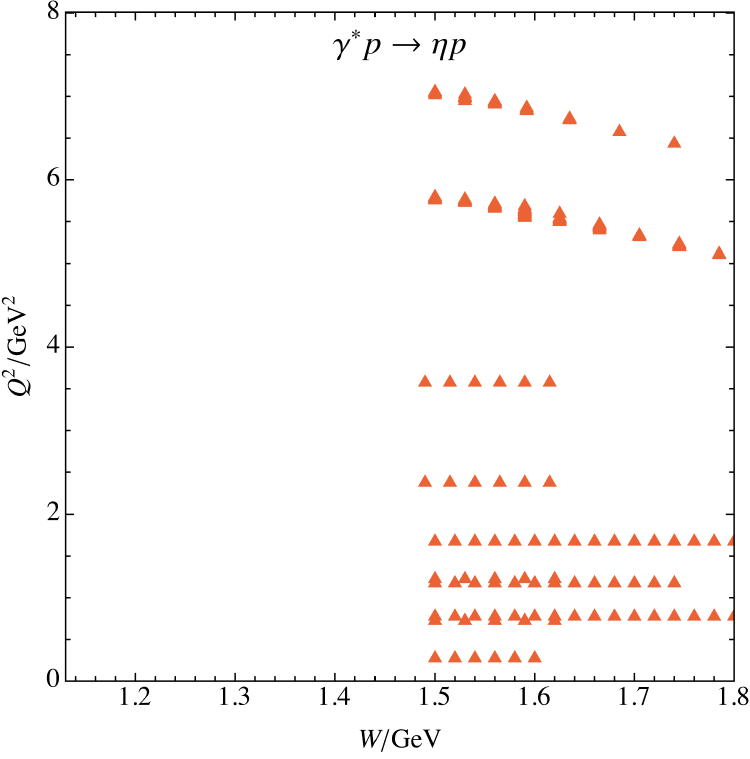}
    ~~~
    \includegraphics[width=0.41\linewidth]{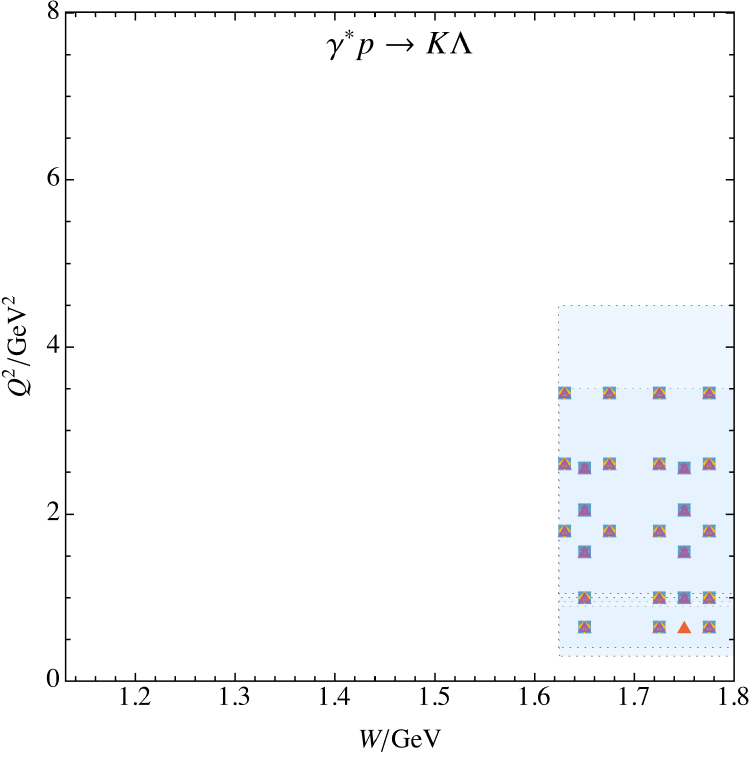}\\~\\
\footnotesize
\begin{tabularx}{\linewidth}{clXXll}
    \hline\hline
    \multicolumn{2}{c}{Type}&$N_{\rm data}^{\pi^0p}$&$N_{\rm data}^{\pi^+n}$&$N_{\rm data}^{\eta p}$&$N_{\rm data}^{K\Lambda}$\BBB\TT\\
    \hline
    ${\color{Cerulean}\bullet}$&$\rho_{LT}$
    &45~\cite{Mertz:1999hp,Elsner:2005cz}
    &--
    &--
    &--\TT\\
    ${\color{Dandelion}\blacksquare}$&$\rho_{LT'}$
    &2768~\cite{Joo:2003uc, Sparveris:2002fh, Kelly:2005jj, Bartsch:2001ea,Bensafa:2006wr}
    &5068~\cite{Joo:2004mi, PARK-pc-08-2007}
    &-- 
    &-- \\
    ${\color{ForestGreen}\blacklozenge}$&$\sigma_L$
    &--
    &2~\cite{Gaskell:2001fn}
    &--
    &--\\
    ${\color{BurntOrange}\blacktriangle}$&$d\sigma/d\Omega$
    &48135~\cite{Laveissiere:2003jf, Ungaro:2006df, Gayler:1971zz, May:1971zza, Frolov:1998pw, Hill:1977sy, Joo:2001tw, Siddle:1971ug, Haidan:1979yqa, Sparveris:2002fh, Kelly:2005jj, Kalleicher:1997qf, Baetzner:1974xy, Latham:1979wea, Latham:1980my, Stave:2006jha, Sparveris:2006uk, Alder:1975xt, Afanasev:1975qa, Shuttleworth:1972nw, Blume:1982uh, Rosenberg:1979zm, Gerhardt:1979zz, MONTANA-phd1971}
    &44266~\cite{Egiyan:2006ks, Breuker:1977vy, PARK-pc-08-2007, Bardin:1975oea, Bardin:1977zu, Gerhardt:1979zz, DavenportMartyn1980Eona, Vapenikova:1988fd, Hill:1977sy, Alder:1975na, Evangelides:1973gg, Breuker:1982nw, Breuker:1982um, Bebek:1974ww, Brown:1973wr, Litt:1971xx}
    &3665~\cite{JeffersonLabE94014:1998czy, CLAS:2007bvs, CLAS:2000mbw, Dalton:2008aa}
    &2055~\cite{Carman:2012qj, CLAS:2006ogr}\\
    ${\color{CadetBlue}\blacktriangledown}$&$\sigma_T+\epsilon \sigma_L$
    &384~\cite{Laveissiere:2003jf, Mertz:1999hp, Sparveris:2002fh, Kunz:2003we, Stave:2006ea, Sparveris:2006uk, Sparveris:2004jn, Alder:1975xt, Laveissiere:2003jf}
    &182~\cite{Breuker:1977vy, Alder:1975na}
    &--
    &204~\cite{Carman:2012qj, CLAS:2006ogr}\\
    ${\color{Maroon}\circ}$&$\sigma_{T}$
    &30~\cite{Blume:1982uh}
    &2~\cite{Gaskell:2001fn}
    &--
    &--\\
    ${\color{Cerulean}\square}$&$\sigma_{LT}$
    &373~\cite{Laveissiere:2003jf, Sparveris:2002fh, Mertz:1999hp, Kunz:2003we, Stave:2006ea, Sparveris:2006uk, Sparveris:2004jn, Alder:1975xt, Laveissiere:2003jf}
    &138~\cite{Breuker:1977vy, Alder:1975na}
    &--
    &204~\cite{Carman:2012qj, CLAS:2006ogr}\\
    ${\color{Dandelion}\Diamond}$&$\sigma_{LT'}$
    &214~\cite{Joo:2003uc, Kunz:2003we, Stave:2006ea, Sparveris:2006uk}
    &208~\cite{Joo:2003uc}
    &--
    &156~\cite{Carman:2012qj, Nasseripour:2008aa}\\
    ${\color{Mulberry}\triangle}$&$\sigma_{TT}$
    &327~\cite{Laveissiere:2003jf, Stave:2006ea, Sparveris:2006uk, Sparveris:2004jn, Alder:1975xt, Laveissiere:2003jf}
    &123~\cite{Breuker:1977vy, Alder:1975na}
    &--
    &204~\cite{Carman:2012qj, CLAS:2006ogr}\\
    ${\color{JungleGreen}\nabla}$&$K_{D1}$
    &1527~\cite{Kelly:2005jj}
    &--
    &--
    &--\\
    ${\color{BrickRed}\bullet}$&$P_Y$
    &--
    &2~\cite{Warren:1999pq, Pospischil:2000ad}
    &--
    &--\BB\\
    \hline
    \multicolumn{2}{c}{Total}
    &53804
    &49989
    &3665
    &2823\TT\BB\\
    \hline\hline
\end{tabularx}
    \caption{World data and data types for meson-electroproduction process. The kinematical region covered by the recent beam-recoil transferred polarization measurement of Ref.~\cite{CLAS:2022yzd} is represented by the blue shaded area. Figure and table adapted from Ref.~\cite{Mai:2023cbp}.}
    \label{fig:DATA}   
\end{figure}

The sparse data situation is another motivation for the simultaneous analysis of $\pi N,\eta N, K\Lambda, K\Sigma$ electroproduction data, apart from the above mentioned property of amplitudes factorizing at resonance poles, improving the reaction-independent extraction of TFFs. The available world data consists mostly of cross sections but includes also some polarization observables. A summary of the data is provided in the table in \cref{fig:DATA}, most of them being included in the latest JBW analysis~\cite{Mai:2023cbp}. Sometimes structure functions are extracted from cross sections in which case JBW analyzes only the underlying cross sections. There are also data on $K\Sigma$ electroproduction~\cite{CLAS:2022yzd, Carman:2012qj, CLAS:2009sbn, CLAS:2006ogr}, but they are not yet included in the JBW analysis.

The JBW group analyzed $\pi N$~\cite{Mai:2021vsw}, $\eta N$~\cite{Mai:2021aui}, and, most recently, the combined $\pi N$, $\eta N$, and $K\Lambda$ electroproduction data base~\cite{Mai:2023cbp}. Each time the maximal energy was increased requiring the inclusion of more and more partial waves, currently up to $F$-waves. The approach represents the first coupled-channel electroproduction analysis, while data at the photon point ($Q^2=0$) are also included as a boundary condition from the JBW approach to pion~\cite{Ronchen:2014cna}, eta~\cite{Ronchen:2015vfa}, and $K\Lambda$~\cite{Ronchen:2018ury} photoproduction~\cite{Ronchen:2022hqk} and pion-induced reactions~\cite{Doring:2010ap, Ronchen:2012eg}. Specifically, the JBW electroproduction multipoles are constructed in close analogy to the photoproduction multipoles of Eq.~\eqref{m2},
\begin{align}
    {\cal M}_{\ell,\mu\gamma^*}(p',E,Q^2)=
    &
    R_{\ell'}(\lambda, q/q_\gamma)
    \left(V_{\ell,\mu\gamma^*}(p',E,Q^2)
    +\sum_\kappa\int\limits_0^\infty \dv{q}\, q^2\, T_{\ell,\mu\kappa}(p',q;E)G_\kappa(q,E)V_{\ell,\kappa\gamma^*}(q,E,Q^2)
    \right) \ ,
    \label{ampl_2}
\end{align}
for outgoing meson momentum $p'$, total energy $E=\sqrt{s}$, and  $\kappa,\, \mu,\,\nu\in\{\pi N,\eta N, K\Lambda, \pi\Delta, \rho N\}$ meson baryon channels. Note that there are no real or virtual photon couplings to the $K\Sigma , \,\sigma N$ states included, while the hadronic rescattering encoded in $T$ does contain these channels. As no $K\Sigma$ electroproduction data are analyzed yet, the input from the photoamplitude is taken from Ref.~\cite{Ronchen:2018ury} which does not contain the analysis of $K\Sigma$ photoproduction, either. The multipoles in Eq.~\eqref{ampl_2} are formulated for a given angular momentum $\ell$, ${\cal M}_{\ell,\mu\gamma^*}\in\{E_\ell,M_\ell,L_\ell\}$ are the electroproduction multipoles. The  $G_\kappa$ are meson-baryon propagators (see Sects.~\ref{sec:pwa} and \ref{sec:tauiso}); $R_\ell'$ is a barrier factor ensuring correct threshold behavior while taming its high-momentum behavior. It depends on the photon momentum $q$ and $q_\gamma=q(Q^2=0)$~\cite{Mai:2023cbp}. For the connection of $\ell$ and $\ell'$ see also Ref.~\cite{Mai:2023cbp}. The $V_{\mu\gamma^*}$ are the $Q^2$-dependent photon transitions such that for $Q^2=0$ they exactly match the JB2017 photoproduction solution~\cite{Ronchen:2018ury}. Note that longitudinal multipoles cannot be  matched to photoproduction data. Instead, a valuable constraint is provided through the long-wavelength limit encoded in the so-called Siegert's condition~\cite{Siegert:1937yt, Tiator:2016kbr}.

Apart from $\pi$ and $\eta$ electroproduction data, the JBW analyses~\cite{Mai:2021vsw,Mai:2021aui,Mai:2023cbp} fit cross section data for $\gamma^* p\to K\Lambda$ from CLAS. Notably, the data base was recently enlarged through the addition of beam-recoil transfer polarization data from CLAS~\cite{CLAS:2022yzd}. This data is confronted (no fit) with the predictions of the JBW approach in Ref.~\cite{Mai:2023cbp}, demonstrating the need for further analysis.
While the comparison of data and fit solutions of pion- and real-photon-induced reactions with JB have been collected on a website~\cite{Juelichmodel:online}, the JBW electroproduction solutions are collected on another interactive website~\cite{JBW-homepage}. 

\begin{figure}[tb]
    \raisebox{-.5\height}{\includegraphics[height=5.5cm]{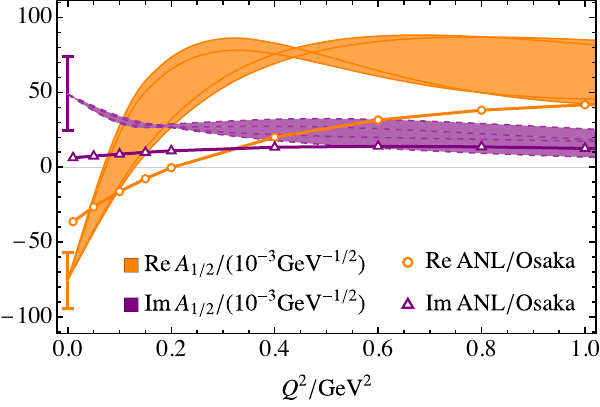}}
    ~~~~~~
    \raisebox{-.5\height}{\includegraphics[height=6.5cm]{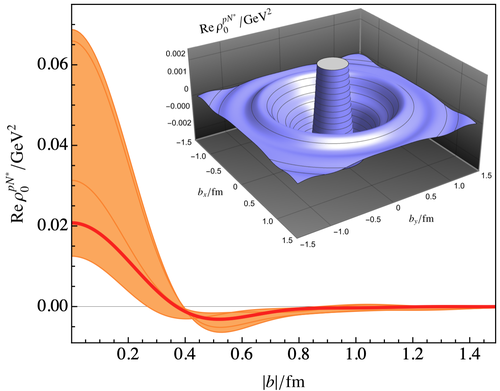}}
	 \caption{
     {\bf Left}: $N(1440)$ transition form factors at small $Q^2$ from the JBW analysis~\cite{Wang:2024byt} in comparison with the ANL-Osaka solutions~\cite{Kamano:2018sfb}. The error bars at $Q^2=0\,\GeV^2$ depict the uncertainties of the photoproduction solution at the pole from Ref.~\cite{Ronchen:2018ury}. A zero in the real part of the TFF is observed.
     {\bf Right}:     
     Unpolarized transverse charge density $\rho_0^{pN^*}$ of the $p\to N(1440)$ transition as a function of the transverse position $b$ in the $xy$-plane from Refs.~\cite{Mai:2023cbp, Wang:2024byt}. The orange band (thick red line) depicts the uncertainty band of the determination (the result using the MAID 2007 helicity couplings~\cite{Drechsel:2007if}). The inset shows the corresponding representation in position space.
     }
     \label{fig:ChargeDensityRoper} 
\end{figure}

After adjusting the amplitude parameters in the simultaneous to $\pi N$, $\eta N$, and $K\Lambda$ electroproduction data, results for TFFs can be calculated. In the JBW coupled-channel approach, twelve $N^*$ and $\Delta$ transition form factors at the pole were extracted recently in Ref.~\cite{Wang:2024byt} using data with the center-of-mass energy from $\pi N$ threshold to $1.8\,\GeV$, and the photon virtuality $0\leq Q^2/(\GeV/c)^2\leq 8$. For the $\Delta(1232)$ and $N(1440)$ states, the results are in qualitative agreement with previous studies, while the transition form factors at the poles of some higher excited states are estimated for the first time. 
Since quarks are charged, the electroproduction probe can provide a spatial scan of the charge density $\rho$ of, e.g., $p\to N^*$ transition in a certain reference frame~\cite{Tiator:2009mt,Tiator:2008kd,Ramalho:2023hqd}.
This is shown in \cref{fig:ChargeDensityRoper}.

In the  ANL-Osaka model the electroproduction data  are analyzed by extending the model of the pion- and photon-induced reactions. The parameters for this analysis are the $Q^2$ dependence of the bare helicity  amplitudes $\Gamma_{\gamma N,j}$ of Eq.~\eqref{eq:dressf-gam} and the new parameters on the transition charge form factors $S_{l\pm}^{N^*}$. The first analysis addressed the pion electroproduction reaction around the $\Delta(1232)$ resonance region~\cite{Sato:2000jf}. Analyzing the available data at that time for $Q^2=2.8$ and $Q^2=4$~(GeV/c)$^2$~\cite{Frolov:1998pw}, three transition form factors $G_M, G_E, G_C$, which are related to the helicity amplitudes, were obtained. The magnitude of the Coulomb quadrupole coupling $G_C$ at $Q^2=0$ was obtained from $G_E$ assuming the relation between the electric and Coulomb form factors in the long wave length limit, i. e., Siegert's theorem. The model predicted a characteristic enhancement of form factors due to the meson cloud illustrated in Fig.~\ref{fig:ANL-photoproduction}. These predictions were later confirmed in the low-$Q^2$ region. The extensive data on pion electroproduction from CLAS in the $\Delta(1232)$ region are re-analyzed in Ref.~\cite{Julia-Diaz:2006ios}. Based on  the $\pi N,\eta N,\pi\Delta,\sigma N$ and $\rho N$ coupled-channel JLMS model~\cite{Julia-Diaz:2007qtz}, single pion electroproduction data for $\sqrt{s} < 1.6 $~GeV and  $Q^2 < 1.45$~(GeV/c)$^2$ are studied in Ref.~\cite{JuliaDiaz:2009ww}. The $\gamma^* N \rightarrow N^*$ helicity amplitudes for $P_{33},P_{11}.S_{11},D_{13}$ resonances are extracted and their $Q^2$ dependence and meson cloud effects are discussed. 

\begin{figure}[tbh]
\begin{center}
    \includegraphics[width=6.5cm]{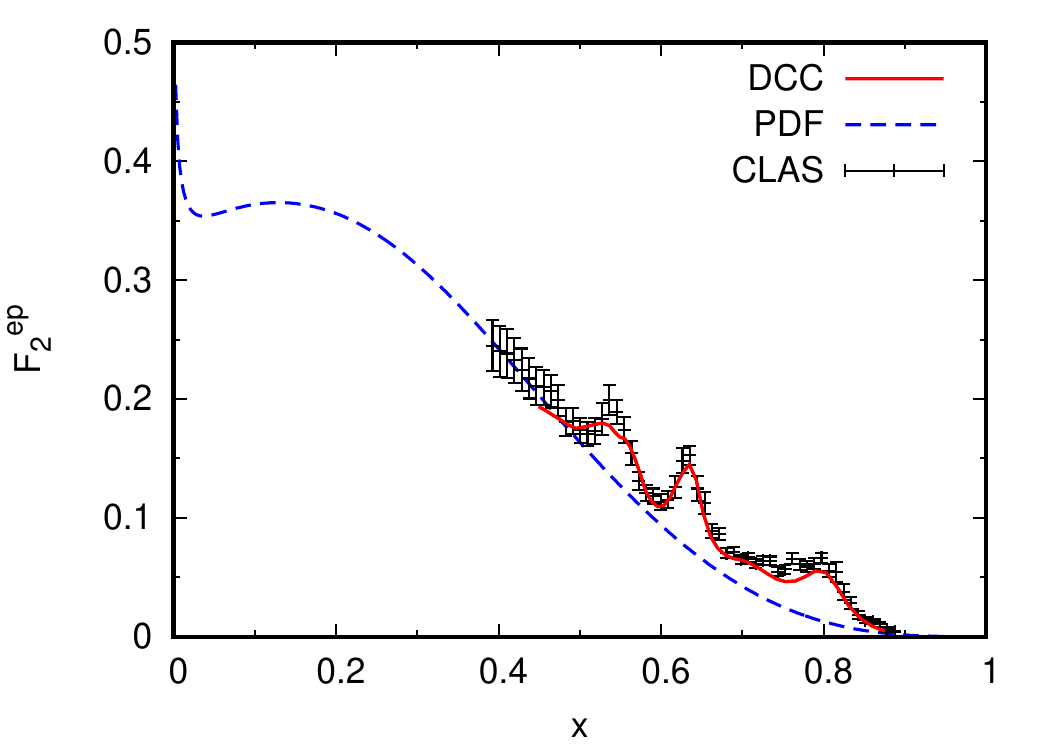}
    \end{center}
    \caption{The proton structure function $F_2$ for inclusive electron-proton scattering at $Q^2=2.425$(GeV/c)$^2$ with the ANL-Osaka model (red solid curve) and with the nucleonic PDF (blue dashed curve). The horizontal axis shows the Bjorken scaling variable $x$. The data are from Ref.~\cite{Osipenko:2003ua}.}
\label{fig:f2ep}
\end{figure}

In Ref.~\cite{Suzuki:2010yn},   $\gamma N \rightarrow N^*$ helicity amplitudes are studied as the residue of the scattering amplitude for the first time at finite $Q^2$. In terms of the pole expansion of the Green's function, helicity amplitudes can be understood as nucleon to resonance transition form factor $\braket{N^*|J_\mu^{EM}(Q)|N}$. Within the JLMS model, the helicity amplitudes are extracted for the $P_{33}(1232), D_{13}(1521), P_{11}(1357), P_{11}(1820)$ resonances up to $Q^2 < 1.45$~(GeV/c)$^2$. Using the ANL-Osaka model~\cite{Kamano:2013iva}, $N^*$ helicity amplitudes are determined in Ref.~\cite{Nakamura:2015rta} for the $Q^2 < 3$~(GeV/c)$^2$ region. The parametrization of the bare helicity amplitudes of the vector current are given in Appendix C of Ref.~\cite{Nakamura:2015rta}. In this analysis, all available data of the single pion electroproduction data  and also empirical parametrization of the structure function $W_1^{em},W_2^{em}$~\cite{Christy:2007ve} are used. The data on $\eta$ and $K$ electroproduction are not included in the analysis. The model is constructed for the neutrino reaction which is supposed to cover the resonance region below the DIS region. The model including two-pion production is reasonably consistent with the nucleonic PDF at $x \sim 0.45$ (where $\sqrt{s} \sim 2$GeV), as shown in Fig.~\ref{fig:f2ep} taken from Ref.~\cite{Nakamura:2016cnn}. The extracted helicity amplitudes  $A_{1/2}, A_{3/2}$ of the $P_{33}(1232), D_{13}(1520)$ resonances from the ANL-Osaka model agree reasonably well with those from the JLMS model for $Q^2 < 1.45$~(GeV/c)$^2$~\cite{Sato:2016dav}. The $A_{1/2}$ helicity amplitude of the Roper resonance reported in Ref.~\cite{Kamano:2018sfb} is shown in Fig.~\ref{fig:ChargeDensityRoper}.

\subsection{Applications in neutrino physics}
\label{sec:neutrinoproduction}
Neutrino-induced reactions have recently received much attention in the context of long base-line  experiments such as T2K, HK, DUNE~\cite{T2K:2011qtm,Hyper-Kamiokande:2016srs,DUNE:2015lol} and atmospheric neutrino data. To extract neutrino properties such as CP violation and the neutrino mixing angle, the precise understanding of neutrino reactions in the GeV neutrino energy range plays a crucial role~\cite{NuSTEC:2017hzk, Nakamura:2016cnn, NuSTEC:2019lqd, NuSTEC:2020nsl}. For neutrino energies from around 1 GeV to a few GeV, meson production is the main mechanism of the neutrino-nucleus reaction.

Isobar models provide simple and useful parametrizations of neutrino reactions in the resonance region. The Rein-Sehgal model~\cite{Rein:1980wg,Berger:2007rq} has often been used in neutrino reaction generators. In this model, the amplitude is constructed using Breit-Wigner forms for nucleon resonances with the masses, widths, and branching ratios from the PDG. Coupling constants have been estimated with a
relativistic quark model. Theoretical analyses from the 1970's, mainly for the $\Delta(1232)$ region using isobar models and dispersion theory, are summarized in  Ref.~\cite{LlewellynSmith:1971uhs}.

The model of Hern\'andez et al.~\cite{Hernandez:2007qq,Hernandez:2010bx, Hernandez:2016yfb} (HNV model) includes the $\Delta(1232)$  and non-resonant mechanisms from the chiral Lagrangian. The vector form factors of the resonances are determined from the analysis of electron scattering data. The model is improved to satisfy the requirement of unitarity using  Watson's theorem for the single-pion production region~\cite{Alvarez-Ruso:2015eva}. Several extensions of HNV including higher resonances such as  $N(1535)1/2^-$, $N(1440)1/2^+$ and $N(1520)3/2^-$ were developed,  for a detailed discussion see Ref.~\cite{Sato:2021pco}. The model of neutrino-induced single-pion production~\cite{Gonzalez-Jimenez:2016qqq} is a hybrid model based on the HNV approach for the low-$W$ region and a reggeized, non-resonant amplitude based on the chiral Lagrangian for the higher $W > 2\,\GeV$ region.

For energies above the $\Delta(1232)$, many meson-baryon channels such as $ \pi N, \pi\pi N, \eta N, K\Lambda, K\Sigma$ open up, rendering the construction of a model for the multi-resonance and multi-channel neutrino reaction very challenging. The coupled-channel approach is a way to construct such a model, which is constrained by the extensive data of pion-, photo- and electro-production data. The first attempt along this line was made in Refs.~\cite{Sato:2003rq, Matsui:2005ns} for the $\Delta$ resonance region, and then the model was extended for $W < 2\,\GeV$ and $Q^2 < 3 \,(\GeV/c)^2$ in Refs.~\cite{Kamano:2012id, Nakamura:2015rta} based on the ANL-Osaka model. A coupled-channel approach has also been used to study the strangeness-changing production of $\Lambda(1405)$ in antineutrino-proton interactions~\cite{Ren:2015bsa}. 
Below, we briefly review the current status of the ANL-Osaka coupled-channel model for  neutrino reactions in the resonance region  for strangeness-conserving processes.

In the Standard Model, the hadronic weak charged current (CC) and neutral current (NC) together with the electromagnetic current (EM) are given in terms of the vector ($V_\mu$) and axial ($A_\mu$) currents as
\begin{align}
    \label{eq:jcc}
    J_\mu^{\rm CC}(x) & = V_{ud} (V_\mu^\pm(x) - A_\mu^\pm(x))  
    \ , \\
    \label{eq:jnc}
    J_\mu^{\rm NC}(x)
    & =   (V_\mu^3(x) - V_\mu^s(x)) - (A_\mu^3(x)  - A_\mu^s(x))
    - 2 \sin^2\theta_W J^{\rm EM}_\mu(x) 
    \ , \\
    J_\mu^{\rm EM}(x) & =  V^3_\mu(x) + V_\mu^{\rm IS}(x)
    \ ,
\end{align}
where the superscript $\pm$ indicates the isospin raising (lowering) current, '3' is the third component of the isovector, and 'IS' is the isoscalar current. In the model, strange currents $V^s$ and $A^s$ are not taken into account.

The starting point of the neutrino-nucleon reaction model~\cite{Nakamura:2015rta} is the ANL-Osaka model~\cite{Kamano:2013iva}. There, all model parameters that govern hadronic interactions are determined through the analysis of pion-, photon-, and electron-induced meson-production reactions. The contributions of the individual isovector and isoscalar 
vector currents are obtained from the analysis of the proton and deuteron reactions~\cite{Kamano:2016bgm}. However, one still needs to fix the remaining unknown axial current. The non-resonant axial current can be derived from a chiral Lagrangian on which the $\pi N$ interaction potentials are based. By construction, the non-resonant axial current and the $\pi N$ potentials are related by the PCAC relation at $Q^2=0$. The neutrino-induced forward reactions for various meson production channels, which are related to the $\pi N \to X$ reaction are studied in Ref.~\cite{Kamano:2012id}. The axial $N$-$N^*$ transition strengths at $Q^2=0$ are related to the corresponding $\pi NN^*$ couplings via the PCAC relation. The advantage of the approach is that one can uniquely fix not only the axial coupling strengths but also their phases. The $Q^2$ dependencies of the axial couplings are difficult to determine because of the lack of experimental information. Here, one assumes that all of the axial couplings have the dipole $Q^2$ dependence of $1/ (1+Q^2/M_A^2)^2$ with $M_A=1.026\,\GeV$.

\begin{figure}[t]
    \centering
    \includegraphics[height=0.2\textwidth]{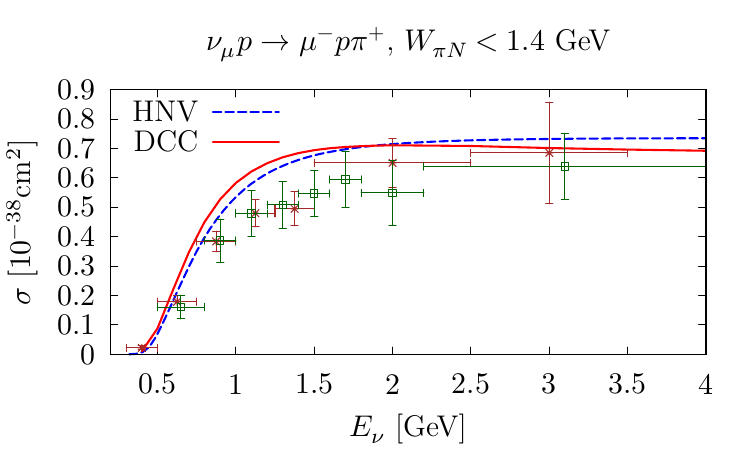}
    \hspace{-5mm}
    \includegraphics[height=0.2\textwidth]{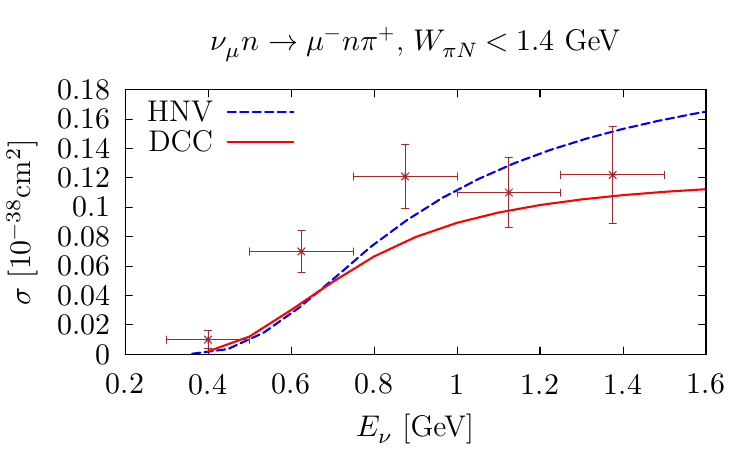}
    \hspace{-5mm}
    \includegraphics[height=0.2\textwidth]{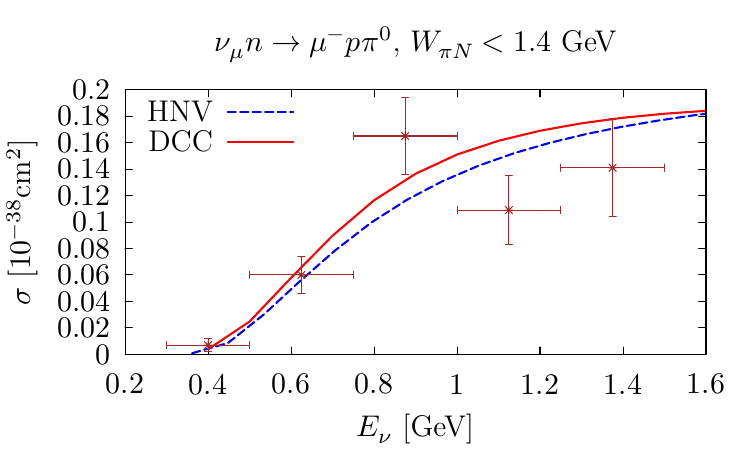}
    \caption{
    The $\nu_\mu N \to \mu^- \pi N$ total cross section as a function of the neutrino energy. The experimental data have been taken from the reanalysis of old ANL (crosses) and BNL (open square) data~\cite{Kitagaki:1986ct} 
    done in Ref.~\cite{Rodrigues:2016xjj}.
    The $W_{\pi N}$ cut is $W_{\pi N} < 1.4\,\GeV$.}
    \label{fig:neut-tot-1pi}
\end{figure}

The total cross sections of $\nu_\mu$ induced CC single pion production on proton and neutron of the ANL-Osaka model and the HNV model are shown in Fig.~\ref{fig:neut-tot-1pi} in comparison with the ANL~\cite{Barish:1978pj,Rodrigues:2016xjj} and BNL~\cite{Kitagaki:1986ct} data. The final state of $\nu p \to l^- \pi^+ p$ is a purely isospin $3/2$ amplitude, therefore, a large part of the cross section is due to the $\Delta(1232)$, while $\nu n \to l^- \pi^+ n$ and $\nu n \to l^- \pi^0 p$ contain the contribution
of the $I=1/2$ amplitude. Both models agree reasonably well, while they both  lie slightly above the data in the $1-2\,\GeV$ region. For $\nu n \to l^- \pi^+ n$, the discrepancies between the two models are larger in the high-energy region. The experimental data are obtained from the analysis of $\nu d$ reactions. Effects of the final state interaction and deuteron wave function within the ANL-Osaka model are studied in Refs.~\cite{Wu:2014rga,Nakamura:2018ntd}, and are found to enhance the cross section at the quasi-elastic peak by 10\% to 30\% for the $\nu n \to \mu^- \pi^+ n$ channel.

\begin{figure}[tb]
    \centering
    \includegraphics[width=0.95\textwidth]{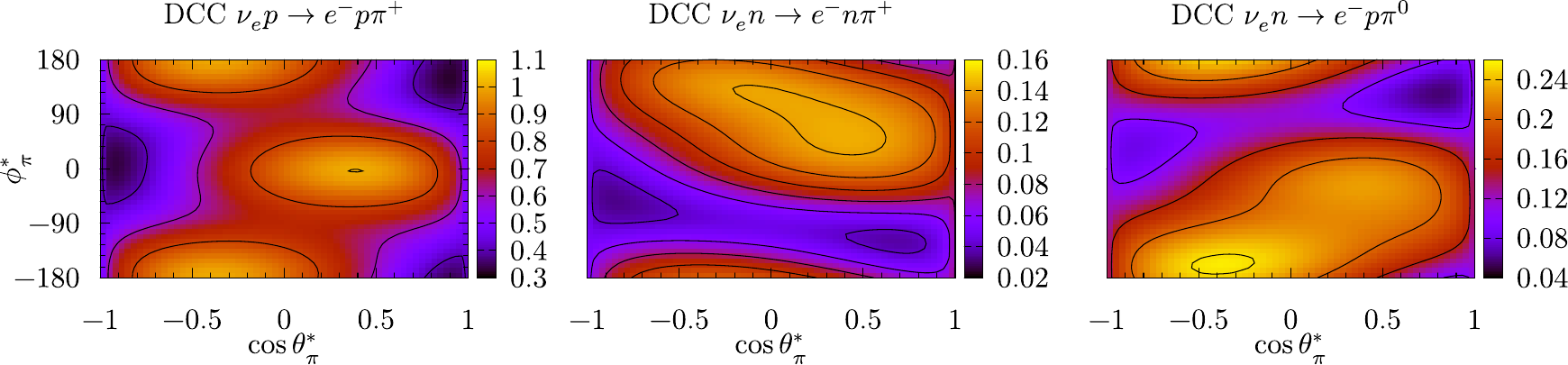}
    \caption{Neutrino-nucleon cross section $d\sigma/dQ^2/dW_{\pi N}/d\Omega^*_\pi$   in units of $10^{-38} {\rm cm}^2/\GeV^3$ for $E_\nu = 1\,\GeV$, $W_{\pi N}=1.23\,\GeV$, and $Q^2=0.1\,\GeV^2$.}
    \label{fig:neut-abcde}
\end{figure}

The mechanism of pion production can be closely examined by the angular distribution of the pion. In the massless lepton limit, the formula of the neutrino-induced pion production can be written in a similar way as pion electroproduction,
\begin{eqnarray}
    \frac{d \sigma_{CC}}{d\Omega' dE_l' d\Omega^*_\pi}
    = \frac{|\bm{k}'|}{|\bm{k}|}\frac{G_F^2}{4\pi^2}
    ( A^* + B^* \cos\phi_\pi^* + C^* \cos 2\phi_\pi^*
          + D^* \sin\phi_\pi^* + E^* \sin 2\phi_\pi^*)\ .
\end{eqnarray}
Here, the pion angular distribution is expressed in the $\pi N$ center of mass system and $\phi_\pi^*$ is the angle between the lepton scattering plane and 
the $\pi N$ production plane. Comparing with the well known formula for pion electroproduction, we note that there is no $E^*$ term in electron scattering. In \cref{fig:neut-abcde}, the contour plot of the neutrino-nucleon cross section at $W_{\pi N}=1.232\,\GeV$ is shown. The cross section is calculated with the ANL-Osaka model.
The parity violation effect $d\sigma(\phi_\pi^*) \neq d\sigma(-\phi_\pi^*)$  is apparent. Therefore, the $D^*$ and $E^*$ terms  should have a similar size as the $A^*,B^*,C^*$ terms.

\begin{figure}[tb]
    \centering
    \includegraphics[width=0.3\textwidth]{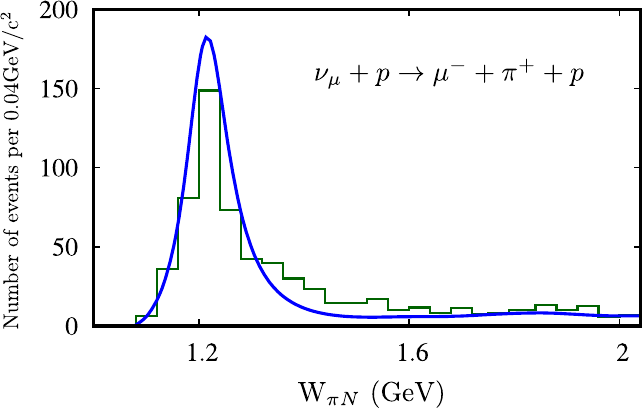}
    \includegraphics[width=0.3\textwidth]{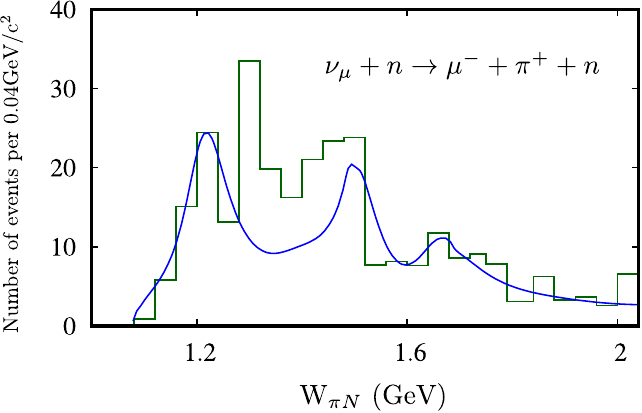}
    \includegraphics[width=0.3\textwidth]{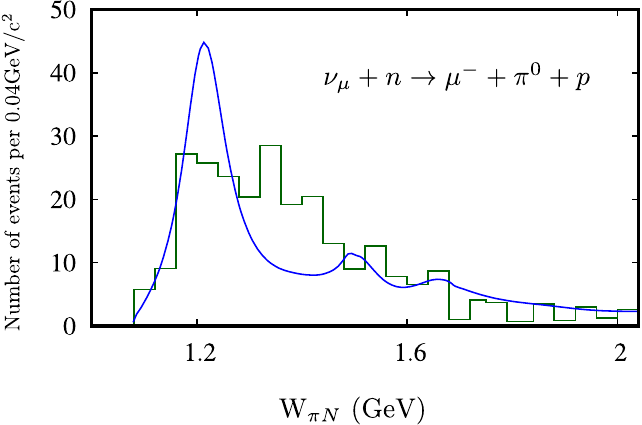}
    \caption{Cross section $d\sigma/dW_{\pi N}$ of the single-pion production reaction.}
    \label{fig-neut-dsdw}
\end{figure}

The excitation spectrum of the neutrino-induced single-pion production reaction $d\sigma/dW_{\pi N}$ is shown in Fig.~\ref{fig-neut-dsdw}. The data are taken from the CERN bubble chamber BEBC~\cite{Allasia:1990uy}. For comparison, the cross section is calculated using the ANL-Osaka model at $E_\nu=50\,\GeV$ and the normalization is adjusted for the $\nu_\mu p \to \mu^- \pi^+ p$ reaction. The reaction $\nu_\mu p \to \mu^- \pi^+ p$ is dominated by the  $\Delta(1232)$, while for the other reactions isospin $I=1/2$ structures are important in the high $W_{\pi N}$ region. Also note that for the $\nu_\mu n \to \mu^- \pi^+ n$ channel, the strength of $N^*$ resonances is significant.

\begin{figure}[tb]
    \centering
    \includegraphics[width=0.3\textwidth]{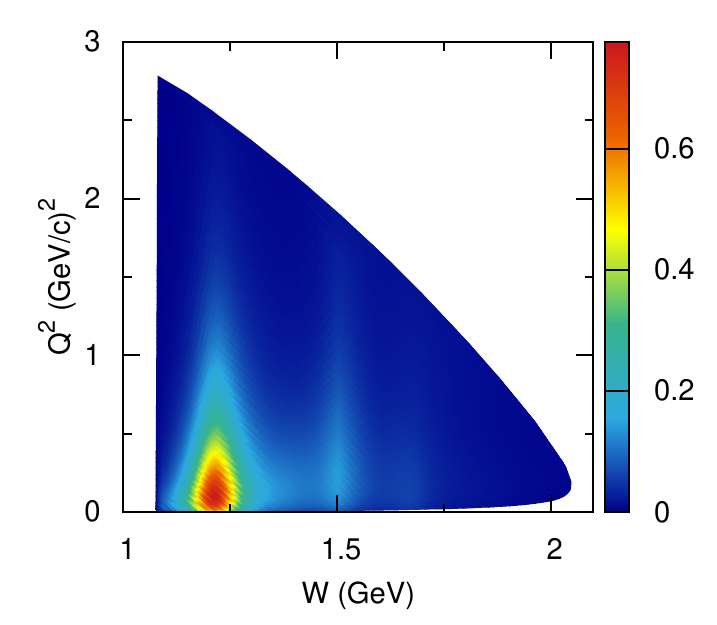}
    \includegraphics[width=0.3\textwidth]{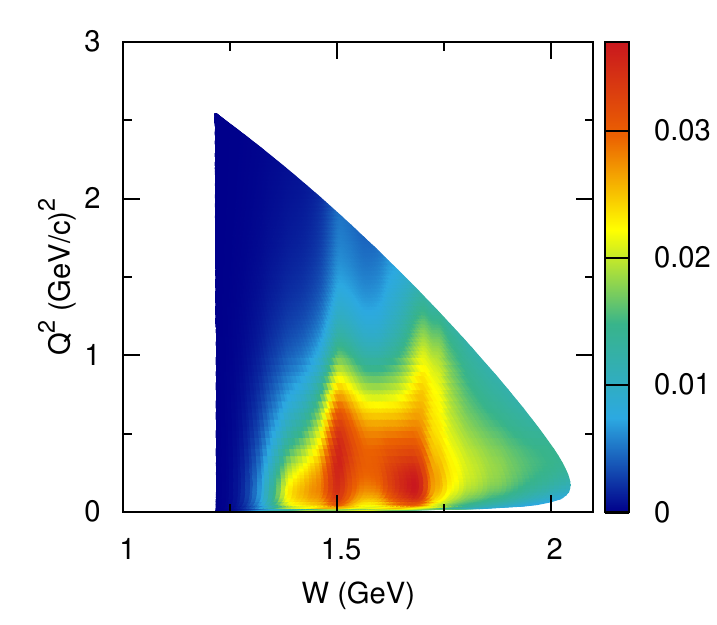}
    \caption{Contour plots of $d\sigma/dW/dQ^2$ for $\nu_\mu n \to \mu^-\pi N$ (left)
    and $\nu_\mu n \to \mu^-\pi^+\pi^- p$ (right) at $E_\nu=2\,\GeV$.}
    \label{fig-neut-dsdwsq}
\end{figure}

Inclusive double-differential cross sections $d\sigma/dW/dQ^2$ for the neutron target carry important physical information. The result is shown in Fig.~\ref{fig-neut-dsdwsq}. There, the prominent peak due to the $\Delta(1232)$ has a long tail toward the high $Q^2$ region. Contributions of the  second resonance region are also seen in the single-pion production cross section. A similar contour plot is shown for double-pion productions. The situation looks very different from the single-pion production case. The main contributions are from the second and the third resonance region.

The ANL-Osaka model of the neutrino-induced meson production reaction is constrained from the experimental data of pion-, photon-, and electron-induced reactions as far as possible. However, there is a limitation on  the axial vector current to data-driven phenomenological models because of the lack of experimental data. Except for the $\Delta(1232)$ region, all of the existing models for the neutrino-induced meson production reactions use assumptions on the $Q^2$ dependence of the axial vector current such as the dipole form factor discussed before. The predicted CC structure functions $F_i$ for high $Q^2$ of the reaction model are known to miss strength in comparison with the parton model even though the model predicts reasonably well for EM structure functions. Possible effects to improve the situation by modifying the axial form factors are discussed in Ref.~\cite{Sato:2021pco}.
Further progress in the theoretical modeling of weak meson production reactions requires new measurement of (anti)neutrino inelastic scattering on hydrogen and deuterium targets.

\subsection{DCC approaches for strangeness systems}
\label{sec:strangebaryons}

Generally, the spectrum of excited hyperons ($Y^*$) is much less known than 
that of non-strange excited nucleons ($N^*$ and $\Delta$). This is mostly owed to the fact that experiments with open strangeness are harder to conduct because the reaction participants decay quickly. However, some options exist, such as using secondary Kaon beams, as proposed for example in the upcoming 
KLF experiment~\cite{KLF:2020gai}, and in ongoing and planed experiments at 
J-PARC~\cite{Ohnishi:2019cif,Aoki:2021cqa,jparc:online}. 
Regarding the latter, for example, spectroscopy of $\Xi$ hyperons (P97), 
of $\Omega$ baryons (P86), access to the region 
below $\bar KN$ threshold ($\Lambda(1405)$)~\cite{J-PARCE31:2022plu} (E31), 
measurements of a missing-mass spectrum near the $\Si N$ threshold 
(``$\Si N$ cusp'') \cite{Ichikawa:2022pjm} (E90), 
as well as a study of the $\eta\Lambda$ final state~\cite{JPARCE72}
(E72) are on the agenda. 
Future measurements in this direction are also discussed for the AMBER experiment~\cite{Quintans:2022utc, Friedrich:2024ylw} and as an opportunity for SIS100@FAIR~\cite{Spiller:2020wtm}. Another alternative is to use photo-induced reactions and separating a third particle (spectator) carrying one unit of strangeness away as in the CLAS experiment~\cite{CLAS:2013rjt}. Also, multi-step decays of charmed states as in the Belle analysis of the $\Xi^-\pi^+$ invariant mass spectrum~\cite{Belle:2018lws} carry new information. See the recent review by Crede and Yelton~\cite{Crede:2024hur}. There is also the femtoscopic methodology employed for example by the ALICE collaboration~\cite{ALICE:2023wjz}, see the review~\cite{Tolos:2020aln}. In the latter cases the theoretical analysis is complicated by the less known systematics in the creation mechanism of the open-strangeness configurations. For a critical discussion see Ref.~\cite{Epelbaum:2025aan}. Obviously, the situation becomes even more complicated if one is interested in higher-strangeness states, and even less is known there regarding the resonance spectrum. For more details and other options see recent reviews~\cite{Hyodo:2020czb, Mai:2020ltx, Wang:2024jyk, Hyodo:2011ur}. 

Theoretically, the by far largest interest in this field was generated from chiral perturbation theory and unitarization techniques such as described in \cref{subsec:S11}. Specifically, CHPT is derived from QCD by integrating out heavy degrees of freedom. Interaction patterns of this Effective Field Theory for the meson-baryon scattering are used then in a unitarized approach to extend the region of applicability to a resonance region. Overall, the inclusion of chiral symmetry and S-matrix principles constrains the amplitudes such that predictions can be made outside of the data region leading for example to a prediction of the second isoscalar pole of the $\Lambda(1405)$, for more details see reviews~\cite{Hyodo:2020czb, Mai:2020ltx, Wang:2024jyk}. We restrict ourselves in the following to the context of DCC models only.

Most of the past partial wave analyses for investigating $Y^*$ resonances
were performed using the Breit-Wigner parametrization. An analysis aiming at
the extraction of the poles and their residue of the partial wave amplitudes was presented only more recently~\cite{Qiang:2010ve}. In the multi-channel 
K-matrix method, the $K^- p$ reactions were analyzed in Refs.~\cite{Zhang:2013cua, Zhang:2013sva} (KSU). In Ref.~\cite{Fernandez-Ramirez:2015tfa}, the amplitudes of Refs.~\cite{Zhang:2013cua,Zhang:2013sva} are used to extract resonance parameters. The Bonn-Gatchina (BnGa) group analyzed the strangeness sector with their K-matrix approach, describing not only two but also three-body final states. Resonance parameters were extracted in Refs.~\cite{Anisovich:2020lec, Sarantsev:2019xxm, Matveev:2019igl}.

In DCC approaches, two-body meson-baryon interactions consist of $s$-, $t$- and $u$-channel hadron exchange mechanism as described in section~\cref{sec:DCC-formalism}.
The data on $K^- p$ reactions were analyzed by the Juelich group~\cite{Mueller-Groeling:1990uxr, Haidenbauer:2010ch} and, 
recently, by the ANL-Osaka group~\cite{Kamano:2014zba, Kamano:2015hxa}. 
In the Juelich approach to the $KN$~\cite{Buettgen:1990yw, Hoffmann:1995ie}
and $\bar KN$~\cite{Mueller-Groeling:1990uxr} interactions, most of 
the parameters of the model are constrained from SU(3) flavor symmetry. 
Additional parameters are fixed from the analysis of $K N$ and 
$\bar{K} N$ elastic and inelastic cross section data that existed at the time. 
The channels included for negative strangeness are $\pi\Lambda$, $\pi\Sigma$, and $\bar KN$. The three-body channels $\bar K\Delta$, $\bar K^* N$, and $\bar K^*\Delta$ are perturbatively included in the form of box diagrams. 
The $\Lambda(1405)$ is predicted as a $\bar{K}N$ quasi-bound state and exhibits a two-pole structure~\cite{Haidenbauer:2010ch} similar to that
observed in works performed in chiral unitary approaches \cite{Mai:2020ltx}.
The validity of 
the model is limited to energies below the $\Lambda$(1520) resonance. 

\begin{figure}[t]
    \centering
    \raisebox{-.5\height}{\includegraphics[height=5cm,trim=0 0 9cm 0,clip]{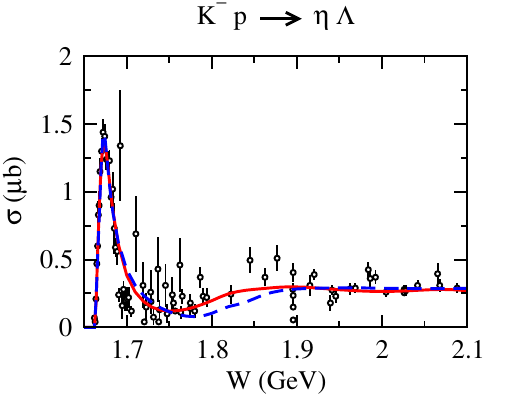}}
    \raisebox{-.5\height}{\includegraphics[height=6cm]{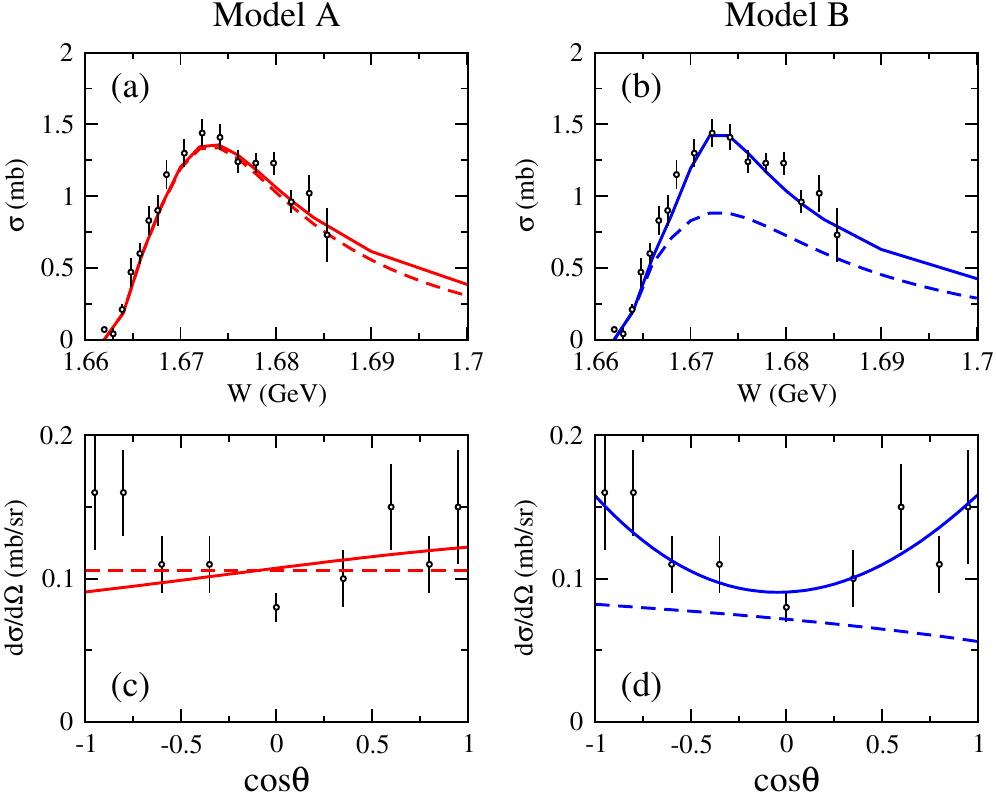}}
    \caption{Total cross section for $K^- p\to \eta \Lambda$ (left).
    Total cross section near the threshold (right top) and differential cross section (right bottom) at $\sqrt{s}=1672\,\MeV$ for $K^- p \rightarrow \eta\Lambda$ (right). Red (blue) curves are results of model A (B) in the ANL-Osaka approach~\cite{Kamano:2014zba}. Solid curves are the full results, while the dashed curves are results obtained by turning off the $P_{03}$ partial wave.}
    \label{fig:ANL-etaLambda}
\end{figure}

In the ANL-Osaka model~\cite{Kamano:2014zba,Kamano:2015hxa}, coupled channel equations for the two-body states $MB = \bar K N, \pi \Sigma, \pi \Lambda , \eta\Lambda, K\Xi$,  $\pi \Sigma^*$, $\bar{K}^* N$ and the three-body states $\pi\pi \Lambda$ and $\pi\pi \Sigma$ are formulated as given in~\cref{sec:DCC-formalism}. The data used in the analyses of the ANL-Osaka model~\cite{Kamano:2014zba} and the KSU model~\cite{Zhang:2013cua} are unpolarized differential and total cross sections and polarization observables of the $K^- p \rightarrow \bar{K}N, \pi\Sigma, \pi \Lambda, \eta \Lambda, K\Xi$ reactions. 
In addition, the ANL-Osaka group also studied strangeness production on the deuteron~\cite{Kamano:2016djv}.
In Ref.~\cite{Matveev:2019igl} by the BnGa group,  $\omega \Lambda$ and the three-body final states $K^- p \rightarrow \pi^0 \pi^0 \Lambda, \pi^0\pi^0 \Sigma$ are included in the fits as well. 
In the work of the ANL-Osaka group~\cite{Kamano:2014zba}, two solutions (model A and model B) have been obtained which have quite different values for the model parameters, while both models fitting the $K^-p$ data only have similar quality of data description. Results for the energies of $Y^*$ resonances from model A, B~\cite{Kamano:2015hxa} and  KSU~\cite{Zhang:2013cua} show excellent agreement for several states. However, in some cases large discrepancies are also seen, mirroring the fact that the existing $K^- p$ data are not yet sufficient to constrain the mass spectrum of $Y^*$.  The $S_{01}(1/2^-)$ resonance around $1.67$~GeV has a large contribution 
to the total cross section of $K^- p \rightarrow \eta \Lambda$ as shown in the left panel of Fig.~\ref{fig:ANL-etaLambda}. The situation is similar to the total cross section of $\pi^- p \rightarrow \eta n$ and the role of $S_{11}(1535)$. Around the $\eta\Lambda$ threshold region, model B additionally predicts a narrow $P$-wave resonance $P_{03}(3/2^+)$ at $E=1671^{+2}_{-8}-i(5^{+11}_{-2})\,\MeV$. This resonance couples strongly with $\eta \Lambda$. About 40\% of the peak of the total cross section is due to this $P_{03}$ resonance in model B, while the $S_{01}(1670)$ resonance dominates the peak of the cross section in model A. The difference between the model A and model B shows up in the angular distribution of $\eta \Lambda$ as shown in Fig.~\ref{fig:ANL-etaLambda} suggesting that the data of angular distribution seems to favor the p-wave contribution and resonance in model B.

\begin{figure}[t]
    \centering
    \includegraphics[height=6cm]{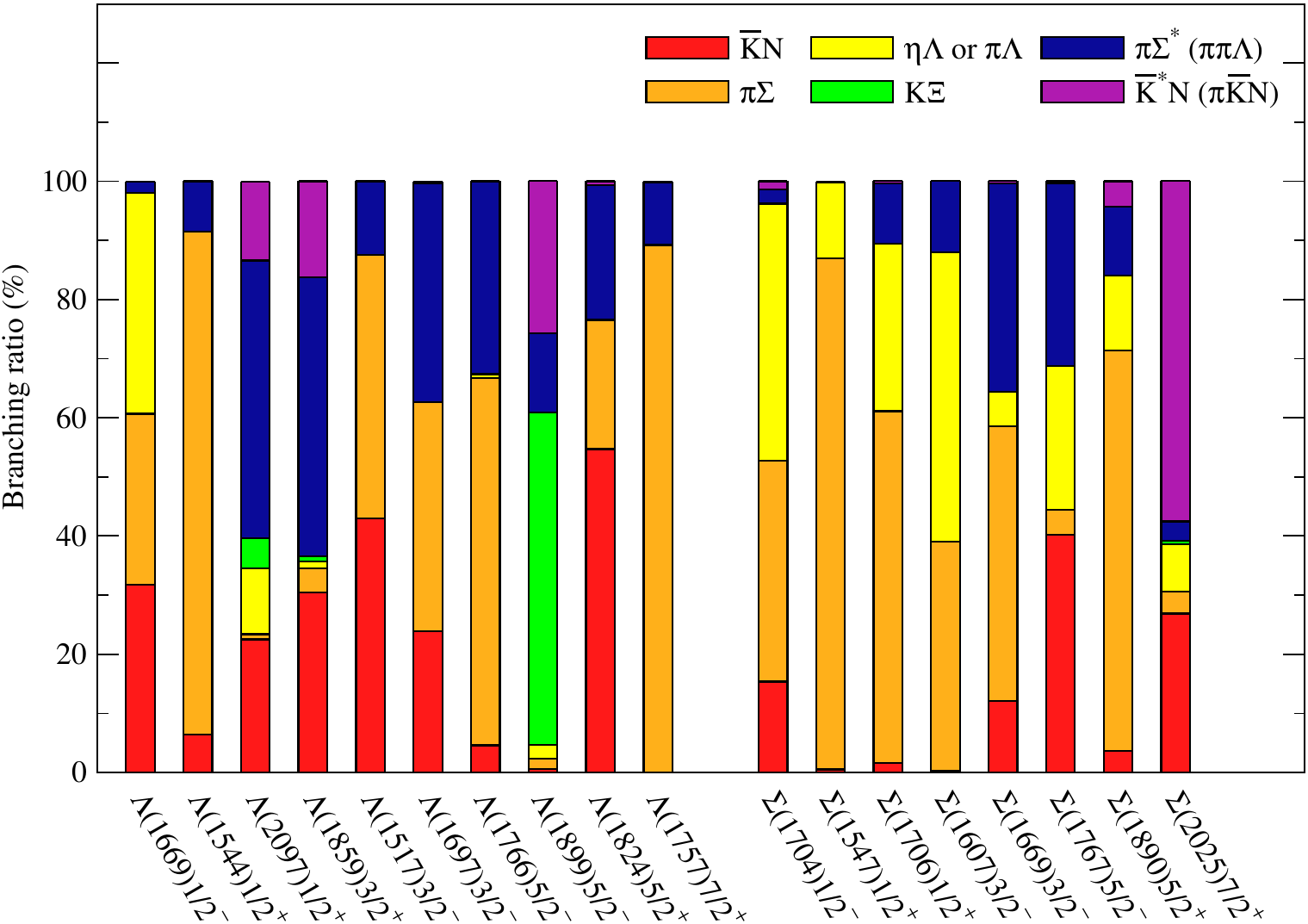}
    ~~~
    \includegraphics[height=6cm]{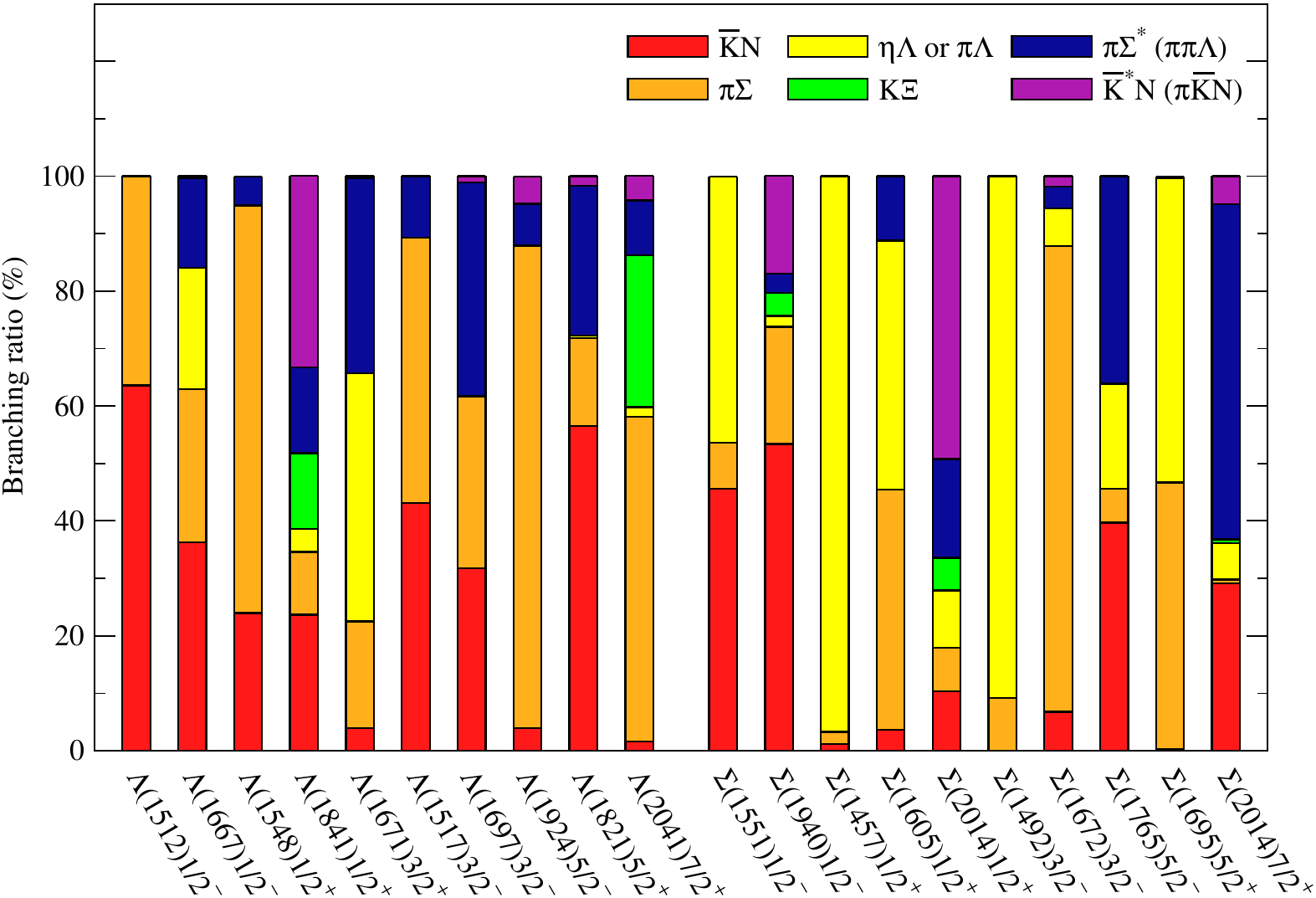}
    \caption{Branching ratio for the decay of $\Lambda^*$ and $\Sigma^*$
    in model A (left) and model B (right). Figure adapted from Ref.~\cite{Kamano:2015hxa}.
    }
    \label{fig:ANL-KY-br}
\end{figure}

For both models also the branching ratios of the $Y^*$ resonances are determined. They are depicted in \cref{fig:ANL-KY-br}. The results show large branching ratios to $\pi \Sigma^*$ and $\bar{K}^*N$ states for higher-energy resonances, suggesting the importance to establish those resonance from the data of three-body reactions $\bar KN \rightarrow \pi\pi \Lambda,\pi\bar{K}N$. For some cases, see, e.g., the lowest $\Lambda$ ${1}/{2}^-$ 
or ${1}/{2}^+$ states, both models predict not only similar masses but also
similar coupling to different channels. Less pronounced agreement in some other cases indicates the need for more precise data to pin down the properties of the hyperons. This again emphasizes the importance of experimental updates~\cite{KLF:2020gai,Quintans:2022utc, Friedrich:2024ylw,Spiller:2020wtm} discussed at the beginning of 
this section.

\section{Three-meson systems: the ANL-Osaka and IVU models}
\label{sec:threemesons}

The previous discussion of excited baryons has shown that the elastic energy window is usually small, making it necessary to consider coupled channels and three or more body dynamics. Indeed, already the first excited state of the nucleon, the Roper-resonance $N^*(1440)1/2^+$, is above the $\pi\pi N$ threshold, decaying substantially into effective three-body channels $f_0(500)N$  and $\pi\Delta$ causing the unusual line shape of that resonance~\cite{Arndt:1995bj, Alvarez-Ruso:2010ayw, Ronchen:2012eg, MartinezTorres:2008kh}. Some resonances like the $a_1(1260)$ meson decay overwhelmingly into three particles~\cite{ParticleDataGroup:2024cfk}. The same is expected for the lightest meson with exotic spin-parity quantum numbers, $J^{PC} = 1^{-+}$, the $\pi_1(1600)$ that was found by COMPASS~\cite{COMPASS:2009xrl}. This state is currently a subject of extensive experimental and theoretical investigations~\cite{COMPASS:2021ogp, Chen:2022asf}, see also  Refs.~\cite{Liu:2024uxn, JPAC:2021rxu} for recent reviews. Three-body effects can also be relevant for the other exotic candidates,  
like the $X(3872)$~\cite{Baru:2011rs}, the $Y(4260)$~\cite{Cleven:2013mka}, or the $T^+_{cc}$(3875)~\cite{Du:2021zzh,Lin:2022wmj} observed in the charmonium spectrum close to various three-body thresholds. 

In experimental analyses, the description of three-body final states with unitary isobar amplitudes has made substantial progress recently. 
See Refs.~\cite{Mai:2017vot, Jackura:2018xnx, Jackura:2019bmu, Albaladejo:2019huw, Mikhasenko:2019vhk, Jackura:2020bsk, MartinezTorres:2020hus, Zhang:2021hcl, Jackura:2023qtp, Zhang:2024dth, Leupold:2008bp} for formal developments. In close conceptual connection to Refs.~\cite{Sadasivan:2020syi, Sadasivan:2021emk} highlighted in this section, in Ref.~\cite{Nakamura:2023hbt} a three-body unitary coupled-channel framework was used to analyze $K_SK_S\pi^0$ Dalitz plot pseudodata for the $J^{PC}=0^{-+}$ amplitude in the $J/\psi\to \gamma K_SK_S\pi^0$ decay related to the controversial $\eta(1405/1475)$, extracting also the associated resonance poles. 
That framework is conceptually related to the pioneering work by the ANL-Osaka group~\cite{Kamano:2011ih, Nakamura:2012xx}. 
The final-state interaction of three pions in the  decays of $a_1(1260),\pi_2(1670), \pi_2(2100)$ and $D^0$ and also the photon-induced three pion production is investigated in DCC. The role of particle-exchange interaction (Z diagram) needed for three-body unitarity is demonstrated for the Dalitz plot, the phase of the decay amplitudes, and the extraction of resonance pole and residue. See also Refs.~\cite{Nakamura:2023obk, Zhang:2024dth} for other unitary three-body coupled-channel applications.  Ref.~\cite{Nakamura:2023obk} is an attempt of a unified description of the various final states in $e^+ e^- \rightarrow c \bar{c}$ similar  to the $N^*$ analysis with DCC. 

Three-body rescattering effects can also be quantified using Khuri-Treimann (KT) equations and in similar frameworks by the Bonn group, JPAC, and others for light meson decays~\cite{Pasquier:1968zz, Aitchison:1976nk, Colangelo:2009db, Kubis:2009sb, Schneider:2010hs,Kampf:2011wr, Niecknig:2012sj, Guo:2011aa,Danilkin:2014cra, Guo:2014vya, Guo:2015zqa,  Daub:2015xja, Niecknig:2015ija, Guo:2016wsi, Isken:2017dkw, Albaladejo:2017hhj,  Niecknig:2017ylb, Dax:2018rvs, Jackura:2018xnx, Gasser:2018qtg,  Albaladejo:2019huw, JPAC:2019ufm, Mikhasenko:2019vhk, Akdag:2021efj}. Faddeev-type arrangements of chiral two-body amplitudes were used to predict resonance states and to study known ones~\cite{MartinezTorres:2007sr,  MartinezTorres:2008kh, MartinezTorres:2011vh,
Magalhaes:2011sh, Magalhaes:2015fva,
Aoude:2018zty}. See also Ref.~\cite{Aitchison:2015jxa} for a pedagogical introduction into dispersive methods and Ref.~\cite{Aitchison:1966lpz} for connections between Khuri-Treiman equations and three-body unitary methods.
We also highlight Ref.~\cite{Stamen:2022eda} in which KT and Omn\`es-functions for the two-body part were used to study  $\pi\rho$ rescattering in close relation to the physical systems discussed in this review.

In experiments producing excited baryons there is often a well-defined initial state such as $\pi N$, $\bar K N$, or $\gamma^{(*)}N$. This usually simplifies modeling the production mechanisms. For example, in elastic $\pi N$ scattering or pion-induced meson-baryon production, there is no need to model the production mechanism separately from the coupled-channel rescattering, at all. Sometimes, however, the production mechanism is not fully known as in the study of excited mesons with three-body channels. In some cases, like semi-leptonic $\tau$ decays, the production via a highly virtual gauge boson filters quantum numbers of the three-meson system, like in case of the $a_1$~\cite{Sadasivan:2020syi}. In  high-energy experiments, on the other hand, many partial waves are simultaneously excited through complicated production mechanisms. One way of dealing with the dependence of results on the production mechanism is to kinematically divide it and check if the results depend, e.g., on the Mandelstam variable $t$ bins~\cite{COMPASS:2018uzl}. See Ref.~\cite{Ketzer:2019wmd} for a review on COMPASS physics.

In the pertinent experimental analyses of three-body final states, the traditional ``isobar model'' employs line shapes (isobars) for the final 2-body correlations in the decay, without explicit three-body  rescattering effects, e.g., at Crystal Barrel~\cite{CrystalBarrel:2019zqh} (includes coupled channels), BESIII~\cite{BESIII:2023qgj}, Belle~\cite{Rabusov:2022woa}, and LHCb~\cite{LHCb:2022lja}. Diagrammatically, this would correspond (but not be equal) to the first term in the second line of Fig.~\ref{fig:ccillu}: There is a complex ``partial-wave amplitude'' that is fit to events, and that couples to spectators and isobars which subsequently decay into asymptotically stable states - no explicit final-state interaction and no coupled channels. Both effects are, at least partially, absorbed into the complex partial wave.
Recently, the COMPASS collaboration significantly advanced the analysis of the three-pion spectrum~\cite{Ketzer:2019wmd, COMPASS:2018uzl}  by performing independent partial-wave decompositions in sub-energy bins of the two-pion subsystems, i.e., without using fixed isobar line shapes~\cite{Krinner:2017dba}. If the isobar is ``freed'' in this way, the amplitude is general enough that three-body effects like rescattering and coupled channels can be accommodated. In contrast, the three-body methods discussed in this review parametrize these effects explicitly, with the consequence that the production vertex $D$ (see Fig.~\ref{fig:ccillu}) is always real and does not contain physical resonance information by itself. 

From a theoretical point of view one can argue that manifestly implementing coupled channels and unitarity corrections in the amplitude allows for a better understanding of these effects, and, in the future, these amplitudes might even be used to analyze data directly. In addition, there are kinematic effects like triangle singularities (TS) which cannot easily be accommodated in simple isobar models. However, these effects naturally arise as coupled-channel effect (with unequal masses) in DCC approaches~\cite{Sakthivasan:2024uwd} as illustrated in the last diagram of Fig.~\ref{fig:ccillu}. We briefly review recent developments in this respect in \cref{sec:triangles}.

In this review, we report on the development and applications of unitary DCC amplitude with explicit rescattering and coupled channels related in concept and formalism to the previously discussed DCC approaches. The previously discussed baryonic JBW models focus on phenomenological aspects and data analysis from experiment by starting from some phenomenological Lagrangian. 
In contrast to this, mesonic three-body DCC approaches are sometimes related to finite-volume methods for the mapping of lattice-QCD energy eigenvalues to the physical scattering amplitudes in infinite volume. In particular, the Finite-Volume Approach (FVU) discussed in Sect.~\ref{3BQC} uses the concept of three-body unitarity~\cite{Mai:2017vot} discussed in \cref{sec:unitarity} to construct a finite-volume amplitude in which spurious singularities cancel systematically in an interplay of disconnected and connected contributions, and exchange singularities with propagation singularities~\cite{Mai:2017bge}. We, therefore, call the corresponding amplitudes discussed here infinite-volume unitary (IVU) approaches. While finite-volume physics is not the focus of this review~\cite{Hansen:2019nir,Mai:2021lwb, Blanton:2020jnm}, it should be mentioned that alternative method exists, referred to as the RFT approach~\cite{Blanton:2019igq, Hansen:2024ffk}. Here, several three-body systems of excited mesons were formulated~\cite{Hansen:2020zhy, Blanton:2020gmf} including the unequal mass case of Refs.~\cite{Draper:2023boj, Dawid:2025doq} and Ref.~\cite{Draper:2024qeh} ($\eta\pi\pi$ and $K\bar K\pi$ channels), as well the solution of integral equations in this context~\cite{Hansen:2020otl}. 

In \cref{sec:a1} we  report  the two-channel approaches of Refs.~\cite{Sadasivan:2021emk, Mai:2021nul} in which, for the first time, a three-body resonance pole was extracted from data with a unitary amplitude. In \cref{sec:nine} a recent extension to the unequal mass case is discussed for a nine-channel amplitude containing pions and kaons~\cite{Feng:2024wyg}.
The unequal-mass case can also lead to the formation of so-called triangle singularities in the physical region. While this is a long-known issue evaluated at the one-loop level, triangle singularities can also be understood as off-diagonal channel transitions of a three-body unitarized amplitude. Recent developments in this direction are discussed in \cref{sec:triangles}.

\subsection{A two-channel model for the \texorpdfstring{$a_1(1260)$}{a1(1260)} meson}
\label{sec:a1}
A DCC amplitude with manifest three-body unitarity  (referred to as infinite-volume unitary, IVU) was developed in Ref.~\cite{Sadasivan:2020syi} for the two coupled partial waves $(\pi\rho)_S$ and $(\pi\rho)_D$ to study the $a_1(1260)$ resonance. This amplitude was later reformulated for the finite volume to perform the first extraction of a three-body resonance from lattice QCD data~\cite{Mai:2021nul}. The analytic continuation of the amplitude to the $a_1(1260)$ resonance pole was explained in detail in Ref.~\cite{Sadasivan:2021emk} and, for the first time, the pole of that resonance was determined from experimental data using the continuation techniques discussed in Sect.~\ref{sec:cont3B}. 

See also Ref.~\cite{JPAC:2018zwp} for an extraction of the $a_1(1260)$ resonance pole with an approximately unitarity amplitude. The $a_1(1260)$ resonance can also decay to the $\pi\sigma$ channel and the branching ratio to this subdominant channel was estimated in Ref.~\cite{Molina:2021awn}. Different parametrizations for the $a_1$ meson and consequences for observables were studied in Ref.~\cite{Wagner:2008gz}. A model that systematically includes $S$- and $P$-wave isobars and also strangeness channels was formulated in Ref.~\cite{Feng:2024wyg} and is discussed in \cref{sec:nine}, but it has not yet been applied to a physical system.

\begin{figure}[tb]
    \centering
    \includegraphics[width=0.4\linewidth,trim=0cm 17cm 17cm 0cm,clip]{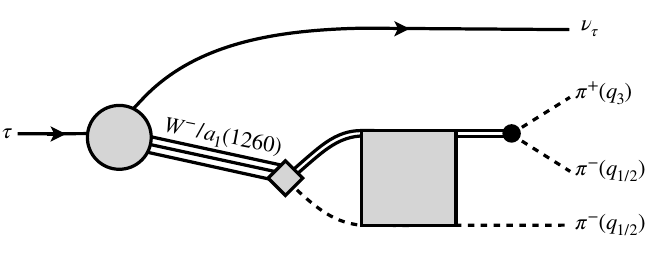}
    \caption{Factorization of the (weak) production mechanism and (hadronic) final state three-body interaction in semileptonic $\tau$ decays. Full-directed, dashed, and double lines denote leptons, mesons and auxiliary $\rho$ fields, respectively. The initial production mechanism is shown by a shaded circle and diamond. The three-body unitary isobar-spectator interaction in the gray square corresponds to $\tilde T$ from Eq.~\eqref{eq:tildeTnoprob}.
    Note that there is also a ``disconnected'' production of the final isobar-spectator pair in the IVU parametrization as indicated in Fig.~\ref{fig:ccillu}.
    The Figure is taken from Ref.~\cite{Sadasivan:2021emk}.
    }
    \label{fig:process}
\end{figure}

Coming back to the  $a_1(1260)$ resonance, it  couples to  three-pion states in the $I^G(J^{PC})=1^-(1^{++})$ channel that can be decomposed as $\pi\rho$ in $S$/$D$-wave, $\pi \sigma$ and $\pi (\pi\pi)_{I=2}$ in $P$-waves and other channels. Phenomenologically, the channel $(\pi\rho)_{\rm S}$ is dominant~\cite{E852:2004gpn}, with the branching ratios into other channels being quite uncertain~\cite{ParticleDataGroup:2024cfk}. The semileptonic $\tau$ decay leading to the $a_1$ formation is shown in Fig.~\ref{fig:process}.
The starting point to formulate the hadronic final-state interaction is, again, the connected production amplitude $\tilde \Gamma$ from Eq.~\eqref{sec:direct} plus the disconnected part $D_j$ that we refer to as a ``bare'' production mechanism in the sense that it is a real, energy and momentum-dependent function in the physical region. The index $j$ stands for the two channels $(\pi\rho)_S$ and $(\pi\rho)_D$.
The production vertices $D_j$  contain a regular piece and a first-order pole in the three-body energy $\sqrt{s}$~\cite{Sadasivan:2021emk}, similar to Eq.~\eqref{vg} for photoproduction. That pole cancels with a corresponding zero in the rescattering part just like for the photoproduction case. Also, there is now a nonzero contact term in rescattering, $C$ from Eq.~\eqref{eq:T}. It contains the same singularity. This corresponds to $V^\po$ in baryon amplitudes, see Eq.~\eqref{blubb}. Together, the matching singularities in $D$ and $C$ provide a flexible way to parametrize the $a_1(1260)$. Also, both $D$ and $C$ contain regular, real-valued terms that can depend on three-body energy $\sqrt{s}$ and spectator momentum $p$. All these parts contain fit parameters that can be adjusted to data. Note the formal similarity of the overall amplitude to the photo and electroproduction amplitude in DCC model, see Eq.~\eqref{ampl_2}. In particular, the production amplitude has the same structure as the ``dressed resonance creation vertex'' in Eq.~\eqref{dressed}, consisting of a direct/disconnected piece and a rescattering/connected piece.

To connect the production amplitude to the observable $\pi^-\pi^-\pi^+$ state in $\tau$ decays, the angular dependence is introduced into the production process $\widebreve\Gamma$ related to Eq.~\eqref{eq:gammabrev} with the help of Wigner-D functions, 
\begin{align}
\Gamma_{\Lambda\lambda} ({\bf q}_1,{\bf q}_2,{\bf q}_3)&=
\sqrt{\frac{3}{4\pi}}{\mathfrak D}^{1*}_{\Lambda\lambda}(\phi_1,\theta_1,-\phi_1)\, U^{T}_{L'\lambda'}\widebreve\Gamma_{L'L}(q_1)U_{L\lambda}v_{\lambda}({\bf q}_2,{\bf q}_3) \ ,
\label{eq:Gammaangle}
\end{align}
with transformation matrices $U$ between helicity and JLS basis~\cite{Chung:1971ri} implying sums over $\lambda'$, $L$, and $L'$ and 
where $\mathfrak{D}^{J}_{\lambda'\lambda}(\phi_1,\theta_1,-\phi_1)$ are the capital Wigner-D function with angles $\theta_1$ and $\phi_1$ giving the polar and azimuthal angles of $\bf{q}_1$. Note the explicit appearance of the last isobar decay vertex $v_\lambda$ here that contains an additional angular dependence due to the $P$-wave decay of the $\rho$ (defined in its rest frame). For consistency, it should be mentioned that in Eq.~\eqref{eq:gammabrev} and in Ref.~\cite{Feng:2024wyg} that vertex is absorbed in the definition of $\breve\Gamma$ while in $\widebreve{\Gamma}$ of Eq.~\eqref{eq:Gammaangle} it is not, following the notation of Ref.~\cite{Sadasivan:2021emk}.

Finally, the amplitude has to be symmetrized to account for indistinguishable particles. The amplitude $\hat\Gamma_{\Lambda\lambda}$, describing the decay of the axial $a_1(1260)$-resonance at rest with helicity $\Lambda$ measured along the $z$-axis into a $\pi^-$ and a $\rho_\lambda^0\to \pi^+\pi^-$ with helicity $\lambda$, is given by 
\begin{align}
\label{eq:decayrate}
&\hat \Gamma_{\Lambda\lambda} (\boldsymbol{q}_1,\boldsymbol{q}_2,\boldsymbol{q}_3)
=\frac{1}{\sqrt{2}}\big[\Gamma_{\Lambda\lambda} (\boldsymbol{q}_1,\boldsymbol{q}_2,\boldsymbol{q}_3)
-\left(\boldsymbol{q}_1\leftrightarrow\boldsymbol{q}_2\right)\big] \ ,
\end{align}
where $\bm{q}_1$, and $\bm{q}_2$ are outgoing $\pi^-$ momenta, and $\bm{q}_3$ is the outgoing $\pi^+$ momentum, see also Fig.~\ref{fig:process}. 

\begin{figure}[tb]
\centering
\includegraphics[width=0.465\linewidth,trim=0.1cm 1.6cm 0 0,clip]{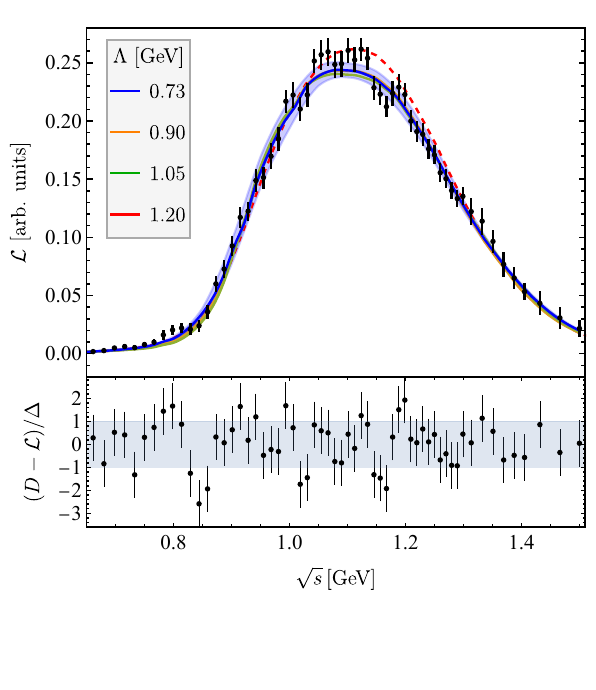}
\includegraphics[width=0.525\linewidth,trim=0.cm 0cm 0 0,clip]{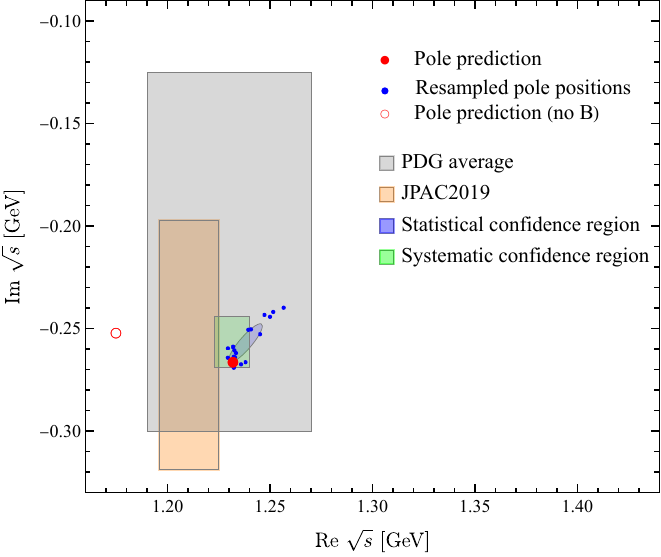}
\caption{
{\bf Left}: Fit to the line shape data from the ALEPH experiment~\cite{Davier:2013sfa} for different cutoff values (top) and the normalized residuals for the $\Lambda=0.73\,\GeV$ case (bottom), where $D-\mathcal{L}$ is a given residual and $\Delta$ the pertinent data uncertainty. In the upper figure, the blue band shows the statistical uncertainties of the $\Lambda=0.73\,\GeV$ fit, multiplied by ten for visibility. The red dashed line shows the $\Lambda=0.73\,\GeV$ fit with all D-wave terms set to 0. {\bf Right}: Compilation of pole positions determined in Ref.~\cite{Sadasivan:2021emk} including statistical and systematic uncertainties. For convenience, the PDG~\cite{ParticleDataGroup:2024cfk} average and a result by JPAC~\cite{JPAC:2018zwp} are quoted as well. See text for further explanations. Figures taken from Ref.~\cite{Sadasivan:2021emk}.
}
\label{fig:phantom}
\end{figure}

In Ref.~\cite{Sadasivan:2020syi} the connection of $\hat\Gamma$ to different observables is evaluated: The most detailed information is contained in Dalitz plots at a fixed three-body energy $\sqrt{s}$; there are Dalitz plot projections, and there is the line shape that is the integral over the full phase space as a function of $\sqrt{s}$, 
\begin{align}
\label{eq:lineshape}
&\mathcal{L}(\sqrt{s})=N(m_\tau^2-s)^2
\int \frac{\dif^3{\bm q}_1}{(2\pi)^3}
\frac{\dif^3{\bm q}_2}{(2\pi)^3}
\frac{\dif^3{\bm q}_3}{(2\pi)^3}
\frac{(2\pi)^4\,\delta^4(P_3-q_1-q_2-q_3)}{8E_{q_1}E_{q_2}E_{q_3}}
\left(
\big|\sum_\lambda\hat\Gamma_{-1\lambda}\big|^2
+\frac{m_\tau^2}{s}\big|\sum_\lambda\hat\Gamma_{0\lambda}\big|^2
+\big|\sum_\lambda\hat\Gamma_{+1\lambda}\big|^2
\right) \ ,
\end{align}
with arbitrary normalization $N$. See also Ref.~\cite{JPAC:2018zwp} for the derivation of the weak decay part. The Monte-Carlo integration over phase space and other numerical details are explained in Ref.~\cite{Sadasivan:2020syi}. Note that the analytic continuation of the amplitude to real spectator momenta using Pad\'e approximants, as proposed in that paper, is less rigorous than the exact methods explained in Sects.~\ref{subsec:Cahill-Triangle} and \ref{sec:direct}. The line shape fit from Ref.~\cite{Sadasivan:2021emk} is shown together with data and fit residuals in Fig.~\ref{fig:phantom} to the left. As the line shape data themselves have been extracted from raw data, they contain correlations that are included in the fit. The cutoff $\Lambda$ refers to the production amplitude in Eq.~\eqref{dirprod}.

As for the analytic continuation to complex energies onto the unphysical Riemann sheet, see \cref{sec:cont3B}. The resulting pole positions and their variations under changes of the parametrization are shown in Fig.~\ref{fig:phantom} to the right. The $\pi\rho$ branch point is outside the shown area. The most probable pole position is shown with the red dot. Data resampling results in the small blue dots that can be used to determine an error ellipse shown in light blue, even though the fit problem is obviously not very linear (i.e., the most probable value is not in the center but at the margin of the ellipse). Systematic uncertainties, shown with the light green rectangle, were determined by varying the cutoff $\Lambda$ in a wide range while re-fitting all other parameters. This shows that the parametrization is consistent and cutoff changes can be well absorbed in changing the contact $C$-terms.

In Ref.~\cite{Sadasivan:2021emk} the role of the pion exchange term was also tested  (see also Eq.~\eqref{btilde} below). The fit quality without this term changes little, but the most probable pole position, shown with the open red circle in Fig.~\ref{fig:phantom}, changes significantly compared to the uncertainties. This demonstrates that the exchange process, demanded by three-body unitarity~\cite{Mai:2017vot}, does have an influence and should be taken into account when determining resonance poles from data.  

\subsection{A unitary amplitude with pions and kaons}
\label{sec:nine}
Reference~\cite{Feng:2024wyg} builds on these works and allows for pions and kaons in the amplitude. Nine channels are included that not only contain canonical $S$- and $P$-wave resonances in the two-body subsystems --the $f_0(500)$ ``$\sigma$'', $f_0(980)$, $\rho(770)$, $K_0^*(700)$ ``$\kappa$'', $K^*(892)$-- but also repulsive $\pi\pi$ (``$\pi_2$'') and $\pi K$ (``$K_{\nicefrac{3}{2}}$'') subsystems at maximal isospin. Amplitude transitions are formulated for both $G$-parities $\eta_G$ and all total isospins $I\in\{0,\dots,3\}$ with zero overall strangeness. An overview of the channel space is shown in Table~\ref{tab:channels}. The last row shows the partial waves for the $a_1$ quantum numbers. Other total angular momenta are easily obtained following Sec.~\ref{sec:pwa}.

\begin{table}[b!]
    \caption{Channels for $\eta_G=-1$ from Ref.~\cite{Feng:2024wyg}. The spin $\ell$ and isospin $I_I$ of the isobar is indicated in the first line for clarity. The second line shows the eleven channels in helicity basis (HB). The third line indicates the nine channels in the JLS basis for total angular momentum $J=1$. The outermost subscripts indicate the relative angular momentum $L$ between isobar and spectator. The abbreviation $\pi_2$ stands for the $S$-wave isospin $I_I=2$ repulsive $\pi\pi$ channel and $K_{3/2}$ for the repulsive $S$-wave $I_I=\nicefrac{3}{2}$ $\pi K$ channel.}
    \center
    \renewcommand{\arraystretch}{1.5}
        \begin{tabularx}{\linewidth}{X|l|l|l|l|l|l|l|l|l}
        \hline\hline
        Isobar $(\ell,I_I)$
        &
        \multicolumn{2}{c|}{$(1,1)$}
        &
        \multicolumn{2}{c|}{$(1,\nicefrac{1}{2})$}
        &
        \multicolumn{2}{c|}{$(0,0)$}
        &
        $(0,2)$ & $(0,\nicefrac{1}{2})$ & $(0,\nicefrac{3}{2})$
        \\
        \hline
        HB basis (11 Ch.)
        & 
        \multicolumn{2}{c|}{$\pi\rho_{\lambda=\pm 1, 0}$}
        & 
        \multicolumn{2}{c|}{$KK^*_{\lambda=\pm 1, 0}$}
        &$\pi\sigma$ & $\pi (K\bar K)_S$
        & $\pi\pi_2$ & $K\kappa$ & $KK_{\nicefrac{3}{2}}$
        \\
        \hline
        JLS basis (9 Ch.)
        &
        $(\pi\rho)_S$ & $(\pi\rho)_D$
        &
        $(KK^*)_S$ & $(KK^*)_D$
        &$(\pi\sigma)_P$ & $(\pi (K\bar K)_S)_P $
        & $(\pi\pi_2)_P$ & $(K\kappa)_P$ & $(KK_{\nicefrac{3}{2}})_P$
        \\
        \hline\hline
        \end{tabularx}
    \label{tab:channels}
\end{table}

All channels contribute to total isospin $I=1$, but only some would contribute to isospins $I=0$, $I=2$, and $I=3$. Note that positive/negative $G$-parity requires the channels containing kaons to be combined according to
\begin{align}
    \eta_+=\frac{1}{\sqrt{2}}\left(\bar K_{I_I} K+K_{I_I} \bar K\right) \,\quad
    \eta_-=\frac{1}{\sqrt{2}}\left(\bar K_{I_I} K-K_{I_I}\bar K\right) \ ,
    \label{gparity}
\end{align}
where $K_{I_I}$ stands for the $S$-wave or $P$-wave isobars in isospin $I_I=\nicefrac{1}{2}$ or $I_I=\nicefrac{3}{2}$. In Table~\ref{tab:channels}, we indicate the combinations $\eta_-$ as $KK^*$, $K\kappa$, or $KK_{\nicefrac{3}{2}}$. 
The included channels represent all possible isobars up to $P$-wave, except for a very small $P$-wave $I=\nicefrac{3}{2}$, $\pi K$ isobar and an $S$-wave, $I=1$, $K\bar K$ isobar. There are 11 partially redundant channels in the helicity basis (HB) and 9 channels in the standard JLS basis. The transformations are carried out via the partial-wave projection discussed in \cref{sec:pwa}.

The projected scattering equation reads
\begin{align}
    \tilde T_{ji}(s,p',p)=\tilde B_{ji}(s,p',p) + \int_\Gamma \frac{\dv{q}\,q^2}{(2\pi)^3\,2E_q} \tilde B_{jk}(s,p',p)\tilde \tau_{kk'}(\sigma(q))\tilde T_{k'i}(s,q,p) 
    \label{eq:tildeTnoprob}
\end{align}
with block-diagonal isobar-spectator propagation $\tau_{kk'}$ allowing for channel transitions during the isobar propagation, see also Eq.~\eqref{dirprod} for the corresponding production amplitude. Currently, these transitions are only allowed for the $f_0$ isobar quantum numbers with $\pi\pi\leftrightarrow K\bar K$ as illustrated in ~\cref{fig:ccillu}. In the above equation, the $\tilde B$ are the exchange (``re-arrangement'') processes that become complex above the three-meson thresholds as required by unitarity, see ~\cref{sec:unitarity}. The $\tilde T$-matrix serves as input for the production reaction according to 
\begin{align}
    \tilde\Gamma_j (s,p')=\int_{\Gamma}\frac{\dv{p}\,p^2}{(2\pi)^3\,2E_p}\,\tilde T_{ji}(s,p',p)\tilde\tau_{ik}(\sigma(p))D_k(s,p)\,,
    \label{eq:normalgamma}
\end{align}
but $\tilde\Gamma$ plus the disconnected piece $D$ can also be obtained via Eq.~\eqref{dirprod} without the need to solve for $\tilde T$ first. The full production amplitude is obtained from Eq.~\eqref{eq:gammabrev}.

\begin{figure}[tb]
\begin{center}
\includegraphics[width=0.35\textwidth]{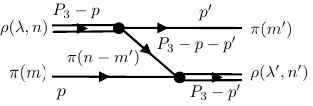}
\end{center}
\caption{The $\pi\rho\to\pi\rho$ transition. The labels $n^{(\prime)},\,m^{(\prime)}$ stand for third components of isospin of the incoming (outgoing) states.}
\label{fig:exlabels}
\end{figure}

In the previous equations as well as in Ref.~\cite{Feng:2024wyg}, a ``tilde'' notation for $\tilde B,\,\tilde\tau$, and $\tilde T$ is introduced to allow for a  clear separation of the isospin structure associated with the exchange $\tilde B$ from the propagation $\tilde \tau$, cast into simple isospin recoupling factors for both quantities. This is in contrast to previous IVU works for excited mesons~\cite{Sadasivan:2020syi, Mai:2021nul, Sadasivan:2021emk} but also the previously discussed JBW approach for excited mesons. There, Lagrangians were used to calculate isospin coefficients for transitions $B$ that partly contain an isospin structure from the isobar propagation $\tau$, see the discussion in Appendix A of Ref.~\cite{Feng:2024wyg}. In particular, the isospin coefficients in Ref.~\cite{Feng:2024wyg} are obtained simply by 
\begin{align}
    \tilde I_F&=\sum_{\substack{m,n\\m',n'} }
    \cg{I_In}{I_Sm}{II_3}
    \cg{I_{I'}n'}{I_{S'}m'}{II_3}
    \cg{I_xn-m'}{I_{S}m}{I_{I'}n'} 
    \cg{I_xn-m'}{I_{S'}m'}{I_In}
    \,,\quad \text{ for $\pi\pi\pi$ only,} 
    \label{eq:isofac}
\end{align}
where $I_{I}$ ($I_{I'}$) is the isospin of the incoming (outgoing) isobar,  $I_S$ ($I_{S'}$) is the isospin of the incoming (outgoing) spectator, and $I_x$ is the isospin of the exchanged particle. Figure~\ref{fig:exlabels} shows the label assignments for the example of $\pi\rho\to\pi\rho$ transition.
The calculation is slightly more intricate in the presence of kaons as one has to adopt a convention of how  particles with strangeness are ordered. The logic of \cref{eq:isofac} can be illustrated as follows. The two-body input of the current formalism are isobar amplitudes in the isospin basis. To construct the particle exchange, one needs to first ``unfold'' these amplitudes to the particle/charge basis (third and fourth coefficient). One then has to sum over all possible particle exchanges. Finally, one needs to construct total isospin states (first and second coefficient). The particle phases~\cite{deSwart:1963pdg} associated with $\pi^+$ and $\rho^+$ always cancel in the process. To demonstrate the angular dependence of isobars with spin, the exchange transitions  with the labeling of Fig.~\ref{fig:exlabels} read in the helicity basis:
\begin{align} 
    \tilde B_{ji}(s,\bm{p}',\bm{p})&
    =\frac{(\tilde {\bm I}_F)_{ji}\,
    v_j^*(p,P_3-p-p')
    v_i(p',P_3-p-p')}
    {2E_{p'+p}(\sqrt{s}-E_{p}-E_{p'}-E_{p'+p})+i\epsilon}
    \quad    \text{for} \quad
    i,j\in {\rm HB}
    \,, 
    \label{btilde}
\end{align}
with the angular structures of $P$- and $S$-wave isobars,
\begin{align}
    v_{i}(p,q)
    =
    \begin{cases}
        g_i\,(p-q)^{\mu}\,\epsilon_{\mu}(\bm{p}+\bm{q}, \lambda)
        \hfill &\quad \text{for $\ell=1$ isobars,}\\
        1&\quad \text{for $\ell=0$ isobars,}
    \end{cases}
    \label{eq:v}
\end{align}
with polarization state vectors $\epsilon$, for exact definitions see Ref.~\cite{Chung:1971ri}. Note that the presence of the vector meson coupling to pseudoscalar mesons, $g_i$, is due to the particular representation of the two-body amplitude with explicit isobars~\cite{Feng:2024wyg}. In contrast, the $\ell=0$ isobars are not parametrized with isobars. While the same is possible for $\ell=1$ isobars, the angular dependence of the vector isobar channel is always moved inside the definition of $\tilde B$ because that is the object undergoing partial-wave projection according to \cref{sec:pwa}.

Another issue concerns symmetry factors as explained in Sect.~II~D of Ref.~\cite{Feng:2024wyg}. The amplitude entering the propagation $\tilde \tau$, together with the adjacent angular structure in case of $P$-wave isobars corresponds to the full, plane-wave, 2-body amplitude that no longer absorbs any symmetry factors in its definition. This is potentially an issue, because it is often customary (e.g., in many chiral unitary approaches) to absorb symmetry factors from the two-body propagation of identical particles -- pions count as identical in the isospin basis -- into the definition of two-body states (e.g., Ref.~\cite{Oller:1998hw}), or the partial-wave projected amplitudes in case of the Inverse Amplitude Method (e.g., Ref.~\cite{GomezNicola:2001as}). In contrast, note that the self-energy integral of Eq.~\eqref{eq:tau} explicitly contains the symmetry factor of $1/2$ for identical particles, where it belongs.
The detailed procedure to match different two-body amplitudes from the literature to $\tilde \tau$ as input for three-body calculations is given in Ref.~\cite{Feng:2024wyg}.

The next step in the amplitude construction consists in determining the isobars from two-body scattering data. In the scalar-isoscalar ($f_0$) channel, there are not only data for the $\pi\pi$ phase shifts, but also for the $\pi\pi\to K\bar K$ transition and the inelasticity $\eta_0^0$ that determine the coupled-channel isobar amplitude. With all input available the integral equation for the production amplitude can be calculated either by direct inversion (\cref{sec:direct}), contour deformation (\cref{sec:realmom}), or continued fractions~\cite{Sakthivasan:2024uwd}. For contour deformation, one needs to analytically continue the amplitude to real spectator momenta as described in \cref{subsec:Cahill-Triangle}. In Ref.~\cite{Feng:2024wyg} it is checked that this method and direct inversion lead to the same result.

\begin{figure}[tb]
    \begin{center}
    \includegraphics[width=0.45\textwidth]{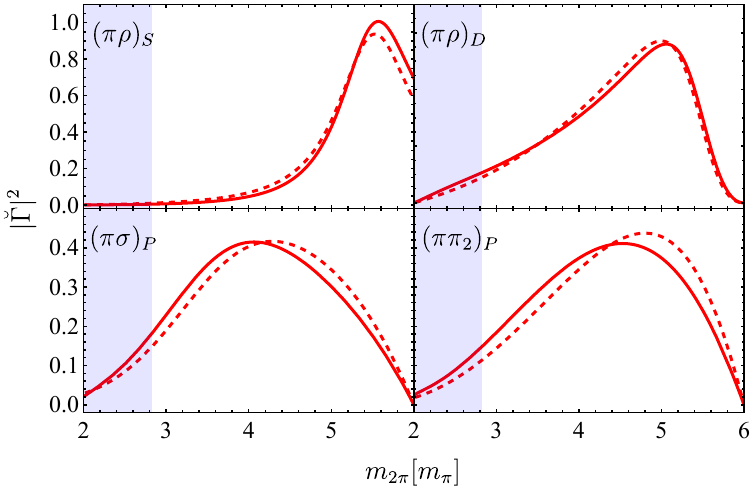}
    \includegraphics[width=0.45\textwidth]{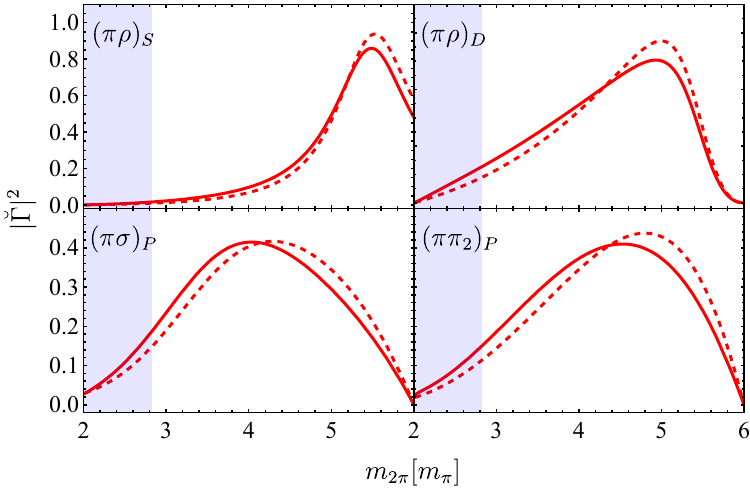}
    \end{center}
    \caption{Production amplitude $|\breve{\Gamma}|^2$ (unitless for all channels) in terms of the isobar invariant mass. The left figure shows the result with the standard choice for the production vertex $D$, while the right figure corresponds to modified input.
    The red dashed lines show the disconnected contributions that correspond to  traditional isobar amplitudes/line shapes, while the red solid lines correspond to the full, three-body unitary amplitude including coupled channels and meson exchanges. The areas under all curves are normalized to one to facilitate the comparison of the line shapes. The light blue area indicates the critical region, see the discussion of Fig.~\ref{fig:illustration} for further explanations. Figure taken from Ref.~\cite{Feng:2024wyg}.}
    \label{fig:normalized production amplitude}
\end{figure}

In Fig.~\ref{fig:normalized production amplitude} the production amplitude $\breve{\Gamma}$ from Eq.~\eqref{eq:gammabrev} is shown for different choices of ``bare'' production vertices $D$ in different channels. The blue highlighted regions correspond to the ``critical region'' shown in Fig.~\ref{fig:illustration} to the left. The fact that the amplitude exhibits a smooth behavior and transition to the noncritical region demonstrates the validity of the  Cahill and Sloan method discussed in \cref{sec:realmom}.

In Fig.~\ref{fig:normalized production amplitude}, to the left, we show the complete production amplitude $\breve\Gamma$ from Eq.~\eqref{eq:gammabrev} including the disconnected part,  the final isobar propagation, and its decay. Consequently, one observes the $\rho$ and $\sigma$ line shapes, but not the $f_0(980)$ line shape, due to the chosen energy being $\sqrt{s}=7\,m_\pi$. 
The dashed lines show the disconnected part alone. This would correspond to the traditional isobar model, in which line shapes are related to two-body scattering amplitudes only. The solid lines show the full, three-body unitary amplitude $|\breve\Gamma|^2$, including coupled channels and rescattering.  
There is a noticeable modifications of the line shapes by these effects, at some invariant masses. We observe a distortion of the $\sigma$ meson towards smaller invariant masses and similarly moderate changes for the other channels. 

If one chooses the input for the production vertex $D_{f}$ differently, one obtains different line shapes as shown in Fig.~\ref{fig:normalized production amplitude} to the right.  Of course, there is additional freedom in the amplitude that will further modify the line shapes. This refers to the isobar-spectator contact terms $ C$ in Eq.~\eqref{eq:T}. These terms encode three-body resonance dynamics~\cite{Sadasivan:2021emk} and other QCD effects, and they have all been set to zero in Ref.~\cite{Feng:2024wyg} for simplicity. In summary, the work from Ref.~\cite{Feng:2024wyg} demonstrates how rescattering, coupled channels, three-body forces $C$, and the relative strengths of  the production processes $D$ all lead to modifications of the traditional isobar line shapes.

\subsection{Triangle singularities as channel transitions in DCC approaches}
\label{sec:triangles}
Three-body effects can be enhanced by a kinematical effect known as triangle singularity (TS)~\cite{Guo:2019twa}. It  sometimes occurs in the decay of a resonance to an intermediate   resonance  and a spectator. In a specific kinematic region one of the decay products of the intermediate resonance can ``catch up'' with the spectator. This on-shell condition produces a bump in the invariant  mass spectrum of these two particles, without any resonance being responsible for it~\cite{Coleman:1965xm, Karplus:1958zz,Landau:1959fi,Booth:1961zz,Anisovich:1964ikk}. See also Refs.~\cite{Liu:2015taa,Bayar:2016ftu} for the conceptual treatment of TS. The $a_1(1420)$, was at first claimed to be a new resonance while posterior analyses suggested that a TS effect can take place in the decay chain $a_1(1260) \to K^* \bar K \to K\bar K \pi \to f_0 \pi \to 3\pi$ resulting in a peak in the $3\pi$ mass $0.2\,\GeV$ above the nominal $a_1$ mass~\cite{Mikhasenko:2015oxp, Aceti:2016yeb, Guo:2019twa, COMPASS:2020yhb}, see also Refs.~\cite{Bayar:2016ftu,Debastiani:2018xoi, Guo:2019twa, Isken:2023xfo}  for general discussions.  
See Refs.~\cite{Dai:2018rra, Liang:2019jtr, Jing:2019cbw} for further examples of TS. The exact position of the TS can also be used to determine the mass of an exotic state close to a three-body threshold when the decay channel of the exotic particle is mediated by a TS~\cite{Guo:2019qcn,Molina:2020kyu}.

In the context of DCC approaches, triangle singularities are nothing but enhanced hadron exchanges between channels of unequal masses. An example is given by the diagram in Fig.~\ref{fig:ccillu} to the lower right: In a system of pions and kaons that interact strongly, the above-mentioned $K^*\bar K\to K\bar K\pi\to f_0\pi$ transition is clearly visible as kaon exchange in an infinite rescattering series. Qualitatively, at around $\sqrt{s}\approx 1.42\,\GeV$ the quasi-on-shell $K^*$ and the kaon spectator are on-sell at the momentum of the exchanged kaon  flying in backwards direction, thus being able to ``catch up'' with the kaon spectator and form an $f_0$ as if it were a classical kinematic process. That kinematic point is given as $p_P$ from Eq.~\eqref{eq:trianglepoint}. This enhancement of processes that are allowed in classical kinematics is referred to as Coleman-Norton theorem~\cite{Coleman:1965xm}.

The role of TS in dynamical coupled channel approaches has been recently discussed in depth in Ref.~\cite{Sakthivasan:2024uwd}. Using the IVU formalism~\cite{Mai:2017vot} the question was studied whether the appearance of triangle singularities is affected by an infinite series of rescattering of final two- and three-body final states. It was found that the position of the triangle singularity is not affected through the final-state rescattering. This is in full agreement with the argument provided by the generalization of the Landau
equations~\cite{Landau:1959fi,Polkinghorne:1960cjb,Polkinghorne:1960udx} to include rescattering effects. Specifically, it was found that inclusion of higher topologies 
(triangle\,$\times$\,box, triangle\,$\times$\,box\,$\times$\,box etc.)  leads either to non-singular configurations or provides a sub-leading singularity through a factorization into the triangle diagram and the rest that is non-singular.
A first estimate of the size of the effect due to rescattering was carried out and found to be quite small. Clearly, some approximations were made, neglecting the full spin/isospin structure for the $a_1(1420)$ system, but the driving effect for the smallness of the rescattering effect seems to lie in the particular size of the coupling constants, fixed mostly by the mass and the width of the $K^*$ and $f_0(980)$ resonances. This explains a different observation of Ref.~\cite{Nakamura:2023hbt} where the rescattering effects for the $\eta(1405/1475)$ are found to be more important. It seems that the general path forward is to always include all rescattering effects and let data decide whether rescattering is relevant or not.

\section{Coupled-channel baryon-baryon systems}
\label{sec:Baryon-Baryon}

\begin{figure}[tb] 
\begin{center}
    \includegraphics[width=0.60\linewidth,trim= 1.5cm 3.5cm 2.1cm 1.5cm, clip]{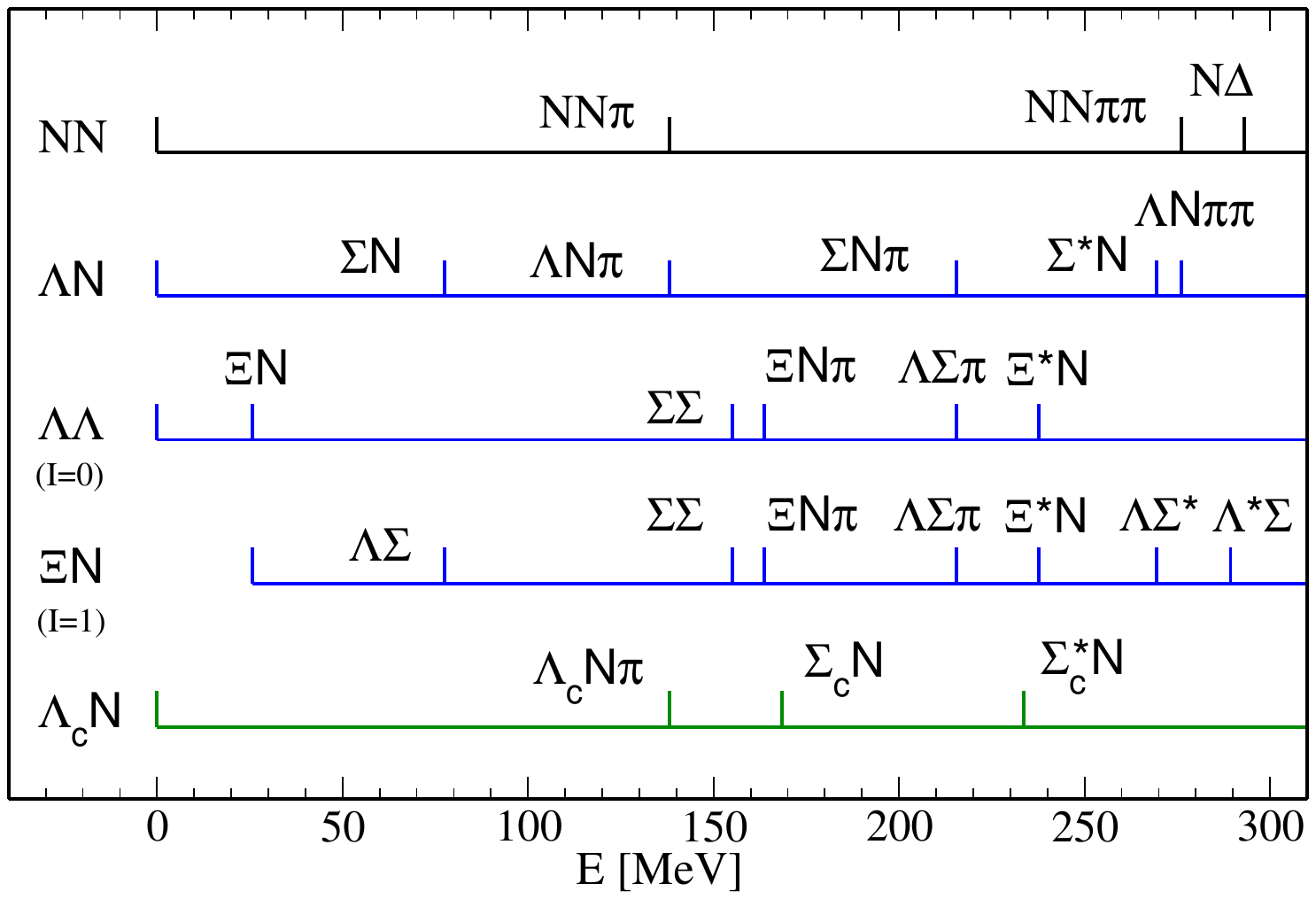}
    \end{center}
    \caption{Thresholds of baryon-baryon systems with strangeness
    $S=0,-1,-2$ and with charm. $E$ is the kinetic energy relative
    to the lowest threshold. $\Si^*$, $\La^*$, $\Xi^*$, and $\Sigma_c^*$
    refer to the $\Si^*$(1385), $\La^*$(1405), $\Xi^*$(1530), and 
    $\Sigma_c^*$(2520) resonances.}
    \label{fig:bb_th}
\end{figure} 

The by far best studied baryon-baryon ($BB$) system, experimentally 
as well as theoretically, is the nucleon-nucleon ($NN$)
system. However, in most theoretical investigations coupled-channel 
effects do not really play an role. It has to do with
the fact that the threshold of the next $BB$ channel ($N\Delta(1232)$) 
is separated by roughly $300\,\MeV$, see Fig.~\ref{fig:bb_th}.
Moreover, there is no coupling between $S$-wave states.
Such a coupling can only occur with the $\Delta\Delta$ channel
whose threshold is another $300\,\MeV$ higher up. 
Indeed, the first channels that opens is that of the three-body system $NN\pi$ and $\pi d$ at about $140\,\MeV$. Near threshold a perturbative treatment of the $\pi$ production is possible \cite{Hanhart:2003pg} because at low momenta the $\pi N$ 
interaction is weak due to constraints from chiral symmetry. However, at intermediate energies one faces the challenge of treating the $NN\pi$ three-body system (and eventually the $NN\pi\pi$ system) and its coupling to $NN$ rigorously \cite{Garcilazo:1990gc, Blankleider:1992pc}. Thus, most of the $NN$ potentials in the literature are designed to describe and reproduce the $NN$ scattering data only up to the pion-production threshold. 
Nonetheless, there is a considerable number of
$NN$ potentials where the $\Delta$ resonance is included
as explicit degree of freedom and where the coupling
to the $N\Delta$ and/or $\Delta\Delta$ systems is 
taken into account, for example in 
Refs.~\cite{Lomon:1981su, vanFaassen:1984jt, Wiringa:1984tg,
Haidenbauer:1993pw}. An extensive overview and references
to the pertinent literature is given in Ref.~\cite{Cattapan:2002rx}.
In some of the potentials established
within the meson-exchange approach $\Delta$'s are only included in 
order to achieve more realistic values for model parameters like the 
meson-baryon coupling constants~\cite{Machleidt:1987hj},
whereas in others the motivation is indeed to describe $NN$ phenomenology
up to and even beyond the thresholds involving the $\Delta$ resonance \cite{Lomon:1981su,vanFaassen:1984jt,Elster:1988zu}. 
As a matter of facts, the influence of the coupling of the $NN$ system
to the $N\Delta$ channel on some $NN$ phase shifts
like the $^1D_2$ or $^3F_3$
has been long established by phase-shift analyses but
also by model calculations.

Also in modern approaches to the $BB$ interaction like chiral effective 
field theory (EFT) the inclusion of the $\Delta$ has been considered
from the very beginning \cite{Ordonez:1995rz}.
Among others, it has been found to improve the 
convergence in peripheral $NN$ phases
\cite{Kaiser:1998wa,Krebs:2007rh}. But like in
some phenomenological $NN$ potentials mentioned above, 
$\Delta$'s are not an active degree of freedom and
the calculations are restricted to energies below 
the $NN\pi$ threshold. Anyway, the dynamics around
the $N\Delta$ threshold and possible coupled-channels effects are 
beyond the validity of the present chiral potentials and cannot 
be addressed. 

The kinematical situation is different for $BB$ systems involving strange 
baryons. In case of strangeness $S=-1$ there are two channels involving
octet baryons, $\La N$ and $\Si N$, while for $S=-2$ there are even four
channels, $\La \La$, $\Xi N$, $\La\Si$ and $\Si\Si$.
In the limit of SU(3) flavor symmetry their thresholds coincide.
SU(3) symmetry is, of course, broken, but even for physical 
masses the separation between the thresholds remains small as can be
seen from the overview provided in Fig.~\ref{fig:bb_th}.
Thus, for a realistic description of the dynamics of $YN$ systems the 
explicit inclusion of the channel coupling is indispensable.  
Indeed, already in the first attempts to establish a potential for the
$S=-1$ sector the coupling between $\La N$ and $\Si N$ was taken
into account~\cite{deSwart:1962}.
An early review on the hyperon-nucleon interaction is provided in Ref.~\cite{DeSwart:1971dr}. An overview of the work of the Nijmegen group can be found in Ref.~\cite{Rijken:2010zzb}, while the tools for describing the $YN$ interaction within chiral EFT, pursued by the Juelich-Bonn group, 
are summarized in Ref.~\cite{Petschauer:2020urh}.

Contrary to the meson-baryon systems discussed above,  
excited hadrons in form of conventional $q\bar q$ or $qqq$ states
cannot occur in $BB$ systems. However, it is possible to find  
exotic hadrons, the so-called dibaryons. Historically, one has tried
to distinguish between genuine (compact) $6\,q$ states, with masses
well below the lowest $BB$ threshold, and bound states/resonances close
to the physical region and generated by the (coupled-channel) dynamics. 
Whether that distinction is meaningful on a practical level is a matter
of believe. In any case, predictions like the $H$-dibaryon (a state 
with $S=-2$, $J=0$, $I=0$) \cite{Jaffe:1976yi} have definitely stimulated 
the interest in the $S=-2$ sector for decades. In recent times there 
has been some excitement about a dibaryon in the $NN$ sector, the
$d^*(2380)$ discovered at the COSY facility in 
Juelich \cite{WASA-at-COSY:2011bjg}, which is associated with a possible
$\Delta\Delta$ bound state (with $J=3$, $I=0$). 
For a detailed discussion of this structure, and 
a more general review of dibaryons, see 
Ref.~\cite{Clement:2016vnl}.
For an alternative explanation of the 
COSY data see Refs.~\cite{Ikeno:2021frl,Molina:2021bwp}.

The next two subsections provide an introduction to the 
theoretical frameworks most commonly applied for constructing the
interactions in the $BB$ system, chiral EFT and meson-exchange. 
In \cref{sec:LNSN} the $\La N$-$\Si N$ system is discussed. An overview
of results for the $\La p$, $\Si^- p$, and $\Si^+p$ channels is given. These 
are the channels which are experimentally reasonably well explored. In 
addition, the situation regarding a possible bound state (dibaryon)
near the $\Si N$ threshold in the $^3S_1$-$^3D_1$ partial wave is
examined. Its appearance and structure is closely related with the strength
of the $\La N$-$\Si N$ channel coupling.
~\cref{jsec:LL} is devoted to the $S=-2$ sector. One of the aspects 
summarized is the present situation regarding the $H$-dibaryon. 
Furthermore the overall experimental situation is discussed and 
results for recent measurements of two-particle momentum correlation
functions are shown. Those allow conclusions on the properties of 
the $\La\La$ and $\Xi^- p$ interaction. 
Finally, in \cref{jsec:Others} the situation in other $BB$ systems 
is briefly reviewed.

\subsection{Baryon-baryon interaction in chiral effective field theory}
\label{subsec:BBEFT}

The application of chiral EFT to the $NN$ system is
thoroughly documented in various reviews, see e.g. 
\cite{Epelbaum:2008ga,Machleidt:2011zz}. Potentials up to
the fifth order in the chiral expansion have been presented
\cite{Reinert:2017usi,Entem:2017gor}. 
Chiral potentials for the $BB$ interaction involving strange baryons are so
far available up to the third order \cite{Haidenbauer:2023qhf}. Details on 
the derivation of the chiral $BB$ potentials for the strangeness sector 
at leading order (LO) using the Weinberg power counting and assuming 
approximate SU(3) flavor symmetry can be found 
in Refs.~\cite{Polinder:2006zh} and \cite{Haidenbauer:2007ra},
and for up to next-to-leading order (NLO) in Ref.~\cite{Haidenbauer:2013oca}.
Here we provide only a brief summary of the essential features. 
 
\begin{figure}[tb]
    \centering
    \begin{tabularx}{\linewidth}{Xl}
        \textbf{\Large LO} &
            \raisebox{-.5\height}{\includegraphics[height=2.5cm]{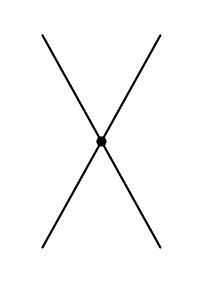}}
            \raisebox{-.5\height}{\includegraphics[height=2.5cm]{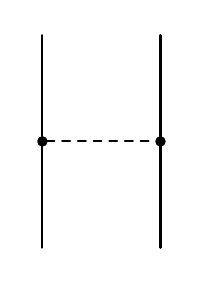}} \\
    \hline
        \textbf{\Large NLO} &
            \raisebox{-.5\height}{\includegraphics[height=2.5cm]{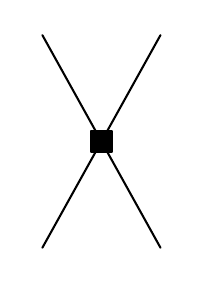}}
            \raisebox{-.5\height}{\includegraphics[height=2.5cm]{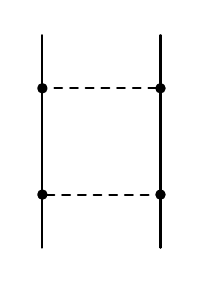}}
            \raisebox{-.5\height}{\includegraphics[height=2.5cm]{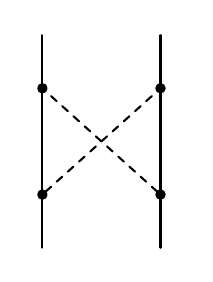}}
            \raisebox{-.5\height}{\includegraphics[height=2.5cm]{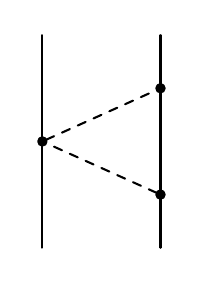}}
            \raisebox{-.5\height}{\includegraphics[height=2.5cm]{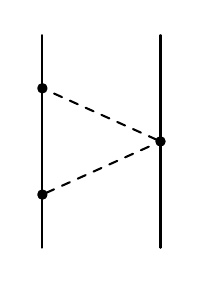}}
            \raisebox{-.5\height}{\includegraphics[height=2.5cm]{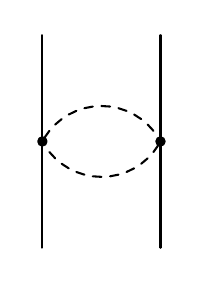}} \\  
    \hline
        \textbf{\Large NNLO} &
            \raisebox{-.5\height}{\includegraphics[height=2.5cm]{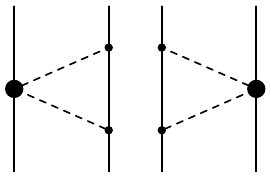}}
            \hspace{1em}
            \raisebox{-.5\height}{\includegraphics[height=2.5cm]{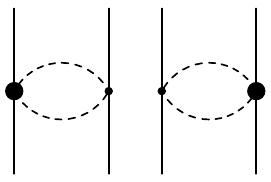}} \\
    \end{tabularx}
    \caption{
        Diagrams contributing at LO (top), NLO (center), and N$^2$LO (bottom) in chiral EFT. Solid and dashed lines denote octet baryons and pseudoscalar mesons, respectively.
    }
    \label{fig:chEFT}
\end{figure}
The LO potential consists of four-baryon contact terms without derivatives 
and of one-pseudoscalar-meson exchanges, see Fig.~\ref{fig:chEFT} (top).
As described in Ref.~\cite{Polinder:2006zh}, imposing SU(3) symmetry there
are six independent contact interactions at LO whose strength can be given in 
terms of six corresponding low-energy coefficients (LECs). Not surprisingly,
the number of terms corresponds to the number of irreducible representations
of SU(3) for the octet baryons: $8 \times 8 = 27 + 10 + 10^* + 8_s + 8_a + 1$
\cite{deSwart:1963pdg}. 
The LECs contribute to the $^1S_0$ and $^3S_1$ partial waves 
(using the standard partial-wave notation $^{2S+1}L_J$ for $BB$ systems)
and need to be determined by a fit to experimental data. 

The lowest order ${\rm SU(3)}_{\rm f}$ invariant pseudoscalar-meson--baryon
interaction Lagrangian with the appropriate symmetries was discussed in~\cite{Polinder:2006zh, Haidenbauer:2013oca}. 
It leads to a one-pseudoscalar-meson-exchange potential similar to the 
static one-pion-exchange potential in \cite{Epelbaum:1998ka} (recoil 
and relativistic corrections give higher order contributions),
\begin{eqnarray}
    V_{B_1B_2\to B'_1B'_2}&=&-f_{B_1B_1'P}f_{B_2B_2'P}\frac{\left(\mbox{\boldmath $\sigma$}_1\cdot{\bf q}\right)\left(\mbox{\boldmath $\sigma$}_2\cdot{\bf q}\right)}{{\bf q}^2+m^2_P}\ ,
    \label{jeq:14}
\end{eqnarray}
where $m_P$ is the mass of the exchanged pseudoscalar meson. The transferred and average momentum, ${\bf q}$ and ${\bf k}$, are defined in terms of the final and initial c.m. momenta of the baryons, ${\bf p}'$ and ${\bf p}$, as ${\bf q}={\bf p}'-{\bf p}$ and ${\bf k}=({\bf p}'+{\bf p})/2$. 
Assuming ${\rm SU(3)}_{\rm f}$ symmetry imposes specific relations between 
the coupling constants $f_{B_iB_j'P}$ so that they can be expressed in terms 
of a coupling constant $f$ and the $F/(F+D)$-ratio $\alpha$
\cite{deSwart:1963pdg}.
Here $f\equiv g_A/2f_0$ where $f_0$ is the pseudoscalar-meson decay constant in the chiral limit, $g_A$ is the axial-vector strength measured in neutron 
$\beta$-decay, and 
$F$ and $D$ are coupling constants which satisfy the relation $F+D=g_A\simeq 1.26$.
In the $YN$ potentials of the Juelich-Bonn group $f_0$ is fixed to
the weak pion decay constant, $f_\pi =  92.4\,\MeV$. 
In the extension to next-to-next-to-leading order
(N$^2$LO) the explicit SU(3) breaking of the decay constant
was taken into account by using the empirical values for $f_\pi$, $f_K$, 
and f$_\eta$ \cite{Haidenbauer:2023qhf}. 
For the $F/(F+D)$-ratio the SU(6) value ($\alpha=0.4$) was adopted in 
all $BB$ studies.

In next-to-leading order (NLO), contributions from (non-iterative) 
two-pseudoscalar-meson exchange diagrams arise  \cite{Epelbaum:2004fk,Haidenbauer:2013oca}. 
Those involve the leading meson-baryon Lagrangian 
which describes a (Weinberg-Tomozawa) vertex between two baryons and two mesons.
Explicit expressions for the resulting NLO potential can be found in the appendix
of Ref.~\cite{Haidenbauer:2013oca}. A graphical representation of the 
contributions is provided in the middle line of Fig.~\ref{fig:chEFT} and
consists of (irreducible) planar box, crossed box, triangle, and 
football diagrams. 
In addition there are four-baryon contact terms with two derivatives, involving
additional LECs. Performing a partial wave projection for the latter and
imposing again ${\rm SU(3)}_{\rm f}$ symmetry one finds that in case of the $YN$ interaction there are ten new coefficients entering the $S$ waves and $S$-$D$ transitions, respectively, and eight coefficients in the $P$-waves
\cite{Haidenbauer:2013oca}.
Finally, at N$^2$LO contributions involving the
subleading meson-baryon Lagrangian \cite{Haidenbauer:2023qhf} arise. 
Those are depicted at the bottom of Fig.~\ref{fig:chEFT}. 

The reaction amplitudes are obtained in a DCC approach by solving a coupled-channel Lippmann-Schwinger (LS) equation for the interaction potentials: 
\begin{eqnarray}
    T_{\rho''\rho'}^{\nu''\nu',J}(p'',p';E)=V_{\rho''\rho'}^{\nu''\nu',J}(p'',p')+
    \sum_{\rho,\nu}\int_0^\infty \frac{dpp^2}{(2\pi)^3} \, V_{\rho''\rho}^{\nu''\nu,J}(p'',p)
    \frac{2\mu_{\nu}}{q_{\nu}^2-p^2+i\eta}T_{\rho\rho'}^{\nu\nu',J}(p,p';E)\ .
    \label{jeq:LS}
\end{eqnarray}
The label $\nu$ indicates the particle channels and the label $\rho$ the partial wave. The pertinent reduced mass is denoted by $\mu_\nu$. The on-shell momentum in the intermediate state, $q_{\nu}$, is defined by 
$E=E_{B_{1,\nu}} + E_{B_{2,\nu}} =
\sqrt{M^2_{B_{1,\nu}}+q_{\nu}^2}+\sqrt{M^2_{B_{2,\nu}}+q_{\nu}^2}$. Relativistic kinematics is used for relating the laboratory energy $T_{{\rm lab}}$ of the hyperons to the c.m. momentum. In principle, one could
also use a relativistic propagator like in Eq.~(\ref{eq:lse}), which in 
fact has been done in the first $YN$ potential developed in
Juelich~\cite{Holzenkamp:1989tq}. 
However, then one can no longer straightforwardly apply the potentials in
standard few-body calculations of hypernuclei 
\cite{Le:2021gxa,Le:2023bfj,Le:2024rkd,Haidenbauer:2025zrr}
or in nuclear matter studies \cite{Haidenbauer:2014uua, Petschauer:2015nea,Jinno:2025vgm}, which all require non-relativistic kinematics. 

The LS equation is solved in the particle basis, in order to incorporate the correct physical thresholds. Depending on the specific values of strangeness and charge up to six baryon-baryon
channels can couple. For the $S=-1$ sector where a comparison with scattering data is possible the Coulomb interaction is taken into account appropriately. 
The potentials in the LS 
equation are cut off with a regulator function, $\exp\left[-\left(p'^4+p^4\right)/\Lambda^4\right]$, 
in order to remove high-energy components of the baryon and pseudoscalar meson fields \cite{Epelbaum:2004fk}.
Cut-off values in the range $500, ..., 650\,\MeV$ are considered, similar to 
what was used for chiral $NN$ potentials \cite{Epelbaum:2004fk}.
In the extension to N$^2$LO \cite{Haidenbauer:2023qhf} a novel
scheme is used, called semilocal momentum-space (SMS) regularization
\cite{Reinert:2017usi}. 

Alternative chiral EFT approaches to the $BB$ interaction involving 
strange baryons have been pursued in Ref.~\cite{Korpa:2001au},
where the so-called Kaplan-Savage-Weise (KSW) resummation 
scheme \cite{Kaplan:1998we} has been applied, 
and in Refs.~\cite{Li:2016paq,Ren:2019qow,Song:2021yab} where calculations
are performed in covariant chiral EFT, but so far only at LO. 

\subsection{Baryon-baryon interaction in the meson-exchange picture} 
\label{subsec:BBmex}

For a long time investigations of the $BB$ interactions have been done primarily within phenomenological meson-exchange
potentials such as the Juelich \cite{Holzenkamp:1989tq,Reuber:1993ip,Haidenbauer:2005zh},
Nijmegen \cite{Rijken:1998yy,Rijken:2010zzb,Nagels:2015lfa}, 
or Ehime \cite{Tominaga:1998iy,Tominaga:2001ra} potentials.
Below we give a brief introduction to these type of models.

Conventional meson-exchange models of the $YN$ interaction are 
usually also based on the assumption of SU(3) flavor symmetry for the  occurring coupling constants, and in some cases even on the SU(6) symmetry of the quark model, see Refs.~\cite{Holzenkamp:1989tq,Reuber:1993ip}. 
In the derivation of the meson-exchange contributions one follows 
essentially the same procedure as outlined in the previous subsection,   
for the case of pseudoscalar mesons. 
Besides the lowest pseudoscalar-meson multiplet also the 
exchanges of vector mesons ($\rho$, $\omega$, $K^*$), of scalar mesons 
($\sigma$ ($f_0(500)$), ...), or even of axial-vector mesons ($a_1(1260)$, ...)~\cite{Rijken:2010zzb, Nagels:2015lfa} are included. 
The spin-space structure of the corresponding meson-baryon Lagrangians differ from 
that for pseudo-scalar mesons and, accordingly, the final 
expressions for the corresponding contributions to the $YN$ interaction potentials differ as well. Details can be found in Refs.~\cite{Holzenkamp:1989tq, Rijken:1998yy, Rijken:2010zzb}.
We want to emphasize that even for pseudoscalar mesons the final result for the interaction potentials differs, in general, from the expression given in Eq.~(\ref{jeq:14}). Contrary to the chiral EFT
approach, recoil and relativistic corrections are often kept in meson-exchange models because no power counting rules are applied. Moreover, in case of the Juelich potential pseudoscalar coupling is assumed for the meson-baryon interaction Lagrangian for the pseudoscalar mesons instead of the  pseudovector coupling
dictated by chiral symmetry.
Note that in some $YN$ potentials of the Juelich group \cite{Holzenkamp:1989tq,Reuber:1993ip}
contributions from two-meson exchanges are included. The ESC08 and ESC16 potentials 
\cite{Rijken:2010zzb,Nagels:2015lfa} include likewise contributions from two-meson exchange, 
in particular, so-called meson-pair diagrams analog to the last three diagrams shown in the center of
Fig.~\ref{fig:chEFT}.
Moreover, there are some $YN$ potentials which include the coupling
to channels with decuplet baryons
($\Si^*N$, $\La \Delta$, $\Si \Delta$) \cite{Holzenkamp:1989tq, Reuber:1993ip,Greenberg:1992fp}. 

The major conceptual difference between the various meson-exchange
models consists in the treatment of the scalar-meson sector. This simply reflects the fact that, unlike for pseudoscalar and vector mesons, 
there is no general agreement about the internal structure of the light 
scalar mesons, see the review \cite{Scalar:2025} in 
Ref.~\cite{ParticleDataGroup:2024cfk}, 
so that the concrete application of SU(3) relations remains ambiguous. 
Therefore, besides the question of the actual masses of the exchange
particles it remains unclear how the relations for the coupling constants should be specified. As a consequence, different prescriptions for describing the scalar sector, whose contributions play a crucial role in any $BB$ interaction at intermediate ranges, were adopted by the various authors who published meson-exchange models of the
$YN$ interaction.
For example, the Nijmegen group views this interaction as being
generated by genuine scalar-meson exchange. In their models NSC97
\cite{Rijken:1998yy} and ESC08 (ESC16) \cite{Rijken:2010zzb,Nagels:2015lfa}
a scalar SU(3) nonet is exchanged - namely, two 
isospin-$0$ mesons (an $\epsilon$(760) and the $f_0(980)$) 
an isospin-$1$ meson ($a_0(980)$) and an isospin-1/2 strange meson
$\kappa$ with a mass of $1000\,\MeV$. 
In the initial $YN$ models of the Juelich group \cite{Holzenkamp:1989tq,Reuber:1993ip}
a $\sigma$ meson with a mass of $\approx 550\,\MeV$
is included which is viewed as arising 
from correlated $\pi\pi$ exchange. In practice, however, the coupling strength of this fictitious $\sigma$ to the baryons is treated as a free parameter and fitted to the data. In the latest meson-exchange $YN$ potential presented by the Juelich group \cite{Haidenbauer:2005zh} a microscopic model of correlated $\pi\pi$ and $K\bar K$ exchange \cite{Reuber:1995vc} is utilized to fix the contributions in the scalar-isoscalar ($\sigma$) and vector-isovector ($\rho$) channels.

Let us mention for completeness that meson-exchange models are typically
equipped with phenomenological form factors in order to cut off the potential
for large momenta (short distances). 
For example, in case of the $YN$ models of the Juelich group the
interaction is supplemented with form factors for each meson-baryon-baryon 
vertex, c.f. Refs.~\cite{Holzenkamp:1989tq,Reuber:1993ip} for details. 
Those form factors are meant to take into account the extended
hadron structure and are parametrized in the conventional monopole or dipole
form. In the case of the Nijmegen potentials, a Gaussian form factor is used. In addition, there is some intricate short-range phenomenology that controls the interaction at short distances \cite{Rijken:2010zzb,Nagels:2015lfa}. 
Finally, we note that realistic models for the $YN$ interaction have been
also established within the constituent quark model
\cite{Fujiwara:2006yh}. 

\begin{figure}[tb]
    \centering
    \includegraphics[width=0.32\linewidth]{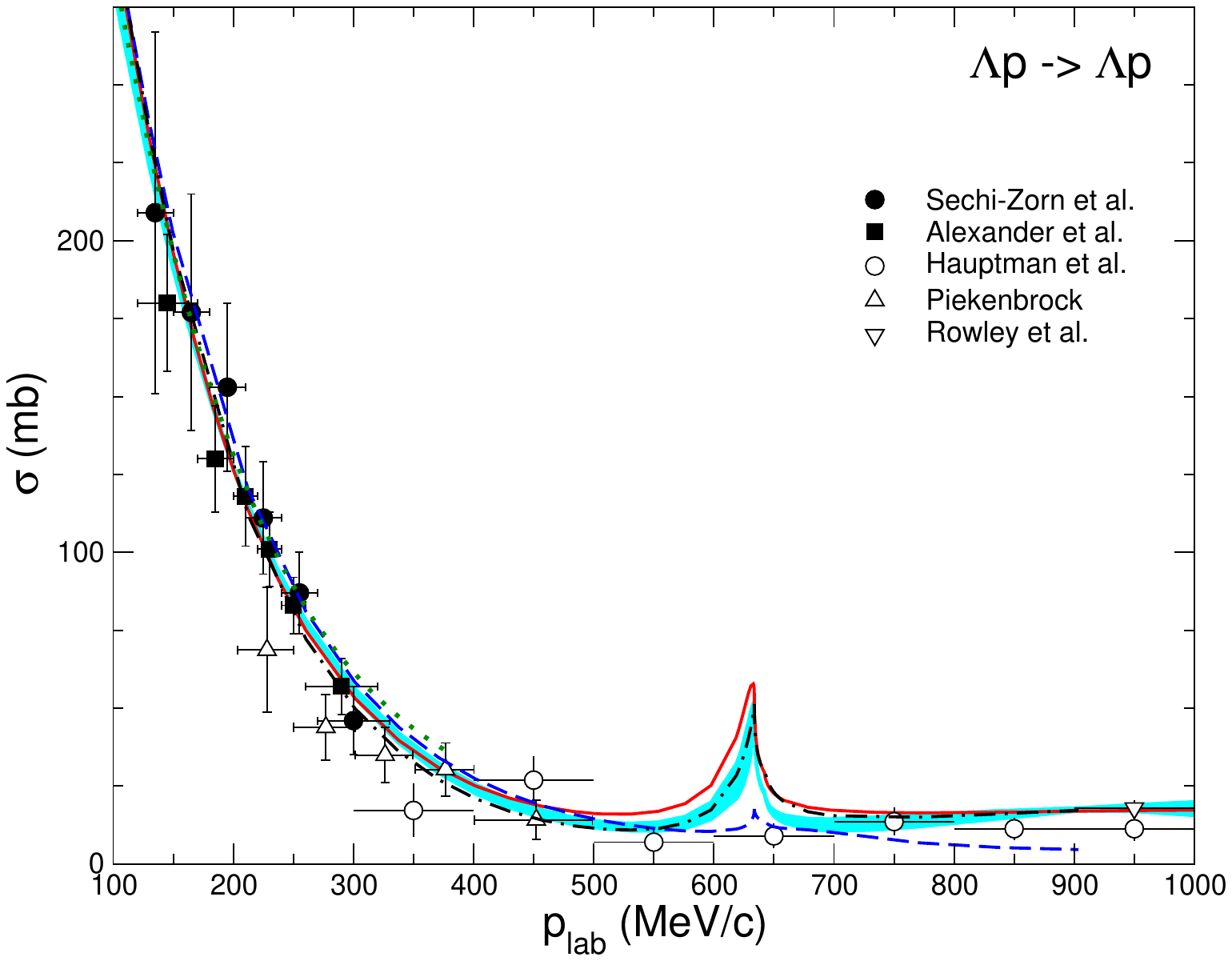}
    \includegraphics[width=0.32\linewidth]{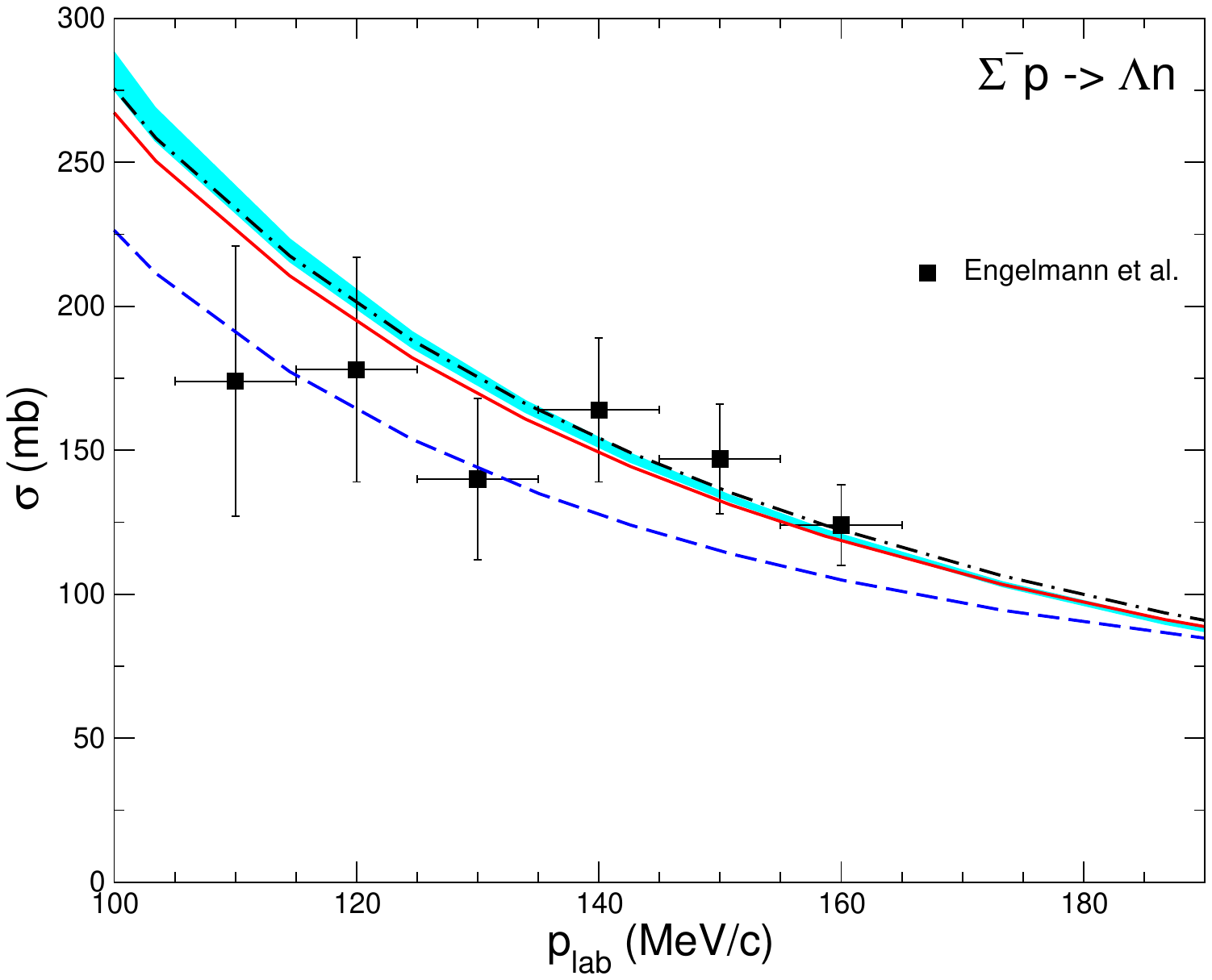}
    \includegraphics[width=0.32\linewidth]{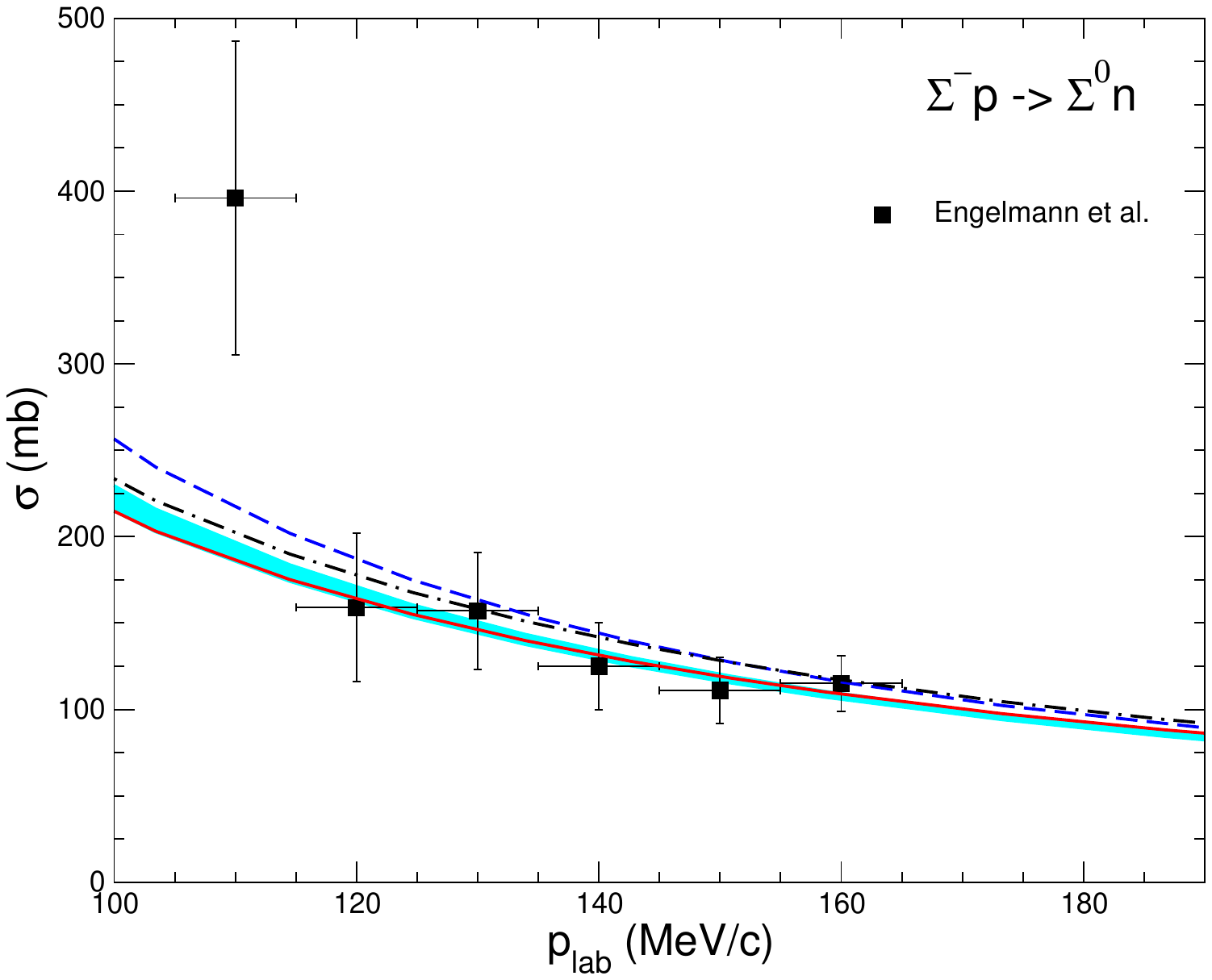}
    \includegraphics[width=0.32\linewidth]{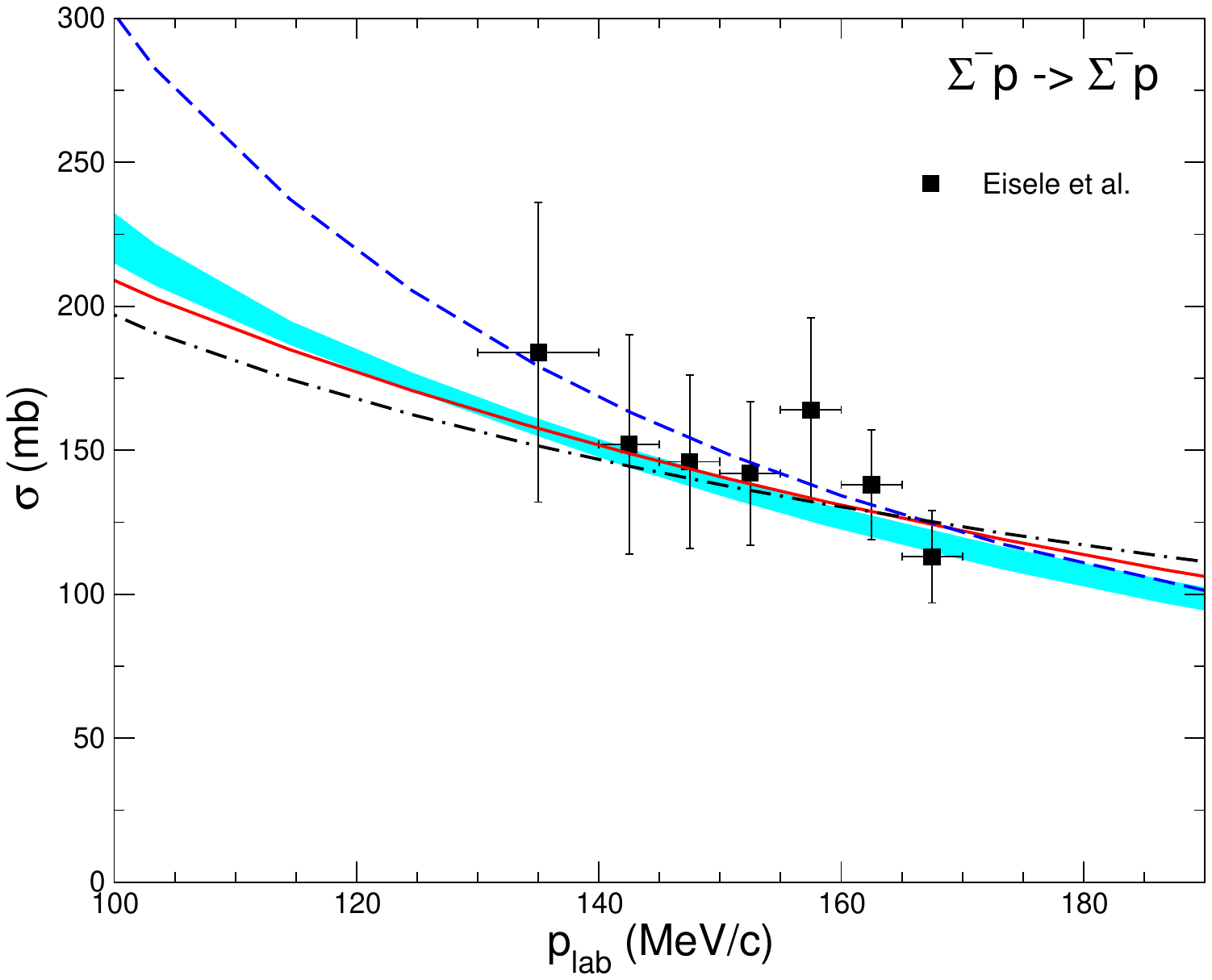}
    \includegraphics[width=0.32\linewidth]{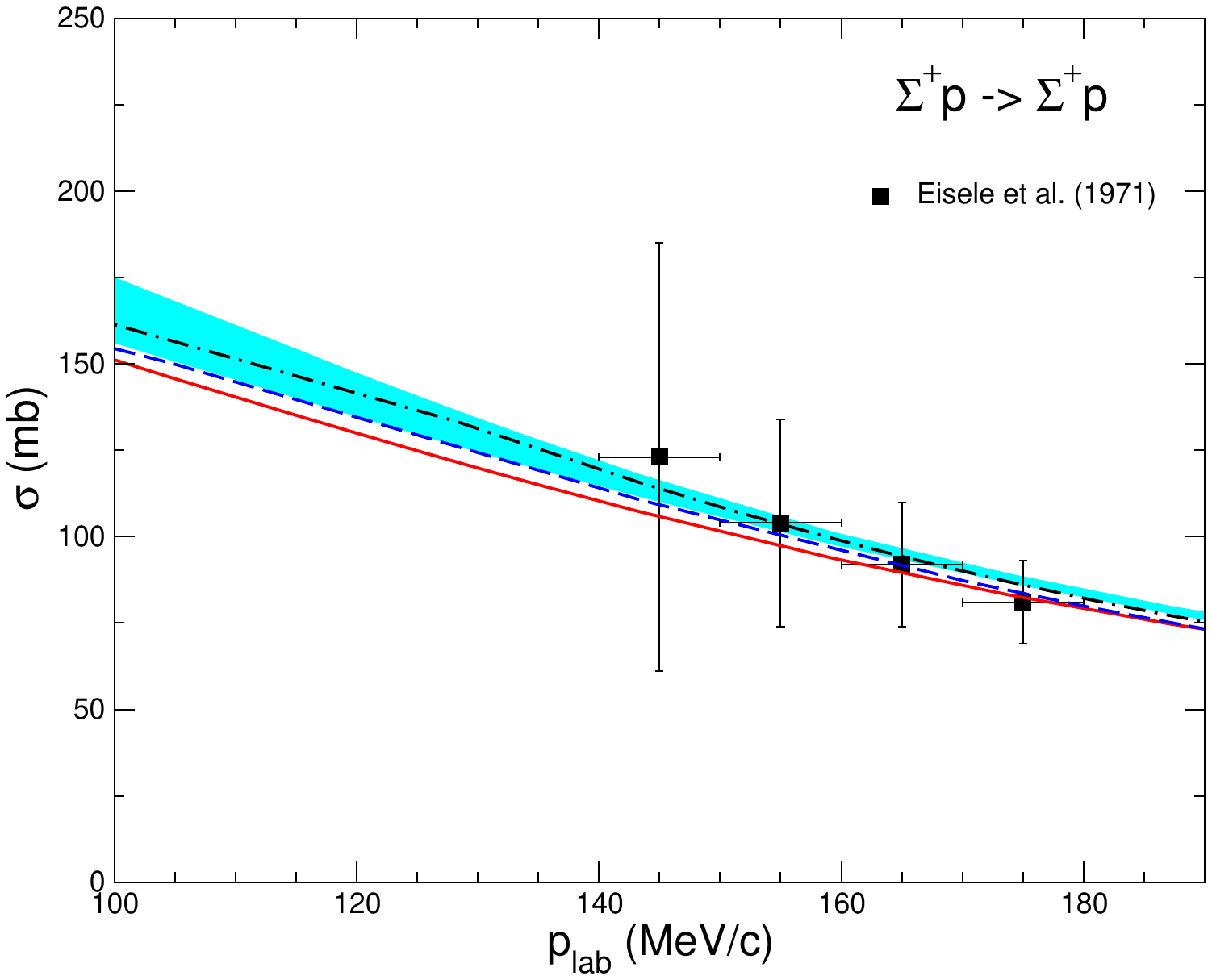}
    \vskip -0.5cm 
    \caption{$\La p$, $\Si^- p$, and $\Si^+p$ cross sections. Results are shown for the chiral $YN$ potentials NLO19 \cite{Haidenbauer:2019boi} (cyan band) and SMS N$^2$LO (550)
    \cite{Haidenbauer:2023qhf} (red line). In 
    addition cross, sections of meson-exchange potentials by the 
    Juelich group (Juelich '04 \cite{Haidenbauer:2005zh}, 
    dashed blue line) and the Nijmegen group 
    (NSC97f \cite{Rijken:1998yy}, dash-dotted black line) 
    are included.
    }
    \label{fig:YNcross}
\end{figure} 

\subsection{The \texorpdfstring{$\La N$-$\Si N$}{LambdaN-SigmaK} system}
\label{sec:LNSN}

Let us first focus on two channels of the $\La N$-$\Si N$ 
coupled-channel system, namely on $\La p$, on the one hand,
which couples to the $\Si^+ n$ and $\Si^0 p$ channels,
and $\Si^- p$, which couples to $\La n$ and $\Si^0 n$.
This choice is dictated by the fact that for
those systems scattering data for low momenta are
available. Indeed, in case of $\Si^- p$ besides the cross section
for elastic scattering also the transition cross sections to
the other two channels have been established. It should
be noted that there are data for $\La p \to \Si^0 p$ too
\cite{Kadyk:1971tc},
which are, however, incompatible with the measurements
for $\Si^- p \to \La n$ when considering isospin symmetry and the 
principle of detailed balance. Moreover, there are data on the $\Si^0 p$
two-particle momentum correlation function
\cite{ALICE:2019buq} which, however, suffer from low statistics. 
There are also data on $\Si^+ p$, which, however, is an
isospin $I=3/2$ state and does not couple to $\La N$.
Thus, we do not discuss the pertinent results here and
refer the interested reader to the original works
\cite{Rijken:1998yy,Haidenbauer:2013oca, Haidenbauer:2019boi,Haidenbauer:2023qhf}.
Nonetheless, it should be emphasized that the interaction
in the $I=3/2$ channel is not disconnected from the results
for the other ones. Because of the underlying isospin symmetry 
the potentials for $\Si^- p$, $\Si^0 p$, etc. are combinations of the $I=1/2$ and $I=3/2$ components. 

Cross sections are presented in Fig.~\ref{fig:YNcross}
for a selection of $YN$ potentials. In case of the chiral potentials
from 2019~\cite{Haidenbauer:2019boi}
bands are shown which reflect the residual cutoff dependence, i.e.,
the variation of the results for $\La = 500,...,650\,\MeV$
\cite{Haidenbauer:2019boi}. 
For completeness, we list also results for the
effective range parameters in Table~\ref{tab:ereF}.
As one can see from the figure 
the data at low energy are rather well reproduced,
though one should certainly keep in mind that those
data are the ones included in the fitting procedure
to determine the free parameters of the interaction. 
The channel coupling, i.e., its effect on $\La p$, is
easily recognizable by the large cusp-like structure
at the $\Si N$ threshold which is predicted by all 
considered potentials and, actually, in all 
coupled-channel treatments which provide a realistic
description of the $\La p$ and $\Si^- p$ data 
including the $\Si^- p$ capture ratio at rest. 
Unfortunately, the few $\La p$ data points available in the pertinent momentum region suffer from low statics
and poor momentum resolution and do not allow any
conclusions. 

\begin{table}[b!]
    \caption{Scattering lengths ($a$) and effective ranges ($r$) for singlet (s) and triplet (t) $S$-waves (in fm), for $\Lambda N$, $\Si N$ with isospin $I=1/2,\, 3/2$. The results are for the chiral $YN$ 
    potentials NLO13~\cite{Haidenbauer:2013oca}, 
    NLO19~\cite{Haidenbauer:2019boi}, and SMS NLO (500) and N$^2$LO (550)
    \cite{Haidenbauer:2023qhf}. Results from the meson-exchange potentials 
    by the Juelich group (Juelich '04 \cite{Haidenbauer:2005zh}), 
    and the Nijmegen group (NSC97f \cite{Rijken:1998yy})
    are included too.
      }
     \label{tab:ereF}
    \begin{center}
    \renewcommand{\arraystretch}{1.45}
    \begin{tabular}{l||l|l||l|l||l||l}
    \hline\hline
    & {SMS NLO} & {SMS N$^2$LO} & NLO13 & NLO19 & Juelich '04 & Nijmegen \\
    \hline
    ${\Lambda}$ [MeV] 
    & 550     & 550      & 600     & 600 & & NSC97f \\
    \hline
    \hline
    $a^{\La N}_s$ 
   & $-2.79$  & $-2.79$  & $-2.91$ & $-2.91$ &-2.56 & -2.60 \\
    $r^{\La N}_s$ 
   & $ 2.72$  & $ 2.89$   & $ 2.78$ & $ 2.78$  &2.74 &3.05\\
    \hline
    $a^{\La N}_t$ 
  & $-1.57$  & $-1.58$  & $-1.54$ & $-1.41$ &-1.67&-1.72\\
    $r^{\La N}_t$ 
   & $ 2.99$  & $ 3.09$  & $ 2.72$  & $ 2.53$ &2.93&3.32\\
    \hline
    \hline
    Re\,$a^{\Si N \ (I=1/2)}_{s}$
    & $ 1.15$  & $ 1.12$  & $ 0.90$ & $ 0.90$ &0.90&1.16\\
    Im\,$a^{\Si N}_{s}$ 
 & $ 0.00$  & $ 0.00$   & $ 0.00$ & $ 0.00$ &-0.13&0.00\\
    \hline
    Re\,$a^{\Si N \ (I=1/2)}_{t}$
 & $ 2.42$ &  $ 2.38$   & $ 2.27$ & $ 2.29$  &-3.83&1.68\\
    Im\,$a^{\Si N}_{t}$ 
    & $-2.95$  & $-3.26$   & $-3.29$ & $-3.39$  &-3.01&-2.35\\
    \hline
    $a^{\Si N \ (I=3/2)}_s$
   & $-4.05$  & $-4.19$   & $-4.45$ & $-4.55$  &-4.71& -6.11\\
    $r^{\Si N}_s$ 
    & $ 3.89$  & $ 3.89$   & $ 3.68$ & $ 3.65$  &3.31& 3.24\\
    \hline
    $a^{\Si N \ (I=3/2)}_t$
& $ 0.47$  & $ 0.44$   & $ 0.44$ & $ 0.43$  &0.29&-0.29\\
    $r^{\Si N}_t$ 
    & $-4.74$  & $-4.96$   & $-4.59$ & $-5.27$  &-11.6&-16.7\\
    \hline\hline
    \end{tabular}
    \end{center}
\end{table}

\begin{figure}[tb] 
    \begin{center}
    \raisebox{-.5\height}{
        \includegraphics[width=0.32\linewidth]{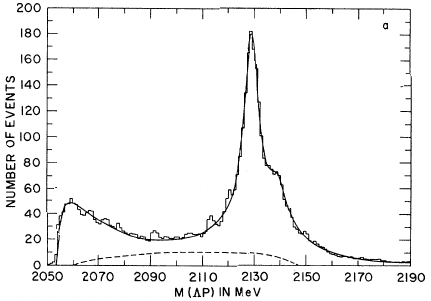}
        \includegraphics[width=0.32\linewidth]{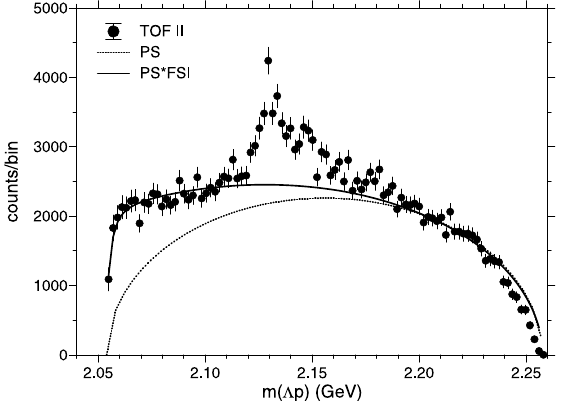}}
    \raisebox{-.5\height}{
        \includegraphics[width=0.32\linewidth]{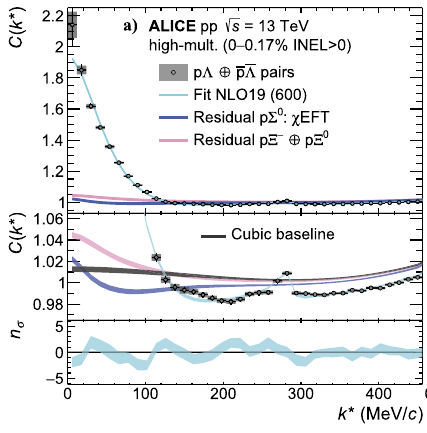}
        }
    \end{center}
    \vskip -0.5cm 
    \caption{Cusp-like structure at the $\Si N$ threshold
    seen in the invariant-mass spectra of the reactions 
    $K^- d\to \pi^- \La p$ \cite{Tan:1969jq} (left),
    $pp \to K^+ \La p$ \cite{COSY-TOF:2012inb} (middle),
    and in the $\La p$ two-particle momentum correlation
    function measured in high energy $pp$ collisions
    \cite{ALICE:2021njx} (right).
    The results on the right-hand side are shown as function of 
    the $\La p$ center-of-mass momentum $k^*$. 
    }
    \label{fig:LNcusp}
\end{figure}

Nonetheless, there is solid information on the existence
of a threshold effect from measurements of the $\La p$
invariant-mass spectrum in reactions like 
$K^- d\to \pi^- \La p$, $\pi^+ d \to K^+ \La p$, or
$pp \to K^+ \La p$, but also from
studies of the $\La p$ two-particle momentum correlation
function in high energy $pp$ collisions. Indeed, the first evidence for a cusp-like structure was already reported as early as 1961 \cite{Dahl:1961zzb}, but the first convincing signal and still one of the most prominent examples is from the measurement of the reaction $K^- d \to \pi^- \La p$ by Tan in 1969~\cite{Tan:1969jq}. 
A review of other early observations can be found in 
Ref.~\cite{Dalitz:1979qv} and an overview of later measurements
is provided by Machner et al.~\cite{Machner:2013hs}. 
More recent examples for the presence of a $\Si N$ threshold effect, in the reaction 
$pp \to K^+ \La p$, can be found in 
Refs.~\cite{COSY-TOF:2012inb,COSY-TOF:2013uqx,COSY-TOF:2016qxd}. 
Very recently evidence of the threshold structure has been also observed 
in measurements of  the $\La p$ correlation function in $pp$ collisions at 
$\sqrt{s}=13$~GeV by the 
ALICE Collaboration \cite{ALICE:2021njx}. For illustration we
present a selection of measurements in Fig.~\ref{fig:LNcusp}.
Note that there are plans at J-PARC for performing a dedicated 
experiment to pin down the threshold cusp more quantitatively
\cite{Ichikawa:2022pjm}. 

\begin{figure}[tb] 
    \begin{center}
    \includegraphics[width=0.32\linewidth]{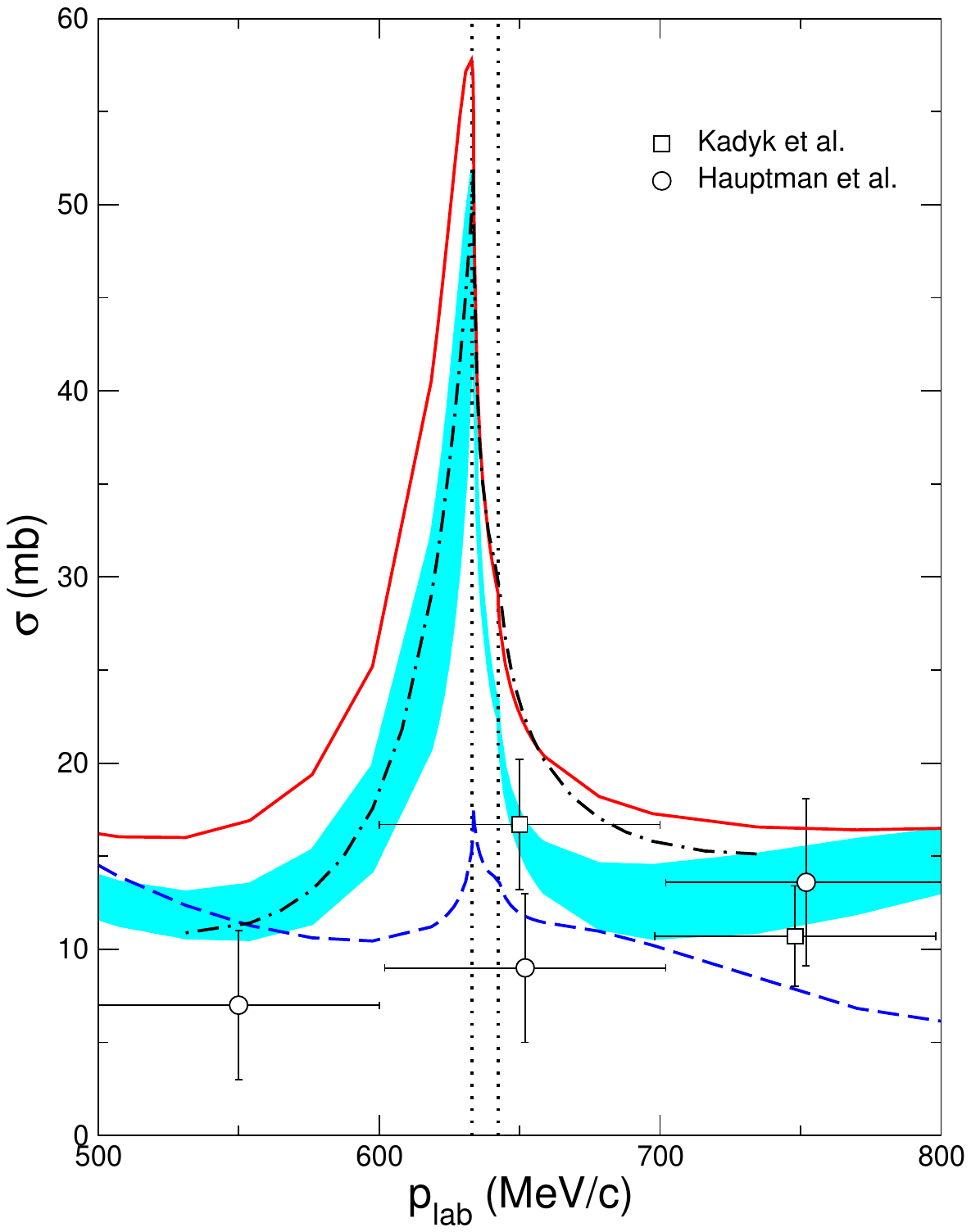}
    \includegraphics[width=0.32\linewidth]{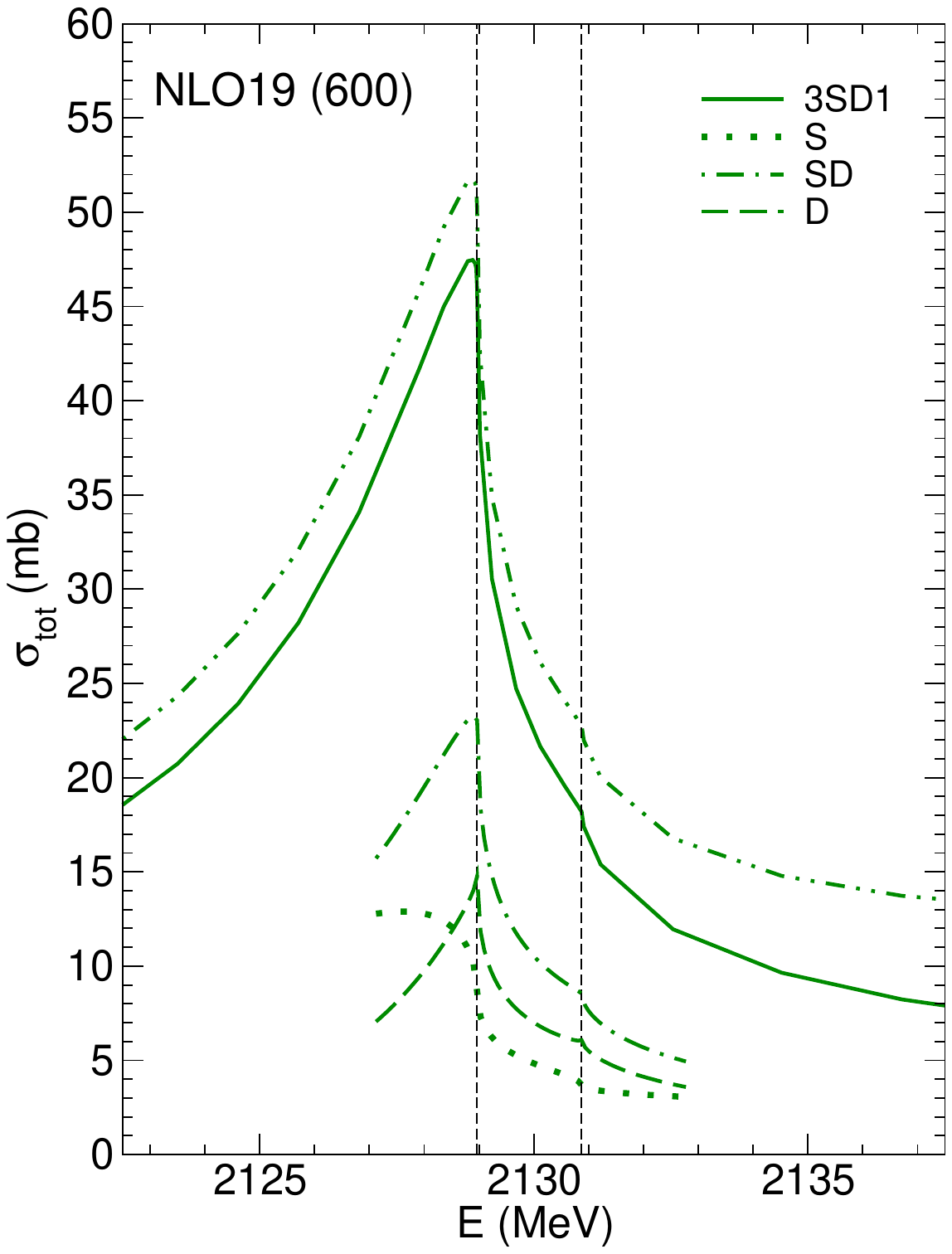}
    \includegraphics[width=0.32\linewidth]{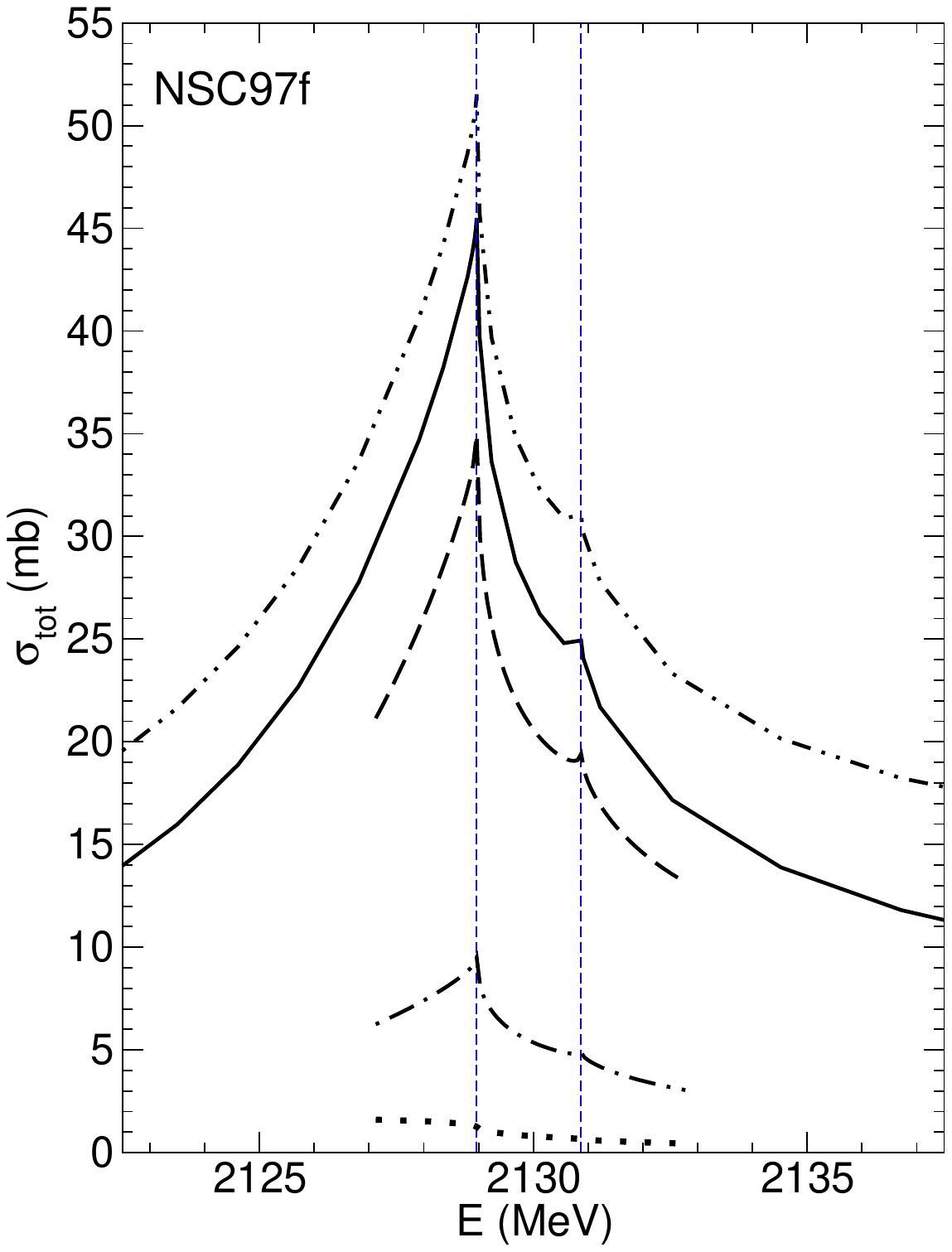}
    \end{center}
    \vskip -0.5cm 
    \caption{A detailed view on the cusp-like structure at the $\Si N$ threshold. On the left the $\La p$ cross section for selected $YN$ potentials is shown as a function of $p_{lab}$. Same description of curves as in 
    Fig.~\ref{fig:YNcross}. 
    In the other two graphs a break down of individual partial waves for 
    the chiral $YN$ potential NLO19 with cutoff 
    $\La = 600\,\MeV$ and the Nijmegen NSC97f potential 
    is presented as a function of the total
    energy. The dashed-doubled-dotted lines are the full results,
    including the $^1S_0$ and $P$- and $D$-waves. 
    }
    \label{fig:LNcusp2}
\end{figure} 

As already mentioned in the introduction to this section,
the interesting question is, of course, whether the observed threshold effect is a signal for a dibaryon - in form of a deuteron-like $\Si N$ bound state -  or whether it is simply a regular cusp that is enhanced
due to the presence of a near-by virtual state. 
In the latter case there would be a sharp peak 
directly at the $\Si N$ threshold while in the former ideally one expects to see a peak below (and well separated from) the threshold.
However, there are two issues that complicate any 
calculation/interpretation. First, while there are only
two channels in isospin space there are three in the
physical world and, thus, there are two $\Si N$ thresholds, 
namely those of $\Si^+ n$ and $\Si^0 p$. 
Second, and more important, because of the tensor
coupling in the $^3S_1$-$^3D_1$ partial wave there
are three amplitudes that are responsible for 
generating the threshold effect, namely the $S$- and
$D$ wave and also the $S\leftrightarrow D$ transition.
The two aspects are illustrated in 
Fig.~\ref{fig:LNcusp2}. 
On the left hand side is a close-up of the $\La p$
cross section around the $\Si N$ thresholds predicted
by the considered potentials. One can see that in all
cases the peak of the produced cusp-like structure 
coincides with the lower ($\Si^+ n$) threshold. At 
the $\Si^0 p$ threshold there is only a kink. 
On the right hand side one can see the different
partial-wave contributions to the $\La p$ cross section
around the $\Si N$ thresholds. 

Anyway, given the experimental situation, in the past the possibility 
of a dibaryon was discussed predominantly in the context of the 
reaction $K^- d \to \pi^- \La p$ owing to the 
prominent experimental signal 
\cite{Dosch:1977at,Dosch:1979hf,Toker:1979pnx,Toker:1981zh,Dalitz:1979qv,Dalitz:1983cu,Torres:1986mr,Deloff:1989uj}. While initial investigations were more or less inconclusive, in the latest works~\cite{Toker:1981zh, Torres:1986mr,Deloff:1989uj} the unanimous conclusion has been drawn 
that a $\Si N$ bound state does not exist near the $\Si N$ threshold. 
However, one must keep in mind that in those studies simplified models of 
the $YN$ interaction were employed. Specifically, with regard to the $^3S_1$-$^3D_1$ partial wave where a possible dibaryon occurs, the tensor coupling mediated by the long-ranged one-pion exchange was ignored and usually only the $S$-wave component was taken into account. Realistic $YN$ potentials suggest that the $\La N \to \Si N$ transition occurs predominantly from the $\La N$ $^3D_1$ state~\cite{Haidenbauer:2013oca, Haidenbauer:2019boi,Rijken:1998yy, Polinder:2006zh, Nagels:1976xq, Nagels:1978sc,Holzenkamp:1989tq, Reuber:1993ip, Haidenbauer:2005zh}.
Furthermore, in those early works often no constraints from SU(3) flavor symmetry were implemented. Though SU(3) symmetry is certainly broken, presumably on the level of 
20 - 30\,\% \cite{Dover:1991sh}, it is likely to end up with unrealistic
results if one ignores it altogether.

In Ref.~\cite{Machner:2013hs} a systematic but purely 
phenomenological attempt was made to determine the
position of the peak. In that work a combined analysis
of the structure as seen in the older data from
$K^- d \to \pi^- \La p$ with the new measurements of
$pp \to K^+ \La p$ was performed. For that as tools a
Breit-Wigner parameterization as well as a Flatt\'e
approach was employed to determine the position of the
peak. Values that agree quite well for the various 
measurements were found and the reported mean value
amounts to $2128.7\pm 0.3\,\MeV$, which pretty much
coincides with the threshold energy of the $\Si^+n$
channel. The authors concluded that the structure
should be interpreted as a genuine cusp, signaling an
inelastic virtual state in the $^3S_1$-$^3D_1$ partial
wave of the $\Si N$ $I=1/2$ channel.

An entirely different strategy was followed in the
works of \cite{Miyagawa:1999zz} and \cite{Haidenbauer:2021smk}.
Those studies focused on the $\La N$-$\Si N$ system itself and
explored directly the pole structure of the coupled-channel
amplitude in the $^3S_1$-$^3D_1$ partial wave in the region of 
the $\Si N$ threshold, see also the more educational analysis that 
has been presented in Ref.~\cite{Pearce:1988rk}. 
Though there is no useful empirical information on the $\La p$ 
cross section around the $\Si N$ threshold, as mentioned above, in 
principle the threshold region is well constrained by cross section data 
on the $\Si^-p$ elastic and inelastic channels. 
In addition the so-called capture ratio at rest has been determined with 
high precision \cite{Hepp:1968zza,Stephen:1970}. 
The latter is defined by \cite{deSwart:1962}
\begin{eqnarray}
    \nonumber
    r_R
      =\frac{1}{4}\,\frac{\sigma_s(\Sigma^-p\rightarrow\Sigma^0n)}
                    {\sigma_s(\Sigma^-p\rightarrow\Lambda n)
                    +\sigma_s(\Sigma^-p\rightarrow\Sigma^0n)} 
      +\frac{3}{4}\,\frac{\sigma_t(\Sigma^-p\rightarrow\Sigma^0n)}
                    {\sigma_t(\Sigma^-p\rightarrow\Lambda n)
                    +\sigma_t(\Sigma^-p\rightarrow\Sigma^0n)}\ ,
    \label{rR}
\end{eqnarray}
where $\sigma_s$ ($\sigma_t$) is the total reaction cross section in the singlet $^1S_0$ (triplet $^3S_1$-$^3D_1$) partial wave. The cross sections are the ones at zero momentum, but in calculations it is common practice \cite{Rijken:1998yy}  to evaluate the cross sections at a small non-zero momentum, namely $p_{\rm lab}=10\,\MeV/c$.
The essential question is, of course, in how far those data pin down the 
coupled-channel amplitude and, in turn, allow one to determine the location
of the poles reliably.

In Ref.~\cite{Miyagawa:1999zz} a variety of meson-exchange potentials
of the Nijmegen group has been explored, NSC97f \cite{Rijken:1998yy} being
the most modern one at that time. The poles were searched in the complex 
planes of the relative momenta in the $\La N$ and $\Si N$ channels, 
$q_{\La N}$ and $q_{\Si N}$, respectively. It turned
out that for two of the considered potentials the poles are on the 
second quadrant of the $q_{\Si N}$ plane while for the other two they are 
on the third quadrant.
(For details on various conventions used for the classification of the 
pole structure we refer to \cite{Badalian:1981xj,Pearce:1988rk}.)
In the first case this should indicate an unstable (dibaryon-like) bound
state while the second implies an inelastic virtual state. The resulting
shape of the $\La N$ cross section around the $\Si N$ threshold is also
illustrated in \cite{Miyagawa:1999zz} and should consist of a round peak,
similar to a Breit-Wigner form, below the threshold in the first case,
and a sharp peak (cusp) at the threshold in the second case, according
to text-book knowledge \cite{Newton:1966}. However, as is
obvious from the concrete results reported in Ref.~\cite{Miyagawa:1999zz}
for the considered potentials, not all show the idealized shape. In 
particular, for NSC97f that produces an unstable bound state, in the nomenclature of Ref.~\cite{Badalian:1981xj}, 
the resulting shape is ambiguous \cite{Miyagawa:1999zz}, see also Fig.~\ref{fig:LNcusp2}. 

\begin{figure}[tb] 
    \begin{center}
    \includegraphics[width=0.32\linewidth]{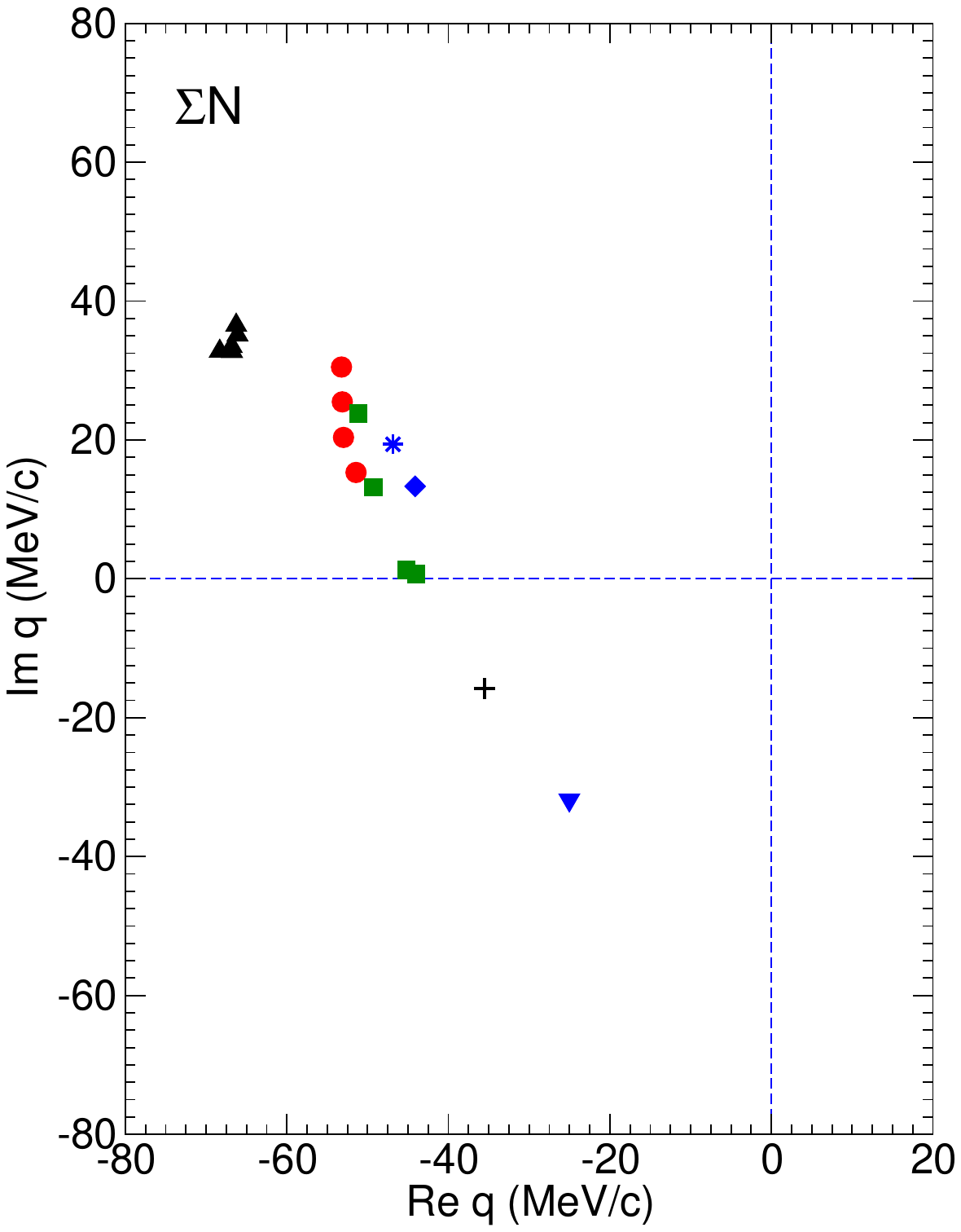}
    \includegraphics[width=0.32\linewidth]{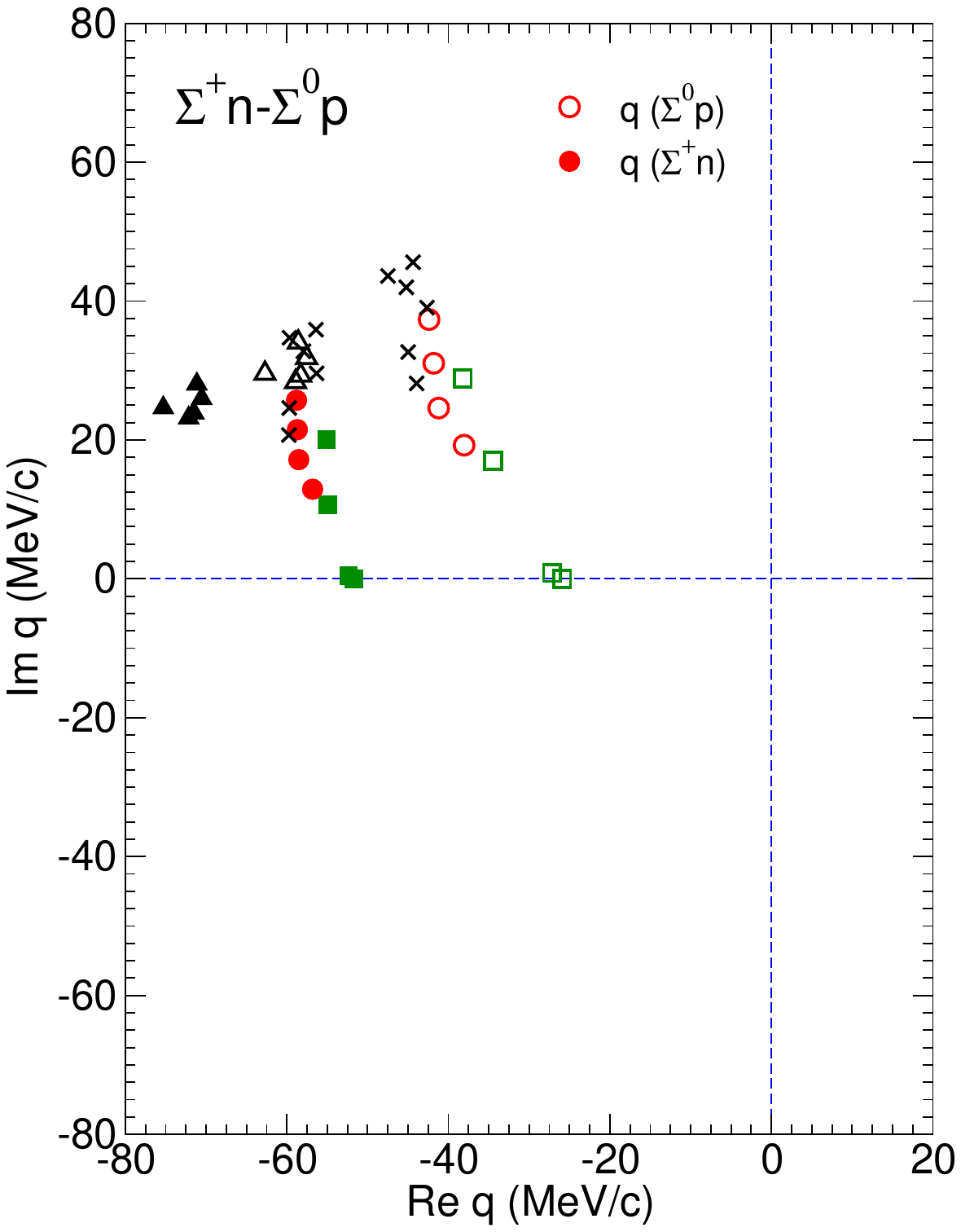}
    \includegraphics[width=0.32\linewidth]{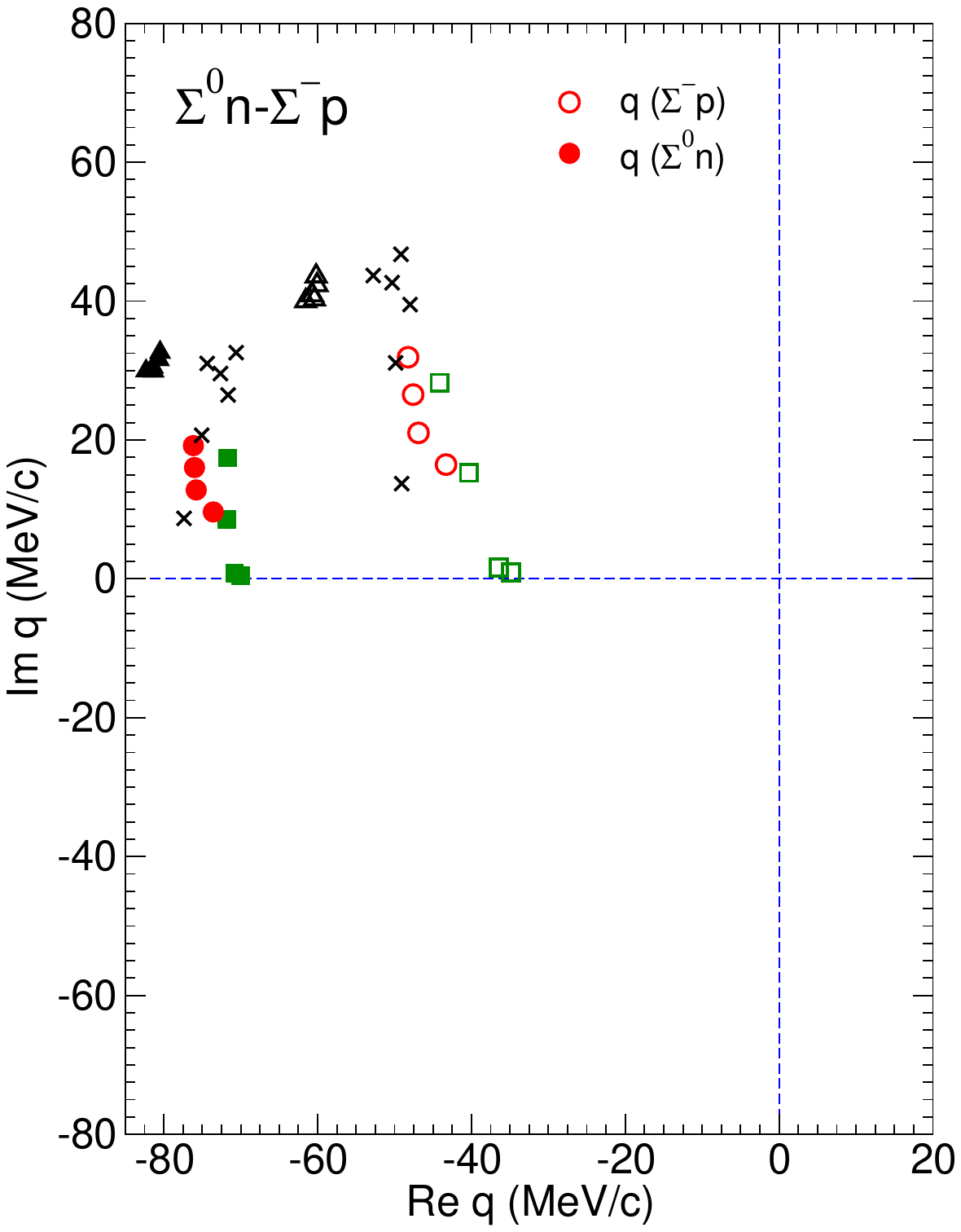}
    \end{center}
    \vskip -0.5cm 
    \caption{Pole positions for the Nijmegen NSC97f model
    (triangles), and the chiral $YN$ potentials 
    NLO13 (squares) and NLO19 (circles), 
    and the SMS potentials from
    Ref.~\cite{Haidenbauer:2023qhf} (crosses)
    in the complex $q$ plane. Results are
    presented for the isospin $I=1/2$ channel and for the
    $\La p - \Si^+ n - \Si^0 p$ and $\La n - \Si^0 n - \Si^- p$
    systems. In the latter cases the poles for the two $\Si N$ channels
    are shown in the same plot. In case of isospin averaged masses 
    also results for the Juelich meson-exchange potentials ’04 
    \cite{Haidenbauer:2005zh} (blue inverted triangle), 
    \~A~\cite{Reuber:1993ip} (blue diamond), 
    and A~\cite{Holzenkamp:1989tq} (blue star), 
    and for the Nijmegen potential ND \cite{Nagels:1976xq} (plus) are included.
    }
    \label{fig:LNpoles}
\end{figure} 

In Ref.~\cite{Haidenbauer:2021smk} the poles and the resulting structure
were investigated more systematically by considering a larger set of 
potentials, including the new chiral $YN$ potentials, and by 
performing the calculations not only in isospin basis but also for 
the charge $Q=1$ ($\La p -\Si^+ n-\Si^0 p$) as well as for the
$Q=0$ ($\La n -\Si^0 n-\Si^- p$) coupled-channel systems. 
The found pole positions are summarized in Fig.~\ref{fig:LNpoles}. One
can see that most of the poles are located in the 2nd quadrant (sheet II). The few exceptions where the poles are either in the 3rd quadrant (sheet IV) or very close to it concern potentials that do not describe the near-threshold $\Si^- p$ data well, as emphasized in Ref.~\cite{Haidenbauer:2021smk}.
For example, the Juelich '04 potential (blue inverted triangle in the left plot) fails to reproduce the capture ratio and it also underestimates the $\Si^- p\to \La n$ cross section, see Fig.~\ref{fig:YNcross}. 

As one can see in Fig.~\ref{fig:LNpoles} the poles are in the lower 
half of the 2nd  quadrant, i.e., ${\rm Im}\, q_{\Si N} \le |{\rm Re}\, q_{\Si N}|$. A Breit-Wigner type signal can be only expected, however, when the poles are sufficiently close to the $\Si N$ threshold and specifically close to the 
negative or positive Im~$q_{\Si N}$ axis. Obviously, the second condition 
is not fulfilled by any of the $YN$ potentials. Thus, one is far from 
the mentioned ideal case and that means, citing Pearce and Gibson \cite{Pearce:1988rk}, in ``a gray area where it is not obvious whether the effect will be cusp or peak''. 

Anyway, the conclusion drawn from the analysis in
Ref.~\cite{Haidenbauer:2021smk} is that, if one takes the presently 
available low-energy $\Si N$ data serious and aims at their best possible 
reproduction, then the appearance of a deuteron-like dibaryon in form of 
a (unstable) $\Si N$ bound state close to the $\Si N$ threshold seems to 
be practically unavoidable. On the other and, there is also strong 
evidence that it might be wishful thinking to expect a truly convincing 
and unambiguous signal for a strangeness $S=-1$ dibaryon, i.e., 
a peak that is well separated from the (and unambiguously below the) 
$\Si N$ threshold.  

\subsection{Coupled channels with strangeness \texorpdfstring{$S=-2$}{S=-2}}
\label{jsec:LL}

As one can see from Fig.~\ref{fig:bb_th} the strangeness
$S=-2$ sector is extremely rich as far as coupled channels
are concerned. Moreover, the separation of the 
thresholds is, in general, fairly small, with the ones 
between $\La\La$ and $\Xi N$ being just about $25\,\MeV$.
However, contrary to the strangeness $S=-1$ sector discussed above experimental
information on $BB$ channels with $S=-2$ is really very much limited. Indeed
for many of the reactions only upper limits on the cross sections have been
established. A summary of available data is provided in Ref.~\cite{Haidenbauer:2015zqb}. Recently, a new source of information has opened up by measurements of two-particle momentum correlation functions by the ALICE and STAR collaborations. So far data on the $\La\La$ \cite{STAR:2014dcy,ALICE:2019eol} and $\Xi^- p$ \cite{ALICE:2019hdt,ALICE:2020mfd,Fu:2024btw} channels have been published.

\subsubsection{The \texorpdfstring{$H$}{H}-dibaryon}
\label{subsec:H-dibaryon}

Certainly the most interesting aspect related to the $S=-2$ sector and 
one of the main motivations for pertinent investigations has been the 
possible existence of the $H$-dibaryon. Initially proposed as 
a deeply bound 6-quark state with $J=0$ and $I=0$ by Jaffe, based on 
the bag model~\cite{Jaffe:1976yi}, it has been suggested as
a dark-matter candidate in recent times~\cite{Farrar:2002ic,Doser:2023gls}.
Given its quantum numbers it should couple to the $\La\La - \Xi N - \Si\Si$ 
system. With its proposed energy of $2150\,\MeV$ it would lie
around $80\,\MeV$ below the $\La\La$ threshold. 
 
\begin{figure}[tb] 
    \begin{center}
    \includegraphics[width=0.60\linewidth]{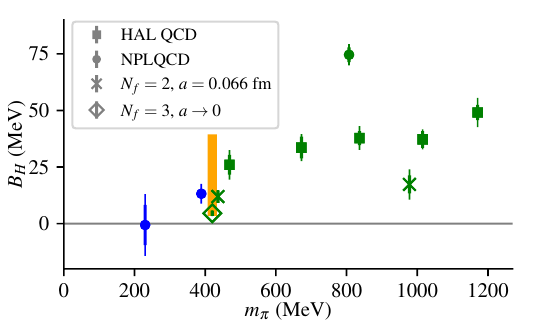}
    \end{center}
    \vskip -0.5cm
    \caption{Lattice QCD results for the H-dibaryon binding energy,
    by the NPLQCD and HAL QCD Collaborations and the Mainz group. The variation of lattice spacing from the calculations of Ref.~\cite{Green:2021qol} is shown by the orange band. The figure is taken from Ref.~\cite{Green:2021qol}.}
    \label{fig:H_dib}
\end{figure}

None of the past experimental searches for the $H$-dibaryon has let 
to convincing signals, see e.g. \cite{Yoon:2007aq}. 
However, in 2010 the interest in that state was revitalized by 
evidence for a bound $H$-dibaryon based on lattice QCD calculations 
\cite{NPLQCD:2010ocs,Inoue:2010es,Inoue:2011ai}. 
The calculations were performed for quark masses
that correspond to unphysical meson and baryon masses,
specifically to pion masses of
$230$ and $389\,\MeV$ in case of the NPLQCD Collaboration and 
$469$-$1171\,\MeV$ by the HAL QCD Collaboration.  
The most recent lattice calculations of the 
$H$ are from the Mainz group~\cite{Green:2021qol} for 
$m_\pi\approx 420\,\MeV$ and the HAL QCD calculation near the physical point~\cite{HALQCD:2019wsz}. The latter work concludes that the 
$\Lambda\Lambda$ attraction is too weak to generate a bound or resonant dihyperon, see the discussion below. In the former work and in \cite{Beane:2011zpa} one can also find a review of earlier lattice calculations. A compilation of available predictions for the $H$-dibaryon is given in Fig.~\ref{fig:H_dib}. As shown in this figure (from Ref.~\cite{Green:2021qol}), systematics due to continuum extrapolations also may play an important role in identifying this illusive state from QCD.

Immediately after the publication of the lattice results several 
attempts have been made to extrapolate those results to the physical 
pion mass. Those extrapolations, performed with different methods and under different assumptions suggested that the $H$-dibaryon could be either loosely bound or move into the continuum \cite{Beane:2011zpa,Shanahan:2011su}.
In Ref.~\cite{Haidenbauer:2011ah,Haidenbauer:2011za} chiral EFT at 
leading order was employed to study the fate of the $H$-dibaryon for
physical masses. In that work not only the effect of the pion mass 
dependence of the interaction (which at LO consists of pion exchange and 
a contact term) was explored, but also the influence of the baryon masses 
which depend likewise on the quark mass and 
which enter the scattering or bound-state equations (\ref{jeq:LS}). 
It was found that the mass dependence of the latter has a much more 
dramatic effect on the binding energy of the $H$ than the pion-mass 
dependence of the potential. The effect is caused by the relative shift 
of the three thresholds $\La\La$, 
$\Si\Si$ and $\Xi N$ due to the SU(3) breaking in the masses. 
Specifically, with physical values the binding energy of the $H$ is 
reduced by as much as $60\,\MeV$ as compared to a calculation based on degenerate (i.e. SU(3) symmetric) $BB$ thresholds. Translating this observation to the situation in the HAL QCD calculation, implied that the bound state should disappear at the physical point. For the case of the NPLQCD 
calculation, a resonance in the $\La\La$ system might survive.

\begin{figure}[tb] 
    \begin{center}
    \includegraphics[width=0.45\linewidth]{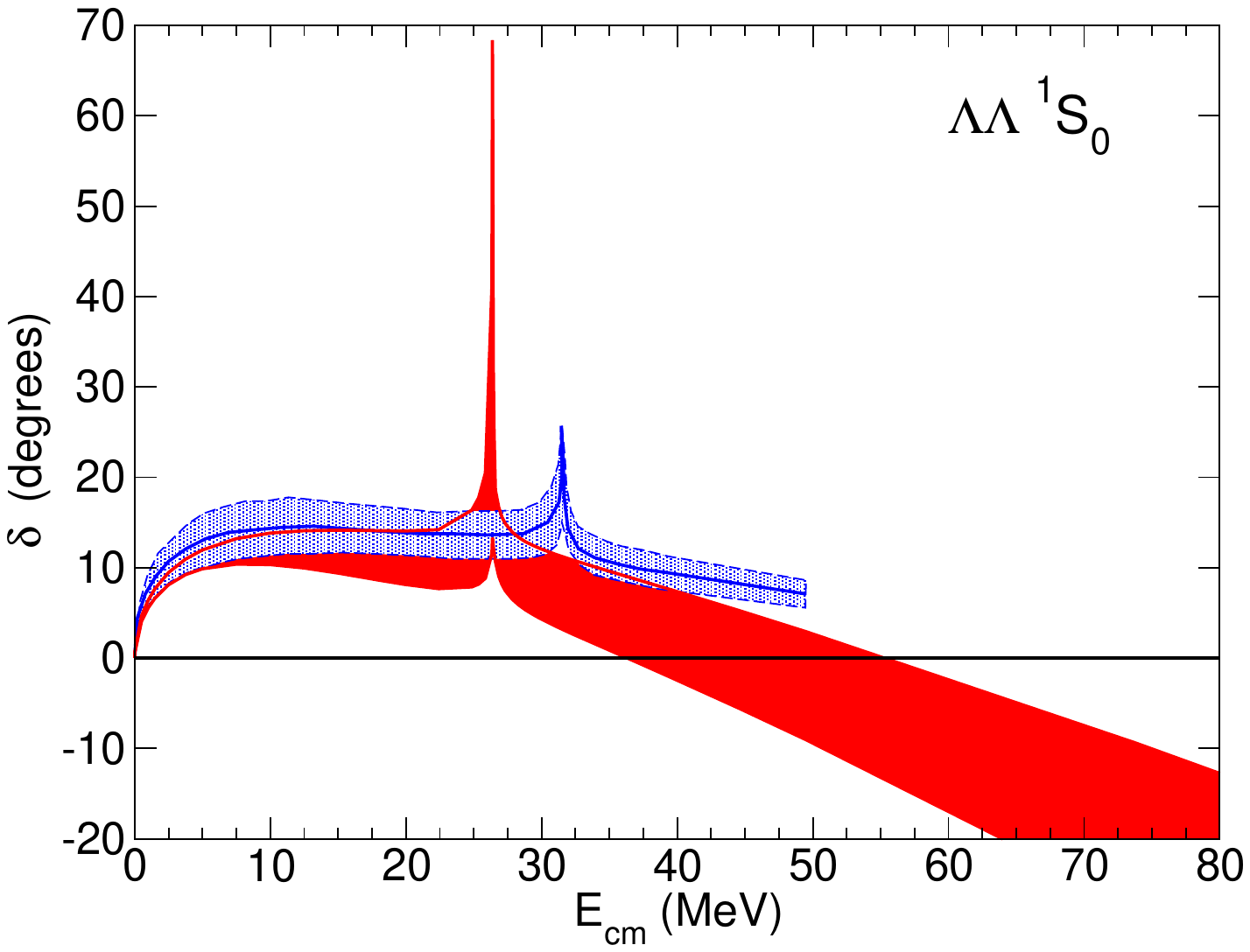}
    \includegraphics[width=0.45\linewidth]{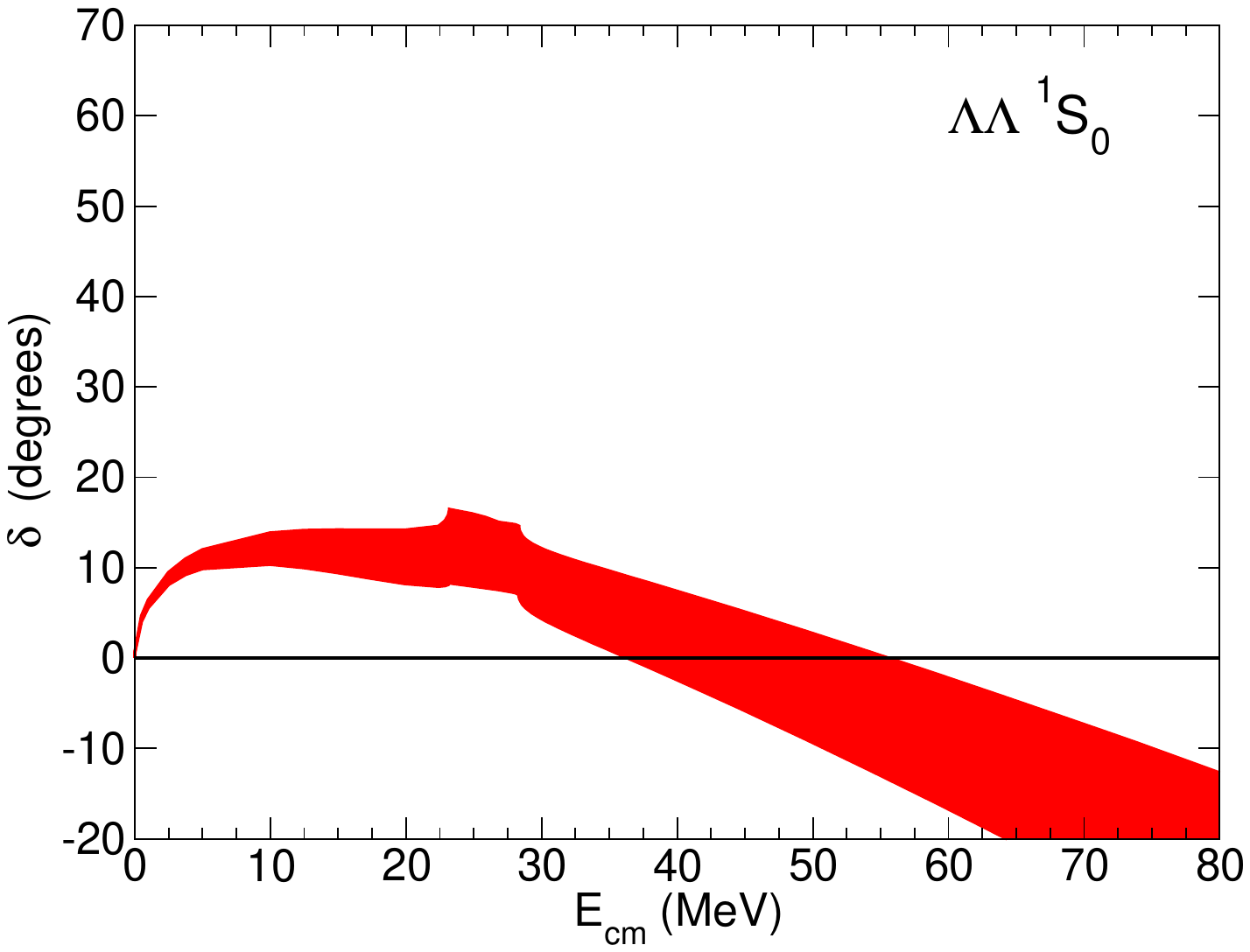}
    \end{center}
    \vskip -0.8cm 
    \caption{Phase shift for the $\La\La$ $^1S_0$ partial wave.
    The results are from the lattice simulation 
    by the HAL QCD Collaboration \cite{HALQCD:2019wsz} (blue band) 
    and the chiral NLO potential \cite{Haidenbauer:2015zqb} (red band).  
    The former calculation corresponds to (meson and baryon) masses which 
    are slightly larger than the physical ones and, therefore, the 
    $\Xi N$ threshold appears at a higher energy. The curves on the left
    are for isospin averaged masses, while on the right-hand side physical
    masses have been used. 
    }
    \label{fig:LL1s0}
\end{figure}

Of course, nowadays lattice calculations at (almost) the physical point
($m_\pi\simeq 146\,\MeV$, $m_K\simeq 525\,\MeV$) are available, thanks
to the work of the HAL QCD Collaboration~\cite{HALQCD:2019wsz}. 
Their results for the $\La\La$ phase-shift in the
the $^1S_0$ partial wave are shown in Fig.~\ref{fig:LL1s0} for the 
$t/a=12$ case with $t$ the sink-source time separation
and $a$ the lattice spacing. In that calculation no bound state (nor a resonances) was found around the $\La\La$ threshold. However, as
one can see in the figure, there is still a remnant
of the $H$-dibaryon, in form of a virtual state in
the $\Xi N$ channel close to its threshold. It produces
a sizable cusp-like structure in the $\La\La$ phase.
Interestingly, the chiral potential at NLO, established a few years earlier by the Juelich group \cite{Haidenbauer:2015zqb}, produces such a cusp too.
(More details of that interaction will be discussed
below.) On the other hand, the results of both groups indicate
that the actual location of the virtual state is still uncertain. 
For example, in the calculation with the chiral potential the large cusp
appears only when solving the coupled-channel equations with isospin 
averaged $\Xi N$ masses (left side). Once the mass difference
between the $\Xi^0$ and $\Xi^-$ is taken into account,
i.e., the splitting between the thresholds for $\Xi^0 n$
and $\Xi^- p$, the virtual state is no longer close to
either of the thresholds and the cusp structure is 
strongly reduced (right side).

\subsubsection{\texorpdfstring{$\Xi N$}{XiN} scattering}

\begin{figure}[thb]
    \begin{center}
    \includegraphics[height=4.6cm,keepaspectratio]{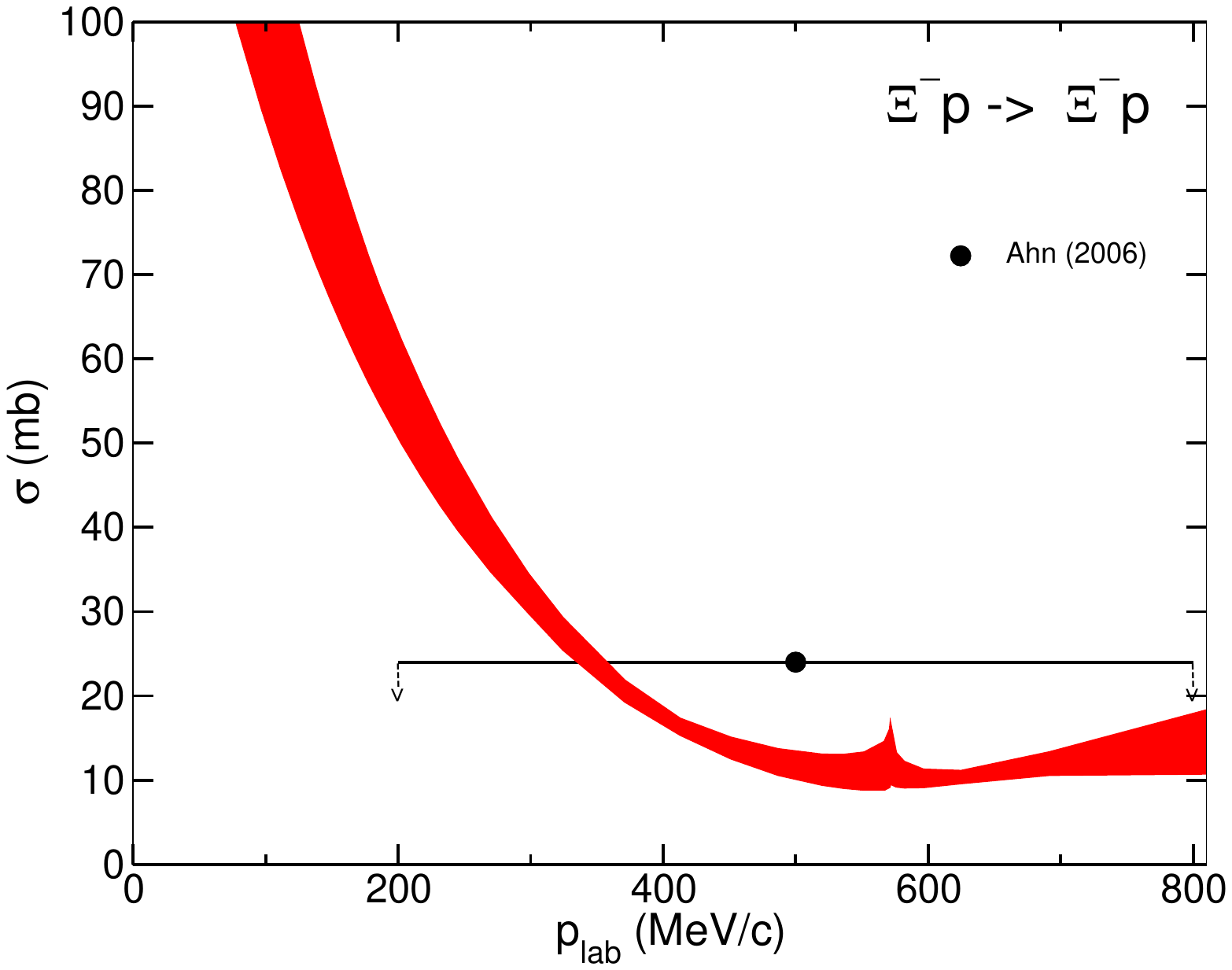}
    \includegraphics[height=4.6cm,keepaspectratio]{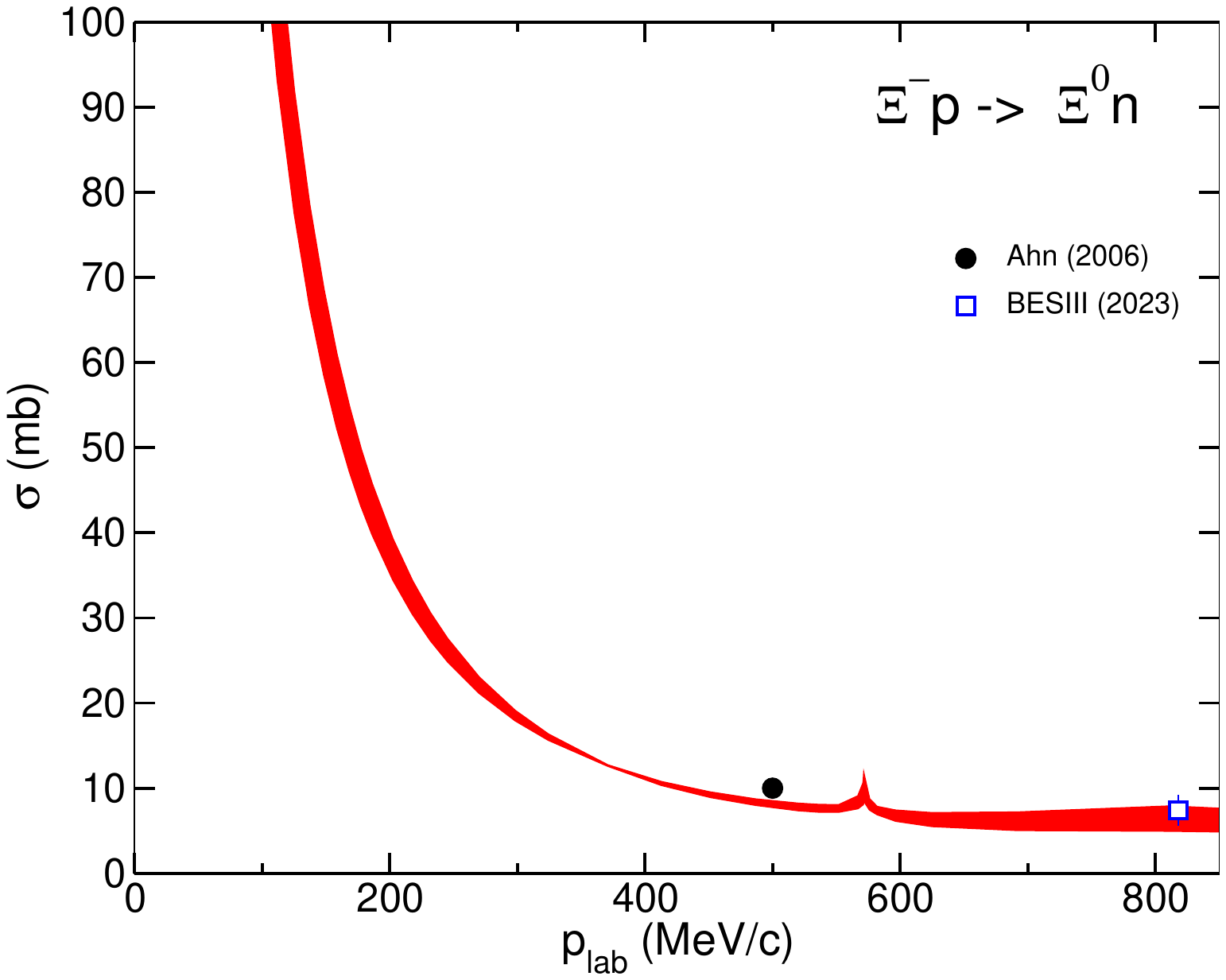}
    \includegraphics[height=4.6cm,keepaspectratio]{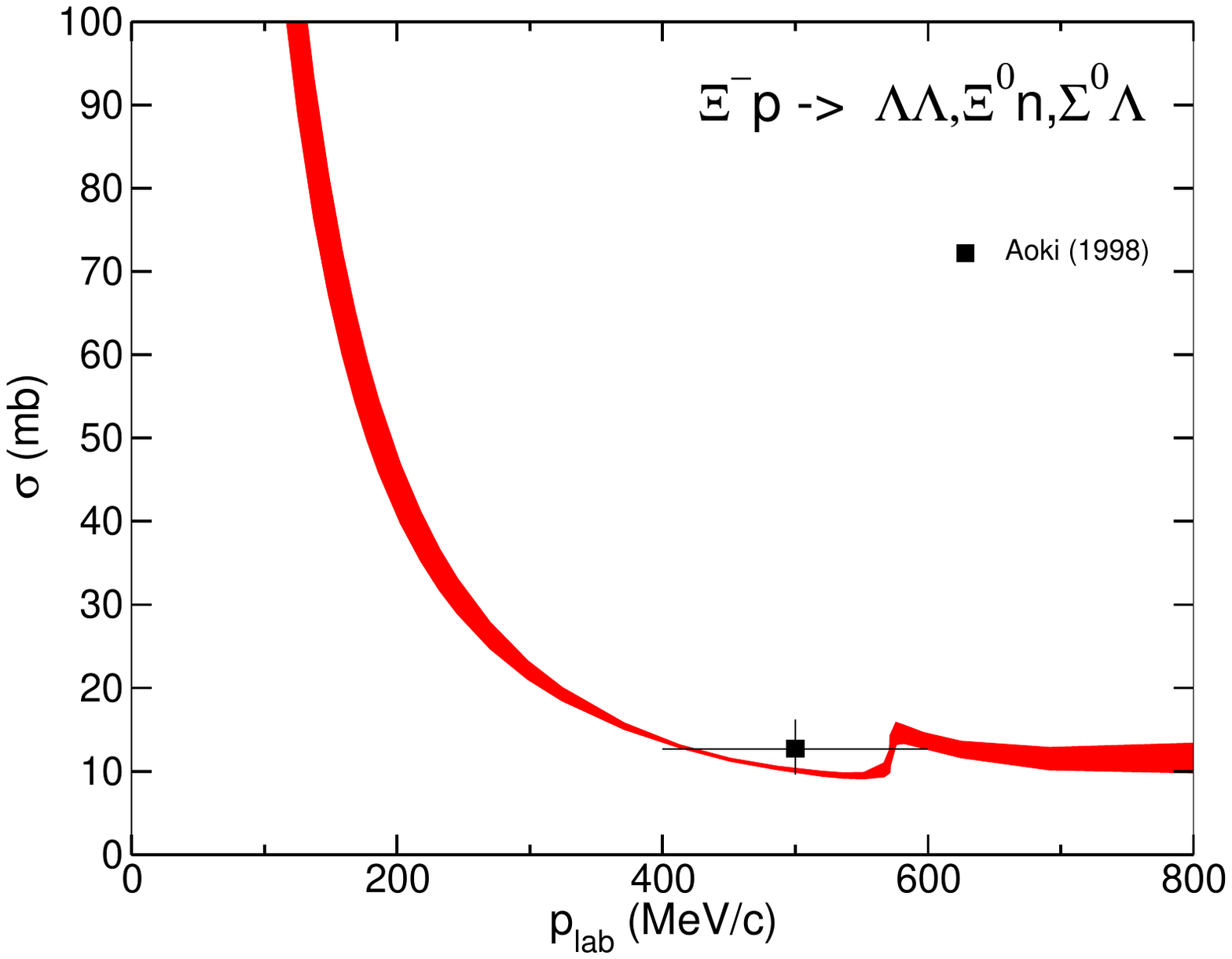}
    \end{center}
    \vspace*{-0.3cm}
    \caption{Results for the $\Xi N$ cross sections of the NLO potential \cite{Haidenbauer:2018gvg}. 
    Data are from Refs.~\cite{Ahn:2005jz,Aoki:1998sv,BESIII:2023clq}. 
    \vspace*{-0.6cm}
    } 
    \label{fig:xmpX}
\end{figure}

\begin{figure}[tb]
    \centering
    \includegraphics[width=0.39\textwidth]{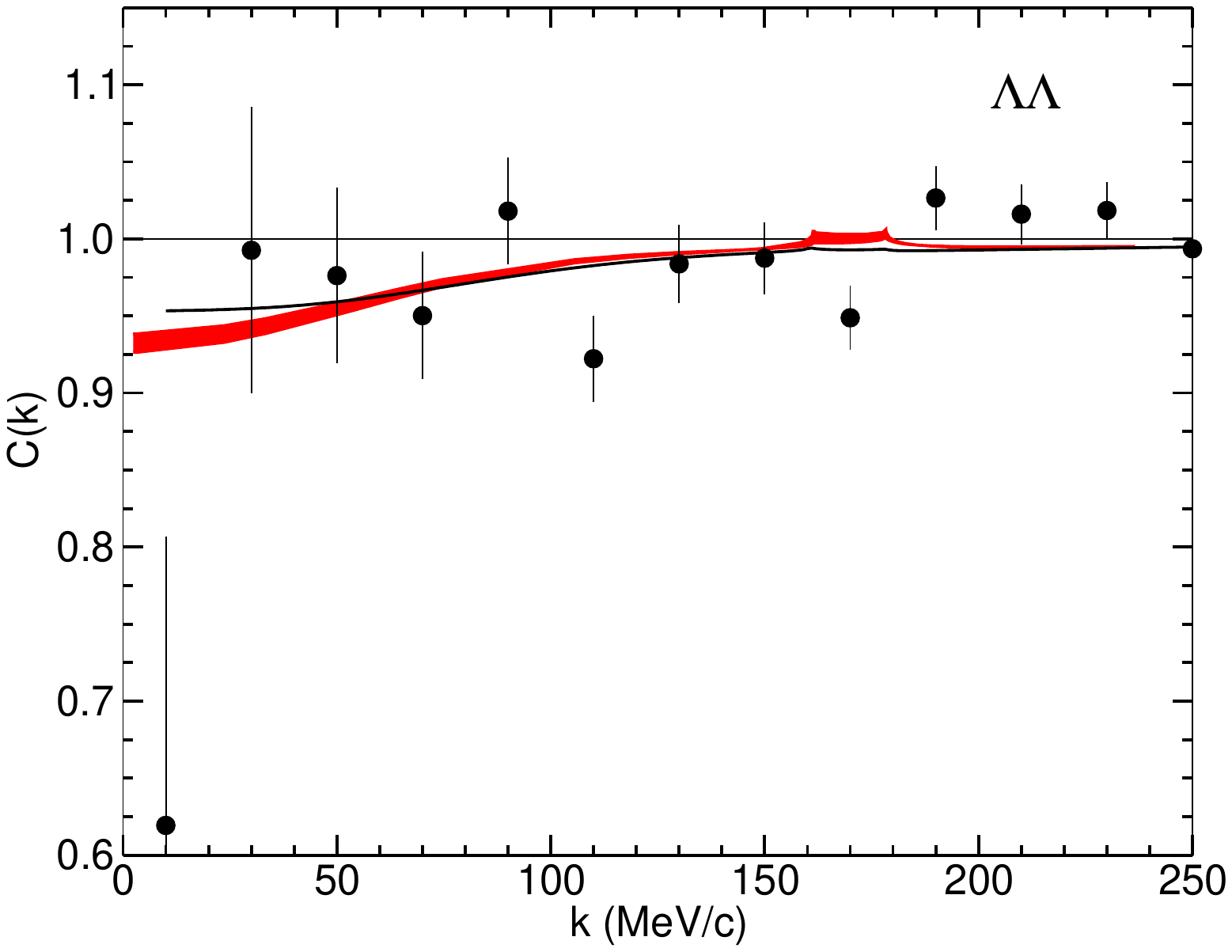}
    \includegraphics[width=0.39\textwidth]{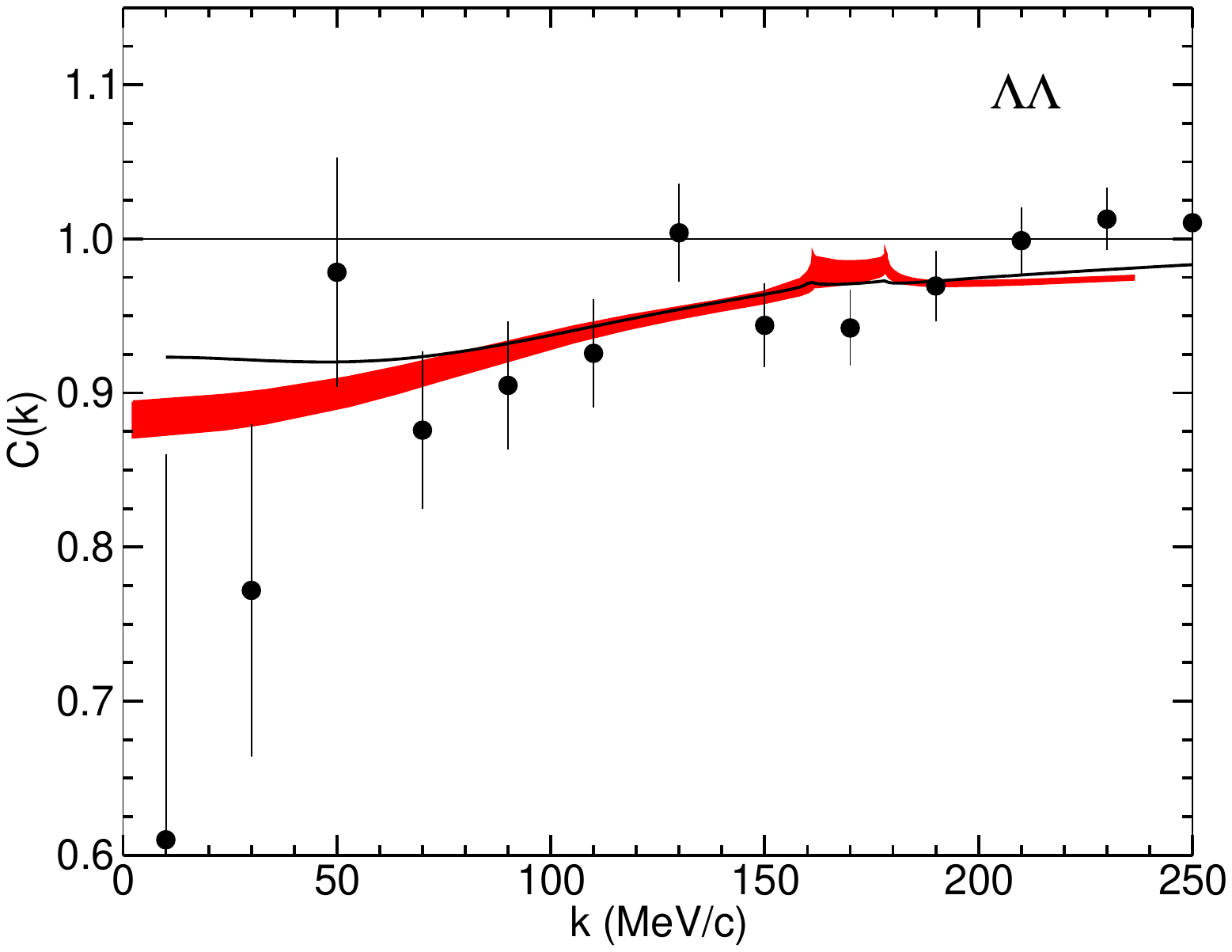}
    
    \includegraphics[width=0.39\textwidth]{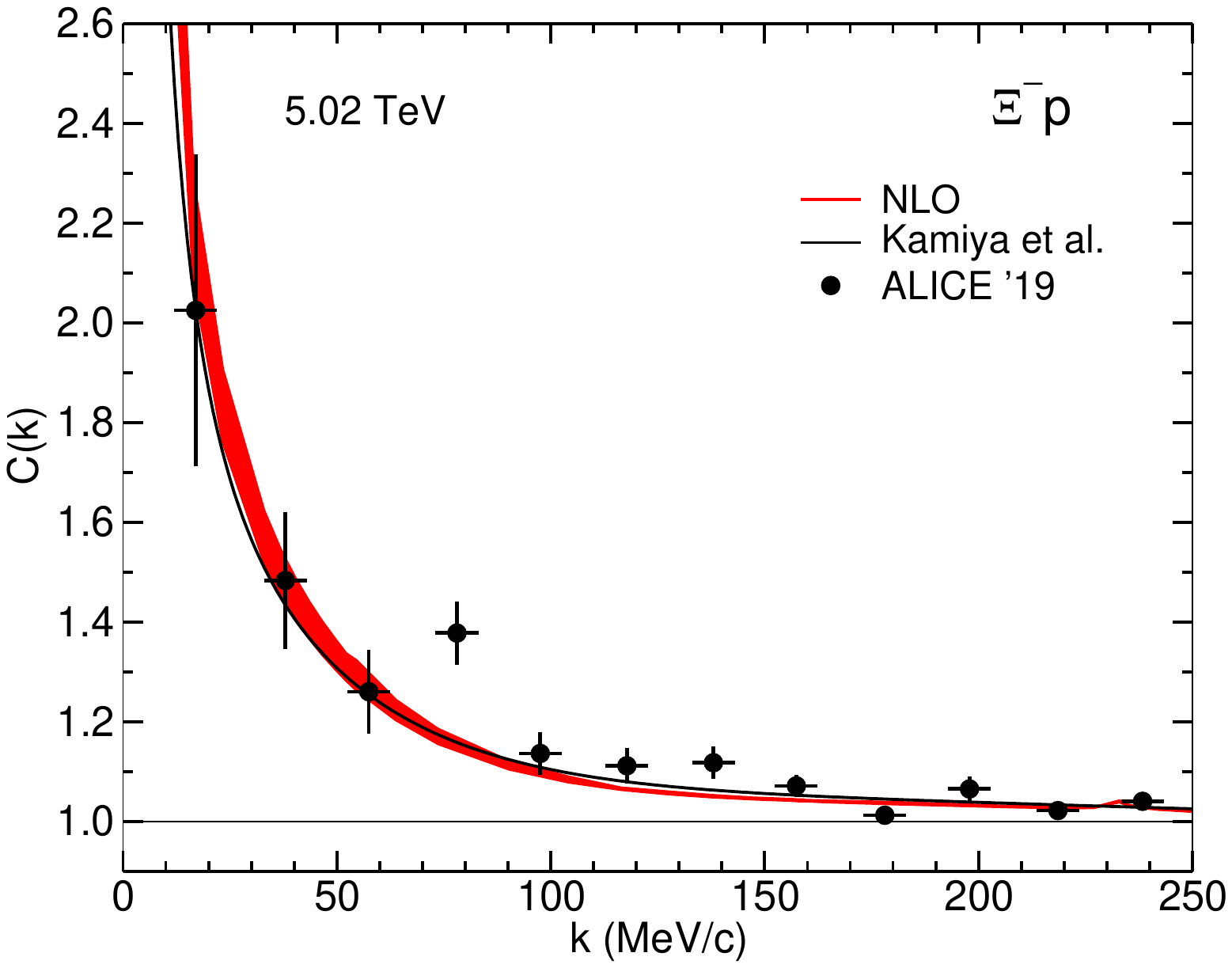}
    \includegraphics[width=0.39\textwidth]{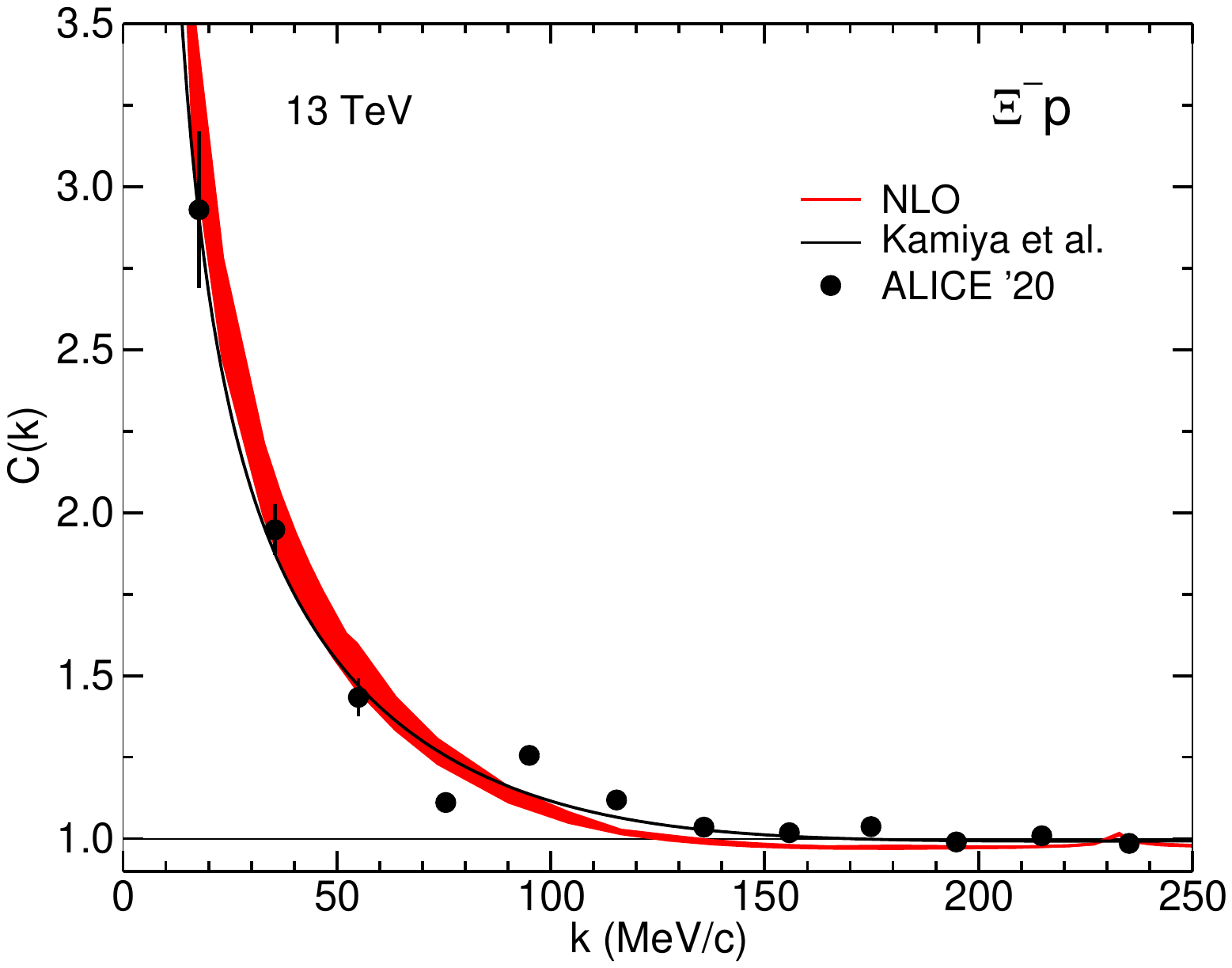}
    \caption{$\La\La$ and $\Xi^-p$ two-particle momentum correlation function
    calculated from the NLO potential \cite{Haidenbauer:2018gvg} (red bands). 
    Results for the HAL QCD potential, taken from 
    Ref.~\cite{Kamiya:2021hdb}, are shown by the solid (black) line. 
    Data from $p$Pb collisions at $5.02$~TeV (left) and $pp$ collisions 
    at $13$~TeV (right) are by the ALICE Collaboration \cite{ALICE:2019eol,ALICE:2019hdt,ALICE:2020mfd}.
    }
    \label{fig:xmp}
\end{figure}

For illustrating the experimental situation in the $S=-2$ sector, as examples 
$\Xi^- p$ elastic and inelastic cross sections are presented in
Fig.~\ref{fig:xmpX}.
The theory results for the $\Xi N$ cross sections are based on a $\Xi N$ potential established within chiral EFT in 2019 \cite{Haidenbauer:2018gvg}.
Again the bands reflect the residual cutoff dependence. In the construction
of the potential constraints from the $\La\La$ scattering length in the $^1S_0$ state
together with experimental upper bounds on the cross sections for $\Xi N$
scattering and for the transition $\Xi N \to \La\La$ have been exploited.
This allowed to fix the additional LECs that arise in the $\{1\}$
irreducible representation of SU(3) \cite{Haidenbauer:2015zqb},
which contribute only to the $BB$ interaction in the $S=-2$
sector.
Furthermore, the consideration of those empirical constraints necessitated
to add SU(3) symmetry breaking contact terms in other irreps
($\{27\}$, $\{10\}$, $\{10^*\}$, $\{8_s\}$, $\{8_a\}$),
with regard to those determined from the $\La N$ and $\Si N$ data.
This is anyway expected and fully in line with the power counting of
SU(3) chiral EFT, as discussed in Ref.~\cite{Haidenbauer:2014rna}. 
It is interesting to see that the theory predicts coupled-channel effects 
at the opening of the $\La \Si$ channel. But these are just standard (moderate) cusp structures and it will be difficult to verify their existence experimentally.  Note that the recent BESIII data \cite{BESIII:2023clq} shown in Fig.~\ref{fig:xmp} (middle) are for $\Xi^0 n\to \Xi^-p$, 
and actually deduced from the $\Xi^0 + ^9$Be reaction.

At present, the only more quantitative test for the $\Xi N$ interaction is 
provided by two-particle momentum correlations. For $\Xi^- p$, correlation functions  have been measured recently by the ALICE Collaboration in $p$-Pb collisions at $5.02$ TeV \cite{ALICE:2019hdt} and in $pp$ collisions at $13$ TeV \cite{ALICE:2020mfd}. There are also new and still preliminary results from Au+Au collisions at $200\,\GeV$ 
by the STAR Collaboration \cite{Fu:2024btw}.
In Fig.~\ref{fig:xmp} predictions for $C(k)$ for the chiral $S=-2$ interaction
from \cite{Haidenbauer:2018gvg} are presented. Details on the evaluation of such correlation functions 
can be found, e.g., in Refs.~\cite{Haidenbauer:2018jvl,Kamiya:2021hdb}. 
Clearly, the correlation functions, evaluated for the source radii
$R$ taken from the corresponding $pp$ fits by ALICE~\cite{ALICE:2019eol}
($1.43$~fm for $5.02$~TeV and $1.18$~fm for $13$~TeV), agree nicely with
the measurements. For a more thorough discussion on the choice of $R$ and of the other parameters that enter into the calculation, see the work of 
Kamiya et al.~\cite{Kamiya:2021hdb}. In this context,
note also the critics on the ``universality
assumption'' for the 
source (radius) in Ref.~\cite{Epelbaum:2025aan}.
Comparable results have been also achieved by
Liu et al.~\cite{Liu:2022nec} based on a full 
coupled-channel potential ($\La\La$, $\Xi N$, $\La\Si$,
$\Si\Si$)
established within covariant chiral EFT, whose 
LECs were fitted to the $\La\La$ and $\Xi N$ phase 
shifts of the HAL QCD potential \cite{HALQCD:2019wsz}.

For completeness, results for the correlation function in the $\La\La$ channel 
are shown too. Exploratory calculations~\cite{Haidenbauer:2018jvl, Kamiya:2021hdb} suggested that there could be a pronounced threshold effect at the opening of the $\Xi N$ channel, cf. Fig.~\ref{fig:LL1s0}.
However, the data from ALICE themselves do not exhibit any 
structure, though admittedly the statistics is low and the momentum resolution too poor.

\subsection{Other coupled baryon-baryon systems}
\label{jsec:Others}

The interaction of octet baryons in $BB$ channels with strangeness $S=-3,-4$ have been studied in the meson-exchange picture 
\cite{Nagels:2023zky} and in chiral EFT \cite{Haidenbauer:2009qn,Liu:2020uxi}
by exploiting the underlying SU(3) symmetry. However,
there are large variations/ambiguities in the predictions. 
First but still limited experimental constraints have
emerged only recently for the $\La\Xi^-$ system, again in form 
of momentum correlation functions~\cite{ALICE:2022uso}. 
The present statistics is not sufficient to see any 
coupled-channel effects, say in form of a structure at the opening 
of the $\Si \Xi$ channels. There are also results from lattice QCD 
simulations by the HAL QCD collaboration for quark masses close to 
the physical point \cite{Ishii:2018ddn}. 
Finally, there are some results on the 
$N\Omega$ interaction from lattice QCD \cite{HALQCD:2018qyu} 
and empirical correlation functions \cite{STAR:2018uho,ALICE:2020mfd}. 
This channel couples to $\Xi\La$ and $\Xi \Si$. 

Regarding $BB$ systems with charmed baryons, there are studies on the
$\Lambda_c^+ N$-$\Sigma_c N$ system. Again there are results from 
lattice QCD simulations by the HAL QCD collaboration 
\cite{Miyamoto:2017tjs,Miyamoto:2017ynx}, though in 
this case the masses are still away from the physical point. 
An extrapolation to the physical
point based on (conventional as well as covariant) chiral EFT
has been attempted \cite{Haidenbauer:2017dua,Song:2020isu,Haidenbauer:2021tlk}
- with in part contradictory results. There are also attempts to construct potential models based on meson exchange relying on the assumption of SU(4) flavor symmetry \cite{Liu:2011xc,Vidana:2019amb}.
In any case, as one can see from Fig.~\ref{fig:bb_th}, the 
thresholds in the charm sector are farther apart and actually  
the $\Lambda_c^+ N\pi$ threshold lies below that of $\Sigma_c N$.
Thus, more pronounced coupled-channel effects are unlikely to 
occur, specifically if the overall strength of the interaction is
in the order of magnitude as suggested by the HAL QCD lattice simulations. 

\section{Heavy sector}
\label{sec:heavy-sector}

\subsection{The \texorpdfstring{$DN$}{DN} interaction and the charm counterpart of the \texorpdfstring{$\La$}{Lambda}(1405)}
\label{sec:DN}
In 1993 first evidence for the $\Lambda_c(2595)^+$ 
resonance was reported
by the CLEO collaboration~\cite{CLEO:1994oxm} and its mass 
was subsequently confirmed by
other experiments~\cite{E687:1995srl,ARGUS:1997snv,CDF:2011zbc}. 
It is now generally believed that this resonance is the 
charm counterpart of 
the $\Lambda$(1405) \cite{ParticleDataGroup:2024cfk}. 
As discussed in \cref{sec:strangebaryons}, there is 
strong theoretical evidence that the $\La$(1405) is a 
$\bar K N$ 
bound state (molecule) and very likely even has a two-pole 
structure~\cite{Mai:2020ltx,Hyodo:2020czb,Meissner:2020khl}. This circumstance triggered several studies under
the premise that the $\Lambda_c(2595)^+$ could be likewise
a dynamically generated state. Those studies build again on
flavor symmetry and on chiral symmetry, and several of them
also on heavy-quark spin symmetry (HQSS). In addition, in most 
of the works, the experimentally known mass of the 
$\Lambda_c(2595)^+$ resonance has been utilized as 
constraint for fixing inherent parameters of the interaction. 

\begin{figure}[tb] 
    \begin{center}
    \includegraphics[width=0.80\linewidth]{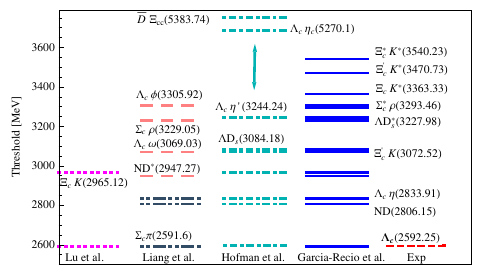}
    \end{center}
    \vskip -0.5cm
    \caption{Thresholds of channels included in different
    studies: Lu et al.~\cite{Lu:2014ina}, 
    Liang et al.~\cite{Liang:2014kra}, 
    Hofmann et al.~\cite{Hofmann:2005sw}, 
    Garcia-Recio et al.~\cite{Garcia-Recio:2008rjt}.
    The figures is adapted from Ref.~\cite{Lu:2014ina}.
    }
    \label{fig:dn_th}
\end{figure} 

The connection between the $\bar KN$ and $DN$ interactions 
can be easily established within chiral  unitary (and related) approaches, by extending the scheme 
from SU(3) to SU(4) in flavor space. Thereby, in analogy to 
the $\Lambda (1405)$ resonance, states in the corresponding charm 
sector are generated by the dynamics in the coupled-channel 
system that involves channels like $\pi\Lambda_c^+$, $\pi\Sigma_c$, or $DN$,
i.e., where the strange quark in the hadrons is simply replaced by the one with charm 
\cite{Lutz:2003jw, Hofmann:2005sw, Mizutani:2006vq,Lutz:2005vx, Garcia-Recio:2008rjt, Haidenbauer:2010ch, Jimenez-Tejero:2011dif, Romanets:2012hm, Liang:2014kra}. 
There is some dispute on the question how one should deal with the flavor 
and chiral symmetries in detail and what additional symmetries should 
be imposed. In the simplest extension based on the SU(3) subgroup
one has again an octet of pseudoscalar mesons and an octet 
(nonet) of ($J^P=1/2^+$) baryons. The number of 
channels that couple is then comparable to those in the 
strangeness $S=-1$ sector.  When keeping only Goldstone bosons in the chiral SU(3) Lagrangian, 
potentially important channels like $DN$ are excluded 
\cite{Lutz:2003jw,Lu:2014ina}. 
On the other hand, when implementing also 
HQSS~\cite{Garcia-Recio:2008rjt,Romanets:2012hm}, vector 
mesons and also $J^P=3/2^+$ baryons enter so that 
the number of coupled channels increases significantly. 
Then one has also 
$D^*N$, $D^*\Delta$, $\omega \La_c$, etc. -- in fact
a total of $16$ channels for $I=0$ and $22$ for $I=1$
that all couple to $DN$ \cite{Garcia-Recio:2008rjt}.
Channels involving vector mesons appear also in the extended
local hidden gauge approach followed in Ref.~\cite{Liang:2014kra}. 
A visual impression of the situation 
is provided in Fig.~\ref{fig:dn_th}. 
Clearly, with the increasing number of channels there is a
simultaneous increase in the number of dynamically 
generated states, as spelled out in Ref.~\cite{Lu:2014ina}:
\emph{``it seems that the number of states generated is proportional
to the number of coupled channels considered''}.
It should be noted that most of the states found in the 
calculations have so far not been confirmed experimentally. 

In the aforementioned approaches the dynamics is  provided by 
vector-meson exchange
\cite{Hofmann:2005sw,Jimenez-Tejero:2011dif}, 
by the Weinberg-Tomozawa (WT) term 
\cite{Lutz:2003jw,Mizutani:2006vq,Lutz:2005vx}, 
or by an extension of the WT interaction to an SU(8) 
spin-flavor scheme 
\cite{Garcia-Recio:2008rjt,Romanets:2012hm}. 
The model presented in Ref.~\cite{Haidenbauer:2010ch} is
derived within the conventional meson-exchange 
picture. There, besides contributions from vector-meson exchanges
also scalar-mesons are considered. The included
channels are $\pi\Lambda_c^+$, $\pi\Sigma_c$, $DN$ but also 
$D^*N$, $D\Delta$, and $D^*\Delta$, in close analogy to 
the Juelich $\bar KN$ potential model \cite{Mueller-Groeling:1990uxr} 
which forms the basis for the extension to the charm sector via 
SU(4) symmetry. 

There are also additional works which focus on specific aspects. For
example, Refs.~\cite{Wang:2018jaj,Wang:2020dhf} consider mainly 
the physics around the $\Sigma_c(2800)$ resonance, 
which is located close to the $DN$ threshold. 
A recent overview of the various investigations can be 
found in Ref.~\cite{Du:2022rbf}.

Let us review in more detail the situation for the $\Lambda_c(2595)^+$ 
and the $\Sigma_c (2800)$. 
First, we want to emphasize that there is an essential difference 
between the situation for the $\La$(1405) and for the  
$\Lambda_c(2595)^+$: The former is located about $30$~MeV below
the $\bar KN$ threshold while the latter is about $200$~MeV below 
the $DN$ threshold. Thus, if the
$\Lambda_c(2595)^+$ is indeed a $DN$ bound state it would be 
rather deeply bound. Accordingly, the interactions have to be 
fairly strong.
Since the mass of the $\Lambda_c (2595)^+$ is commonly used as input, 
as mentioned above, it is in usually well reproduced in such studies.
There is a bit more variation when it comes to the width $\Gamma$, 
where the present experimental value is 
$2.6\pm 0.6$~MeV~\cite{ParticleDataGroup:2024cfk}. 
The theoretical predictions range from $0.6$~MeV 
\cite{Garcia-Recio:2008rjt} to $4.8$~MeV \cite{Haidenbauer:2010ch}.
However, given the delicate situation that the $\Lambda_c (2595)^+$
coincides with the $\pi\Sigma_c$ thresholds and in view of the fact that 
the three-body decay channel $\pi\pi\Lambda_c^+$ is not included in any
of the calculations the agreement is still remarkably good. Anyway, one
has to keep in mind that in calculations typically the pole position
is determined whereas the experimental value is the result of a 
Breit-Wigner type fit. 

One of the characteristic features of the $\Lambda(1405)$ in chiral 
unitary approaches is its two-pole structure, 
see, e.g., the review by Mai~\cite{Mai:2020ltx}.
Indeed such a two-pole
structure is also found for the $\Lambda(2595)^+$ in several studies,
notably in Refs.~\cite{Mizutani:2006vq, Garcia-Recio:2008rjt,Romanets:2012hm,Liang:2014kra,Haidenbauer:2010ch}. 
In the majority of the calculations, the second pole is very close to
that of the $\Lambda(2595)^+$, say within $10$ to $15$~MeV but with
a considerably larger width. The latter reflects the fact that the
second pole is already well above the $\pi\Sigma_c$ threshold and it is
dominated by the $\pi\Sigma_c$ component.
Compared to the $\bar K N$ ($\Lambda(1405)$) case, 
there is even less concrete experimental
evidence for the partner state. Initial works in the chiral 
unitary approach associated the second pole with the 
$\Lambda_c (2625)^+$ but the predicted width is much too large
as compared to the experiment. Anyway, the latter is now
commonly assumed to be a $J^P=3/2^-$ state, i.e., the 
counterpart of the $\Lambda$(1520). 

\begin{figure}[tb]
    \centering
    \includegraphics[width=0.39\textwidth]{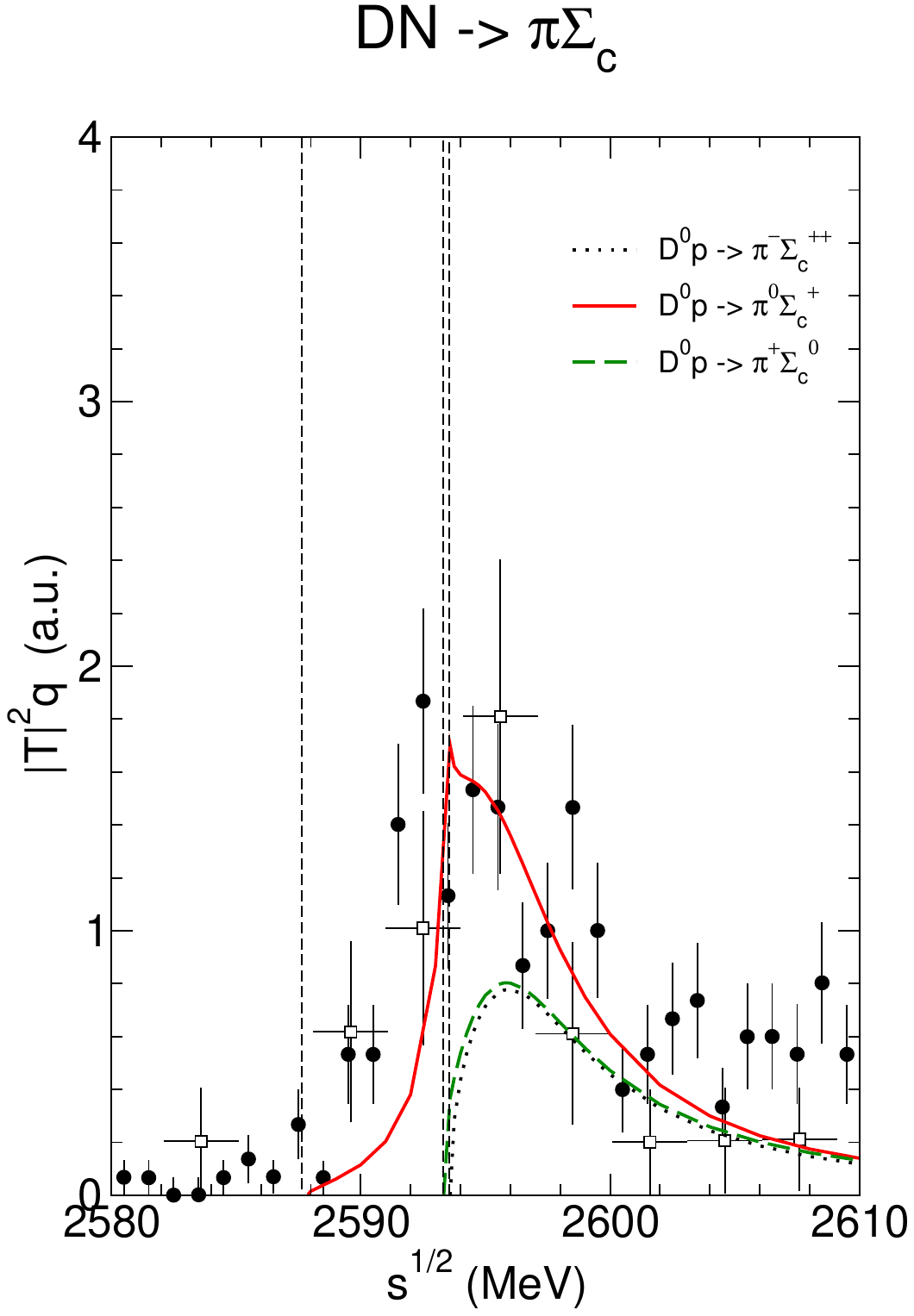}
    \includegraphics[width=0.39\textwidth]{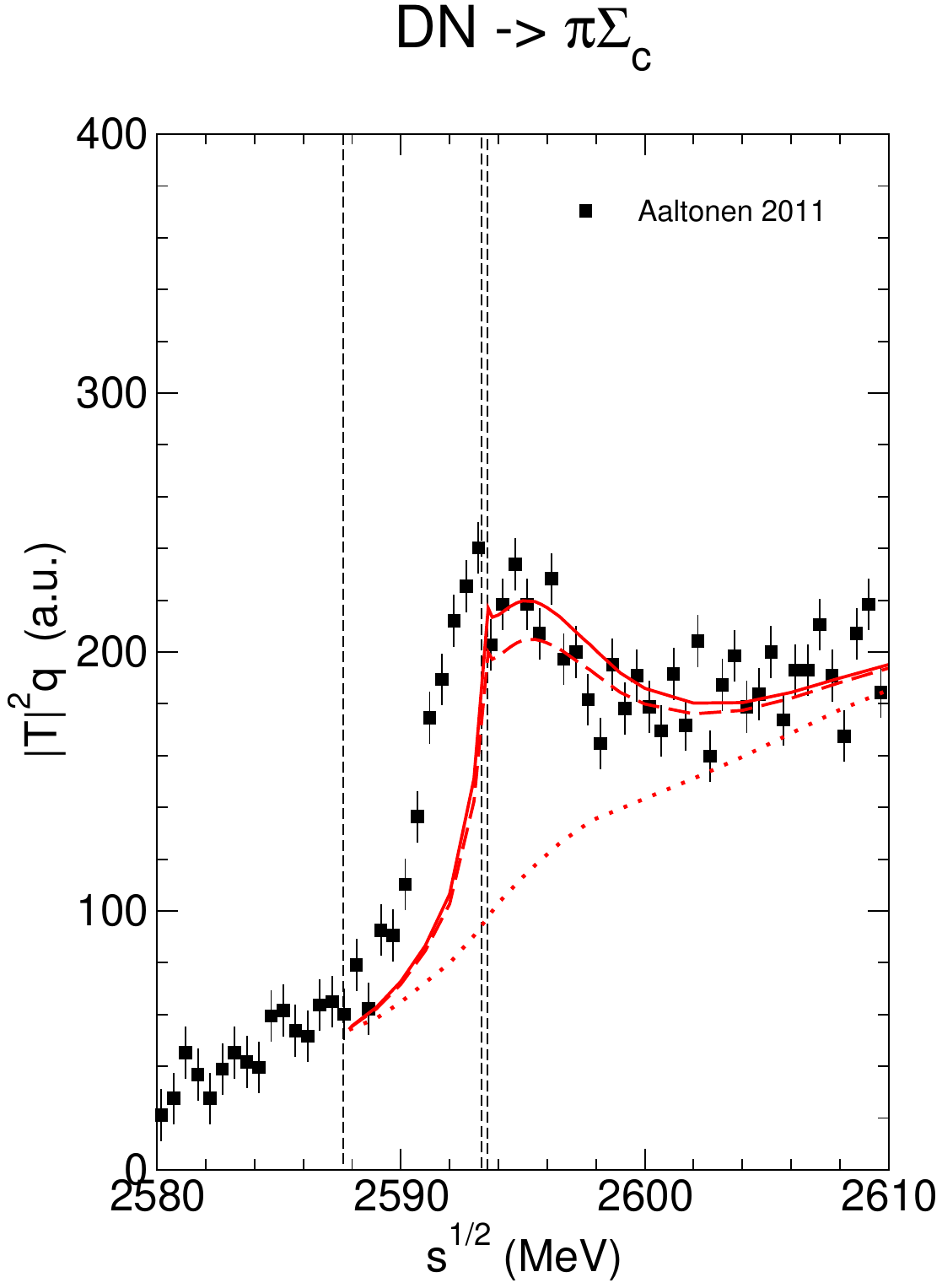}
    \vspace*{-0.2cm}
    \caption{$\pi\Sigma_c$ invariant-mass spectrum predicted
    by the $DN$ potential of the Juelich group \cite{Haidenbauer:2010ch}. For comparison experimental 
    information on the $\pi^+\pi^-\Lambda_c^+$ mass distribution 
    in the $\Lambda_c(2593)^+$ region is shown. The
    data are from ARGUS~\cite{ARGUS:1997snv} (open squares), 
    Zheng (CLEO)~\cite{Zheng:1999} (circles) 
    and CDF~\cite{CDF:2011zbc} (filled squares). The nominal thresholds 
    of the $\pi^0\Sigma_c^+$, $\pi^+\Sigma_c^0$, and $\pi^-\Sigma_c^{++}$
    channels are indicated by vertical lines.
    }
    \label{fig:mdn}
\end{figure}

Predictions for the $\pi\Sigma_c$ invariant-mass spectrum in 
the region of the $\Lambda_c (2595)^+$ resonance were presented 
in Ref.~\cite{Haidenbauer:2010ch}, obtained from the 
$\pi\Sigma_c\to \pi\Sigma_c$ and $DN\to \pi\Sigma_c$
reaction amplitudes multiplied with the appropriate 
phase-space factors. 
Since there is no qualitative difference between the two cases,
we reproduce here only the results for the latter, see Fig.~\ref{fig:mdn}. 
The invariant-mass distributions are given in arbitrary units 
but the relative normalization between the different charge 
channels as predicted by the model has been kept fixed. 
Then the drastic effect due the splitting of the
thresholds in the $\pi\Sigma_c$ channels caused by the mass differences 
between $\pi^\pm$ and $\pi^0$ and between the $\Sigma_c$'s 
is preserved. 
Indeed, the threshold of the $\pi^0\Sigma_c^+$ channel is about $6$~MeV
lower than those of the other two charge channels and, as a consequence, 
the predicted invariant-mass distribution is about twice as large.
Moreover, there is a clear cusp in the $\pi^0 \Sigma_c^+$ amplitude
at the opening of the $\pi^+\Sigma_c^0$ channel. The threshold of
$\pi^-\Sigma_c^{++}$ is just about $0.3$~MeV above the one for $\pi^+\Sigma_c^0$.
This produces a noticeable kink in the $\pi^0 \Sigma_c^+$
invariant-mass distribution.

For illustration, the experimental $\pi^+\pi^-\Lambda^+_c$
mass distributions where the signal for the $\Lambda_c(2595)^+$
is seen \cite{ARGUS:1997snv,Zheng:1999} are also displayed in the left panel of the Fig.~\ref{fig:mdn}. 
The results for the $D^0p \to \pi^0\Sigma_c^+$ channel resemble 
very much the measured signal and one can imagine that smearing out the results by the 
width of the $\Sigma_c^+$, which is around $2.3$~MeV 
\cite{ParticleDataGroup:2024cfk}, could further improve the already good agreement with the data. 
However, experimentally it was found that the $\Lambda_c(2595)^+$ decays predominantly 
into the $\pi^+\Sigma_c^0$ and $\pi^-\Sigma_c^{++}$ channels with 
a combined branching fraction in the range of 
50\% \cite{ParticleDataGroup:2024cfk}.
Smearing out the corresponding results with the widths of the $\Sigma_c^0$ 
and $\Sigma_c^{++}$, which are likewise around $2$~MeV 
\cite{ParticleDataGroup:2024cfk}, would still 
leave a large fraction of the events found below the nominal 
$\pi^+\Sigma_c^0$ and 
$\pi^-\Sigma_c^{++}$ threshold unexplained, especially for the 
data from Ref.~\cite{Zheng:1999}.
Thus, it seems that the position of the $\Lambda_c(2595)^+$ resonance
being so close to or even below the nominal $\pi^+\Sigma_c^0$/$\pi^-\Sigma_c^{++}$
thresholds and the found large branching ratios into those channels
are difficult to reconcile. In any case, one has to keep in mind that
there is also a branching fraction of 20\% directly into the three-body 
channel $\pi^+\pi^-\Lambda^+_c$ 
\cite{ParticleDataGroup:2024cfk}, which is not included in the  
theoretical curve. 
For a dedicated study of the situation of the $\Lambda_c (2595)^+$ 
being so close to the $\pi\Sigma_c$ threshold see \cite{Guo:2016wpy}. 

A comparison with the data from the CDF experiment~\cite{CDF:2011zbc} is provided in
the right panel of Fig.~\ref{fig:mdn}. In this case a background contribution has been
established/added in the analysis of the data (cf. dotted line) and we add 
this contribution to the $\pi\Sigma_c$ invariant-mass spectrum evaluated in
Ref.~\cite{CDF:2011zbc}. Once that is done, there is again a 
fairly good agreement with the experimental spectrum, but again an underestimation below the
nominal $\pi^+\Sigma_c^{0}$ and $\pi^-\Sigma_c^{++}$ thresholds.

Concerning the actual structure of the $\Lambda_c (2595)^+$ the conclusions drawn from the various studies remain ambiguous. While there is consent that the resonance contains a large
meson-baryon component, it remains unclear which of those components might
be the dominant one. Actually, it seems that the compositeness depends 
strongly on the number of considered coupled channels and also on the 
regularization schemes employed in the course of the unitarization \cite{Lu:2014ina}. 
Specifically, in approaches where only (approximate) flavor symmetry is imposed one
finds either a dominant $\pi\Sigma_c$ component 
or a dominant $DN$ component. 
On the other hand, models that implement also HQSS often find 
a large $D^*N$ component. 
For example, in the SU(8) spin-flavor scheme of 
Ref.~\cite{Garcia-Recio:2008rjt} the 
$\Lambda_c(2595)^+$ becomes predominantly a $D^*N$ 
quasi-bound state -- though
still with important additional binding effects from the $DN$ channel.
Finally, in a recent work by Nieves et al.~\cite{Nieves:2024dcz} even
the analogy between the $\Lambda_c(2595)^+$ and the $\Lambda(1405)$ is
called into question. The study is based on the standard Weinberg-Tomozawa
(WT) term and incorporates HQSS. However, in addition, it includes also 
the exchange of resonances from the constituent quark 
model in the $s$-channel. 

Now to the $\Sigma_c (2800)$, a state which is located very close 
to the $DN$ threshold and whose quantum numbers are still not
determined \cite{ParticleDataGroup:2024cfk}.
The Juelich potential predicts a dynamically generated 
$S$-wave state with mass $2797$~MeV \cite{Haidenbauer:2010ch} 
and isospin $I=1$, and it is quite tempting to identify it with 
the $\Sigma_c (2800)$ \cite{ParticleDataGroup:2024cfk}. 
However, there is a caveat because the predicted 
width is just about $12$~MeV, significantly smaller than 
the value of roughly $75\pm 20$~MeV given by the PDG. Since 
that resonance decays primarily to $\pi\Lambda_c^+$ the failure 
to reproduce the width could be an indication
that the $DN\to\pi\Lambda_c^+$ transition potential is too 
small in that model. Most of the other studies do not find states with suitable quantum number in that region. Actually, the majority of the predictions suggest masses around or even below $2700$~MeV for $S$-wave states with $I=1$. Only in \cite{Jimenez-Tejero:2011dif} an isovector state
with $2800$~MeV is predicted, interestingly with $t$-channel 
vector-meson exchange as driving force just as in the Juelich model, 
and again with a somewhat too small width.   

In this context, let us mention that recently evidence against 
the bound-state nature of the $\Sigma_c (2800)$ was reported by 
Wang et al.~\cite{Wang:2020dhf}, who studied the state within 
chiral EFT to NLO, including, however, only the $DN$ and $D^*N$ 
channels. In contrast to that, a theoretical analysis of the $D^0p$ 
invariant-mass spectrum, measured by the LHCb Collaboration 
in the reaction $\Lambda_b\to \pi^- D^0p$ \cite{LHCb:2017jym},  
supports the existence of a bound state right below the
$D^0p$ threshold \cite{Sakai:2020psu}. In that work also 
the $DN$ scattering lengths were extracted. Interestingly, 
but not really surprising, the obtained values are fairly close 
to those predicted by the Juelich model \cite{Haidenbauer:2010ch},
thus, giving support to the strength of the $DN$ 
interaction generated by that potential.

\subsection{\texorpdfstring{$XYZ$}{XYZ} and pentaquark states} 

DCC approaches play also an essential role in studies of the
structure of the so-called $XYZ$ states and of possible pentaquark 
states, see the reviews  
\cite{Lebed:2016hpi, Esposito:2016noz,Ali:2017jda,A2:2018doh, Guo:2017jvc, Olsen:2017bmm,Liu:2019zoy, Brambilla:2019esw,Guo:2019twa,Dong:2021bvy, Chen:2022asf,Meng:2022ozq}.
Many of the new states seen in the charm or bottom sectors, which
do not fit into the standard quark-antiquark or three-quark
classification, have been observed close to thresholds. The first 
and perhaps still most famous one is the $\chi_{c1}$(3872),
formerly known as $X$(3872) \cite{Belle:2003nnu}, whose mass coincides 
with the $D\bar D^*/D^*\bar D$ threshold within the experimental 
uncertainty. Indeed channel coupling and aspects associated
with it like threshold anomalies have been revisited in various 
works since they are considered as possible explanation for 
the nature and structure of some $XYZ$ and other exotic states, 
see, e.g., Refs.~\cite{Matuschek:2020gqe,Dong:2020hxe,Zhang:2024qkg} 
and the review \cite{Guo:2019twa} for very recent examples.

Since the focus of the present review are DCC approaches we do not have 
the ambition and the space to provide an adequate overview over the vast 
amount of studies and publications dedicated to exotic states in
the heavy sector. Anyway, as mentioned above, there are already 
various excellent and comprehensive reviews available. Accordingly, here we want to highlight a few specific and illustrative systems/works, selected by the criterion that channel coupling plays 
a decisive role for the physical conclusions drawn in those studies. 
Accordingly, our emphasis will be on specific meson-baryon systems 
involving heavy hadrons where potentially many channels are coupled. 
In this context the (approximate) heavy-quark spin symmetry (HQSS)
\cite{Neubert:1993mb}, a symmetry which requires the dynamics of heavy 
hadrons to became independent of the spin of the constituent heavy quark 
($Q=c,\, b$) in the limit $m_Q \to \infty$,  
plays an important role in a two-fold way. 
On the one hand side, it results in mass splittings
$\bar D^*-\bar D$, $\Sigma_c^* - \Sigma_c$, etc. 
that are significantly smaller 
than those between the strange counterparts $K^*-K$ and $\Sigma^* - \Sigma$,   
\begin{equation}
    M_{K^*}-M_K = 396~{\rm MeV} \quad M_{D^*}-M_D = 141~{\rm MeV} \quad 
    M_{B^*}-M_B = 45~{\rm MeV} \nonumber
\end{equation}
\begin{equation}
    M_{\Sigma^*}-M_\Sigma = 191~{\rm MeV} \quad M_{\Sigma_c^*}-M_{\Sigma_c} = 65~{\rm MeV} \quad 
    M_{\Sigma_b^*}-M_{\Sigma_b} = 19~{\rm MeV}
    \label{eq:masses}
\end{equation}
so that the thresholds of the meson-baryon channels that can be formed are
much closer together. In addition it provides constraints on the interaction between
the pertinent hadrons and it singles out the relevant channels,
which would all have degenerate thresholds in the strict HQSS
limit. 

DCC approaches in the heavy sector are characterized by the 
intention to interpret observed exotic states in terms of
bound states, commonly referred to as molecules or dynamically generated states.
This strategy is motivated by the fact that many of the exotic states
seen in experiments are located conspicuously close to thresholds, as 
mentioned above, which suggests that their nature could be analogous to that of the deuteron.
The coupling between channels whose thresholds are fairly close together
enhances effectively the overall strength of the interaction in the 
system, and specifically in channels where the underlying direct interaction 
is attractive, and, consequently, favors the formation of bound states. 
For a specific review from a molecular perspective see Ref.~\cite{Guo:2017jvc}.

\subsubsection{Hidden-charm pentaquarks}

\begin{figure}[tb]
    \centering
    \includegraphics[width=0.387\textwidth]{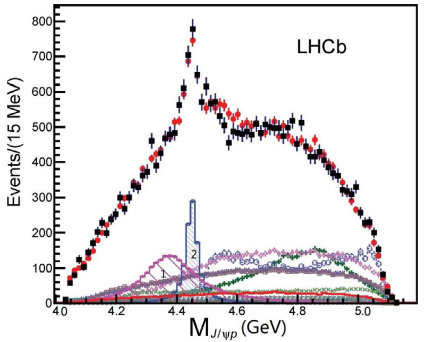}
    \includegraphics[width=0.39\textwidth,height=0.33\textwidth]{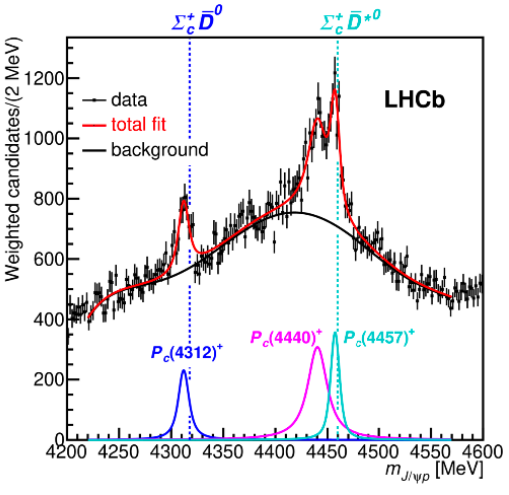}
    \vspace*{-0.2cm}
    \caption{Hidden-charm pentaquarks established by the LHCb Collaboration in the $\Lambda_b^0 \to J/\psi pK^-$ decays in 2015 \cite{LHCb:2015yax} (left) and 
    (2019) \cite{LHCb:2019kea} (right). 
    Left: Squares represent the data and circles the
    result of a fit. The histograms labeled ``1'' and ``2''
    indicate the contributions from the $P_c(4380)$ and $P_c(4450)$ states. 
    Right: The fit to the data is based on a sixth-order
    polynomial background and three Breit-Wigner functions
    representing the pentaquark candidates. 
    }
    \label{fig:PentLHCb}
\end{figure}

First evidence for two possible pentaquark candidates, the
$P_c(4380)$ and $P_c(4450)$ ($P_{c\bar c}$ in the new convention
by the PDG~\cite{ParticleDataGroup:2024cfk}), 
were reported by the LHCb Collaboration
in 2015 based on a measurement of the $J/\psi p$ invariant mass
spectrum via the weak decay $\Lambda^0_b\to J/\psi p K^-$~\cite{LHCb:2015yax}.
Four years later the LHCb Collaboration presented new results by exploiting 
a much larger data sample~\cite{LHCb:2019kea}. In that work a new structure,
the $P_c(4312)$, was found and, in addition, the (former) $P_c(4450)$
resolved into two fairly narrow states, the 
$P_c(4440)$ and $P_c(4457)$, respectively. The mass spectra from the two
LHCb publications are shown in Fig.~\ref{fig:PentLHCb} while the properties
of the states are summarized in \cref{tab:Penta}, together with
there theoretically favored quantum numbers. 
Another state, the $P_c(4338)$, was identified in the 
decay $B^0_s\to J/\psi p\bar p$~\cite{LHCb:2021chn}. 
Note that 
a detailed overview on pentaquarks, by Karliner and Skwarnicki, is included in the latest review of the PDG~\cite{ParticleDataGroup:2024cfk}.

\begin{table}[b!]
    \centering
    \caption{Properties of the $P_c(4312)$, $P_c(4440)$, $P_c(4457)$~\cite{LHCb:2019kea}, the
    $P_c(4380)$~\cite{LHCb:2015yax}, and the $P_c(4338)$~\cite{LHCb:2021chn} (in MeV). 
    The $I(J^P)$ quantum numbers are the theoretically favored ones; experimentally they are not yet established. Adapted from Ref.~\cite{Wang:2019ato}. 
    }
    \label{tab:Penta}
    \begin{tabularx}{\linewidth}{c|XXXX|c}
        \hline\hline
        States& Mass &Width&Threshold&Binding energy&$I(J^P)$
        \TT\BB
        \\
        \hline
        $P_c(4312)$&$4311.9\pm0.7^{+6.8}_{-0.6}$&$9.8\pm2.7^{+3.7}_{-4.5}$&$\Sigma_c^+\bar{D}^0$&$-5.83\pm0.7^{+6.8}_{-0.6}$&$\frac{1}{2}\left(\frac{1}{2}^-\right)$\\
        $P_c(4440)$&$4440.3\pm1.3^{+4.1}_{-4.7}$&$20.6\pm2.7^{+8.7}_{-10.1}$&$\Sigma_c^+\bar{D}^{\ast0}$&$-19.45\pm1.3^{+4.1}_{-4.7}$&$\frac{1}{2}\left(\frac{1}{2}^-\right)$\\
        $P_c(4457)$&$4457.3\pm0.6^{+4.1}_{-1.7}$&$6.4\pm2.0^{+5.7}_{-1.9}$&$\Sigma_c^+\bar{D}^{\ast0}$&$-2.45\pm0.6^{+4.1}_{-1.7}$&$\frac{1}{2}\left(\frac{3}{2}^-\right)$\\
        $P_c(4380)$&$4380\pm8\pm29$&$205\pm18\pm86$&$\Sigma_c^{\ast+}\bar{D}^{0}$&$-2.33\pm8\pm29$&$\frac{1}{2}\left(\frac{3}{2}^-\right)$\\
        $P_c(4338)$&$4337^{+7}_{-4}$$^{+2}_{-2}$&$29^{+26}_{-12}$$^{+14}_{-14}$&
        & & \BBB \\
        \hline\hline
    \end{tabularx}
\end{table}

The plot on the right-hand side of Fig.~\ref{fig:PentLHCb} 
reveals already the most interesting aspect,
the $P_c(4312)$, $P_c(4440)$ and $P_c(4457)$ are just below the 
$\bar D^0 \Sigma^+_c$ and $\bar D^{*0} \Sigma^+_c$ thresholds of
4318 and 4460 MeV, respectively. 
Those channels are two of the four primary channels, 
$\D\Sigma_c$, $\D\Sigma_c^*$, $\D^*\Sigma_c$, and 
$\D^*\Sigma_c^*$, which are closely connected by HQSS.
Other channels of relevance are $\D\Lambda_c^+$ and
$\D^*\Lambda_c^+$, and the hidden-charm channels $\eta_c N$ and specifically 
$J/\psi N$, since the latter system is the one whose invariant mass 
spectrum was measured in the LHCb experiments. 

Besides those obvious channels, further and also more special channels
have been considered.
For example, $\D\Lambda_c(2595)^+$ could be of relevance because its threshold, at $4457$~MeV,  
coincides exactly with the mass of one of the LHCb pentaquarks
\cite{Burns:2019iih,Yalikun:2021bfm,Wu:2024bvl}. 
Of course, since the $\Lambda_c(2595)^+$ has $J^P=\frac{1}{2}^-$, then 
the $\D\Lambda_c(2595)^+$ $S$-wave state
can couple only to $P$-wave states of $J/\psi N$ and of 
the primary channels mentioned above and the pentaquark
in question should have $J^P=\frac{1}{2}^+$ instead
of $\frac{1}{2}^-$ or $\frac{3}{2}^-$. 
There have been also speculations about the role of the $\chi_{c1} p$
channel whose threshold is at 4449~MeV, i.e., right in the
pentaquark region \cite{Meissner:2015mza}. Also in this case the $S$-wave is in $J^P=\frac{1}{2}^-$, just 
like for the $D\Lambda_c(2595)^+$ channel. Another unusual channel 
is considered in the work of
Wu et al.~\cite{Wu:2024bvl}. Here, the $P_c(4457)$ pentaquark is interpreted
as a $J^P = 1/2^+$ state in the $\bar D^0\Lambda_c(2595)^+$–$\pi^0P_c(4312)$
{DCC} system. 
Finally, let us mention that, in principle, there are also three-body 
channels like $\D\Lambda_c^+\pi$ that open in the relevant energy region. 

While by far the majority of the works on the pentaquarks that have 
been published list only the poles, there are a few which provide results 
for the $J/\psi p$ invariant-mass spectrum which then can be 
compared directly with the measurements~\cite{Du:2019pij,  Du:2021fmf,Wang:2022oof, Shen:2024nck}.
We will focus on those works in the following, starting with a discussion 
of the publication of Du et al.~\cite{Du:2021fmf}. In their study, 
all four $\D^{(*)}\Sigma^{(*)}_c$ channels are included. 
Their thresholds are all close to the discovered $P_c$ states and 
they are all related by HQSS so that their consistent
inclusion is anyway demanded by that symmetry. Those are called
elastic channels by Du et al. In addition, $J/\psi N$ is included as
so-called inelastic channel \cite{Du:2019pij} and, 
in Ref.~\cite{Du:2021fmf}, also $\eta_c N$ and $\D^{(*)}\Lambda_c^+$. 
The latter step is necessary from a HQSS perspective since 
$(\eta_c,J/\psi)$ form likewise a heavy quark spin doublet, just as 
$(\D,\D^*)$. 
In the actual calculations three different schemes are considered~\cite{Du:2021fmf}. 
In scheme~I the leading-order (LO) contact potentials in the
$\D^{(*)}\Sigma^{(*)}_c$ channels are taken into account  
and in the same way the coupling to the $S-$ and $D-$wave $J/\psi N$ 
and $\eta_c N$ channels. 
In scheme~II, the contributions from one pion exchange (OPE) are added.
The OPE potential is derived in line with chiral effective field theory
and the involved coupling constants are fixed from the measured width
of the decay $D^{*+} \to D^0\pi^+$ on one hand side, and from
results by lattice QCD calculations regarding the coupling to the
$\Sigma^{(*)}_c$s \cite{Detmold:2012ge}. Furthermore, an NLO
$S$-$D$ contact term for the $\D^{(*)}\Sigma^{(*)}_c$ channels is 
added to absorb undesired artifacts from the regularization of 
the OPE potential~\cite{Wang:2018jlv,Baru:2019xnh}. 
Finally, in scheme~III, the coupling to the $\D^{(*)}\Lambda_c^+$ 
channels is added.

\begin{figure}[tb]
    \centering
    \includegraphics[width=0.35\textwidth]{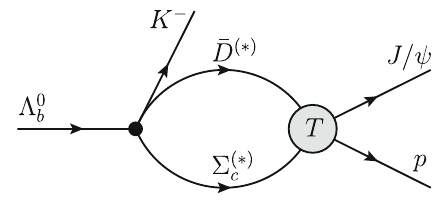}
    \vspace*{-0.2cm}
    \caption{Illustration of the treatment of the reaction 
    $\Lambda^0_b\to J/\psi p K^-$ in Refs.~\cite{Du:2019pij,Du:2021fmf}
    and \cite{Shen:2024nck}.
    }
    \label{fig:PentFSI}
\end{figure}

\begin{figure}[tb]
    \centering
    \includegraphics[width=0.445\textwidth,height=0.34\textwidth]{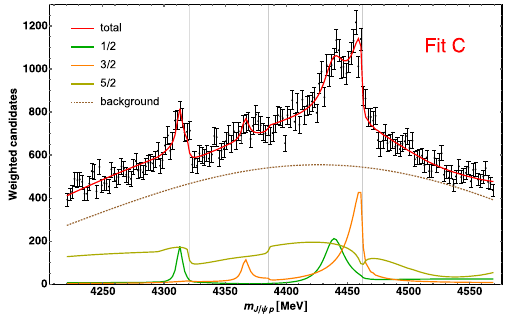}
    \includegraphics[width=0.49\textwidth]{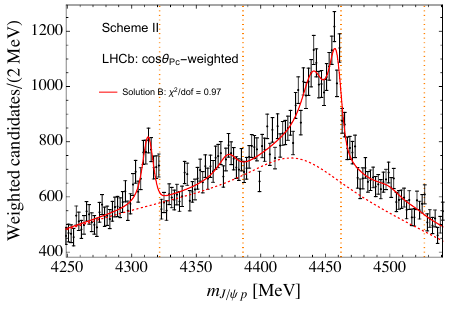}
    \vspace*{-0.2cm}
    \caption{$J/\psi p$ invariant-mass spectrum measured by the LHCb Collaboration \cite{LHCb:2019kea}. Left: Calculation taken from Ref.~\cite{Shen:2024nck}. Right: Calculation taken from Ref.~\cite{Du:2021fmf}. The vertical lines indicate the $\bar D\Sigma_c$, $\bar D\Sigma_c^*$, $\bar D^*\Sigma_c$, and $\bar D^*\Sigma_c^*$ thresholds, respectively, from left to right.
    }
    \label{fig:Penta1T}
\end{figure}

The evaluation of the $J/\psi p$ invariant mass is illustrated schematically 
in Fig.~\ref{fig:PentFSI}. Its calculation involves the reaction
amplitudes (T-matrices) for the meson-baryon channels, which 
are obtained by solving a Lippmann-Schwinger type equation. 
As indicated in the figure, in the intermediate state only the 
elastic channels are kept. 
The free parameters of the different schemes are fixed in a fit to the 
$J/\psi N$ invariant mass reported by the LHCb collaboration. In fact, 
that work contains three
sets of $m_{J/\psi p}$ distributions, one that includes all $m_{Kp}$
events, one where only events with $m_{Kp}> 1.9$~GeV are included,
and a third set where $\cos \theta_{Pc}$-dependent weights are applied
to each event candidate \cite{LHCb:2019kea} so that the $P_c$ signals
are enhanced. In Ref.~\cite{Du:2021fmf} Du et al. present a variety
of fits, based on the different schemes and for the different data sets, 
where all of them reproduce the data with fairly similar quality 
(i.e. similar $\chi^2$). 
For illustration, we pick out the one in scheme II and for the 
$\cos \theta_{Pc}$ weighted case, see Fig.~\ref{fig:Penta1T} (right side). 
One can see that the mass spectrum is very well reproduced, including the
structure associated with the $P_c$ states. Note that besides the theoretical
$J/\psi p$ invariant-mass distribution, evaluated via the mechanism 
displayed in Fig.~\ref{fig:PentFSI}, a smooth incoherent background has 
to be added. The latter is indicated by the dotted line. 

Exploring the pole structure of the reaction amplitude, the authors found
a total of seven $P_c$ states, for all fits/schemes considered, and in 
line with other investigations where HQSS was imposed on the employed
interaction models~\cite{Liu:2019tjn}. To be concrete, they identified 
three states with
$J^P=\frac{1}{2}^-$, three states with $J^P=\frac{3}{2}^-$, and one with
$J^P=\frac{5}{2}^-$. The lowest one corresponds to the $P_c(4312)$ with
$\frac{1}{2}^-$ and is a $\bar D\Sigma_c$ bound state. There are two 
$\bar D^*\Sigma_c$ bound states, with $\frac{1}{2}^-$ and $\frac{3}{2}^-$,
respectively, which in the solution shown in Fig.~\ref{fig:Penta1T}
correspond to a reversed order for $P_c(4457)$ and $P_c(4440)$.  
It should be said that in scheme I  the ``theoretically favored" ordering 
$P_c(4440)$, $P_c(4457)$, see Table~\ref{tab:Penta}, can 
be realized too~\cite{Du:2019pij,Du:2021fmf}.
In addition, there is a pole around 4380~MeV, with a dominant 
$\bar D\Sigma^*_c$ component. This mass is close to the structure reported
in the first LHCb paper \cite{LHCb:2015yax}, however, it is much narrower
than the state reported by LHCb. Finally, there are three $P_c$ states 
(with $\frac{1}{2}^-$, $\frac{3}{2}^-$, and $\frac{5}{2}^-$, respectively) which are 
$\D^*\Sigma^*_c$ bound states and which are located close
to the threshold of that channel. Their interpretation is unclear so far,
since the mass spectrum from the LHCb measurement does not show any 
unambiguous structure in that invariant-mass region.

The other study of the LHCb pentaquarks {reviewed here} is performed by members of the 
{Juelich-Bonn (JB)} group \cite{Wang:2022oof, Shen:2024nck}, who
likewise presented results for the $J/\psi p$ invariant mass. This DCC study is performed within the traditional meson-exchange picture 
and can be viewed as an extension of the JB model~\cite{Ronchen:2012eg} (see \cref{sec:JB})
to the heavy sector. It includes six channels, namely 
$\D\Lambda_c^+$, $\D\Sigma_c$, $\D^*\Lambda_c^+$, $\D^*\Sigma_c$,
$\D\Sigma^*_c$, and $J/\psi N$. The potentials acting 
in the various channels are derived from an effective interaction
based on the Lagrangians of Wess and Zumino~\cite{Wess:1967jq}
using time-ordered
perturbation theory.
They consist of contributions from $t$-channel meson-exchange processes 
and also from $u$-channel baryon exchanges. The former comprise the pseudoscalar
mesons $\pi$, $\eta$, $\eta'$ and the vector mesons $\rho$, $\omega$,
depending on the quantum numbers of the involved charmed mesons
and baryons. The transition  $\D^{(*)}\Sigma^{(*)}_c,\D^{(*)}\Lambda_c^+\to J/\psi N$ is
mediated by $\D$ and $\D^*$ exchanges. The coupling constants at the various meson-meson-meson- and baryon-baryon-meson vertices are fixed by assuming SU(3) and SU(4) flavor symmetry. The free parameters in the model are the cutoff masses associated with 
the phenomenological form factors that are attached to each vertex. Some of them are determined in the fitting procedure while others are kept fixed, see Table~5 in Ref.~\cite{Shen:2024nck} for details.

The evaluation of the $J/\psi p$ invariant mass is similar to that of Du et al. described above, see also Fig.~\ref{fig:PentFSI}. 
Again, the employed reaction amplitudes for the meson-baryon channels are obtained by solving a Lippmann-Schwinger type equation. And again only $\D^{(*)}\Sigma^{(*)}$ are kept in the intermediate state. Finally, as in \cite{Du:2021fmf} different solutions are found which describe the $J/\psi p$ invariant-mass spectrum almost equally well. Out of the three fits presented in Ref.~\cite{Shen:2024nck} we show here exemplary Fit C, see the left-hand side of Fig.~\ref{fig:Penta1T}. 

As one can see from the figure, also here a smooth incoherent background was added, which is indicated by the brown solid line. In addition the contributions from the different partial waves are shown, specifically those for $J=\frac{1}{2}$ (green), $\frac{3}{2}$ (orange), and $\frac{5}{2}$ (olive). Evidently, the three $P_c$ states from the recent LHCb measurement are reproduced where the $P_c(4312)$ appears as a $\D\Sigma_c$ bound state with $J^P=\frac{1}{2}^-$ and the $P_c(4440)$ and $P_c(4457)$ are states with $\frac{1}{2}^-$ and $\frac{3}{2}^-$, respectively, associated mainly with the $\D^*\Sigma_c$ channel. Interestingly, (and different from Fit A and B) in Fit C the pole for the latter is above the threshold, a situation similar to that observed for  a possibly strange dibaryon in the $\Lambda N$-$\Sigma N$ system \cite{Haidenbauer:2021smk}, see \cref{sec:LNSN}. The pole around 4366~MeV with $\frac{3}{2}^-$, a $\D\Sigma_c^*$ bound state, is in line with the structure reported in the first LHCb paper, but again it is much narrower than what has been deduced by the experimentalists. Note that, besides the states discussed here, the JB model predicts many  more states, from $\frac{1}{2}^-$ up to $\frac{7}{2}^-$, however, with fairly large widths. Future and more precise experiments might shed light on the question whether these exist or not. 

\begin{figure}[tb]
    \centering
    \includegraphics[width=0.90\textwidth]{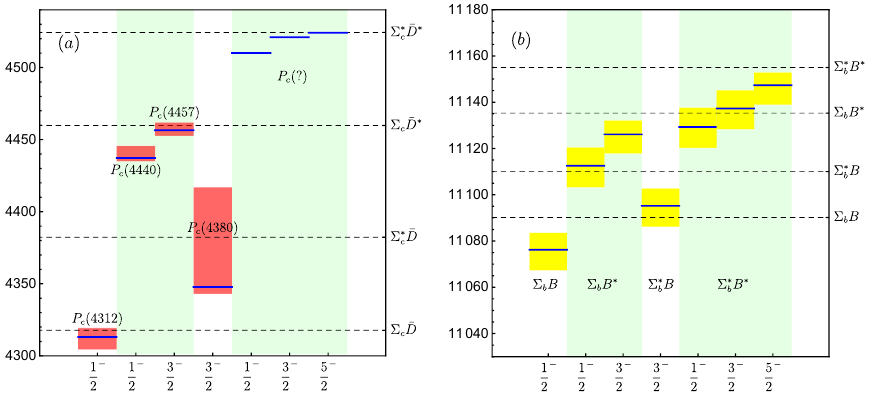}
    \vspace*{-0.2cm}
    \caption{Hidden-charm (a) and hidden-bottom (b) pentaquarks. Results are taken from \cite{Wang:2019ato}. 
    Experimentally established hidden-charm states are indicated by the red bands. The yellow bands for the
    hidden-bottom predictions are an estimate of the
    theoretical uncertainty. 
    }
    \label{fig:ZhouBB}
\end{figure}

As already said above, there are many more theoretical studies of the LHCb pentaquarks employing various approaches. However, as far as we could see, they are all restricted to the determination of the poles and no comparison with the observed $J/\psi p$ mass spectrum is provided. Moreover, several of those works are not performed in a factual coupled-channel framework. Nonetheless, we want to mention selectively some of them, namely  
three of the earliest works~\cite{Wu:2010jy, Wu:2010vk, Wu:2012md}, published well before experimental evidence for pentaquark candidates were found, the studies of Meng et al.~\cite{Meng:2019ilv}, where the interaction is derived within chiral perturbation theory, including contact terms, and one-pion and two-pion exchange, 
a calculation by Xiao et al.~\cite{Xiao:2019aya}, performed with a 
dynamical input obtained from an extension of the local hidden gauge approach, 
and the works of Wu et al.~\cite{Wu:2019adv} and of Zhang~\cite{Zhang:2024dkm}, which is based on the conventional meson-exchange approach and where predictions for $\gamma p \to J/\psi p$ are provided.
Interesting are also works like Ref.~\cite{Wang:2019ato} where hidden-charm pentaquarks are considered 
and predictions for hidden-bottom pentaquarks are given,
see Fig.~\ref{fig:ZhouBB}. For yet other prediction of hidden-bottom pentaquarks, see e.g., Ref.~\cite{Zhu:2020vto}.

\subsubsection{Hidden-charm pentaquarks with strangeness}

\begin{figure}[tb]
    \centering
    \includegraphics[width=0.44\textwidth]{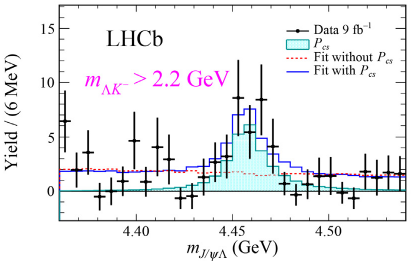}
    \includegraphics[width=0.44\textwidth]{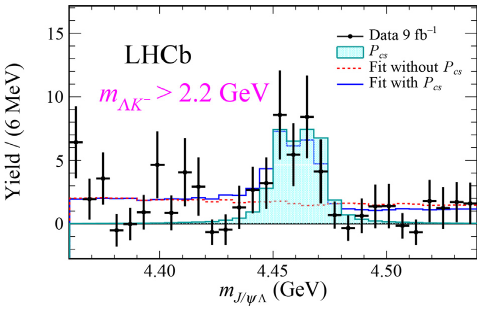}
    \vspace*{-0.2cm}
    \caption{Hidden-charm pentaquark(s) with strangeness established by the LHCb Collaboration in the $\Xi_b^- \to J/\psi\Lambda K^-$ decays
    \cite{LHCb:2020jpq}. Interpretation of the data in terms of one (left) or
    two (right) pentaquarks are shown. 
    }
    \label{fig:PentLHCbs}
\end{figure}

First experimental evidence for a hidden-charm pentaquark candidate
with strangeness, the $P_{cs}(4459)^0$ ($P_{c\bar c s}$ in the new 
convention by the PDG~\cite{ParticleDataGroup:2024cfk}), was reported by 
the LHCb Collaboration in 2020 from an analysis of 
$\Xi_b^- \to J/\psi\Lambda K^-$ decays~\cite{LHCb:2020jpq}. 
The state is located about 18~MeV below the $\D^{*0}\Xi^0_c$
threshold (4477~MeV) and about 15~MeV above the one of the $\D\Xi_c'$ channel 
(4444~MeV). Two years later a second candidate was identified, the $P_{cs}(4338)$, in $B^- \to J/\psi \Lambda \bar p$ decays \cite{LHCb:2022ogu}.
It is located basically at the $\D\Xi_c$ threshold. Very recently, the $P_{cs}(4459)^0$ was also observed by the Belle
Collaboration~\cite{Belle:2025pey}, in inclusive decays of $\Upsilon(1S, 2S)$ to $J/\psi \Lambda$. 
The $J/\psi\Lambda$ invariant-mass spectrum measured by LHCb, where the 
$P_{cs}(4459)^0$ was identified, is reproduced in Fig.~\ref{fig:PentLHCbs}.
As one can see from the figure, besides the single-pentaquark solution
(left), the LHCb Collaboration explored also the possibility that there 
are two states (right), inspired by corresponding model predictions of a calculation based on chiral 
effective field theory up to the next-to-leading order~\cite{Wang:2019nvm}. 
The masses suggested from that analysis for the pentaquark candidates
are $4454.9\pm 2.7$~MeV and $4467.8\pm 3.7$~MeV, respectively. 
The quantum numbers predicted by theory \cite{Wang:2019nvm} are 
$J^P=\frac{1}{2}^-$ and $\frac{3}{2}^-$, respectively. 

\begin{figure}[tb]
    \centering
    \raisebox{-.5\height}{\includegraphics[width=0.48\textwidth]{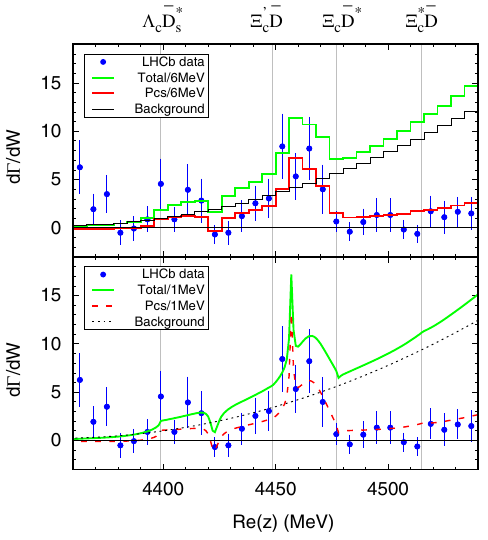}}
    \raisebox{-.5\height}{\includegraphics[width=0.50\textwidth]{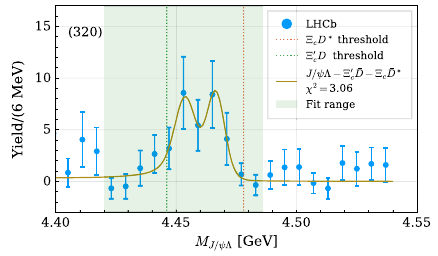}}
    \caption{Results for the $J/\psi\Lambda$ invariant-mass spectrum, taken from Refs.~\cite{Du:2021bgb} (right) and \cite{Zhu:2022wpi} (left).  
    Data are by the LHCb Collaboration from the $\Xi_b^- \to J/\psi\Lambda K^-$
    decays \cite{LHCb:2020jpq}. 
    }
    \label{fig:Pent3}
\end{figure}

A first study of the nature of the $P_{cs}(4459)^0$ within a DCC approach was presented by Du et al.~\cite{Du:2021bgb}. Besides $J/\psi \Lambda$, out of the many  $\D^{(*)}\Xi^{(*)}_c$ and $\D^{(*)}\Xi'_c$ open-charm channels only those whose thresholds are close to the observed structure were taken into account, namely $\D \Xi_c'$ and $\D^* \Xi_c$. 
The calculation was performed in a leading-order EFT-type ansatz with  the potentials represented by constant terms. For the considered channels constraints from HQSS were implemented. The reaction amplitude is established in the on-shell factorization 
approximation where the one-loop two-point Green's function is evaluated in the dimensional regularization scheme \cite{Du:2021bgb}. The 8 parameters of the model, three that specify the interaction in the $\D \Xi_c'$ and $\D^* \Xi_c$ channels, two for the coupling of those channels to $J/\psi\Lambda$, and three related to the elementary decay 
(production) vertices, are fitted to the mass spectrum. 
The result is shown in Fig.~\ref{fig:Pent3} (right). One can see that the potential considered by Du et al. leads indeed to a two-state structure of the observed pentaquark. Both are dominated by the $\D^* \Xi_c$ channel, where the lower state corresponds to $\frac{1}{2}^-$ and the higher one to $\frac{3}{2}^-$. Note that the greenish area indicates the
fitting region. It should be mentioned that in Ref.~\cite{Du:2021bgb} some other scenarios 
are discussed too, with just two coupled channels, and
in some only a single structure appears. In addition the authors considered a perturbative treatment of the
$\D \Xi_c'$ channel. 

DCC results for the invariant-mass spectrum were also presented 
by Zhu et al.~\cite{Zhu:2022wpi}. In that study a total of nine coupled channels are considered:  $J/\psi \Lambda$, $\D^{(*)} \Xi_c'$, $\D^{(*)} \Xi_c$, $\D^{(*)} \Xi^*_c$, $\D_s^{(*)} \Lambda_c^+$. The work is performed within the conventional meson-exchange picture, where the exchanges of pseudoscalar ($\pi$, $\eta$) and vector ($\rho$, $\omega$) mesons are considered, and in addition a scalar-isoscalar ($\sigma$) meson. Furthermore, the transition to the $\D^{(*)}_s\Lambda_c^+$ channels is mediated by $K$ and $K^*$ exchange and the one to the $J/\psi \Lambda$ channel
by $D,D^*$ and $D_s,D^*_s$ exchanges, respectively. The coupling constants for all pseudoscalar and vector mesons are 
fixed assuming SU(4) symmetry. 
The reaction amplitude is obtained by solving a LS-type equation
for the coupled-channel potentials, where the latter are furnished 
with Gaussian form factors. Their cutoff masses (one for the light
mesons and one for $D^{(*)},\,D^{(*)}_s$) are free parameters. The other
free parameters are related to the elementary decay (production) 
vertices; there is one for the $J=1/2$ channel and one for $3/2$. 
Note that in this work the invariant-mass spectrum is evaluated in the 
Migdal-Watson approximation \cite{Watson:1952ji,Migdal:1955}, 
i.e., it is assumed to be proportional to the square
of the meson-baryon reaction amplitudes times the phase space factor. 

The resulting invariant-mass spectrum is shown in Fig.~\ref{fig:Pent3} (left). 
In the upper panel the outcome of the calculation is adapted to the experimental bin width 
of 6~MeV, while in the lower panel a bin width of only 1~MeV is
assumed. In both cases from the full calculation (green curves) a 
background is subtracted (black curves) so that the resonance signals
can be isolated and a comparison with the LHCb data can be made
\cite{Zhu:2022wpi}. Zhu et al. report four poles in the energy region 
displayed in Fig.~\ref{fig:Pent3}, namely at 
4423~MeV ($\D\Xi_c'$, $\frac{1}{2}^-$), at 
4457~MeV ($\D^*\Xi_c$, $\frac{3}{2}^-$), at 
4465~MeV ($\D^*\Xi_c$, $\frac{1}{2}^-$), and at  
4512~MeV ($\D\Xi_c^*$, $\frac{3}{2}^-$), where the dominant
channel is indicated in the brackets. Obviously, the experimental signal
of the $P_{cs}(4459)$ pentaquark is primarily reproduced by the 
$\frac{1}{2}^-$ state with 4465~MeV. The near-by $\frac{3}{2}^-$ state
predicted by this model is too narrow to be seen when the bin width is 
6~MeV. The same is true for the $\frac{1}{2}^-$ state with 4423~MeV.
Note that the potential of Zhu et al. produces also a pole around
4336.5~MeV, dominated by the strong coupling between $\D_s\Lambda$
and $\D\Xi_c$, which could be identified with the $P_{cs}(4338)$ pentaquark
of the LHCb collaboration. 

\begin{figure}[tb]
    \centering
    \includegraphics[width=0.80\textwidth]{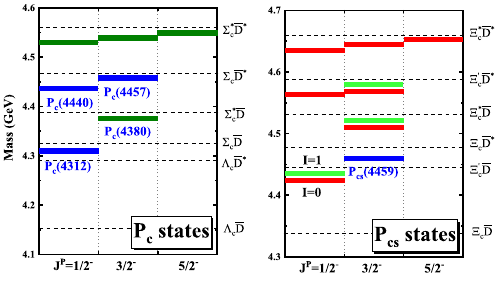}
    \vspace*{-0.2cm}
    \caption{Hidden-charm open-strange pentaquarks predicted in 
    Ref.~\cite{Chen:2022onm} based on an interaction established in 
    the hidden-charm sector. 
    Experimentally established states are indicated by blue bars. 
    }
    \label{fig:PentChen}
\end{figure}

In Ref.~\cite{Chen:2022onm} another coupled-channel calculation based on dynamics generated by one-boson exchange
can be found. The channels included are
$\D^{(*)} \Xi_c'^{(*)}$, $\D^{(*)} \Xi^{(*)}_c$.
The interaction is deduced from the one for
$\D^{(*)} \Sigma_c^{(*)}$ by using heavy quark symmetry and 
SU(3) flavor symmetry, where the latter is fixed from a fit to the
$P_c$ states established by LHCb. Only poles are reported but no
$J/\psi\Lambda$ invariant-mass distribution. Results for the resonance spectrum are shown in Fig.~\ref{fig:PentChen}. A similar coupled-channel 
study has been presented in Ref.~\cite{Clymton:2025hez} very recently. 
Finally, in Ref.~\cite{Feijoo:2022rxf} one can find a coupled-channel
calculation performed with a dynamical input obtained from an extension 
of the local hidden gauge approach, showing the different roles played by the channels involving $\Lambda_c$  in the $P_c$ and $P_{cs}$ states. Also here there are no predictions
for the invariant-mass distribution. 

\subsubsection{Other heavy systems}

Channel coupling plays also a role in the interpretation of the
exotic $XYZ$ states and the doubly charmed tetraquark state $T^+_{cc}$,
observed in meson-meson systems. Most studies in this field are based 
on the chiral unitary approach or comparable formalisms and exploit the 
on-shell factorization to simplify the calculations. 
However, whenever vector mesons like $K^*$, $D^*$, or $B^*$
are involved pion exchange is possible and generates a 
long-ranged interaction that couples the different channels. 
In particular, the strong tensor force induced by pion exchange 
leads to a coupling between the $S$-wave (where molecules/bound states 
are primarily expected) and the $D$-wave. These aspects are considered only in a small number of works because it requires a much more elaborate and tedious computational treatment. 

One state where pion exchange plays an essential and delicate role 
is the aforementioned $\chi_{c1}$(3872)~\cite{Belle:2003nnu},  
discovered in the $\pi^+\pi^- J/\psi$ channel, and located 
extremely close to the $\D^{*0}D^0$ threshold. Since the $D^{*0}$
mass itself is very close to the $D^0\pi^0$ threshold, a pion
exchanged in the coupled $D \D^*$-$D^*\D$ system can go on shell,
in other words the $D\D\pi$ channel opens right in the crucial 
region where the bound state is located. This makes a proper inclusion 
of the three-body 
$D \D \pi$ unitarity cuts mandatory if one wants to obtain
reliable results \cite{Baru:2011rs}. See also the discussion in
\cref{{sec:Three-body unitarity}}.

Pion exchange plays likewise an important role in case of the tetraquark
candidate $T^+_{cc}$, observed in the $D^0D^0\pi^+$ 
spectrum \cite{LHCb:2021vvq}. The structure is
located just about 300~keV below the $D^{*+}D^0$ threshold. 
Also here the exchanged pion can go on shell so that it is
necessary to include three-body effects \cite{Du:2021zzh,Dawid:2024dgy}.

Regarding systems involving the bottom mesons $B^{(*)}$, $\bar B^{(*)}$ 
the overall conditions are similar but differ in important details. 
Here, the $Z_b (10610)$ and $Z_b (10650)$ states have been  
observed by Belle in the invariant-mass distributions of the 
$\Upsilon (nS)\pi^{\pm}$ ($n=1,2,3$) and $h_b(mP)\pi^{pm}$ ($m=1,2$) subsystems of the dipion transitions from $\Upsilon(10860)$ \cite{Belle:2011aa}. They are located
close to the $B\bar B^*$ and $B^*\bar B^*$ thresholds. 
Since the separation between the $\bar BB$, $\bar BB^*$-$\bar B^*B$, 
$\bar B^* B^*$ channels is smaller than the pion mass,
see  Eqs.~(\ref{eq:masses}), the opening of three-body channels involving the pion 
does not play a role. On the other hand, the 
coupling between $\bar BB$ and $\bar BB^*$-$\bar B^*B$ to 
$\bar B^* B^*$ becomes more important as compared to the charmed 
counterpart. Various aspects of the coupled-channel dynamics like 
the role of pion exchange (and also $\eta$ exchange) 
and the coupling between $S$- and $D$-waves were addressed in
Refs.~\cite{Baru:2017gwo,Wang:2018jlv}.

Another state where coupled-channel dynamics
has been advocated for explaining its structure, is the 
$X(6900)$ observed in the invariant-mass spectrum of $J/\psi$ pairs,
produced in proton-proton collisions \cite{LHCb:2020bwg}. 
In the study of Dong et al.~\cite{Dong:2020nwy} the channels
$J/\psi-J/\psi$, $\psi(2S)-J/\psi$, and $\psi(3770)-J/\psi$ 
are considered, motivated by the observation that the 
$\psi(2S)-J/\psi$ threshold (6783~MeV) coincides with 
the dip seen in the invariant mass while the $\psi(3770)-J/\psi$ 
threshold is close to the sharp peak in the experimental spectrum. 
It should be said, however, that the dynamics considered in this work 
is restricted to a pure contact coupled-channel short-range potential. 
Note that there is a recent update \cite{Song:2024ykq}, where additional 
new from ATLAS \cite{ATLAS:2023bft} and CMS \cite{CMS:2023owd} were
considered in combination with the previous LHCb results.

Finally, let us point to the work of Malabarba et al.~\cite{Malabarba:2021taj}, where hidden-charm states with a possible three-body nature have been studied by considering the 
$ND\bar D^*$-$ND^*\bar D$ systems within a Faddeev-type approach based on 
the fixed center approximation. Thereby the dynamics in the $DN$
system, discussed in \cref{sec:DN}, is combined with the assumption that the $D\bar D^*$-$D^*\bar D$ subsystems form bound states corresponding to the $\chi_{c1}(3872)$ and the $Z_c(3900)$ states.

\section{Summary and outlook}
\label{sec:summary}
Dynamical coupled-channel approaches are a versatile tool to study hadron dynamics in different contexts. Historically, they arose as data in wide kinematic ranges and for different reactions became available, calling for a unified analysis and interpretation of different data within one overarching formalism. Uniting principles of S-matrix theory with the option to incorporate microscopic dynamics of underlying theories (strong and electroweak interactions), they became a universal tool in baryon spectroscopy and beyond. As discussed, they allow to describe hadronic reactions with known initial states such as pions, real and virtual photon probes, but also neutrino-induced reactions and baryon-baryon interactions. They also allow to parametrize the final-state interaction in the mesonic sector with three-body decays. 
Owing to their formulation, DCC approaches are particularly suited to incorporate full three-body dynamics while maintaining  three-body unitarity.

In this review the emphasis lies on two out of several DCC approaches, the ANL-Osaka and the Juelich-Bonn-Washington approach (JBW). They have been used in the analysis of light baryons, but they are both members of a larger family of DCC approaches addressing mesons with two and three-body dynamics and baryon-baryon interactions (light, strange, and heavy), as well as the exotic $XYZ$ and pentaquark states, that often require to map out complicated threshold dynamics. DCC approaches also allow to describe and predict hadronic phenomena such as cusps, three-body thresholds, triangle singularities and dynamical resonance generation, all of which have been discussed in this review.

DCC approaches become a bridge connecting various areas of hadron spectroscopy bundling searches for new excited states across different initial states, or studies of electromagnetic transition form factors across different final states. They also connect to  functional methods, or, most recently, infinite volume mappings of finite volume spectra obtained in Lattice QCD.

With new experiments being commissioned exploring various strangeness sectors and higher photon virtualities at BESIII, CERN (AMBER, COMPASS), ELSA, Jefferson Lab, J-PARC, MAMI, SIS100, and other facilities, DCC approaches are  well positioned for coming decades exploring the transition region from perturbative to non-perturbative QCD, and the interplay with electroweak probes, with the option to match to low-energy effective QCD and with perspectives to connect even larger data sets across different flavor sectors. 

DCC approaches deal with new data coming from high-precision modern facilities but, also, include data from  decades ago, often with unknown systematic effects. This calls for utilization of modern statistical tools beyond $\chi^2$ minimization. The input from Bayesian statistics and machine learning for DCC approaches will help in analyzing the large, inhomogeneous data sets that DCC approaches usually deal with. Planned experiments with, e.g.,  meson beams will substantially help in this endeavor and provide further analysis opportunities for DCC approaches.

The neutrino interaction in the $N^*$ region plays a crucial role in neutrino physics, especially for the analysis of long base line neutrino experiments. DCC models well constrained from the electromagnetic reactions are in a good position to describe the electroweak reactions in the $N^*$ region. With further  theoretical or experimental inputs on the axial vector response of nucleon, such as parity-violating asymmetry of electron scattering, DCC approaches will serve as a key tool in neutrino physics research.

When data becomes more and more abundant, novel data driven machine learning and AI tools -- ranging from hands-on tools like LASSO or information criteria, to neural networks or more involved architectures such as transformers -- become more relevant. Application of such techniques and integration thereof into dynamical coupled channel formalisms is an exciting avenue for future explorations.

\section*{Acknowledgment}
The authors thank Luis \'Alvarez Ruso, Vadim Baru, Daniel Carman, Evgeny Epelbaum, Jeremy Green, Feng-Kun Guo, Hiroyuki Kamano, T.-S. H. Lee, Terry Mart, Ulf-G. Mei{\ss}ner, Viktor Mokeev, Osamu Morimatsu, Satoshi Nakamura, Eulogio Oset, Fernando Romero-López, Deborah R\"onchen, Denny Sombillo, Ron Workman, and Wren Yamada for useful discussions and for pointing out literature.
The work of MD is supported by the National Science Foundation under Grant No. PHY-2310036. The work of MD is also supported by the U.S. Department of Energy grant DE-SC0016582 and DOE Office of Science, Office of Nuclear Physics under contract DE-AC05-06OR23177. This work contributes to the aims of the U.S. Department
of Energy ExoHad Topical Collaboration, contract DE-SC0023598.
The work of MM was supported through the Heisenberg Programme (project number: 532635001) and by the Deutsche Forschungsgemeinschaft (DFG, German Research Foundation), the NSFC through the funds provided to the Sino-German Collaborative Research Center CRC 110 “Symmetries and the Emergence of Structure in QCD” (DFG Project-ID 196253076 - TRR 110, NSFC Grant No. 12070131001). 
The work of TS was supported by the JSPS KAKEMHI Grant No. 24K07062.
\bibliographystyle{elsarticle-num} 
\bibliography{bibs/biblio,bibs/jh-bb,bibs/jh-mb,bibs/ppnp-ts,bibs/NON-INSPIRE}
\end{document}